\documentclass[12pt,a4paper]{article}
\pdfoutput=1
\usepackage{color}
\usepackage{amssymb,amsmath,bm,bbold}
\usepackage{epsf}
\usepackage{epsfig}
\usepackage{relsize}
\usepackage[dvipsnames]{xcolor}
\usepackage[linktoc=page,bookmarks=false,colorlinks=false,linkbordercolor=RoyalBlue,citebordercolor=ForestGreen,urlbordercolor=CornflowerBlue]{hyperref}
\usepackage{latexsym,mathrsfs,dsfont}
\usepackage[normalem]{ulem} % for strikeout \sout
\usepackage[compress]{cite}
\usepackage{graphicx}
\usepackage{url}
\usepackage{booktabs}
\usepackage{float}
\usepackage{multirow}
\usepackage{changepage}
\usepackage[hypcap]{caption, subcaption}

\usepackage[titles]{tocloft}

\usepackage{tabularx,colortbl}

\usepackage{fancyhdr}
\advance \headheight by 3.0truept       % for 12pt mandatory...
\allowdisplaybreaks[1]

\definecolor{red}{cmyk}{0,1,1,0.4}
\definecolor{darkgreen}{rgb}{0.0,0.6,0.0}
\definecolor{cDarkGrey}{RGB}{91,91,91}
\definecolor{cGrey}{RGB}{245,243,238}
\definecolor{cBlue}{RGB}{0,110,191}
\definecolor{cLightBlue}{RGB}{214,237,252}
\definecolor{cRed}{RGB}{196,0,100}
\definecolor{cLightRed}{RGB}{254,222,237}
\definecolor{cGreen}{RGB}{0,166,80}
\definecolor{cLightGreen}{RGB}{254,222,237}
\definecolor{cOrange}{RGB}{221,74,44}
\definecolor{cLightOrange}{RGB}{255,215,210}
\definecolor{cPurple}{RGB}{93,35,125}
\definecolor{cLightPurple}{RGB}{241,230,252}
\definecolor{cYellow}{RGB}{252,191,10}
\definecolor{cISSRBlue}{RGB}{0,111,174}
\definecolor{cISSRGrey}{RGB}{167,169,172}

\newcommand{\beq}{\begin{equation}}
\newcommand{\eeq}{\end{equation}}
\newcommand{\be}{\begin{equation}}
\newcommand{\ee}{\end{equation}}
\newcommand{\bi}{\begin{itemize}}
\newcommand{\ei}{\end{itemize}}
\newcommand{\ba}{\begin{array}}
\newcommand{\ea}{\end{array}}
\newcommand{\beqa}{\begin{eqnarray}}
\newcommand{\eeqa}{\end{eqnarray}}
\newcommand{\bea}{\begin{eqnarray}}
\newcommand{\eea}{\end{eqnarray}}
\newcommand{\beqn}{\begin{eqnarray}}
\newcommand{\eeqn}{\end{eqnarray}}

\newcounter{TODO}

\def \refeq#1{(\ref{#1})}

\def \refsec#1{Section~\ref{#1}}

\def \refapp#1{Appendix~\ref{#1}}
\def \reffig#1{Figure~\ref{#1}}
\newcommand{\reffigs}[2]{Figures~\ref{#1}--\ref{#2}}
\def \reftab#1{Table~\ref{#1}}

\DeclareMathOperator{\re}{Re}
\DeclareMathOperator{\im}{Im}

\newcommand{\oL}[1]{\overline{#1}}
\newcommand{\wT}[1]{\widetilde{#1}}

\newcommand{\ord}[1]{\mathcal{O}\left( #1 \right)}

\newcommand{\MSbar}{${\overline{\text{MS}}}$}

\newcommand{\DF}{\Delta F}

\newcommand{\cL}{\mathcal{L}}
\newcommand{\cH}{\mathcal{H}}
\newcommand{\cN}{\mathcal{N}}

\newcommand{\MeV}{\,\text{MeV}}

\newcommand{\GeV}{\,\text{GeV}}
\newcommand{\geV}{\text{GeV}}
\newcommand{\TeV}{\,\text{TeV}}

\newcommand{\epe}{\varepsilon'/\varepsilon}

\newcommand{\KKbar}{K^0-\bar K^0}
\newcommand{\DDbar}{D^0-\bar D^0}
\newcommand{\BBbar}{B_{s,d}-\bar B_{s,d}}
\newcommand{\BBbarD}{B_d-\bar B_d}
\newcommand{\BBbarS}{B_s-\bar B_s}

\newcommand{\alS}{\alpha_s}

\newcommand{\muLow}{{\mu_\text{had}}}
\newcommand{\muEW}{{\mu_\text{ew}}}
\newcommand{\muNP}{{\Lambda}}

\newcommand{\wc}[3][{}]{\big[\mathcal{C}_{#2}^{#1}\big]_{#3}}
\newcommand{\wcup}[3][{}]{[{\hat{\mathcal{C}}}_{#2}^{#1}]_{#3}}
\newcommand{\bwc}[3][{}]{\big[c_{#2}^{#1}\big]_{#3}}
\newcommand{\bwcup}[3][{}]{[{\hat{c}}_{#2}^{#1}]_{#3}}

\newcommand{\Wc}[2][{}]{\mathcal{C}_{#2}^{#1}}
\newcommand{\Wcup}[2][{}]{\hat{\mathcal{C}}_{#2}^{#1}}
\newcommand{\bWc}[2][{}]{c_{#2}^{#1}}

\newcommand{\Op}[2][{}]{\mathcal{O}_{#2}^{#1}}

\newcommand{\wcL}[3][{}]{[C_{#2}^{#1}]_{#3}}

\newcommand{\WcL}[2][{}]{C_{#2}^{#1}}

\newcommand{\OpL}[2][{}]{Q_{#2}^{#1}}
\newcommand{\opL}[3][{}]{[Q_{#2}^{#1}]_{#3}}

\newcommand{\MeL}[2][{}]{\langle Q_{#2}^{#1} \rangle}

\newcommand{\OpLt}[2][{}]{\widetilde{Q}_{#2}^{#1}}

\newcommand{\eps}{\epsilon}

\newcommand{\bra}[1]{\ensuremath{\langle #1 |}}
\newcommand{\ket}[1]{\ensuremath{| #1 \rangle }}

\newcommand\numberthis{\addtocounter{equation}{1}\tag{\theequation}}
\newcommand\nline{\\ & \phantom{\;}}
\newcommand\nlineS[1]{\\[#1] & \phantom{\;}}
\newcommand\cdt[1]{\!\cdot\! 10^{#1}}

\begin{document}

\vspace{-1cm}

\begin{flushright}
  AJB-20-4\\
  TUM-HEP-1283/20  \\
  UdeM-GPP-TH-20-283\\
\end{flushright}

\medskip

\begin{center}
{\LARGE\bf
\boldmath{SMEFT ATLAS of $\DF=2$ Transitions}}
\\[1.2cm]
{\bf
  Jason~Aebischer$^{a}$,
  Christoph~Bobeth$^{b}$,
  Andrzej~J.~Buras$^{c}$ and
  Jacky Kumar$^{d}$
}\\[0.5cm]

{\small
$^a$Department of Physics, University of California at San Diego,
    La Jolla, CA 92093, USA \\[0.2cm]
$^b$Physik Department T31,
    James-Franck-Stra\ss e~1,
    Technische Universit\"at M\"unchen,\\
    D--85748 Garching, Germany \\[0.2cm]
$^c$TUM Institute for Advanced Study,
    Lichtenbergstr. 2a, D--85747 Garching, Germany \\[0.2cm]
$^d$Physique des Particules, Universite de Montreal,
    C.P. 6128, succ. centre-ville, \\
    Montreal, QC, Canada H3C 3J7
}
\\[0.5 cm]
\footnotesize
E-Mail:
\texttt{jaebischer@physics.ucsd.edu},
\texttt{christoph.bobeth@ph.tum.de},
\texttt{andrzej.buras@tum.de},
\texttt{jacky.kumar@umontreal.ca}
\\[0.2 cm]
\end{center}

\vskip 1.0cm

\begin{abstract}
\noindent We present a model-independent anatomy of the $\DF=2$ transitions
$\KKbar$, $\BBbar$ and $\DDbar$ in the context of the Standard Model Effective
Field Theory (SMEFT). We present two master formulae for the mixing amplitude
$\big[M_{12} \big]_\text{BSM}$.
One in terms of the Wilson coefficients (WCs) of the Low-Energy Effective Theory
(LEFT) operators evaluated at the electroweak scale $\muEW$ and one in terms
of the WCs of the SMEFT operators evaluated at the BSM scale $\muNP$. The
coefficients $P_a^{ij}$ entering these formulae contain all the information
below the scales $\muEW$ and $\muNP$, respectively. Renormalization group effects
from the top-quark Yukawa coupling play the most important role.
The collection of the individual contributions of the SMEFT operators to
$\big[M_{12}\big]_\text{BSM}$ can be considered as the SMEFT ATLAS of $\DF=2$
transitions and constitutes a travel guide to such transitions far beyond the
scales explored by the LHC. We emphasize that this ATLAS depends on whether
the down-basis or the up-basis for SMEFT operators is considered. We illustrate
this technology with tree-level exchanges of heavy gauge bosons ($Z^\prime$,
$G^\prime$) and corresponding heavy scalars.
\end{abstract}

\setcounter{page}{0}
\thispagestyle{empty}
\newpage

\setcounter{tocdepth}{2}
\setlength{\cftbeforesecskip}{0.21cm}
\tableofcontents

\newpage

%--------+---------+---------+---------+---------+---------+---------+---------+
%
%
%
%--------+---------+---------+---------+---------+---------+---------+---------+
\section{Introduction}

$\KKbar$, $\BBbar$ and $\DDbar$ mixings have been already for many years the
stars among the flavour-changing neutral-current processes (FCNC)
\cite{Buras:2013ooa, Buras:2020xsm}. This is in particular the case of the
parameter $\varepsilon_K$, of the  $\BBbar$ mass differences $\Delta M_{B_{s,d}}$
and of mixing induced CP asymmetries in the latter systems. The $K_L-K_S$ mass
difference $\Delta M_K$ remained due to large theoretical uncertainties until
recently under the shadow of these observables although it played a very
important role in the past in estimating successfully the charm-quark mass
prior to its discovery \cite{Gaillard:1974hs}. However, recently progress in
evaluating $\Delta M_K$ within the Standard Model (SM) has been made by
the RBC-UKQCD collaboration \cite{Bai:2014cva, Christ:2015pwa, Bai:2018mdv} so
that $\Delta M_K$ begins to play again an important role in phenomenology, not
only to bound effects beyond the SM (BSM) \cite{Gerard:1984bg, Gabbiani:1996hi,
Bona:2007vi, Isidori:2010kg, Silvestrini:2018dos, Calibbi:2019lvs}, but also
to help identifying what this new physics (NP) could be.

These days, the absence of the discovery of new particles at the Large Hadron
Collider (LHC) points towards a mass gap between the electroweak (EW) scale
$\muEW$ and the next threshold scale $\muNP$ of new heavy degrees of freedom.
For such a scenario, NP effects can be discussed for all processes between
scales sufficiently below $\muNP$ down to the EW scale conveniently within
the Standard Model Effective Field Theory (SMEFT), the renormalizable SM
augmented with higher-dimensional operators invariant under the full SM
group $\text{SU(3)}_c \times \text{SU(2)}_L \times \text{U(1)}_Y$. The
flavour-changing processes of light quarks and leptons further below the
EW scale are described by the usual respective Low-Energy Effective Field
Theories (LEFT) that are invariant under $\text{SU(3)}_c \times
\text{U(1)}_\text{em}$. The LEFTs are characterized by disjoint sets of
flavour-changing operators for each process and are conveniently separated from
each other. They have been studied systematically in the SM and generalizations
in terms of the renormalization group (RG) equations also at higher orders in
QCD and in some cases in QED. Currently this program is extended within
the framework of SMEFT and recently the complete matching of SMEFT to LEFTs at
$\muEW$ at tree- and one-loop level has been summarized in \cite{Jenkins:2017jig,
Dekens:2019ept}.
The matching conditions of SMEFT to LEFT together with the RG equations within
SMEFT and LEFTs are the essential ingredients to study phenomenological
correlations of low- and high-energy\footnote{``High-energy'' refers here to
scales sufficiently below $\muNP$.} observables.

Several model-independent SMEFT analyses of FCNC processes can be found in
the literature \cite{Endo:2016tnu, Bobeth:2017xry, Feruglio:2017rjo,
Gonzalez-Alonso:2017iyc, Buttazzo:2017ixm, Kumar:2018kmr, Aebischer:2018iyb, Endo:2018gdn,
Silvestrini:2018dos, Aebischer:2019mlg, Aebischer:2020mkv, Aebischer:2020lsx}.
In particular in \cite{Silvestrini:2018dos} a systematic analysis of the RG
effects above the EW scale has been performed and for the first time the full
set of constraints on all relevant dimension-six operators
resulting from $\Delta F=2$ transitions has been presented in
numerous tables.

In the present paper we want to explore two different avenues involving $\DF=2$
transitions. First we will provide in the spirit of the SMEFT anatomy of the
ratio $\epe$ \cite{Aebischer:2018csl, Aebischer:2018quc} a master formula for
$\DF=2$ processes in terms of the standard LEFT operator basis used already by
many authors for two decades following the expressions presented in
\cite{Buras:2001ra} with Wilson coefficients (WCs) of the LEFT given at the
EW scale $\muEW$. The second avenue leads us to a master formula for $\DF=2$
processes given directly in terms of the SMEFT operator basis with SMEFT WCs,
evaluated at the NP scale $\muNP$. To our knowledge the SMEFT
formula in question is presented here for the first time. It allows to include
automatically SMEFT effects above the EW scale, in particular the ones from
the RG running of the top-Yukawa coupling. Moreover, in contrast to
recent SMEFT analyses found in the literature it includes in addition to
leading order (LO) QCD RG corrections also the next-to-leading order (NLO)
ones below the electroweak scale. This is necessary to have a proper
matching of LEFT WCs to the hadronic matrix elements from lattice QCD (LQCD).

We believe that the collection of the individual contributions of the SMEFT
operators to this master formula that can be considered as
the SMEFT ATLAS for $\DF=2$ transitions will allow model builders in an
efficient manner to obtain predictions for $\DF=2$ processes in a plethora
of NP models that are consistent with the rules of the SMEFT. As we will see
such an analysis is also useful for an analytic insight into model-independent
analyses, which complements the very extensive numerical analysis in
\cite{Silvestrini:2018dos}.\footnote{While the RG analysis in
\cite{Silvestrini:2018dos} is based on the first-leading-log approximation,
we perform full resummation of leading logarithms which allows to include
automatically the mixing between operators that is absent if only the first
leading logarithm is kept. See \cite{Buras:2018gto} for a detailed analysis
of such effects.} To this end our ATLAS exhibits in addition to usually assumed
most important contributions of a given operator to
$\big[M_{12}^{ij}\big]_\text{BSM}$ with $ij = ds, db, sb, cu$ also subleading
ones, which in some NP scenarios could turn out to be the most important ones.
Seeing these contributions to various meson systems side-by-side illustrates
possible correlations generated by RG evolution between $\DF=2$
and $\DF=1$ transitions among the various meson systems
that have to be taken into account. This is in particular the case between
$B^0_d$ and $B^0_s$ as well as $K^0$ and $D^0$ meson systems. Our ATLAS casts
in this manner some doubts on the validity of many analyses present in the literature
that consider only one or two operators at the time and restrict the analyses
to a single meson system. Such analyses can only be considered as a first look
and have to be supplemented eventually by a more complete SMEFT analysis
that optimally includes a concrete UV completion. This is undermined
by the fact that within the SMEFT the results depend on the chosen basis for
SMEFT operators that signals the need for the UV completions that include some
aspects of a theory for Yukawa couplings.

Our paper is organized as follows. In \refsec{sec:2} we recall the LEFT for
the $\DF=2$ processes in any BSM scenario. We pay particular attention to
scheme transformations among the various operator bases in the literature to
enable a consistent use of RG equations, hadronic matrix elements as well as
UV matching conditions to SMEFT at the EW scale. In particular
we discuss the treatment of the evanescent operators, which must be consistent
with the available two-loop anomalous dimensions of the involved operators.

In \refsec{sec:LEFT-master} we present an update of the master formula for
$\DF=2$ processes ($\KKbar$, $\BBbar$ and $\DDbar$) in LEFT valid in any
BSM scenario first presented in \cite{Buras:2001ra}. It depends on the
{\em model-independent} matrix elements evaluated at $\muEW = 160\GeV$
and on Wilson coefficients of these operators evaluated at
the same scale. All the model dependence is collected in the values of
these coefficients. As a byproduct we present in \refsec{HME} a review of
various estimates of hadronic matrix elements found in the literature.

In \refsec{sec:SMEFT} we perform a general SMEFT anatomy of $\DF=2$ processes,
which eventually leads us to the most important formula for $\DF=2$ processes
in our paper, the one given entirely in terms of SMEFT WCs in the Warsaw
basis \cite{Grzadkowski:2010es} at the NP scale $\muNP$. In this context we
stress the importance of the Yukawa RG effects in the evolution from $\muNP$
down to $\muEW$. We also emphasize the differences between results obtained
in {\em{down}}- and {\em{up}}-Warsaw bases. In this section we discuss one-loop
matching of SMEFT onto LEFT in the analytic form and collect the relevant
RG equations accompanied by RG flow charts.
This section culminates in the SMEFT ATLAS, mentioned previously, built out
of numerous formulae for the individual contributions of the relevant
operators. In the main text we present these formulae for $\muNP = 5\TeV$
while in the \refapp{app:100ATLAS} the corresponding expressions are given
for $\muNP = 100\TeV$. The full set of contributions from all
relevant SMEFT operators can be found in an ancillary file
to the archive submission of this article. We also present the
effective sensitivity scales $\muNP_i$ of the Wilson coefficients of the
dominant operators.

In \refsec{sec:5} we illustrate this technology on a number of simplified
models that allow for $\DF=2$ processes at tree-level via heavy spin-zero or
spin-one boson exchange. This includes models with colourless
heavy gauge bosons ($Z^\prime$) and scalars and models with coloured
heavy gauge bosons ($G_a^\prime$) and scalars. Also the cases of
vector-like quarks and leptoquarks are briefly considered.

In \refsec{sec:6} we summarize the main results of our paper and present a
brief outlook for the coming years. In appendices we present the
SMEFT ATLAS for $\muNP = 100\TeV$ and its version for NP scenarios of
\refsec{sec:5} at $\muNP = 5\TeV$ in the up-basis. We list the relations
between various operator bases and we report the one-loop matching
of the SMEFT onto the LEFT in an analytic form  both in the down basis and
the up basis. Finally, we elaborate on the issue of evanescent operators.

%--------+---------+---------+---------+---------+---------+---------+---------+
%
%
%
%--------+---------+---------+---------+---------+---------+---------+---------+
\section{\boldmath LEFT Anatomy of $\DF=2$ Processes}
\label{sec:2}

%
%
%
%--------+---------+---------+---------+---------+---------+---------+---------+
\subsection{Preliminaries}

The $\DF=2$ LEFTs for $\KKbar$, $\BBbar$ and $\DDbar$ mixing arise in the SM
from the decoupling of the heavy electroweak gauge bosons, the Higgs field
and the top quark at $\muEW$ and similarly by the decoupling of heavy degrees
of freedom in any UV completion where additional light degrees of freedom below
$\muEW$ are absent, as is the case for SMEFT. We decompose the corresponding
effective Hamiltonian \cite{Ciuchini:1997bw, Buras:2000if} into the SM and
BSM contribution as follows
\begin{align}
  \label{eq:DF2-Heff}
  \cH_{\DF=2}^{ij} &
  =  \big[\cH_{\DF=2}^{ij}\big]_\text{SM}
  + \sum_a \WcL[ij]{a}(\mu)\, \OpL[ij]{a}
  + \text{h.c.} ,
\end{align}
with $ij = ds$ for $\KKbar$ mixing and $ij = sb,db$ for $\BBbar$ mixing,
respectively. In the SM there is only a single $\DF=2$ operator in each
meson system
\begin{align}
  \label{eq:DF2-Heff-SM}
  \big[\cH_{\DF=2}^{ij}\big]_\text{SM} &
  = \cN \, \big(\lambda^{ij}_t\big)^2 \;
    [\WcL[ij]{\text{VLL}}(\mu)]_\text{SM}\, \OpL[ij]{\text{VLL}}
  + \text{h.c.} ,
\end{align}
with the expression for $\OpL[ij]{\text{VLL}}$ given in \eqref{eq:BMU-basis}.
In the SM analyses it is common to extract a normalisation factor $\cN$ and
the CKM combinations given respectively by
\begin{align}
  \label{eq:DF2:norm-factor}
  \cN & = \frac{G_F^2  m_W^2}{4 \pi^2} ,
&
  \lambda^{ij}_{t} & = V_{ti}^*V_{tj}^{} ,
\end{align}
where we show the case of down-type mixing. Then the effect of the SM one-loop
box diagrams with top-quark exchange for down-type mixing is contained
in\footnote{{Note that whereas $ \WcL[ij]{\text{VLL}}(\muEW)|_\text{SM}$
is dimensionless, the WCs in the BSM part carry dimension $1/\TeV^2$. This will
also be the case of the BSM contributions from $\OpL[ij]{\text{VLL}}$ to
$\DF=2$ processes and will be evident from our results.}}
\begin{align}
  \label{eq:DF2:S0}
  \WcL[ij]{\text{VLL}}(\muEW)|_\text{SM} & = \delta_{ij} \, S_0(x_t) , &
  S_0(x) &
  = \frac{x (4 - 11 x + x^2)}{4\, (x-1)^2}
  + \frac{3 x^3 \ln x}{2\, (x-1)^3} \,,
\end{align}
at the scale $\muEW$. The function $S_0(x_t)$ depends on the ratio of the
top-quark and $W$-boson masses $x_t \equiv m_t^2/m_W^2$. The NLO QCD matching
correction is known from \cite{Buras:1990fn} and NLO EW matching corrections
from \cite{Gambino:1998rt}.

The case of $\Delta C = 2$ is found by exchanging the down-type quarks with
up-type quarks in the operators and using $ij = cu$. The pure short-distance
box diagrams with $W$-boson exchange and light down-type quarks yield vanishing
contribution due to the unitarity of the CKM matrix, i.e. GIM cancellation.
This can be also inferred from $S_0(x)\to 0$ in the limit $x\to 0$ and in
consequence $\big[\cH_{\DF=2}^{cu} \big]_\text{SM} = 0$ in~\eqref{eq:DF2-Heff}.

The most general LEFT operator basis consists of eight operators $\OpL[ij]{a}$,
with different basis conventions in the literature. We anticipate that in SMEFT
only four out of the eight operators are generated. The Wilson coefficients
$\WcL[ij]{a}(\mu)$ depend on the renormalization scale $\mu$. They are obtained
at the matching scale $\muEW$ and can be evolved via the RG equations to a
typical low-energy scale $\muLow$ of the order of a few $\geV$, i.e. the order
of the relevant external scales.

The $\DF=2$ operators contribute to the off-diagonal element of the
mass matrix of neutral meson ($M^0 = K^0, D^0, B_{s,d}$) mixing as follows
\begin{align}
  \label{eq:def-M12}
  M_{12}^{ij} &
  \equiv \big[M_{12}^{ij}\big]_\text{SM}
    +    \big[M_{12}^{ij}\big]_\text{BSM}
  = \frac{\bra{{M}^0} \cH_{\DF=2}^{ij} \ket{\oL{M^0}}}{2 M_{M^0}}
    {\, + \, \ord{\text{dim-8}}} \,,
\end{align}
where
\begin{align}
  \label{eq:def-M12BSM}
  \big[M_{12}^{ij}\big]_\text{BSM} &
  = \frac{1}{2 M_{M^0}} \sum_{a} \WcL[ij]{a}(\mu) \, \MeL[ij]{a} (\mu)
    {\, + \, \ord{\text{dim-8}}} \,,
\end{align}
the central quantity in our paper, is given  in terms of Wilson coefficients
and hadronic matrix elements of the operators,
\begin{align}
  \label{eq:def-hadr-ME}
  \MeL[ij]{a} = \MeL[ij]{a} (\mu) &
  \equiv \bra{{M}^0} \OpL[ij]{a} \ket{\oL{M^0}} (\mu) \,.
\end{align}
The $\mu$ dependence and more generally renormalization-scheme dependences
of Wilson coefficients and hadronic matrix elements cancel in
\eqref{eq:def-M12BSM}, such that observables that depend on the $M_{12}^{ij}$
are independent of the renormalization scheme. We will list the most important
observables in \refsec{sec:LEFT-master}.

The hadronic matrix elements $\MeL[ij]{a} (\mu)$ are calculated with
nonpertubative methods like LQCD or QCD sum rules at the
typical low energy scales $\mu = \muLow$ of a few $\geV$ set by the
masses of the neutral mesons $M_{M^0}$. For $\KKbar$ mixing also the
dual QCD (DQCD) approach is useful \cite{Buras:2018lgu}.

The leading contribution in the expansion of the ratio $\muLow/\muEW \ll 1$ to
the off-diagonal element $M_{12}^{ij}$ is due to the dimension-six contributions
of the $\DF=2$ effective Hamiltonian at the low-energy scales as given in
\eqref{eq:def-M12}. In fact, this is a very good approximation for $\BBbar$
mixing in the SM \cite{Boos:2004xp}.\footnote{Constraints from $\BBbar$ mixing
on NP contributions in current-current operators have been studied in
\cite{Bobeth:2014rda, Lenz:2019lvd}.} However, in the case of $\KKbar$ mixing
the strong CKM hierarchies suppress in the SM this contribution such that
dimension-eight contributions are CKM-enhanced and compete numerically with it.
In the SM the numerically most important contributions at dimension eight are
from double-insertions of $\DF=1$ dimension-six operators. Here the contributions
of $\DF=1$ operators with charm quarks can be still decoupled perturbatively
and actually absorbed as additional $\DF=2$ contributions in \eqref{eq:def-M12}.
However, there is a remaining dispersive contribution to $M_{12}^{ij}$ from
$\DF=1$ operators with light quarks ($q = u,d,s$), which can be evaluated only
with nonperturbative methods and prevents up to now the full prediction of
$\Delta M_K$ in the SM and also beyond. These dimension-eight contributions,
those that can still be decoupled at the charm scale and also the genuine
long-distance ones, are indicated in \eqref{eq:def-M12} and \eqref{eq:def-M12BSM}.
Although our objective is to provide a general anatomy of $\DF=2$, our main
focus will be the anatomy of the leading dimension-six term of \eqref{eq:def-M12}
in SMEFT.

The Wilson coefficients $\WcL[ij]{a}(\mu)$ in \eqref{eq:def-M12BSM}
are calculated perturbatively and are renormalised in
the \MSbar{} scheme because of the particularly simple RG evolution from $\muEW$
to any other scale $\mu$. The necessary anomalous dimension matrices (ADM) are
known at NLO in QCD \cite{Buras:2000if} for the so-called BMU operator basis,
see \eqref{eq:BMU-basis}. It is important to bear in mind that these
higher-order calculations in dimensional regularisation with $D\neq 4$ dimensions
require a generalization of the four-dimensional Levi-Civita tensor, and
specifically $\gamma_5$, which can be conveniently done with the help of
so-called evanescent operators \cite{Buras:1989xd, Dugan:1990df, Herrlich:1994kh}
and naively anti-commuting $\gamma_5$ (NDR). The presence of evanescent operators
besides the physical ones requires for example also a finite renormalization of
those parts of matrix elements of evanescent operators that are proportional to
physical operators. Therefore the Wilson coefficients and ADMs are dependent on
the choice of evanescent operators as well. The hadronic matrix elements
$\langle \OpL[ij]{a} \rangle$ are calculated by LQCD collaborations
employing nonperturbative renormalization schemes. These results are then
converted to the \MSbar{}-NDR including the very same evanescent operators to
guarantee a cancellation of the renormalization scheme dependences in
\eqref{eq:def-M12} to the NLO in QCD. The typical scale $\muLow$ after conversion
is of order of a few $\geV$, depending on the meson type. The NLO QCD corrections
are particularly important at these low scales because the QCD coupling $\alS$
is large. Of course, $\MeL[ij]{a}(\mu)$ obey the very same RG
equations as the Wilson coefficients except that the ADMs used
 in the latter are replaced by the transposed ones.\footnote{Usually in LEFT
the RG evolution of WCs is governed by transposed ADMs of the operators
\cite{Buchalla:1995vs}, but in the SMEFT literature these transposed ADMs
are just called ADMs.} Therefore they can be evolved to any
(perturbative) scale $\mu$, in particular $\muEW$ or even $\muNP$.

Unfortunately there is no unique choice of the operator basis for the
calculation of hadronic matrix elements in the literature, requiring
basis changes at NLO in QCD, that we will explain in more detail now.

%
%
%
%--------+---------+---------+---------+---------+---------+---------+---------+
\subsection[$\DF=2$ Operator Bases]
{\boldmath $\DF=2$ Operator Bases}
\label{sec:LEFT-ops-bases}

Several $\DF=2$ operator bases have been chosen for various reasons in the past.
We begin with the so-called BMU basis \cite{Buras:2000if} for which
the complete ADMs at NLO in QCD have been calculated in \cite{Buras:2000if}.
The BMU basis consists of $(5 + 3) = 8$ physical operators
\begin{equation}
  \label{eq:BMU-basis}
\begin{aligned}
  \OpL[ij]{\text{VLL}} &
  = [\bar{d}_i \gamma_\mu P_L d_j][\bar{d}_i \gamma^\mu P_L d_j] , &
\\[0.2cm]
  \OpL[ij]{\text{LR},1} &
  = [\bar{d}_i \gamma_\mu P_L d_j][\bar{d}_i \gamma^\mu P_R d_j] , & \qquad\quad
  \OpL[ij]{\text{LR},2} &
  = [\bar{d}_i P_L d_j][\bar{d}_i P_R d_j] ,
\\[0.2cm]
  \OpL[ij]{\text{SLL},1} &
  = [\bar{d}_i P_L d_j][\bar{d}_i P_L d_j] , &
  \OpL[ij]{\text{SLL},2} &
  = -[\bar{d}_i \sigma_{\mu\nu} P_L d_j][\bar{d}_i \sigma^{\mu\nu} P_L d_j] ,
\end{aligned}
\end{equation}
which are built exclusively out of colour-singlet currents $[\bar{d}^\alpha_i
\ldots d^\alpha_j] [\bar{d}^\beta_i \ldots d^\beta_j]$, where $\alpha,\, \beta$
denote colour indices. This feature is very useful for calculations in DQCD
\cite{Buras:2018lgu, Aebischer:2018rrz}, because the matrix elements
in the large-$N_c$ limit can be obtained directly without using Fierz identities.
The chirality-flipped sectors VRR and SRR are obtained from interchanging $P_L
\leftrightarrow P_R$ in VLL and SLL. Note that the minus sign in
$\OpL[ij]{\text{SLL},2}$ arises from different definitions of
$\tilde{\sigma}_{\mu\nu} \equiv [\gamma_\mu,\, \gamma_\nu]/2$
in \cite{Buras:2000if} w.r.t. $\sigma_{\mu\nu} = i \tilde{\sigma}_{\mu\nu}$
used here. The ADMs of the five distinct sectors (VLL, SLL, LR, VRR, SRR) have
been calculated at NLO in QCD \cite{Buras:2000if}, and numerical solutions for
$ij = ds, bd, bs$ are given in \cite{Buras:2001ra}.

The so-called SUSY basis \cite{Gabbiani:1996hi, Aoki:2019cca} instead uses
the operators
\begin{equation}
  \label{eq:SUSY-basis}
\begin{aligned}
  \OpL[ij]{1} & = \OpL[ij]{\text{VLL}} \,, &
\\
  \OpL[ij]{4} & = \OpL[ij]{\text{LR},2} \,, \qquad\quad
&
  \OpL[ij]{5} & = [\bar{d}_i^\alpha P_L d_j^\beta][\bar{d}_i^\beta P_R d_j^\alpha]
  = -\frac{1}{2} \OpL[ij]{\text{LR},1} \,,
\\
  \OpL[ij]{2} & = \OpL[ij]{\text{SLL},1} \,,
&
  \OpL[ij]{3} & = [\bar{d}_i^\alpha P_L d_j^\beta][\bar{d}_i^\beta P_L d_j^\alpha]
  = -\frac{1}{2} \OpL[ij]{\text{SLL},1} + \frac{1}{8} \OpL[ij]{\text{SLL},2} \,,
\end{aligned}
\end{equation}
and $\OpL[ij]{1',2',3'}$ obtained from $\OpL[ij]{1,2,3}$ via $P_L \to P_R$.
The relations for $\OpL[ij]{3,5}$ are the usual  Fierz relations valid in $D=4$
only. Beyond the LO evanescent operators must be added to the r.h.s of these
relations. However, as demonstrated in \cite{Buras:1989xd}, and subsequently
discussed in \cite{Dugan:1990df, Herrlich:1994kh}, a particular definition
of these operators can be made so that these operators affect only two-loop
anomalous dimensions, but have no impact on one-loop matching and allow to
use $D=4$ Fierz relations in transforming one operator basis to another one.
In the present paper we use exclusively this definition (BMU) of the evanescent
operators\footnote{There is a second choice of evanescent operators known as
BBGLN \cite{Beneke:1998sy} in the context of SM calculations of life times and
decay width differences for $B_q$ mesons, which does not preserve Fierz
relations beyond LO QCD. The transformation from the BMU to the BBGLN basis at
NLO QCD is given in \cite{Gorbahn:2009pp}.} that is consistent with the
two-loop anomalous dimensions calculated in \cite{Buras:2000if} that are used
in the master formulae of \cite{Buras:2001ra}. Therefore we did not show them
explicitly in the formulae above. However, this is an important issue for the
future NLO SMEFT analyses and we elaborate on it in \refapp{EVO}. The ADMs of
the $\OpL[ij]{1,1'}$ at NLO in QCD are identical to the results of the ones of
the VLL and VRR sectors in the BMU basis.

Yet another basis is relevant for our study, which has been introduced to
facilitate the classification of the complete LEFT operator basis
\cite{Jenkins:2017jig} for the purpose of matching with SMEFT. We will refer
to it as JMS basis. The full one-loop matching of SMEFT to LEFT in the JMS
basis was recently given in \cite{Dekens:2019ept}. The relevant $\DF=2$
operators are
\begin{equation}
  \label{eq:JMS-basis}
\begin{aligned}
  \opL[VLL]{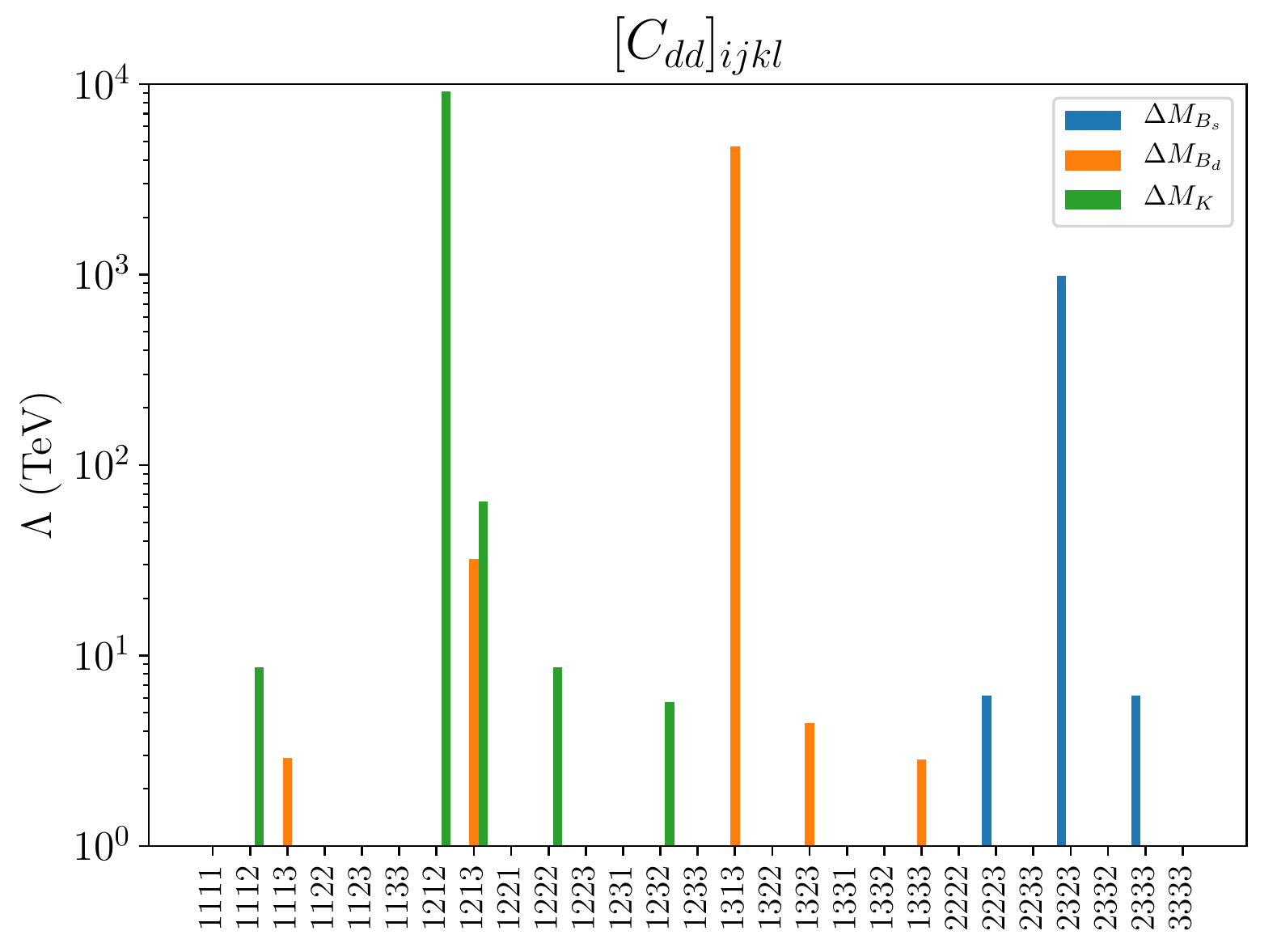}{ijij} & = \OpL[ij]{\text{VLL}} \,, &
\\
  \opL[VRR]{dd}{ijij} & = \OpL[ij]{\text{VRR}} \,, &
\\
  \opL[V1,LR]{dd}{ijij} & = \OpL[ij]{\text{LR},1} \,,
\\
  \opL[V8,LR]{dd}{ijij} &
  = [\bar{d}_i \gamma_\mu P_L T^A d_j][\bar{d}_i \gamma^\mu P_R T^A d_j]
  =-\frac{1}{6}\OpL[ij]{\text{LR},1} -\OpL[ij]{\text{LR},2} \,,
\\
  \opL[S1,RR]{dd}{ijij} & = \OpL[ij]{\text{SRR},1} \,,
\\
  \opL[S8,RR]{dd}{ijij} &
  = [\bar{d}_i P_R T^A d_j][\bar{d}_i P_R T^A d_j]
  =-\frac{5}{12} \OpL[ij]{\text{SRR},1} + \frac{1}{16}\OpL[ij]{\text{SRR},2} \,,
\end{aligned}
\end{equation}
where $T^A$ are $\text{SU(3)}_c$ colour generators of the fundamental representation.
Note that $(\opL[S1,RR]{dd}{jiji})^\dagger = \OpL[ij]{\text{SLL},1}$ etc.
To make use of the one-loop matching results in the JMS basis and connect them
with the hadronic matrix elements from LQCD collaborations it is hence necessary
to transform the Wilson coefficients in the JMS basis to the BMU (or SUSY) basis
in order to cancel the scheme dependence of hadronic matrix elements at NLO in
QCD. We collect these relations in \refapp{Relations}.

Eventually we note that the four operators $Q_{1,2}^\text{SLL}$ and
$Q_{1,2}^\text{SRR}$ violate hypercharge and, although allowed by
$\text{SU(3)}_c \times \text{U(1)}_\text{em}$, cannot be generated
at and above the EW scale in the context of SMEFT and moreover cannot
be generated through RG evolution.

%
%
%
%--------+---------+---------+---------+---------+---------+---------+---------+
\subsection{Hadronic Matrix Elements}
\label{HME}

The matrix elements $\langle \OpL[ij]{a} \rangle$ in \eqref{eq:def-M12BSM}
are provided by LQCD collaborations that present results either for the BMU
or the SUSY basis. The matrix elements of the LR and SLL/SRR sectors are
chirally enhanced compared to the VLL/VRR sector, as the latter vanish in the
chiral limit. The corresponding chiral enhancement factor
\begin{align}
  \label{eq:def-r_chi}
  r_\chi^{ij}(\mu) &
  \equiv
  \frac{(f_{M^0} M_{M^0})^{-2}}{
     \bra{M^0} \oL{d}_i \gamma_5 d_j \ket{0} \,
     \bra{0} \oL{d}_i \gamma_5 d_j \ket{\oL{M^0}}}
  \; \stackrel{\text{(VIA)}}{\approx} \;
     \left(\frac{M_{M^0}}{m_i(\mu) + m_j(\mu)}\right)^2
\end{align}
is related to the meson decay constant $f_{M^0}$ and the overlap of the
scalar densities with the meson states. It is renormalization-scheme
dependent and involves in the vacuum insertion approximation (VIA) and
DQCD approach the \MSbar{} quark masses. It becomes especially large
for $\KKbar$ mixing.

Usually the LQCD collaborations prefer not to calculate directly
$\langle \OpL[ij]{a} \rangle$, but rather ratios, as for example
\cite{Babich:2006bh}
\begin{align}
  \label{eq:def-R_a^ij}
  R_a^{ij} (\mu) &
  \equiv \frac{\MeL[ij]{a} (\mu)}{\MeL[ij]{\text{VLL}} (\mu)} ,
&
  a & \neq \text{VLL} \,,
\end{align}
which exhibit a cancellation of LQCD-specific systematic uncertainties.
The $R_a^{ij}(\mu)$ advantageously include the nonpertubative evaluation of
$r_\chi^{ij}$.

However, for historical reasons the $\MeL[ij]{a}(\mu)$ are expressed often
also in terms of bag factors $B_a^{ij}(\mu)$ that are  unity in the
VIA, i.e. they quantify the deviation from VIA. This allows also for
getting insight in their LQCD values for $\KKbar$ mixing with the help of
DQCD \cite{Buras:2018lgu}. The bag parameters are also
subject to cancellation of systematic uncertainties in LQCD calculations
\cite{Jang:2015sla, Dowdall:2019bea}. There are different conventions for
the various neutral meson systems, but all are for the SUSY basis
\begin{equation}
  \label{eq:rel-ME-bag}
\begin{aligned}
  \MeL[ij]{1} (\mu) &
  = \frac{2}{3} (F_{M^0} M_{M^0})^2 B_1^{ij} (\mu) , &
\\
  \MeL[ij]{a} (\mu) &
  = N_a^{ij} \left(r_\chi^{ij} + d_a^{ij}\right)
    (F_{M^0} M_{M^0})^2 B_a^{ij} (\mu),
&
  \qquad a & \in (2,3,4,5)
\end{aligned}
\end{equation}
with $N_a^{ij} = (-5/12,\, 1/12,\, 1/2,\, 1/6)$. The $B_1^{ds}$
is the well-known $B_K$ parameter of $\KKbar$ mixing. Further, the constants
$d_a^{ij} = 0$ for $\KKbar$ and $\DDbar$ \cite{Carrasco:2015pra} mixing, whereas
$d_a^{ib} =(0,\, 0,\, 1/6,\, 3/2)$ for $\BBbar$ mixing \cite{Bazavov:2016nty,
Dowdall:2019bea}.

It is obvious that the bag factors themselves are not sufficient to calculate
the $\MeL[ij]{a}$, but require the knowledge of $r_\chi^{ij}$. For example,
FNAL/MILC calculates $\MeL[ij]{a}$ for $\BBbar$ \cite{Bazavov:2016nty}
and $\DDbar$ \cite{Bazavov:2017weg} mixing directly and converts then to
bag factors using the VIA form of $r_\chi^{ib}$ in
\eqref{eq:def-r_chi} with fixed numerical values of the quark masses and decay
constant. In this way $r_\chi^{ib}$ can be regarded as a fixed
numerical convenience factor that in principle would not introduce additional
uncertainties related to quark masses and decay constant, despite being scheme
dependent. Consequently, in all applications then strictly the numerical
values of quark masses and decay constants used by FNAL/MILC must be used as
well for the conversion of $B_a^{ij} \to \MeL[ij]{a}$. On the
other hand the Flavour Lattice Averaging Group~(FLAG)~\cite{Aoki:2019cca}
provides currently $\KKbar$ bag factors without $r_\chi^{ds}$, and hence in
phenomenological predictions the applicant is forced to introduce unknown
systematic uncertainties when using the VIA approximation of $r_\chi^{ds}$
together with the parametric uncertainties from the quark masses and the decay
constants. The latter two quantities are usually obtained from different dedicated
LQCD calculations that in principle involve different systematic uncertainties
as well. Therefore it is important that the FLAG provides also the $r_\chi^{ds}$
or alternatively directly averages of the matrix elements $\MeL[ij]{a}$
or the ratios $R_a^{ij}$. The summary of the current status is as follows.
\begin{itemize}
\item
  In the case of $\KKbar$ mixing, FLAG provides a summary \cite{Aoki:2019cca} of
  the available results for the bag factors $B_a^{ds}$ ($a = 1,2,3,4,5$) of
  the SUSY basis.\footnote{Note that we use the convention $ij = ds$
  as opposed to FLAG's $ij=sd$, which complies with the convention in
  \texttt{WCxf} \cite{Aebischer:2017ugx}.} They are from ETM for $N_f = 2$
  \cite{Bertone:2012cu} and for
  $N_f = 2+1+1$ from \cite{Carrasco:2015pra}, as well as the $N_f = 2+1$ average from
  the SWME \cite{Jang:2015sla} and RBC-UKQCD \cite{Garron:2016mva, Boyle:2017skn}
  LQCD collaborations. We will use here the FLAG average for $N_f = 2+1$, for
  further details see also \cite{Boyle:2017ssm, Aoki:2019cca}. We note that
  SWME provides results for the BMU basis, which have been converted to the
  SUSY basis by FLAG. In principle RBC-UKQCD provides also the ratios $R_a^{ds}$,
  which do not require the knowledge of $r_\chi^{ds}$. We will use the
  averages of the $N_f = 2+1$ bag factors from FLAG.
\item
  For the case of $\BBbar$ mixing, FLAG provides only averages for the bag factor
  $B_1^{ib}$ of the SM operator. However, the full set of matrix elements has
  been calculated with LQCD methods for $N_f = 2$ by ETM \cite{Carrasco:2013zta},
  $N_f = 2 + 1$ by FNAL/MILC \cite{Bazavov:2016nty} and for $N_f = 2+ 1 +1$
  by HPQCD \cite{Dowdall:2019bea}. Further, the bag factors were also calculated
  with sum rules \cite{Grozin:2016uqy, Kirk:2017juj, Grozin:2018wtg, King:2019lal}
  and the average of LQCD (except \cite{Carrasco:2013zta}) and sum rule results
  can be found in \cite{DiLuzio:2019jyq}. We use the averages of the
  bag factors from HPQCD and FNAL/MILC, as given in \cite{Dowdall:2019bea}.
\item
  For the case of $\DDbar$ mixing, ETM calculated bag factors for
  $N_f = 2+1+1$~\cite{Carrasco:2015pra} and more recently FNAL/MILC the matrix
  elements for $N_f = 2+1$~\cite{Bazavov:2017weg} for the full set of
  $\Delta C = 2$ operators (BBGLN and BMU). Both results are consistent with
  largest tensions for bag factors $B^{cu}_{4, 5}$. The sum rule determination
  \cite{Kirk:2017juj} suffers from larger uncertainties and is consistent
  with the LQCD determinations. We prefer to use the more recent direct
  determinations of matrix elements from FNAL/MILC \cite{Bazavov:2017weg}
  over the bag factors of ETM \cite{Carrasco:2015pra}.
\end{itemize}
The numerical values of the $\DF=2$ nonperturbative input for $\KKbar$,
$\DDbar$ and $\BBbar$ mixing is collected in \reftab{tab:DF2-me-input}.
There we provide the matrix elements in the \MSbar{}-NDR scheme at
a low-energy scale $\muLow$ for the SUSY basis, which can be converted to
the BMU basis with the help of \eqref{eq:SUSY-basis}. The number of flavours,
$N_f$, in \reftab{tab:DF2-me-input} gives the starting number of active flavours
used in the running of $\alS$ for the RG evolution.

\begin{table}
\centering
\renewcommand{\arraystretch}{1.4}
\resizebox{\columnwidth}{!}{
\begin{tabular}{|lll|lll|}
\hline
  Parameter
& Value
& Ref.
&  Parameter
& Value
& Ref.
\\
\hline \hline
  $\alS^{(5)}(m_Z)$    & $0.1181(11)$           & \cite{Tanabashi:2018oca}
& $m_Z$                & $91.1876(21)$ GeV      & \cite{Tanabashi:2018oca}
\\
\hline
  $M_{K^0}$            & $497.611(13)$ MeV      & \cite{Tanabashi:2018oca}
& $f_K/f_\pi$          & $1.194(5)$             & \cite{Aoki:2019cca}
\\
  $\Delta M_K$         & $3.484(6) \cdt{-15}\GeV$  & \cite{Tanabashi:2018oca}
& $f_\pi$              & $130.41(20)$ \MeV      & \cite{Tanabashi:2018oca}
\\
\hline
  $M_{D^0}$            & $1864.83(5)$ MeV       & \cite{Tanabashi:2018oca}
& $\tau_{D^0}$         & $4.101(15) \cdt{-13}$ s & \cite{Tanabashi:2018oca}
\\
  $\Delta M_D$         & $6.3(^{+1.8}_{-1.9}) \cdt{-15}\GeV$ &
& $x$                  & $0.39(^{+11}_{-12}) \cdt{-2}$ & \cite{Tanabashi:2018oca}
\\
\hline
  $M_{B_s}$            & $5366.88(17)$ MeV      & \cite{Tanabashi:2018oca}
& $M_{B_d}$            & $5279.64(13)$ MeV      & \cite{Tanabashi:2018oca}
\\
  $\Delta M_{B_s}$     & $1.1683(13) \cdt{-11}\GeV$ & \cite{Tanabashi:2018oca}
& $\Delta M_{B_d}$     & $3.334(13)  \cdt{-13}\GeV$ & \cite{Tanabashi:2018oca}
\\
  $f_{B_s}$            & $230.3(1.3)$ MeV       & \cite{Aoki:2019cca}
& $f_{B_d}$            & $190.0(1.3)$ MeV       & \cite{Aoki:2019cca}
\\
  $\oL{m}_s(2\GeV)$    & $92.0(1.1)$ MeV        & \cite{Aoki:2019cca}
& $\oL{m}_d(2\GeV)$    & $4.67(9)$ MeV          & \cite{Aoki:2019cca}
\\
\hline
\end{tabular}
}
\renewcommand{\arraystretch}{1.0}
\caption{\label{tab:LEFT-num-input}
  \small
  Numerical input values for parameters entering the conversion of bag factors
  to matrix elements and the LEFT master formula \eqref{eq:master-M12BSM}. The
  values of the strange- and down-quark masses in the \MSbar{} scheme are the
  $N_f = 2+1$ averages of lattice determinations from the FLAG group from
  \cite{Blum:2014tka, Durr:2010vn, Durr:2010aw, McNeile:2010ji, Bazavov:2009fk,
  Fodor:2016bgu}. The $B_q$-meson decay constants $f_{B_q}$ are averages from
  the FLAG group for $N_f = 2+1+1$ from \cite{Bazavov:2017lyh, Bussone:2016iua,
  Dowdall:2013tga, Hughes:2017spc}. They are almost identical to the single
  determination of FNAL/MILC $f_{B_s} = 230.7(1.3)\MeV$ and $f_{B_d} =
  190.5(1.3)\MeV$ \cite{Bazavov:2017lyh}. We determine $\Delta M_D
  = x/\tau_{D^0}$ with the value of $x$ from a global fit when allowing CP
  violation in all decays \cite{Tanabashi:2018oca}, for comparison it would
  be $x = 0.50(^{+13}_{-14}) \cdt{-2}$ assuming no CP violation.
}
\end{table}

In the absence of any information on the chiral enhancement factor from FLAG,
we use for the conversion of the $\KKbar$ bag factors to matrix elements the
$N_f= 2+1$ \MSbar{} quark masses from FLAG \cite{Aoki:2019cca} together with
the Kaon decay constant and masses \cite{Tanabashi:2018oca} listed in
\reftab{tab:LEFT-num-input}. The chiral enhancement factor is $r_\chi^{ds} =
33.44(78)$ at $\muLow = 3.0 \GeV$ for $N_f = 3$. Note that $B_1^{ds}$ is
given by FLAG at $\mu = 2.0 \GeV$, whereas all the other bag factors
$B_{2,3,4,5}^{ds}$ at $\muLow = 3.0\GeV$, thus we have evolved $\MeL[ds]{1}$
to that scale as well.

The conversion of the averaged bag factors for $\BBbar$ mixing given in
\cite{Dowdall:2019bea} to matrix elements is done with the very same values
of the \MSbar{} quark masses as given in Table 1 of \cite{Dowdall:2019bea}
\begin{align}
  \oL{m}_b(\oL{m}_b) & = 4.162(48) \GeV   , \quad
& \frac{\oL{m}_b}{\oL{m}_s} & = 52.55(55) , \quad
& \frac{\oL{m}_s}{\oL{m}_d} & = 27.18(10) ,
\end{align}
the $B_q$-meson decay constants from \cite{Bazavov:2017lyh} and $B_q$-meson
masses in \reftab{tab:LEFT-num-input}. Thus, we use the very same values as
HPQCD to calculate the chiral enhancement factors $r_\chi^{db} = 1.607(37)$
and $r_\chi^{sb} = 1.601(37)$ at $\muLow = 4.16\GeV$.

\begin{table}
\centering
\renewcommand{\arraystretch}{1.5}
\begin{tabular}{|c||cc|ccccc|}
\hline
  $ij$     & $\muLow$ & $N_f$  & $\MeL[ij]{1}$ & $\MeL[ij]{2}$ & $\MeL[ij]{3}$ & $\MeL[ij]{4}$ & $\MeL[ij]{5}$ \\
           & $[\geV]$ &        & $[\geV^4]$    & $[\geV^4]$    & $[\geV^4]$    & $[\geV^4]$    & $[\geV^4]$    \\
\hline\hline
  $sd$     & $3.0$    & 3      & $0.002156(34)$ & $-0.0420(16)$ & $0.0128(6)$  & $0.0930(30)$  & $0.0241(14)$  \\
\hline
  $cu$     & $3.0$    & 4      & $0.0806(56)$  & $-0.1442(72)$ & $0.0452(31)$  & $0.2745(140)$ & $0.1035(74)$  \\
\hline
  $db$     & $4.16$   & 5      & $0.56(2)$     & $-0.53(3)$    & $0.106(8)$    & $0.96(5)$     & $0.51(2)$     \\
  $sb$     & $4.16$   & 5      & $0.86(3)$     & $-0.85(5)$    & $0.174(11)$   & $1.40(6)$     & $0.74(3)$     \\
\hline
\end{tabular}
\renewcommand{\arraystretch}{1.0}
\caption{\small
  \label{tab:DF2-me-input}
  The values of the matrix elements in the SUSY basis in the \MSbar{}-NDR
  scheme at the low-energy scale $\muLow$ for number of flavours $N_f$.
}
\end{table}

The RG evolution of the matrix elements depends strongly on the running
coupling $\alS$, for which we use the initial value $\alS^{(5)}(m_Z)$ with
$N_f = 5$ given in \reftab{tab:LEFT-num-input} and three-loop equations
for the RG evolution. The quark-threshold crossings to $N_f = 4$ is set
to $\mu_4 = 4.2\GeV$ and for $N_f = 3$ to $\mu_3 = 1.3\GeV$, whereas when
going to $N_f = 6$ we use $\mu_5 = \muEW = 160\GeV$.

Inspecting \reftab{tab:DF2-me-input} we observe the following pattern:
\begin{itemize}
\item
  The matrix elements of LR operators are in the $\KKbar$ system much
  larger than the matrix element of the SM VLL operator due to the chiral
  enhancement. In particular $\MeL[ds]{\text{LR},2} = \MeL[ds]{4}$ is very
  large with significant impact on phenomenology as known already for decades
  \cite{Beall:1981ze, Bagger:1997gg, Ciuchini:1997bw, Buras:2000if}.
\item
  While this pattern is also seen to some extent in the charm system
  it is practically absent at these scales in the $\BBbar$ systems where
  chiral enhancement is absent.
\item
  However, as we will see in the master formulas for $M_{12}^{ij}$
  in LEFT, and in particular in SMEFT, the hierarchy in question is further
  enhanced in the $\KKbar$ system through RG effects. Even in the $\BBbar$
  and charm systems the matrix elements of the LR operators and consequently
  the coefficients in the master formulae are significantly larger
  at these high scales than the SM one. This feature is known from
  \cite{Blanke:2008zb, Buras:2014zga}.
\end{itemize}

%
%
%
%--------+---------+---------+---------+---------+---------+---------+---------+
\subsection{LEFT Master Formula}
\label{sec:LEFT-master}

Hadronic matrix elements from LQCD, DQCD and sum rules are usually obtained at
low energy scales $\muLow$ of a few $\geV$. However, for a transparent study of
NP contributions that are generated at higher scales it is useful to evaluate
these matrix elements at the EW scale $\muEW$, the largest scale of validity
of LEFT. This can be done by means of RG methods as explained in great detail
in \cite{Buras:2001ra}. In this section we adapt these results to derive the
first master formula for $\DF=2$ processes given in terms of LEFT Wilson
coefficients evaluated at the EW scale. The one involving SMEFT Wilson
coefficients will be presented in \refsec{sec:SMEFT-master}.

\begin{table}
\centering
\renewcommand{\arraystretch}{1.3}
\begin{tabular}{|c||ccccc|c|}
\hline
  BMU   & $P_\text{VLL}$ & $P_{\text{SLL},1}$ & $P_{\text{SLL},2}$ & $P_{\text{LR},1}$ & $P_{\text{LR},2}$ & units\\
\hline
  $K^0$ &       0.102(2) &     -4.32(16) &     -7.93(37) &     -8.55(28) &     14.14(82) & $10^{7} \TeV^2$ \\
  $D^0$ &        0.56(4) &     -2.20(11) &     -4.04(28) &     -4.23(22) &      6.18(44) & $10^{7} \TeV^2$ \\
  $B_d$ &       2.67(10) &     -4.99(28) &     -9.05(68) &    -10.29(54) &     12.75(50) & $10^{5} \TeV^2$ \\
  $B_s$ &       1.15(4)  &     -2.24(13) &     -4.08(26) &     -4.20(18) &      5.22(21) & $10^{4} \TeV^2$ \\
\hline
\hline
  SUSY  & $P_1$          & $P_2$         & $P_3$         & $P_4$         & $P_5$         & units \\
\hline
  $K^0$ &       0.102(2) &     -4.32(16) &       1.09(5) &     14.14(82) &      4.28(14) & $10^{7} \TeV^2$ \\
  $D^0$ &        0.56(4) &     -2.20(11) &       0.56(4) &      6.18(44) &      2.12(11) & $10^{7} \TeV^2$ \\
  $B_d$ &       2.67(10) &     -4.99(28) &       1.12(8) &     12.75(50) &      5.15(27) & $10^{5} \TeV^2$ \\
  $B_s$ &       1.15(4)  &     -2.24(13) &       0.51(3) &      5.22(21) &       2.10(9) & $10^{4} \TeV^2$ \\
\hline
\hline
  JMS   & $P_{dd}^{VLL}$ & $P_{dd}^{S1,RR}$ & $P_{dd}^{S8,RR}$   & $P_{dd}^{V1,LR}$ & $P_{dd}^{V8,LR}$ & units \\
\hline
  $K^0$ &       0.102(2) &     -4.32(16) &       1.31(5) &     -8.55(28) &    -12.72(74) & $10^{7} \TeV^2$ \\
  $D^0$ &        0.56(4) &     -2.20(11) &       0.66(3) &     -4.23(22) &      -5.48(39)& $10^{7} \TeV^2$ \\
  $B_d$ &       2.67(10) &     -4.99(28) &       1.51(9) &    -10.29(54) &    -11.04(44) & $10^{5} \TeV^2$ \\
  $B_s$ &       1.15(4)  &     -2.24(13) &       0.68(4) &     -4.20(18) &     -4.52(18) & $10^{4} \TeV^2$ \\
\hline
\end{tabular}
\renewcommand{\arraystretch}{1.0}
\caption{\small
  \label{tab:LEFT-Pa}
  The values of the coefficients $P_a^{ij}(\muEW)$ entering the LEFT master formula
  \eqref{eq:master-M12BSM} at $\muEW = 160 \GeV$ in the BMU, SUSY and JMS bases,
  using as input the \MSbar{}-NDR matrix elements at the low-energy scale $\muLow$
  from \reftab{tab:DF2-me-input}. The shown uncertainties are due to the
  either matrix elements or the bag factors and their corresponding
  chiral enhancement factors.
}
\end{table}

We find it convenient to present the numerical results in the form of
the master formula
\begin{equation}
  \label{eq:master-M12BSM}
\boxed{
  2\big[M_{12}^{ij}\big]_\text{BSM}
  = (\Delta M_{ij})_\text{exp}
    \sum_{a}^{\phantom{x}} P_a^{ij}(\muEW) \, \WcL[ij]{a}(\muEW).
}
\end{equation}
The normalization to the experimental value of $(\Delta M_{ij})_\text{exp}$
allows easily to infer the size of the BSM Wilson coefficients
$\WcL[ij]{a}(\muEW)$ that would generate a certain fraction of this
measured value, in view of the known numerical values of the coefficients
$P_a^{ij}(\muEW)$, which are collected in \reftab{tab:LEFT-Pa}.
The $P_a^{ij}(\muEW) = \MeL[ij]{a} (\muEW)/(M_{M^0} (\Delta M_{ij})_\text{exp})$
are given in terms of the matrix elements at the EW scale as follows
from \eqref{eq:def-M12BSM}. They are related to the $\left[P_a^{ij}(\muEW)
\right]^{\text{BJU}}$ from \cite{Buras:2001ra} as
\begin{align}
  \label{eq:P_a-BJU}
  P_a^{ij}(\muEW) & =
  \frac{2}{3} \frac{M_{M^0} f_{M^0}^2}{(\Delta M_{ij})_\text{exp}}
  \left[P_a^{ij}(\muEW)\right]^{\text{BJU}} .
\end{align}
The expressions of $\left[P_a^{ij}(\muEW)\right]^{\text{BJU}}$ and
$P_a^{ij}(\muEW)$ summarize the RG evolution from the low-energy scale
$\muLow$ and the matrix elements $\MeL[ij]{a}(\muLow)$ from
\reftab{tab:DF2-me-input}, such that the $\muLow$ dependence cancels
\cite{Buras:2001ra}.

\begin{figure}
\centering
  \includegraphics[width=0.45\textwidth]{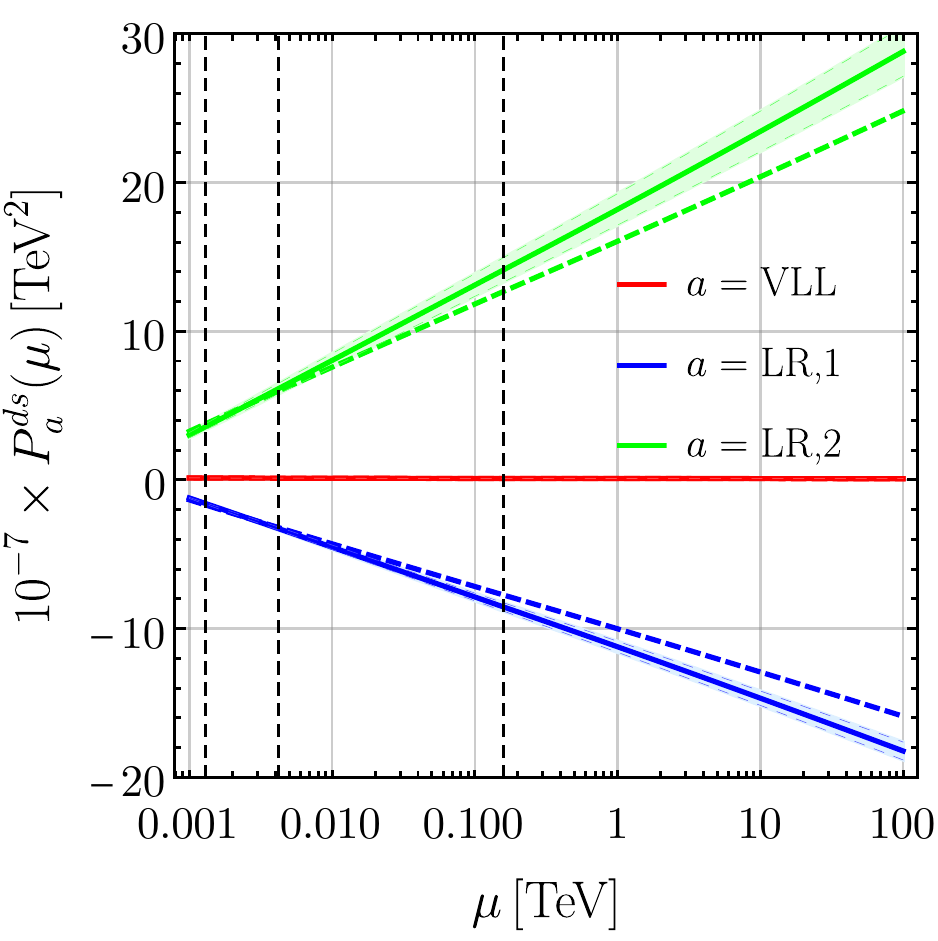}
  \hskip 0.06\textwidth
  \includegraphics[width=0.45\textwidth]{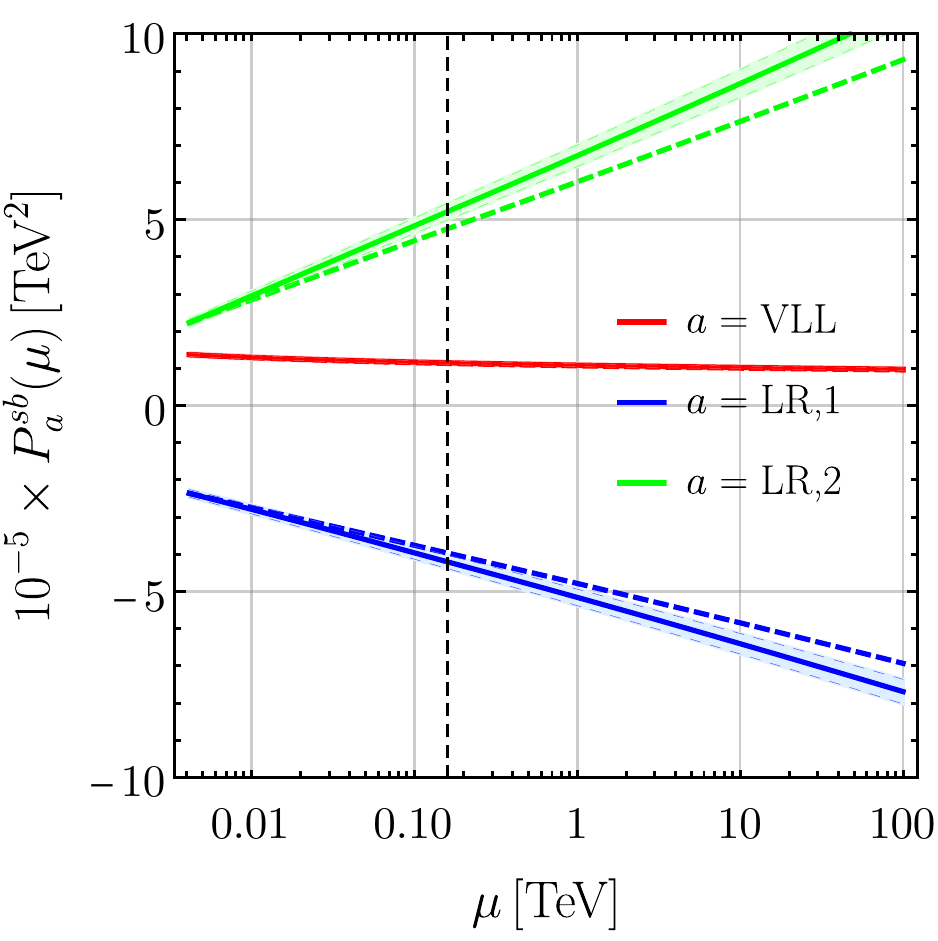}
\caption{\small The QCD RG evolution at NLO [solid] versus LO [dashed]
  for some of the coefficients $P_a^{ij}(\mu)$ for $\KKbar$ [left] and
  $\BBbarS$ mixing [right]. The coloured band around the NLO
  results shows the hadronic uncertainties from the matrix elements.
  The vertical dashed line indicates $\muEW = 160\GeV$ and for $\KKbar$
  mixing also the $N_f = 4, 5$ flavour thresholds. For scales larger
  than $\muEW$ the full SMEFT RG evolution should be used in principle,
  including Yukawa and the full SM gauge sector contributions.
}
\label{fig:qcd-1-2}
\end{figure}

For the numerical evaluation of the RG evolution of the matrix elements at NLO
in QCD, we use for $\BBbar$ mixing the initial scale and $N_f = 5$ as given in
\reftab{tab:DF2-me-input}. In the case of $\DDbar$ mixing the RG evolution
starts at $\mu = 3.0\GeV$ with $N_f = 4$ and is switched to $N_f = 5$ at
$\mu_4 = 4.2\GeV$. In the case of $\KKbar$ mixing the evolution is done first
with $N_f = 3$ from $\mu = 3.0\GeV$ down to $\mu_3 = 1.3\GeV$ and only then we
switch to $N_f = 4$ to evolve up in scale to $\muEW$, with the intermediate
threshold crossing to $N_f = 5$ at $\mu_4$. The NLO QCD corrections always
lead to an increase of the LO results for the $P_a^{ij}$ of about
$\{1.5-1.9,\; 4.7-6.0,\; 2.0-2.7,\; 6.0-10.4,\; 9.7-11.4 \} \%$ for
$a = \{\text{VLL},\, \text{SLL1},\, \text{SLL2},\, \text{LR1},\, \text{LR2}\}$
in the BMU basis, with smallest numbers for $\BBbar$ mixing and largest for
$\KKbar$ mixing. The size of the NLO corrections is up to roughly a
factor of two larger than the current hadronic uncertainties due to the
matrix elements. The effect is also shown in \reffig{fig:qcd-1-2} for
$\KKbar$ and $\BBbarS$ mixing.

The formulae for observables in terms of these matrix elements can be found
in many papers, in particular in the recent book \cite{Buras:2020xsm}.
Here we recall just the general dependence
\begin{align}
  \label{sd}
  ij & = ds :
& \Delta M_K    & =     2 \re\! \big( M_{12}^{ds} \big) ,
& \varepsilon_K & \propto \im\! \big( M_{12}^{ds} \big) ,
\\
  \label{bj}
  ij & = ib :
& \Delta M_{B_i} & = 2 \big|M_{12}^{ib}\big|,
& \phi_i   & = \mbox{Arg} \big(M_{12}^{ib} \big) .
\end{align}
We remind the reader that $\Delta M_{K,D}$ receive also substantial
long-distance corrections. Further, we have assumed that SM QCD penguin
pollution and new physics in $b\to s c\bar{c}$ processes are negligible.

The magnitude of the dimensionfull $P_a^{ij} \sim \mu^2$ must be cancelled by the
$\WcL[ij]{a} \sim (\mu')^{-2}$, such that their ratio $\mu/\mu' < 1$ does not lead
to excessive BSM contributions to $[M_{12}^{ij}]_\text{BSM}$ that would be ruled
out by observations. In the case of BSM contributions at tree-level the
$\mu'$ can be interpreted as the mass scale $\muNP$ of the new physics. It is
instructive to calculate the size of $\WcL[ij]{\text{VLL}}$ in the SM using
\eqref{eq:DF2-Heff-SM}. The normalization factor and the universal one-loop
function entering down-type meson mixing are of order $\cN S_0(x_t) \approx
0.0518 \TeV^{-2}$. The CKM combinations entering $\BBbar$ and $\KKbar$ mixing
found from a SM CKM fit are $|(V_{tb}^{} V_{td}^*)^2| \approx 8.0 \cdot 10^{-5}$,
$|(V_{tb}^{} V_{ts}^*)^2| \approx 1.7 \cdot 10^{-3}$ and
$(V_{td}^* V_{ts}^{})^2 \approx (9.5 + 9.9\, i) \cdot 10^{-8}$.\footnote{The
CKM input is found from the Wolfenstein parameters $\lambda = 0.22453(44)$,
$A = 0.836(15)$, $\oL{\rho} = 0.122(^{+18}_{-17})$ and $\oL{\eta} =
0.355(^{+12}_{-11})$ \cite{Tanabashi:2018oca}, respectively.}
Then the corresponding inverse, now given in the BSM normalization, is found
to be
\begin{align}
  \label{eq:SM-Pa}
  \big(\WcL[ij]{\text{VLL}}\big)_{\text{SM}}^{-1} =
  \frac{1}{\cN S_0(x_t) \big| \big(\lambda_t^{ij} \big)^2 \big|} &
  \approx \left\{
    \begin{array}{ccc}
      2.0 \cdt{8} \TeV^2 &               & ij = ds \\[0.2cm]
      2.4 \cdt{5} \TeV^2 & \phantom{for} & ij = db \\[0.2cm]
      1.1 \cdt{4} \TeV^2 &               & ij = sb
    \end{array}
  \right. .
\end{align}
It shows that for $\BBbar$ mixing the short-distance SM contribution yields
to a good accuracy the experimental value as can be seen by comparison with
$P_\text{VLL}^{db} = 2.67 \cdt{5} \TeV^2$ and
$P_\text{VLL}^{sb} = 1.15 \cdt{4} \TeV^2$ from \reftab{tab:LEFT-Pa}.
On the other hand for $\KKbar$ mixing the ``experimental''
$P_\text{VLL}^{ds} = 0.10 \cdt{7} \TeV^2$ in \reftab{tab:LEFT-Pa} is not
made up by the short-distance SM contribution of the top-quark
in~\eqref{eq:SM-Pa} alone, which is only about 0.5\% of the measured value
of $\Delta M_K = 2 \re(M_{12}^{ds})$. {Similarly $\eps_K \propto
\im(M_{12}^{ds})$ is not nearly reproduced by the top-quark contribution
alone.} Here there are also charm-top and charm-charm contributions, as
reanalysed recently in the SM at NNLO in QCD \cite{Brod:2019rzc}.

We have completed the calculation of the dynamics below the EW scale by
obtaining the values of the coefficients $P_a^{ij}(\muEW)$ accompanying
the LEFT Wilson coefficients $\WcL[ij]{a}(\muEW)$. In the next
section we will present a detailed SMEFT anatomy of the corresponding
coefficients $P_a^{ij}(\muNP)$ that collect the information
about the SMEFT dynamics up to the new physics scale $\muNP$.

%--------+---------+---------+---------+---------+---------+---------+---------+
%
%
%
%--------+---------+---------+---------+---------+---------+---------+---------+
\section{\boldmath SMEFT Anatomy of $\DF=2$ Processes}
\label{sec:SMEFT}

%
%
%
%--------+---------+---------+---------+---------+---------+---------+---------+
\subsection{Preliminaries}

The main assumption inherent to the SMEFT framework is that NP interactions
have been integrated out at some high scale $\muNP \gg \muEW$ above the
electroweak scale. The field content of the SMEFT Lagrangian
\begin{align}
  \label{eq:SMEFT-Leff}
  \cL_\text{SMEFT} &
  = \cL_{\text{SM}}^{(4)} + \sum_a \Wc{a}(\mu)\, \Op{a}
\end{align}
are the SM fields and the interactions are locally invariant under the
SM gauge group $\text{SU(3)}_c \times \text{SU(2)}_L \times \text{U(1)}_Y$.
Here $\cL_{\text{SM}}^{(4)}$ is the renormalizable part known from the SM,
whereas the $\Op{a}$ are higher-dimensional ($\text{dim} = 5,6$) operators
that parametrize the effects of new physics. We use here  the non-redundant
set of operators as classified in \cite{Grzadkowski:2010es}, also known as
the ``Warsaw'' basis. The dimension-four part $\cL_{\text{SM}}^{(4)}$ contains
all the couplings known from the SM, however, their numerical values can
be significantly altered in the presence of higher-dimensional SMEFT operators,
i.e. nonvanishing Wilson coefficients~$\Wc{a}(\mu)$.

In SMEFT it is convenient to work above $\muEW$ in the unbroken
$\text{SU(2)}_L \times \text{U(1)}_Y$ phase, however electroweak symmetry
breaking (EWSB) is taking place at $\muEW$ and it is more convenient to
transform gauge bosons and fermions from the weak to their mass eigenstates,
see details on SMEFT-specific modifications due to the presence of
higher-dimensional operators in \cite{Alonso:2013hga}.
There is some freedom in the choice of the weak eigenstates of fermions,
which allows to choose the mass term of either the down-type quarks or the
up-type quarks to be flavour\footnote{We use here ``flavour'' as synonymous
to ``generation'' of fermions.} diagonal already at the NP scale $\muNP$,
to which we refer as {\em down-basis} and {\em up-basis}, respectively.
Usually the down- or up-basis are defined at the electroweak scale, but our
choice at the new physics scale seems more appropriate in the context of
matching a UV completion on SMEFT. Note that in SMEFT the mass term consists
of dimension-four Yukawa couplings and a dimension-six contribution, and
hence in SMEFT their sum is chosen diagonal. This fixes then also the
definition of all SMEFT Wilson coefficients as explained in more detail in
\cite{Aebischer:2015fzz}. Throughout we denote Wilson coefficients in the
down-basis by $\Wc{a}$ and those of the up-basis are denoted with a hat as
$\Wcup{a}$.

The RG evolution of the dimension-four couplings in $\cL_{\text{SM}}^{(4)}$
and of the Wilson coefficients $\Wc{a}(\mu)$ from the scale $\muNP$ to $\muEW$
is governed by the ADMs in SMEFT.
Given some initial coefficients $\Wc{a}(\muNP)$, they can be evolved down
to $\muEW$, thereby resumming leading logarithmic (LL) effects to all orders
in the quartic Higgs, gauge and Yukawa couplings into $\Wc{b}(\muEW)$.
The formal solution of the coupled ordinary first order differential
equations is given as
\begin{equation}
  \label{eq:SMEFT-RG-sol}
  \Wc{b}(\muEW)
  = \sum_a U_{ba}(\muEW, \muNP)\, \Wc{a}(\muNP) ,
\end{equation}
which can be obtained in the most general case numerically.
The following comments should be made.
\begin{itemize}
\item
  The matrix $U_{ba}(\muEW, \muNP)$ is the RG evolution matrix. It is presently
  known from the one-loop ADMs of the Warsaw basis \cite{Jenkins:2013zja,
  Jenkins:2013wua, Alonso:2013hga}. At NLO it requires the calculation of
  the two-loop ADMs, of which some results, in particular for QCD, are
  scattered over the literature and recent discussions can be found
  in \cite{Bern:2020ikv, deVries:2019nsu}. At NLO $U_{ba}(\muEW, \muNP)$ depends
  on the renormalization scheme used for the evaluation of two-loop ADMs.
\item In order to cancel the renormalization scheme and scale dependences
  in $\Wc{a}(\muNP)$ around the $\muNP$ scale the matching between a given
  NP model (UV completion) and SMEFT has to include tree- and one-loop
  corrections. The $\Wc{a}(\muNP)$ carry then dependence on the fundamental
  parameters of the NP model.
\item
  The RG evolution of the dimension-four couplings has to be performed
  in the presence of nonvanishing $\Wc{a}(\muNP)$.
\end{itemize}
In particular the RG evolution will reintroduce flavour off-diagonal
entries in the mass terms at $\muEW$, which can be undone in principle
with an additional {\em back-rotation} \cite{Aebischer:2020lsx} to the
down- or up-basis at $\muEW$. We will return soon to this issue.

Subsequently, the SMEFT is matched on the LEFT when decoupling the heavy
$W$ and $Z$ bosons, the Higgs boson and the top-quark. In this matching
the LEFT Wilson coefficients $\WcL{d}(\muEW)$ are determined in terms of
the SMEFT Wilson coefficients $\Wc{b}(\muEW)$ at the electroweak scale.
The $\cL_{\text{SM}}^{(4)}$ part is responsible for the known SM expressions
of the LEFT Wilson coefficients, for which often higher order QCD
and partially also EW radiative corrections are known. The effects of NP
parameterized by the higher-dimensional operators in the matching is known
nowadays at tree-level \cite{Jenkins:2017jig} and at one-loop level
\cite{Dekens:2019ept}.\footnote{Numerous partial results have been known
in the literature before.} As mentioned before, this is done in the broken
phase in terms of mass eigenstates of the Higgs and gauge bosons and also fermions.
In the literature the matching is done in terms of SMEFT Wilson coefficients
in either down- or up-basis at $\muEW$, in particular the one-loop results
\cite{Dekens:2019ept} are given for the up-basis. The transformation of the
SMEFT Wilson coefficients between down- and up-basis is governed entirely
by the quark-mixing matrix (CKM matrix) at $\muEW$ \cite{Aebischer:2015fzz}.
As can be seen, the back-rotation at $\muEW$ to either the down- or the
up-basis is required to make use of these matching results. The phenomenological
impact of the back-rotation on $B$-physics observables, including $\BBbar$
mixing, has been discussed in \cite{Aebischer:2020lsx}. The matching
equation for the BSM contribution to a LEFT Wilson coefficient is\footnote{We
have reserved the index ``$a$'' for SMEFT coefficients at $\muNP$ so that we
use in this section  the index ``$d$'' for LEFT coefficients.}
\begin{equation}
  \label{eq:LEFT-SMEFT}
  \WcL{d}(\muEW)
  = \sum_{b\, \in\, B}  M_{db}^{(0)}(\muEW) \, \Wc{b}(\muEW)
  + \sum_{c\, \in\, C}  M_{dc}^{(1)}(\muEW) \, \Wc{c}(\muEW)
  + \ldots,
\end{equation}
where $M^{(0)}(\muEW)$ and $M^{(1)}(\muEW)$ denote the tree-level and one-loop
threshold corrections between SMEFT and LEFT, see \cite{Jenkins:2017jig} and
\cite{Dekens:2019ept} respectively. The following comments are in order:
\begin{itemize}
\item
  The combination of \eqref{eq:SMEFT-RG-sol} with \eqref{eq:LEFT-SMEFT}
  expresses the LEFT coefficients in terms of the SMEFT coefficients at the
  NP scale $\muNP$. Note that the second term in \eqref{eq:LEFT-SMEFT}
  is a NLO correction, which requires in principle to include in the RG
  solution \eqref{eq:SMEFT-RG-sol} the two-loop ADMs, which are not fully
  available yet.
\item
  The one-loop threshold corrections $M^{(1)}_{dc}(\muEW)$ depend on
  logarithms $\ln(m_i/\muEW)$ with $i = W, Z, t, h$ that cancel $\muEW$
  dependences present in the tree-level term $M_{db}^{(0)}(\muEW)\,
  U_{ba}(\muEW, \muNP)\, \Wc{a}(\muNP)$, and replacing them by large
  logarithms $\ln(m_i/\muNP)$, which are resummed in $U_{ba}(m_i, \muNP)$.
\item
  In particular all scheme dependences related to the top-Yukawa
  coupling cancel up to neglected higher order effects. This is the case
  because the top quark is integrated out at $\muEW$ and therefore the
  cancellation has to happen at this scale.
\item
  In the case of $\DF=2$ Wilson coefficients $\WcL{d}(\muEW)$, its
  dependence on the QCD renormalization scheme and $\muEW$, given by
  $M^{(1)}_{dc}(\muEW)$, cancels the one present in $P_d^{ij}(\muEW)$ of
  the LEFT master formula \eqref{eq:master-M12BSM}.
\item
  It is well understood that the set of operators $C$ contains all
  operators that mix at one-loop into set $B$ to guarantee the
  aforementioned renormalization scheme independence. However,
  set $C$ can contain in principle additional operators that do
  not mix into set~$B$.
\end{itemize}

Based on these preliminaries, we explain in the following the extension of
the LEFT master formula \eqref{eq:master-M12BSM} of $[M_{12}^{ij}
\big]_\text{BSM}$ to the one of SMEFT
\begin{equation}
  \label{eq:master-M12SMEFT}
\boxed{
  2 \big [M_{12}^{ij} \big]_\text{BSM}
  = (\Delta M_{ij})_\text{exp} \sum_{a} \,
    P_a^{ij}(\muNP) \; \wc{a}{ij}(\muNP)
  = (\Delta M_{ij})_\text{exp} \sum_{a} \,
    P_a^{ij}(\muNP) \frac{\bwc{a}{ij}(\muNP)}{\muNP^2}
}
\end{equation}
in terms of all relevant SMEFT Wilson coefficients at the NP scale
$\muNP$ of the operators collected in \reftab{tab:OPERATORS}.
Here the $P_a^{ij}(\muNP)$ generalize the $P_a^{ij}(\muEW)$ of
\eqref{eq:master-M12BSM} by including effects in \eqref{eq:SMEFT-RG-sol}
and \eqref{eq:LEFT-SMEFT}. Now the sum over $a$ has to be taken over
SMEFT Wilson coefficients at the scale $\muNP$. For later convenience
we introduce also dimensionless Wilson coefficients $\bwc{a}{ij}(\muNP)$.
To begin with we will use approximate solutions of the
RG evolution to explain the most important contributions for $\DF=2$ processes
in SMEFT. The actual calculation of the $P_a^{ij}(\muNP)$ is based on all
known results and has to be done numerically, as explained in more detail
in \refsec{sec:SMEFT-master}. As an example we will present the ``SMEFT ATLAS''
of $[M_{12}^{ij} \big]_\text{BSM}$ for the specific scale $\muNP = 5\TeV$
in \refsec{sec:SMEFT-ATLAS}. The case of $\muNP = 100\TeV$ is presented
in \refapp{app:100ATLAS}.

\newcommand{\vp}{\phi}
\newcommand{\tvp}{\widetilde{\phi}}
\newcommand{\vpj}{\vp^\dagger i \overleftrightarrow{\mathcal{D}}_{\!\!\!\mu} \vp}
\newcommand{\vpjt}{\vp^\dagger i \overleftrightarrow{\mathcal{D}}^I_{\!\!\!\mu}\vp}

\begin{table}[t]
\renewcommand{\arraystretch}{1.5}
\resizebox{\columnwidth}{!}{
\begin{tabular}{|c|c||c|c||c|c|}
\hline
\multicolumn{2}{|c||}{$(\oL LL)(\oL LL)$} &
\multicolumn{2}{c||}{$(\oL RR)(\oL RR)$} &
\multicolumn{2}{c|}{$(\oL LL)(\oL RR)$}
\\
\hline
  $\Op[(1)]{qq}$  & $(\bar q_p \gamma_\mu q_r)(\bar q_s \gamma^\mu q_t)$
& $\Op{uu}$       & $(\bar u_p \gamma_\mu u_r)(\bar u_s \gamma^\mu u_t)$
& $\Op{lu}$       & $(\bar \ell_p \gamma_\mu \ell_r)(\bar u_s \gamma^\mu u_t)$
\\
  $\Op[(3)]{qq}$  & $(\bar q_p \gamma_\mu \tau^I q_r)(\bar q_s \gamma^\mu \tau^I q_t)$
& $\Op{dd}$       & $(\bar d_p \gamma_\mu d_r)(\bar d_s \gamma^\mu d_t)$
& $\Op{ld}$       & $(\bar \ell_p \gamma_\mu \ell_r)(\bar d_s \gamma^\mu d_t)$
\\
  $\Op[(1)]{lq}$  & $(\bar \ell_p \gamma_\mu \ell_r)(\bar q_s \gamma^\mu q_t)$
& $\Op{eu}$       & $(\bar e_p \gamma_\mu e_r)(\bar u_s \gamma^\mu u_t)$
& $\Op{qe}$       & $(\bar q_p \gamma_\mu q_r)(\bar e_s \gamma^\mu e_t)$
\\
  $\Op[(3)]{lq}$  & $(\bar \ell_p \gamma_\mu \tau^I \ell_r)(\bar q_s \gamma^\mu \tau^I q_t)$
& $\Op{ed}$       & $(\bar e_p \gamma_\mu e_r)(\bar d_s\gamma^\mu d_t)$
& $\Op[(1)]{qu}$  & $(\bar q_p \gamma_\mu q_r)(\bar u_s \gamma^\mu u_t)$
\\
\cline{1-2}\cline{1-2}
  \multicolumn{2}{|c||}{$\psi^2 X\phi$}
& $\Op[(1)]{ud}$  & $(\bar u_p \gamma_\mu u_r)(\bar d_s \gamma^\mu d_t)$
& $\Op[(8)]{qu}$  & $(\bar q_p \gamma_\mu T^A q_r)(\bar u_s \gamma^\mu T^A u_t)$
\\
\cline{1-2}\cline{1-2}
  $\Op[]{uW}$
& $(\bar q_p \sigma^{\mu\nu} u_r)\tau^I \widetilde{\phi}\, W^I_{\mu\nu}$
& $\Op[(8)]{ud}$  & $(\bar u_p \gamma_\mu T^A u_r)(\bar d_s \gamma^\mu T^A d_t)$
& $\Op[(1)]{qd}$  & $(\bar q_p \gamma_\mu q_r)(\bar d_s \gamma^\mu d_t)$
\\
&&&
& $\Op[(8)]{qd}$  & $(\bar q_p \gamma_\mu T^A q_r)(\bar d_s \gamma^\mu T^A d_t)$
\\
\hline\hline
  \multicolumn{2}{|c||}{$(\oL LR)(\oL RL)$ and $(\oL LR)(\oL LR)$} &
  \multicolumn{4}{c|}{$\psi^2\vp^2 D$}
\\
\hline
  $\Op{ledq}$ & $(\bar \ell_p^j e_r)(\bar d_s q_t^j)$
& $\Op[(1)]{\vp q}$ & \multicolumn{3}{c|}{$(\vpj)(\bar q_p \gamma^\mu q_r)$}
\\
  $\Op[(1)]{quqd}$ & $(\bar q_p^j u_r) \eps_{jk} (\bar q_s^k d_t)$
& $\Op[(3)]{\vp q}$  & \multicolumn{3}{c|}{$(\vpjt)(\bar q_p \tau^I \gamma^\mu q_r)$}
\\
  $\Op[(8)]{quqd}$ & $(\bar q_p^j T^A u_r) \eps_{jk} (\bar q_s^k T^A d_t)$
& $\Op{\vp u}$ & \multicolumn{3}{c|}{$(\vpj)(\bar u_p \gamma^\mu u_r)$}
\\
  $\Op[(1)]{lequ}$ & $(\bar \ell_p^j e_r) \eps_{jk} (\bar q_s^k u_t)$
& $\Op{\vp d}$  & \multicolumn{3}{c|}{$(\vpj)(\bar d_p \gamma^\mu d_r)$}
\\
  $\Op[(3)]{lequ}$
& $(\bar \ell_p^j \sigma_{\mu\nu} e_r) \eps_{jk} (\bar q_s^k \sigma^{\mu\nu} u_t)$
& $\Op{\vp u d}$
& \multicolumn{3}{c|}{$(\tvp^\dag i \mathcal{D}_\mu \vp)(\bar u_p \gamma^\mu d_r)$}
\\
\hline
\end{tabular}
}
\renewcommand{\arraystretch}{1.0}
\caption{\label{tab:OPERATORS}
  The SMEFT operators entering the master formulae.
}
\end{table}

%
%
%
%--------+---------+---------+---------+---------+---------+---------+---------+
\subsection[$\DF = 2$ Processes in SMEFT]
{\boldmath $\DF = 2$ Processes in SMEFT}
\label{sec:RGRunning}

The most important Wilson coefficients of SMEFT operators that enter
\eqref{eq:LEFT-SMEFT} for $\DF=2$ processes are
\begin{equation}
  \label{eq:wc_smeft_tree}
  B = \left\{
  \Wc[(1)]{qq}{} \,,\quad
  \Wc[(3)]{qq}{} \,,\quad
  \Wc[(1)]{qa}{} \,,\quad
  \Wc[(8)]{qa}{} \,,\quad
  \Wc[]{aa}{}  \right\}  ,
\end{equation}
in the down $(a=d)$ and up $(a=u)$ sector, respectively \cite{Aebischer:2015fzz}.

At tree-level for $\BBbar$ and $\KKbar$ mixing one finds the following matching
conditions at~$\muEW$ in the {\bf down-basis}
\hfill
\begin{equation}
  \label{eq:left-smeft-down}
\begin{aligned}
  \wcL[V,LL]{dd}{ijij} &
  = - \wc[(1)]{qq}{ijij} - \wc[(3)]{qq}{ijij} \,, \qquad
&
  \wcL[V1,LR]{dd}{ijij} & = - \wc[(1)]{qd}{ijij} \,,
\\
  \wcL[V,RR]{dd}{ijij}  & = - \wc{dd}{ijij}      \,,
&
  \wcL[V8,RR]{dd}{ijij} & = - \wc[(8)]{qd}{ijij} \,,
\end{aligned}
\end{equation}
and for $\DDbar$ mixing in the {\bf up-basis}
\begin{equation}
  \label{eq:left-smeft-up}
\begin{aligned}
  \wcL[V,LL]{uu}{ijij} &
  = - \wcup[(1)]{qq}{ijij} - \wcup[(3)]{qq}{ijij} \,, \qquad\qquad
&
  \wcL[V1,LR]{uu}{ijij} & = - \wcup[(1)]{qu}{ijij} \,,
\\
  \wcL[V,RR]{uu}{ijij}  & = - \wcup{uu}{ijij}      \,,
&
  \wcL[V8,RR]{uu}{ijij} & = - \wcup[(8)]{qu}{ijij} \,.
\end{aligned}
\end{equation}
Here we have chosen the JMS basis in the LEFT.\footnote{Note that we use the Hamiltonian for LEFT to define Wilson coefficients contrary to \cite{Jenkins:2017jig,
Dekens:2019ept}, who use the Lagrangian, in consequence minus signs are present
in the matching conditions.}

The transformations of the SMEFT Wilson coefficients from the down to
the up basis is governed by elements of the CKM matrix
\cite{Aebischer:2015fzz} as follows
\begin{align}
  \wcup[(1,3)]{qq}{ijij} & = \sum_{prst}
  V_{ip}^{} V^*_{jr} V_{is}^{} V^*_{jt} \, \wc[(1,3)]{qq}{prst} \,, &
\\
  \wcup[(1,8)]{qa}{ijij} & = \sum_{pr}
  V_{ip}^{} V^*_{jr} \, \wc[(1,8)]{qa}{prij} \,, &
  (a & = u, d) .
\end{align}
The SMEFT four-quark operators in \eqref{eq:left-smeft-down}
and \eqref{eq:left-smeft-up} form the operator set~$B$ of the first term in
\eqref{eq:LEFT-SMEFT} at $\muEW$. All these operators undergo
self-mixing,\footnote{A strict use of the term ``self-mixing'' implies that
the flavour structure of the $\DF=2$ Wilson coefficient is conserved to be
$\wc{b}{ijij}$.} under gauge-interactions and also Yukawa interactions.
Consequently,  they remain the coefficients with largest $P_a^{ij}(\muNP)$ also
at $\muNP$. Whether large Wilson coefficients are generated for these operators
is then a matter of the flavour structures in the considered
UV completion and whether tree-level or loop mediation occurs.

The RG evolution \eqref{eq:SMEFT-RG-sol} however will also introduce via
mixing all Wilson coefficients that mix into set $B$, which we call
set $B' \subset C$ in the following. Their numerical impact is loop-suppressed
and depends on the size of the ADMs $\gamma_{ba}$, but a large
logarithm $\ln(\muEW/\muNP)$ appears as in the self-mixing of set $B$.
This can be illustrated with the approximate solution of \eqref{eq:SMEFT-RG-sol}
\begin{align}
  \label{eq:SMEFT-RG-sol-appr}
  \Wc{b}(\muEW) &
  \approx \sum_a \left[\delta_{ba} + \gamma_{ba}\, L \right] \, \Wc{a}(\muNP), &
  L &
  \equiv \frac{1}{(4 \pi)^2} \ln \left (\frac{\muEW}{\muNP} \right) \,,
\end{align}
when retaining only the first leading logarithm $L$.

The most sizeable ADMs from Yukawa-mixing is due to the
top-Yukawa coupling $y_t \approx 1$. The contributions $\gamma_{ba}
\propto y_t$ for down-type mixing in \eqref{eq:SMEFT-RG-sol-appr}
(in {\bf the down-basis}) are
\begin{align}
  \wc[(1)]{qq}{ijij}(\muEW) =
  \wc[(1)]{qq}{ijij} + y_t^2 & \bigg[
    \lambda_t^{ik} \wc[(1)]{qq}{kjij} + \lambda_t^{kj} \wc[(1)]{qq}{ikij}
\notag \\ &
    - \lambda_t^{ij} \big( \wc[(1)]{qu}{ij33}
      + \frac{1}{12} \wc[(8)]{qu}{ij33} - \wc[(1)]{\phi q}{ij} \big)
  \bigg] L \,,
  \label{eq:ll-rge1}
\\
  \wc[(3)]{qq}{ijij}(\muEW) =
  \wc[(3)]{qq}{ijij} + y_t^2 & \bigg[
    \lambda_t^{ik} \wc[(3)]{qq}{kjij} + \lambda_t^{kj} \wc[(3)]{qq}{ikij}
\notag \\ &
    - \lambda_t^{ij} \big(
       \frac{1}{4} \wc[(8)]{qu}{ij33} + \wc[(3)]{\phi q}{ij} \big)
    \bigg] L\,,
  \label{eq:ll-rge2}
\\
  \wc[(1)]{qd}{ijij}(\muEW) =
  \wc[(1)]{qd}{ijij} + y_t^2 & \bigg[
      \frac{\lambda_t^{ik}}{2} \wc[(1)]{qd}{kjij}
    + \frac{\lambda_t^{kj}}{2}\wc[(1)]{qd}{ikij}
\notag \\ &
    - \lambda_t^{ij} \big( \wc[(1)]{ud}{33ij} - \wc[]{\phi d}{ij}  \big)
  \bigg] L \,,
  \label{eq:ll-rge3}
\\
  \wc[(8)]{qd}{ijij}(\muEW) =
  \wc[(8)]{qd}{ijij} + y_t^2 & \bigg[
     \frac{\lambda_t^{ik}}{2} \wc[(8)]{qd}{kjij}
   + \frac{\lambda_t^{kj}}{2} \wc[(8)]{qd}{ikij}
   - \lambda_t^{ij}  \wc[(8)]{ud}{33ij}
  \bigg] L \,,
\label{eq:ll-rge4}
\end{align}
where a summation over $k$ is implied.
We have suppressed the argument of the NP scale $\muNP$ in the Wilson
coefficients on the r.h.s to simplify the notation. Note that the up-type
Yukawa matrix in the down-basis is given by\footnote{For illustration
we neglect here the dimension-six terms to the mass matrix, but take
them into account in our numerics.}
\begin{equation}
  Y^U = \frac{\sqrt 2}{v} V_{\rm CKM}^\dagger m_U^{\rm diag} \,.
\end{equation}
The flavour-mixing of the Yukawa couplings is
seen by the presence of CKM elements $\lambda_t^{ij}$. Although at first
sight numerically suppressed, other than top-Yukawa mixings can be
phenomenologically important, depending on the UV completion and also
on the SM suppression factors for the observable under consideration.

Concerning the gauge sector, indeed the most sizeable ADMs are those
due to the strong coupling $4 \pi \alpha_s = g_s^2 \approx 1.4$ and less
sizeable due to $\text{SU(2)}_L \times \text{U}(1)_Y$, however this
mixing is flavour-diagonal. The most important evolution due to
gauge couplings are
\begin{align}
  \label{eq:ll-gauge-rge1}
  \wc[(1)]{qq}{ijij}(\muEW) & =
  \wc[(1)]{qq}{ijij} + \bigg[\!
    \left(\frac{g^{\prime 2}}{3} + g_s^2 \right) \wc[(1)]{qq}{ijij}
  + 9 \left(g^2 + g_s^2 \right) \wc[(3)]{qq}{ijij} \bigg] L \,,
\\
  \label{eq:ll-gauge-rge1b}
  \wc[(3)]{qq}{ijij}(\muEW) & =
  \wc[(3)]{qq}{ijij} + \bigg[\!
    \left(\frac{g^{\prime 2}}{3} - 6 g^2 -5 g_s^2 \right) \wc[(3)]{qq}{ijij}
  + 3 \left(g^2 + g_s^2 \right) \wc[(1)]{qq}{ijij} \bigg] L \,,
\\
  \label{eq:ll-gauge-rge2}
  \wc[(1)]{qd}{ijij}(\muEW) &
  = \wc[(1)]{qd}{ijij} + \frac{2}{3} \bigg[ g^{\prime 2} \wc[(1)]{qd}{ijij}
   -4 g_s^2 \wc[(8)]{qd}{ijij}
   \bigg] L \,,
\\
  \label{eq:ll-gauge-rge3}
  \wc[(8)]{qd}{ijij}(\muEW) & =
  \wc[(8)]{qd}{ijij} + \bigg[
    \left(\frac{2}{3} g^{\prime 2} -14 g_s^2 \right) \wc[(8)]{qd}{ijij}
   - 12 g_s^2 \wc[(1)]{qd}{ijij} \bigg] L \,,
\\
  \label{eq:ll-gauge-rge4}
  \wc[]{dd}{ijij}(\muEW) & =
  \bigg( 1 + 4 \left[\frac{g^{\prime 2}}{3} + g_s^2 \right] L \bigg)
  \wc[]{dd}{ijij} \,.
\end{align}

In {\bf the up-basis}, where the up-type Yukawa matrix $Y^U = \sqrt 2 \,
m_U^\text{diag}/v$ is diagonal, all flavour-changing mixing terms in
\eqref{eq:ll-rge1}--\eqref{eq:ll-rge4} disappear, whereas mixing due
to gauge couplings shown in \eqref{eq:ll-gauge-rge1}--\eqref{eq:ll-gauge-rge4}
remains unaltered.

\begin{figure}
\centering
  \includegraphics[clip, trim=0.5cm 12cm 0.5cm 12cm, width=0.85\textwidth]
  {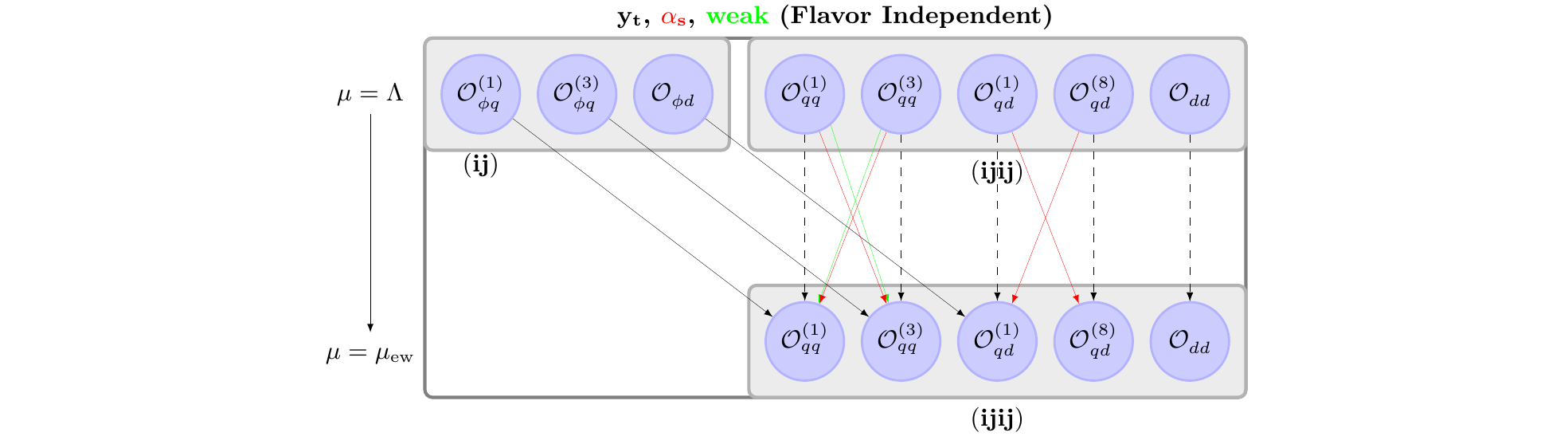}
  \includegraphics[clip, trim=2cm   11cm 0.2cm 11cm, width=0.85\textwidth]
  {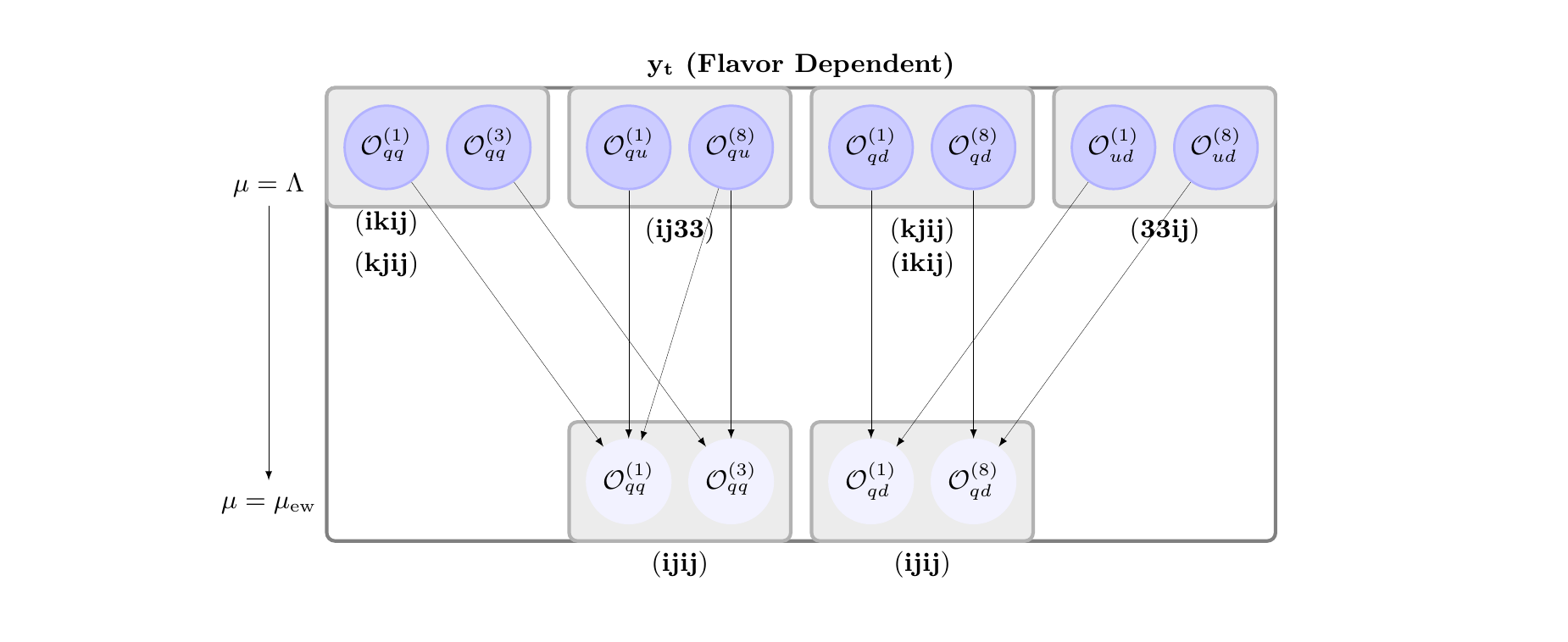}
\caption{\small Mixing of operators relevant for $\DF=2$ observables for
  the $\KKbar$ and $\BBbar$ mixing in the Warsaw down-basis. The red, green
  and black lines indicate the mixing due to strong, electroweak and
  top-Yukawa couplings, respectively. The self-mixing is shown by a dashed black line. Here
  $k=1,2,3$. For the up-basis the mixing due to top-Yukawa,
  i.e. all black lines disappear.}
\label{fig:running:down}
\end{figure}

From the above equations one sees that the operator set $B'$ for down-type
mixing contains at least the following operators
\begin{equation}
  \label{eq:wc_smeft_loop}
  \left\{
  \Wc[(1)]{qu} \,,\quad
  \Wc[(8)]{qu} \,,\quad
  \Wc[(1)]{ud} \,,\quad
  \Wc[(8)]{ud} \,,\quad
  \Wc[(1)]{\phi q} \,,\quad
  \Wc[(3)]{\phi q} \,,\quad
  \Wc{\phi d}
  \right\} .
\end{equation}
They are additional four-quark operators and modified $Z$- and $W$-couplings
of quarks parameterized by the $\psi^2 \phi^2 D$ operators $\Op[(1,3)]{\phi q}$
and $\Op{\phi d}$. The RG running of SMEFT Wilson coefficients from the NP
scale down to the EW scale contributing to $\KKbar$ and $\BBbar$ mixing in
the down-basis is displayed in \reffig{fig:running:down} and for $\DDbar$
mixing in the down-basis in \reffig{fig:running:up}. The following
clarifying comments are in order:

\begin{figure}
\centering
  \includegraphics[clip, trim=0.5cm 12cm 0.5cm 11cm, width=0.85\textwidth]
  {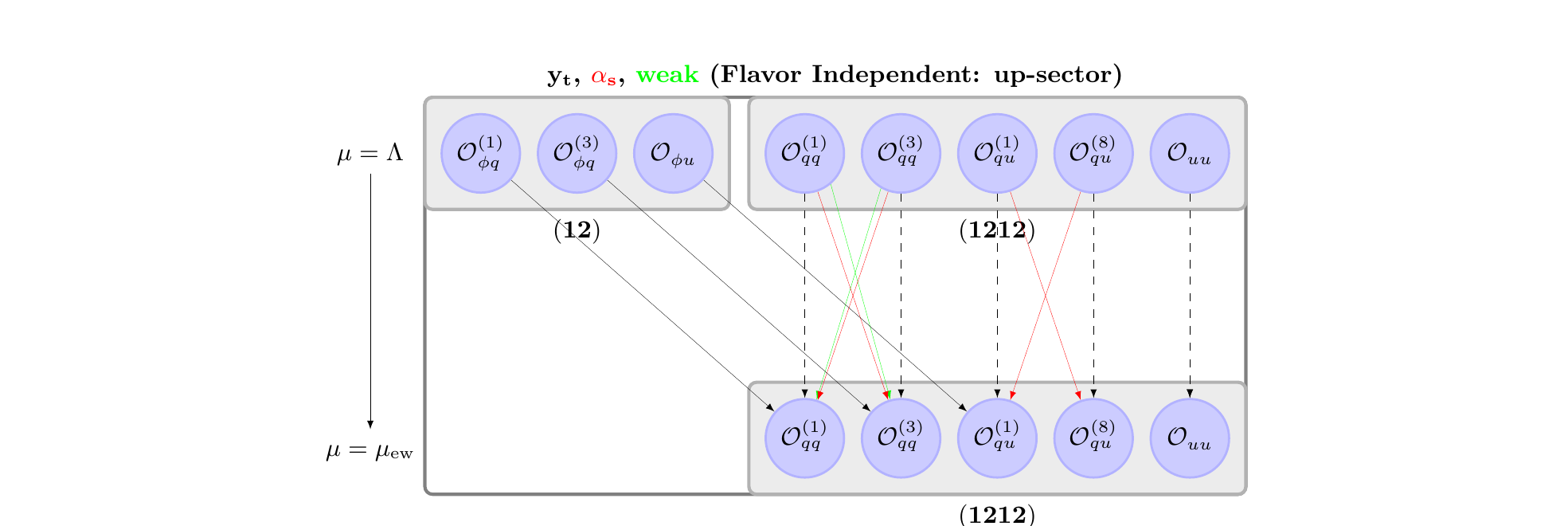}
  \includegraphics[clip, trim=2.0cm 11cm 0.5cm 11cm, width=0.85\textwidth]
  {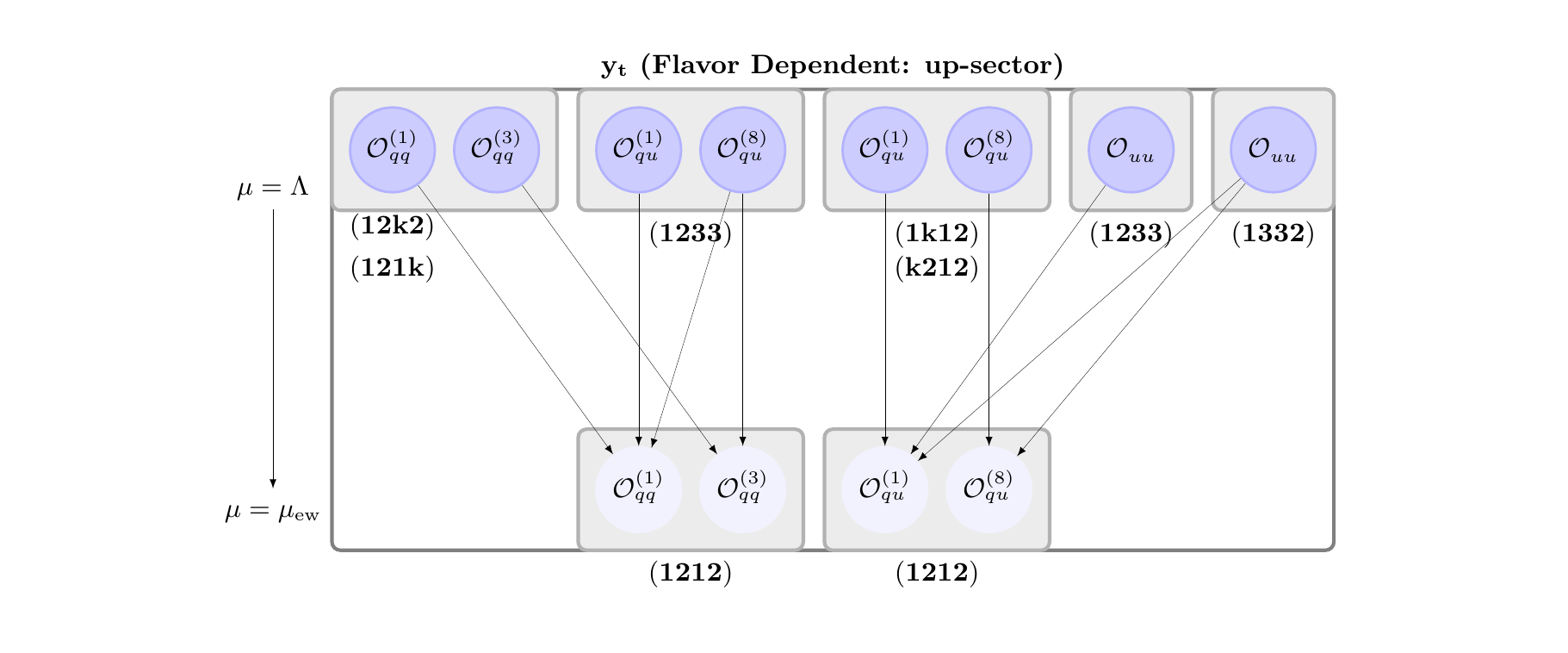}
\caption{\small Mixing of operators relevant for $\DF=2$ observables for
  $\DDbar$-mixing in the Warsaw down-basis. The red, green and black lines
  indicate the mixing due to strong, electroweak and top-Yukawa couplings,
  respectively. Here $k=1,2,3$. The self-mixing is shown by
  a dashed black line. For the up-basis the mixing due to top-Yukawa, i.e.
  all black lines disappear.
  }
\label{fig:running:up}
\end{figure}

\begin{itemize}
\item
  At the scale $\muNP$ operators are listed that contribute to $\DF=2$
  operators at $\muEW$ either directly or through operator mixing in the
  RG evolution. In case the former operators are absent at $\muNP$ but
  are generated at $\muEW$ solely via RG evolution they are placed on a
  lighter background than the original operators.
\item
  The distinction between strong, weak and Yukawa interactions is made with
  the help of colours as described in the figure caption.
\item We split the operator mixing into two parts (a) referred as
  \emph{flavour independent} in which the flavour structure of the Wilson
  coefficients at $\muNP$ remains intact at $\muEW$ and (b) the
  \emph{flavour dependent} part in which the original flavour structure is
  modified due to RG running. Specifically, the operator mixing due to
  gauge couplings is flavour independent whereas the Yukawa couplings
  give rise to both, flavour-dependent and flavour-independent mixing.
  These are shown in the upper and lower panels of \reffig{fig:running:down}
  for $\KKbar$ and $\BBbar$ mixing. For $\DDbar$ the operator mixing is
  shown in \reffig{fig:running:up}, which is flavour independent.
\end{itemize}
In the numerical evaluation we include all Yukawa mixings, which lead
to additional operators in set $B'$ than we listed in
\refeq{eq:wc_smeft_loop}.

So far we have discussed the effects of operator mixing in the RG evolution.
But as already mentioned above, the set $C$ in the second term of
\eqref{eq:LEFT-SMEFT} can contain further operators. Indeed the only
one is the up-type dipole operator $\Op{uW}$, such that
\begin{equation}
  C = B' + \{\Wc{uW}{} \}   \,.
\end{equation}
Its Wilson coefficient at $\muEW$ contributes via $M_{dc}^{(1)}(\muEW)$
in  \eqref{eq:LEFT-SMEFT} and as such it is one-loop suppressed.
Despite being formally a NLO correction, the absence of mixing into
set $B$ guarantees scheme-independence and allows actually to include
this correction without full knowledge of two-loop ADMs.

The complete set of one-loop SMEFT threshold corrections $M_{dc}^{(1)}(\muEW)$
has been calculated in \cite{Dekens:2019ept}, but is provided there only
in electronic form. We have extracted those relevant for $\DF = 2$ processes
and collected them in \refapp{app:one-loop-SMEFT}. With the help of these
results in \eqref{eq:1loop-match1}--\eqref{eq:1loop-match3}, one can verify
the cancellation of the $\muEW$ dependence with the corresponding dependences
in \eqref{eq:ll-rge1}--\eqref{eq:ll-rge3}. This can be seen explicitly by
making the replacement $y_t^2 = {2 \pi x_t\, \alpha}/{s_W^2}$.

%
%
%
%--------+---------+---------+---------+---------+---------+---------+---------+
\subsection{Derivation of the SMEFT Master Formula}
\label{sec:SMEFT-master}

In the previous section we provided an approximate analytic insight into the
one-loop RG evolution relevant for $\DF=2$ processes, with some details on the
cancellation of scale dependences. It also showed that Yukawa couplings are
responsible for a complex flavour-mixing. As also mentioned before, the complete
solution of the RG has to be performed numerically in order to determine the
$P_a^{ij}(\muNP)$ in \eqref{eq:master-M12SMEFT}. Here we provide the details
of their determination before we go to an explicit example with $\muNP = 5\TeV$
in the next section.

We first calculate $\big [M_{12}^{ij} \big]_\text{BSM}$ and determine
then the coefficients $P_a^{ij}(\muNP)$ via \eqref{eq:master-M12SMEFT}
with the following steps:
\begin{itemize}
\item
  From the high scale $\muNP$ down to $\muEW$, the full one-loop RG
  equations \cite{Alonso:2013hga, Jenkins:2013wua, Jenkins:2013zja} are taken
  into account, using the package {\tt wilson} \cite{Aebischer:2018bkb}. As a
  result of operator mixing several secondary WCs are generated at $\muEW$ through
  operator mixing as depicted in the RG charts of \reffig{fig:running:down},
  and \ref{fig:running:up}.
\item
  At $\muEW$ the full one-loop matching onto the LEFT Wilson coefficients
  \refeq{eq:BMU-basis} is considered. Here we have used the results from
  \cite{Dekens:2019ept}, which are implemented in {\tt wilson}. The back-rotation
  \cite{Aebischer:2020lsx} is taken into account automatically at $\muEW$
  when using {\tt wilson}.
\item
  In the next step the complete LO \cite{Aebischer:2017gaw, Jenkins:2017dyc}
  and NLO QCD \cite{Buras:2000if} running of the LEFT Wilson coefficients down
  to lower scales $\muLow$ is taken into account following
  \refsec{sec:LEFT-master}.
\item
  Finally, at the low scale $\muLow$ the LEFT Wilson coefficients are combined
  with the hadronic matrix elements given in \reftab{tab:DF2-me-input} to
  calculate $\big [M_{12}^{ij} \big]_\text{BSM}$.
\end{itemize}
Note that, apart from the set $B'$ defined in \refeq{eq:wc_smeft_loop}, the
following SMEFT Wilson coefficients can contribute to $\DF=2$ observables
at one-loop
\begin{align}
  \text{four-quark} & : \qquad
  \wc[(1)]{quqd}{} ,\; \wc[(8)]{quqd}{} ;
\\
  \text{semileptonic} & : \qquad
  \wc[(1)]{lq} ,\; \wc[(3)]{lq}{} ,\; \wc[]{ld}{} ,\; \wc[]{qe}{} ,\;
  \wc[]{ledq}{} ,\; \wc[(1)]{lequ}{} ,\; \wc[(3)]{lequ}{} .
\end{align}
In the process of obtaining the $P_a^{ij}$ factors at the electroweak scale,
the dimension-four parameters are needed at the NP scale $\muNP$ for the RG
evolution of the Wilson coefficients. They are obtained through an iterative
process explained in \cite{Aebischer:2020mkv}, where we assume no NP
contributions to the CKM parameters stemming from four-quark operators.
In particular as input serve tree-level determinations of $|V_{us}|$, $|V_{ub}|$
and $|V_{cb}|$ from semi-leptonic processes and the determination of the
CKM angle $\gamma$ from hadronic tree-level decays \cite{Aebischer:2018bkb}.

%
%
%
%--------+---------+---------+---------+---------+---------+---------+---------+
\subsection[SMEFT ATLAS at $\muNP = 5\TeV$]
{\boldmath SMEFT ATLAS at $\muNP = 5\TeV$}
\label{sec:SMEFT-ATLAS}

As a numerical example we present here the SMEFT ATLAS at the scale $\muNP = 5\TeV$.
It consists of the master formulae for the contributions of individual operators
to the sum in \eqref{eq:master-M12SMEFT} given in terms of dimensionless Wilson
coefficients $\bWc{a}$ introduced in \refeq{eq:master-M12SMEFT}. The numerical
 coefficients in these formulae are just the central values of the coefficients
$P_a^{ij}(\muNP)$ divided by $\muNP^2$ with $\muNP = 5\TeV$.

The numerical values in the master formulae of this section have been
evaluated at $\muNP = 5\TeV$. Corresponding results for $\muNP = 100\TeV$
are given in \refapp{app:100ATLAS}. While some visible changes are present
in the left-right operators they are fully subdominant relative to
the change of $\muNP$, which amounts to the suppression of BSM contributions
by a factor of 400 relative to the $\muNP = 5\TeV$ case considered here.

Before describing this $\muNP = 5\TeV$ SMEFT ATLAS in more detail
let us make the following observations on the general pattern of these
coefficients
\begin{itemize}
\item
  In contrast to the analogous coefficients entering \eqref{eq:master-M12BSM}
  that were real valued, the ones in \eqref{eq:master-M12SMEFT} are complex
  quantities. The phases in these formulae originate from the complex CKM
  factors $\lambda_t^{ij}$ in the RG equations \eqref{eq:ll-rge1}--\eqref{eq:ll-rge4}
  and from one-loop matching as seen in the formulae in \refapp{app:one-loop-SMEFT}.
  They are often represented by the phase $(-\beta)$ of the CKM element $V_{td}$.
  Its presence is signaled by the values of the phases in the ballpark of
  $\pm 22^\circ$ or $\pm 44^\circ$ often shifted by the small phase of $V_{ts}$.
  But in non-leading contributions also other phases are present. They result
  from the interplay of the complex values of CKM elements.
  We do not include in the numerical coefficients phases smaller than
  three degrees to simplify the formulae.
\item
  The by far largest coefficients are the ones with indices
  (1212), (1313) and (2323) for $K^0$, $B_d$ and $B_s$ systems, respectively,
  in particular for the coefficients $\wc[(1,8)]{qd}{1212}$,
  but also $\wc[(1,8)]{qd}{1313}$ and $\wc[(1,8)]{qd}{2323}$.
  Also for charm (1212) dominates. Yet, in particular in the up-basis
  among the contributions with smaller numerical coefficients multiplying
  $\bWc{a}$ there are several with repeated indices signalizing flavour
  conserving contributions. These are often weaker constrained by $\DF=1$
  observables than the flavour violating ones so that larger size of the
  corresponding $\bWc{a}$ relative to the ones with larger $P_a^{ij}(\muNP)$
  factors could enhance their importance. We will return to this phenomenon
  in \refsec{sec:5} in the context of simplified models.
\item
  Yet, when comparing different meson systems there is a large hierarchy in
  the values of the largest coefficients, which results not only from chiral
  enhancement of hadronic matrix elements of left-right operators in the
  $K^0$ meson system, but also from
  a large hierarchy in the normalization factors $(\Delta M_{ij})_\text{exp}$.
  Consequently, while for $\KKbar$ mixing the largest $P_a^{12}$
  are $\ord{10^8}$, for $\BBbarD$ mixing the largest $P_a^{13}$ are
  $\ord{10^6}$ and for $\BBbarS$ mixing the largest $P_a^{23}$ is in the
  ballpark of $\ord{10^5}$. This clearly confirms the known fact that $\KKbar$
  mixing can probe much larger energy scales than $\BBbarD$
  or $B_s-\bar B_s$ mixing. The largest $P_a^{12}$
  for $\DDbar$ mixing is $\ord{10^8}$ and can probe very short
  distance scales. We will quantify this in \refsec{sec:Lambda-bounds}.
\item
  While at first sight the $\wc[]{\phi X}{ij}$ coefficients with $X=q,d,u$
  would appear irrelevant when compared with $\Wc[(1,8)]{qd}$, in models
  in which the latter are not generated at the high scale, they can play
  a dominant role. This is in particular the case of vector-like quark
  models \cite{Bobeth:2016llm}.
\item
  The $P_a^{ij}(\muNP)$ for semileptonic operators are much smaller than
  those of nonleptonic ones, implying that the corresponding WCs
  are much weaker constrained, if at all, by $\DF=2$ transitions.
\end{itemize}

After these general statements on the SMEFT ATLAS let us look at it a bit closer.
Below we retain only those contributions to the sum in \refeq{eq:master-M12SMEFT},
which amount to at least $5\%$ of $(\Delta M_{ij})_\text{exp}$ when setting
$\bWc{a} = 10$. The $5\%$ cut is in the ballpark of the uncertainties of the
hadronic matrix elements in \reftab{tab:DF2-me-input} that enter the prediction
of $\big [M_{12}^{ij} \big]_\text{BSM}$ linearly. The choice of a maximal
value $\bWc{a} = 10$ is close to the generic value of $4\pi$, which one
would still consider a magnitude that allows perturbative expansions in
couplings. We show only results for the three down-type meson
systems $K^0$, $B_d$ and $B_s$, whereas the results for $D^0$ can be found
in the ancillary files to the arXiv submission of this article, together with
non-vanishing $P_a^{ij}(\muNP)$ that yield numerically subleading contributions
below 5\% to the down-type meson systems.

%
%--------+---------+---------+---------+---------+---------+---------+---------+
\subsubsection{\boldmath $\Wc[(1)]{qq}$ and $\Wc[(3)]{qq}$}

Before presenting the master formulae for the contributions of $\wc[(1,3)]{qq}{ijkl}$
to various meson systems in the down and up bases let us already make general
statements on them on the basis of the size of numerical coefficients multiplying
WCs.
\begin{itemize}
\item
  In the down basis constraints from $\DDbar$ mixing dominate for many
  entries, but $\KKbar$ mixing provides a very important constraint for
  $\wc[(1,3)]{qq}{1212}$. In view of potential poorly known long distance
  effects in the former it appears that presently it is plausible to put
  the constraint on $\wc[(1,3)]{qq}{1212}$ mainly from $\KKbar$ mixing.
\item
  In the up basis constraints from $\KKbar$ mixing dominate for many entries
  but $\DDbar$ mixing could still play the role for $\wcup[(1,3)]{qq}{1212}$
  if long-distance contributions were under better control.
\item
  The entries 1313 and 2323 are at first sight in both down and up
  bases dominantly constrained by $\BBbarD$ and $\BBbarS$ mixing, respectively.
\item
  However, as already mentioned above, it should also be noticed that beyond
  the leading $\DF=2$ entries 1212, 1313, 2323, the $\DF=1$
  entries in particular in the up-basis could play a role dependently on the
  size of WCs in a given model. Indeed, in the up-basis strong
  correlations between $\DF=2$ and $\DF=1$ should be expected.
  This will be the case if in a given NP scenario also
  flavour-conserving couplings can contribute through RG effects to
  $\DF=2$ transitions.
\item
  There is a tendency of correlated constraints from $\KKbar$ and $\DDbar$
  mixing on one hand and $\BBbarD$ and $\BBbarS$ mixings on the other hand.
  This is for instance the case of 1212 for $\KKbar$ and $\DDbar$ mixings
  in both bases and of 1323 for $\BBbarD$ and $\BBbarS$ mixings in the up basis.
  Even if the coefficient for 1323 in $\BBbarD$ mixing is much larger than in
  $\BBbarS$ mixing, their relative importance to the leading terms in these
  mixings is similar.
\item
  The results for $\wc[(3)]{qq}{ijkl}$ are practically equal to the ones for
  $\wc[(1)]{qq}{ijkl}$ for the dominant terms and the differences are only in
  subleading terms. Therefore we present only the terms which have
  different numerical coefficients, although it should be kept in mind that
  the WCs entering here are $\wc[(3)]{qq}{ijkl}$ and not $\wc[(1)]{qq}{ijkl}$.
  They could differ in specific NP scenarios.
\end{itemize}

This pattern is summarized by the following rather accurate formulae for
the sum in~\refeq{eq:master-M12SMEFT} in {\bf the down-basis}
\begin{align*}
  \Sigma^{B_s}_{qq1} & =
  -3.9\cdt{2} \bwc[(1)]{qq}{2323} -5.4\cdt{-1} \bwc[(1)]{qq}{2333}
  \nline
  +3.1\cdt{-1} \bwc[(1)]{qq}{2223}
  -6.8\cdt{-2} e^{i22^\circ}\bwc[(1)]{qq}{1232} \,,
  \numberthis
\\
  \Sigma^{B_d}_{qq1} & =
  -9.1\cdt{3}\bwc[(1)]{qq}{1313} +7.2\bwc[(1)]{qq}{1213}
  +2.7e^{i22^\circ}\bwc[(1)]{qq}{1333} -1.6e^{i22^\circ}\bwc[(1)]{qq}{1113}
  \nline
  -1.2\cdt{-1} e^{i23^\circ}\bwc[(1)]{qq}{1323}
  +2.3\cdt{-2} e^{i21^\circ}\bwc[(1)]{qq}{1332}
  +5.8\cdt{-3} e^{i22^\circ}\bwc[(1)]{qq}{1223}
  \nline
  -5.7\cdt{-3} \bwc[(1)]{qq}{1212}
  -5.1\cdt{-3}  e^{i44^\circ}\bwc[(1)]{qq}{1331} \,,
 \numberthis
\\
  \Sigma^{K}_{qq1} & =
  -3.6\cdt{4} \bwc[(1)]{qq}{1212} -6.0\cdt{1} \bwc[(1)]{qq}{1213}
  +1.3\cdt{1} e^{i22^\circ}\bwc[(1)]{qq}{1232}
  \nline
  -5.7\cdt{-1} e^{i23^\circ}\bwc[(1)]{qq}{1222}
  +2.5\cdt{-1} e^{i23^\circ}\bwc[(1)]{qq}{1112}
  +1.2\cdt{-1} e^{i23^\circ}\bwc[(1)]{qq}{1233}
  \nline
  -1.0\cdt{-1} \bwc[(1)]{qq}{1313}
  +2.6\cdt{-2} e^{i23^\circ}\bwc[(1)]{qq}{1332}
  -5.2\cdt{-3} e^{i24^\circ}\bwc[(1)]{qq}{1223}\,.
  \numberthis
\end{align*}
Before continuing let us explain shortly on the above example the meaning
of these results. Taking the case of $\BBbarS$ mixing, the numerical value
of $\Sigma^{B_s}_{qq1}$ corresponds to its contribution to the ratio
$2 \big [M_{12}^{ij} \big]_\text{BSM} / (\Delta M_{B_s})_\text{exp}$ in
\refeq{eq:master-M12SMEFT}. This choice of normalization allows to read off
the fraction of the NP contribution to the prediction of $\Delta M_{B_s}$
compared to its central experimental value. For illustration, the value
$\Sigma^{B_s}_{qq1} = 0.1$ is a deviation from the SM prediction of the
size of 10\% of the measured experimental value. Since the SM prediction
of $\BBbarS$ mixing and the experimental measurement are in close agreement,
this corresponds also to a 10\% deviation due to NP from the SM prediction
of $\Delta M_{B_s}$. The numerically leading $\DF=2$ contribution
$\Sigma^{B_s}_{qq1} \approx -3.9\cdot 10^{2} \bwc[(1)]{qq}{2323}$ yields
then a constraint on $|\bwc[(1)]{qq}{2323}| \lesssim 2.6\, \cdot 10^{-4}$
for the NP contribution to $\Delta M_{B_s}$ not to exceed $10\%$ deviation
from the SM prediction. The same arguments apply also to $\BBbarD$ mixing,
because the SM prediction is in close agreement with the experimental
measurement. The SM predictions of $\KKbar$ and $\DDbar$ mixing receive
in part unknown nonperturbative contributions and here the fraction refers
to the experimentally measured $(\Delta M_{K,D})_\text{exp}$.

If the numerically leading terms are absent in a UV completion, then the
subleading $\DF=1$ terms like $\Sigma^{B_s}_{qq1} \approx -5.4\cdot 10^{-1}
\bwc[(1)]{qq}{2333}$ would be subject to a much weaker constraint
$|\bwc[(1)]{qq}{2333}| \lesssim 1.9\cdot 10^{-1}$. Complicated
interference between the numerically leading $\DF=2$ term and the subleading
$\DF=1$ terms will arise in UV completions that admit all contributions.
Thus there can be strong correlations to $\DF=1$ processes, depending
on the suppression factors in the SM predictions of the $\DF=1$ observables.
Then it remains to be seen whether $\DF=1$ observables will impose
stronger constraints.

The result for {\bf the up-basis} reads
\begin{align*}
  \hat\Sigma^{B_s}_{qq1} & =
  -3.7\cdt{2}\bwcup[(1)]{qq}{2323}
  -8.5\cdt{1}\bwcup[(1)]{qq}{1323}
  -2.0\cdt{1}\bwcup[(1)]{qq}{1313}
  -1.5\cdt{1}\bwcup[(1)]{qq}{2223}
  \nline
  +1.5\cdt{1}\bwcup[(1)]{qq}{2333}
  -3.5\bwcup[(1)]{qq}{1223}-3.5\bwcup[(1)]{qq}{1322}
  +3.5\bwcup[(1)]{qq}{1333}
  -1.3 e^{-i73^\circ}\bwcup[(1)]{qq}{1232}
  \nline
  -8.2\cdt{-1}\bwcup[(1)]{qq}{1213}
  -6.4\cdt{-1}\bwcup[(1)]{qq}{2222}
  +6.3\cdt{-1}\bwcup[(1)]{qq}{2233}
  +6.3\cdt{-1}\bwcup[(1)]{qq}{2332}
  \nline
  -6.3\cdt{-1}\bwcup[(1)]{qq}{3333}
  -3.0\cdt{-1} e^{-i73^\circ}\bwcup[(1)]{qq}{1231}
  -3.0\cdt{-1} e^{-i73^\circ}\bwcup[(1)]{qq}{1123}
  \nline
  -1.7\cdt{-1}e^{-i18^\circ}\bwcup[(1)]{qq}{1222}
  +1.7\cdt{-1}e^{-i19^\circ}\bwcup[(1)]{qq}{1233}
  +1.7\cdt{-1}e^{-i19^\circ}\bwcup[(1)]{qq}{1332}
  \nline
  -7.0\cdt{-2} e^{-i73^\circ}\bwcup[(1)]{qq}{1113}
  -3.0\cdt{-2} e^{-i5^\circ}\bwcup[(1)]{qq}{1212}
  -1.3\cdt{-2} e^{-i73^\circ}\bwcup[(1)]{qq}{1221}
  \nline
  -1.3\cdt{-2}e^{-i73^\circ}\bwcup[(1)]{qq}{1122}
  +1.2\cdt{-2}e^{-i74^\circ}\bwcup[(1)]{qq}{1133}
  +1.2\cdt{-2}e^{-i74^\circ}\bwcup[(1)]{qq}{1331} \,,
  \numberthis
\\
  \hat\Sigma^{B_d}_{qq1} & =
  -8.6\cdt{3}\bwcup[(1)]{qq}{1313}
  +2.0\cdt{3}\bwcup[(1)]{qq}{1323}
  -4.6\cdt{2}\bwcup[(1)]{qq}{2323}
  -3.6\cdt{2}\bwcup[(1)]{qq}{1213}
  \nline
  +8.2\cdt{1}\bwcup[(1)]{qq}{1223}
  +8.2\cdt{1}\bwcup[(1)]{qq}{1322}
  -7.8\cdt{1}e^{i22^\circ}\bwcup[(1)]{qq}{1333}
  \nline
  -3.1\cdt{1} e^{-i73^\circ}\bwcup[(1)]{qq}{1113}
  -1.9\cdt{1}\bwcup[(1)]{qq}{2223}
  +1.8\cdt{1}e^{i22^\circ}\bwcup[(1)]{qq}{2333}
  -1.5\cdt{1}\bwcup[(1)]{qq}{1212}
  \nline
  +7.0e^{-i73^\circ}\bwcup[(1)]{qq}{1231}
  +7.0e^{-i73^\circ}\bwcup[(1)]{qq}{1123}
  +3.4\bwcup[(1)]{qq}{1222}
  -3.2 e^{i23^\circ}\bwcup[(1)]{qq}{1233}
  \nline
  -3.2 e^{i23^\circ}\bwcup[(1)]{qq}{1332}
  -1.6 e^{-i68^\circ}\bwcup[(1)]{qq}{1232}
  -1.3 e^{-i72^\circ}\bwcup[(1)]{qq}{1112}
  -7.8\cdt{-1}\bwcup[(1)]{qq}{2222}
  \nline
  +7.4\cdt{-1}e^{i22^\circ}\bwcup[(1)]{qq}{2233}
  +7.3\cdt{-1}e^{i22^\circ}\bwcup[(1)]{qq}{2332}
  -7.0\cdt{-1} e^{i44^\circ}\bwcup[(1)]{qq}{3333}
  \nline
  +2.9\cdt{-1}e^{-i73^\circ}\bwcup[(1)]{qq}{1221}
  +2.9\cdt{-1}e^{-i73^\circ}\bwcup[(1)]{qq}{1122}
  -2.8\cdt{-1} e^{-i51^\circ}\bwcup[(1)]{qq}{1133}
  \nline
  -2.7\cdt{-1} e^{-i51^\circ}\bwcup[(1)]{qq}{1331}
  +1.1\cdt{-1}e^{i34^\circ}\bwcup[(1)]{qq}{1111} \,,
  \numberthis
\\
  \hat\Sigma^K_{qq1} & =
  -3.2\cdt{4}\bwcup[(1)]{qq}{1212}
  -7.0\cdt{3}\bwcup[(1)]{qq}{1112}
  +7.0\cdt{3}\bwcup[(1)]{qq}{1222}
  -1.7\cdt{3}\bwcup[(1)]{qq}{1111}
  \nline
  +1.7\cdt{3}\bwcup[(1)]{qq}{1221}
  +1.7\cdt{3}\bwcup[(1)]{qq}{1122}
  -1.7\cdt{3}\bwcup[(1)]{qq}{2222}
  +1.3\cdt{3}\bwcup[(1)]{qq}{1213}
  \nline
  +2.9\cdt{2}\bwcup[(1)]{qq}{1113}
  -2.9\cdt{2}\bwcup[(1)]{qq}{1322}
  -2.9\cdt{2}\bwcup[(1)]{qq}{1223}
  -2.7\cdt{2} e^{i23^\circ}\bwcup[(1)]{qq}{1232}
  \nline
  -1.3\cdt{2} e^{i11^\circ}\bwcup[(1)]{qq}{1123}
  -1.3\cdt{2} e^{i11^\circ}\bwcup[(1)]{qq}{1231}
  +1.3\cdt{2}e^{i11^\circ}\bwcup[(1)]{qq}{2223}
  \nline
  -5.3\cdt{1}\bwcup[(1)]{qq}{1313}
  +1.2\cdt{1}\bwcup[(1)]{qq}{1323}
  +1.1\cdt{1}e^{i23^\circ}\bwcup[(1)]{qq}{1233}
  +1.1\cdt{1}e^{i23^\circ}\bwcup[(1)]{qq}{1332}
  \nline
  -5.0e^{i22^\circ}\bwcup[(1)]{qq}{2323}
  +2.7e^{i23^\circ}\bwcup[(1)]{qq}{1133}
  -2.7e^{i23^\circ}\bwcup[(1)]{qq}{2233}
  +2.7e^{i23^\circ}\bwcup[(1)]{qq}{1331}
  \nline
  -2.7 e^{i23^\circ}\bwcup[(1)]{qq}{2332}
  -4.6\cdt{-1}e^{i23^\circ}\bwcup[(1)]{qq}{1333}
  +2.1\cdt{-1}e^{i34^\circ}\bwcup[(1)]{qq}{2333} \,.
  \numberthis
\end{align*}

For $\Wc[(3)]{qq}$ one finds:
\begin{align*}
  \Sigma^{B_s}_{qq3} & =
  \Sigma^{B_s}_{qq1}|^{1\to 3}
  -8.3\cdt{-3} \bwc[(3)]{qq}{2233} -6.4\cdt{-3} \bwc[(3)]{qq}{1123} \,,
  \numberthis
\\
  \Sigma^{B_d}_{qq3} & =
  \Sigma^{B_d}_{qq1}|^{1\to 3}
  +4.3\cdt{-2} e^{i21^\circ}\bwc[(3)]{qq}{1233}
  -4.1\cdt{-2} e^{i21^\circ}\bwc[(3)]{qq}{1332}
  +3.3\cdt{-2} e^{i22^\circ}\bwc[(3)]{qq}{1322}
  \nline
  -9.2\cdt{-3} e^{i44^\circ}\bwc[(3)]{qq}{1133}
  -5.8\cdt{-3} e^{i22^\circ}\bwc[(3)]{qq}{1223}
  +5.1\cdt{-3} e^{i44^\circ}\bwc[(3)]{qq}{1331} \,,
  \numberthis
\\
  \Sigma^{K}_{qq3} & =
\Sigma^{K}_{qq1}|^{1\to 3}
  -0.5\cdt{-1} e^{i23^\circ}\bwc[(3)]{qq}{1332}
  +8.7\cdt{-2} e^{i23^\circ}\bwc[(3)]{qq}{1233}
  \nline
  -8.7\cdt{-3} e^{i24^\circ}\bwc[(3)]{qq}{1322}
  +5.2\cdt{-3} e^{i24^\circ}\bwc[(3)]{qq}{1223} \,,
  \numberthis
\end{align*}
and in the up-basis
\begin{align}
  \hat\Sigma^{B_s}_{qq3} & =
  \hat\Sigma^{B_s}_{qq1}|^{1\to 3} \,,
&
  \hat\Sigma^{B_d}_{qq3} & =
  \hat\Sigma^{B_d}_{qq1}|^{1\to 3} \,,
&
 \hat\Sigma^{K}_{qq3} & =
 \hat\Sigma^{K}_{qq1}|^{1\to 3} \,.
\end{align}

%
%--------+---------+---------+---------+---------+---------+---------+---------+
\subsubsection{\boldmath $\Wc[(1)]{qd}$ and $\Wc[(8)]{qd}$}

We next present the results for $\wc[(1)]{qd}{ijkl}$ and $\wc[(8)]{qd}{ijkl}$
in the down and up bases. Evidently these coefficients provide very strong
constraints for $\DF=2$ processes. We observe
\begin{itemize}
\item
  The $\KKbar$ mixing is most constraining except for 1313 and 2323 elements
  for which constraints from $\BBbarD$ and $\BBbarS$ mixing dominate. Yet,
  inspection of the formulae below shows that contributions from other WCs
  with different indices could also play a role in some UV completions.
\item
  In certain cases there is a large difference between down and up bases.
\item
  The results for $\wc[(1)]{qd}{ijkl}$ and $\wc[(8)]{qd}{ijkl}$ are similar.
\item
  Constraints from $\DDbar$ mixing are irrelevant for both bases.
\end{itemize}

This pattern can be summarized by the following rather accurate formulae for
the sum in \refeq{eq:master-M12SMEFT} in {\bf the down-basis}
\begin{align*}
  \Sigma^{B_s}_{qd1} & =
   2.3\cdt{3} \bwc[(1)]{qd}{2323} +3.3\bwc[(1)]{qd}{3323}
  -1.8\bwc[(1)]{qd}{2223}  +4.0\cdt{-1} e^{i22^\circ}\bwc[(1)]{qd}{1232}
  \nline
  -1.1\cdt{-1} \bwc[(1)]{qd}{2322} +1.1\cdt{-1} \bwc[(1)]{qd}{2333}
  +2.5\cdt{-2} \bwc[(1)]{qd}{2332}
  \nline
  -2.3\cdt{-2} e^{-i23^\circ}\bwc[(1)]{qd}{1323}
  -5.5\cdt{-3} e^{i21^\circ}\bwc[(1)]{qd}{1332}\,,
  \numberthis
\\
  \Sigma^{B_d}_{qd1} & =
  5.7\cdt{4} \bwc[(1)]{qd}{1313} -4.5\cdt{1} \bwc[(1)]{qd}{1213}
  -1.8\cdt{1} e^{i22^\circ}\bwc[(1)]{qd}{3313}
  \nline
  +9.8e^{i22^\circ}\bwc[(1)]{qd}{1113} -2.8\bwc[(1)]{qd}{1312}
  +7.6\cdt{-1} e^{i23^\circ}\bwc[(1)]{qd}{2313}
  \nline
  -1.4\cdt{-1} e^{i21^\circ}\bwc[(1)]{qd}{2331}
  +6.5\cdt{-2} e^{i23^\circ}\bwc[(1)]{qd}{1323}
  -3.1\cdt{-2} e^{i21^\circ}\bwc[(1)]{qd}{1333}
  \nline
  +3.1\cdt{-2} e^{i21^\circ}\bwc[(1)]{qd}{1311}
  +3.0\cdt{-2} e^{i44^\circ}\bwc[(1)]{qd}{1331}
  +7.2\cdt{-3} e^{i22^\circ}\bwc[(1)]{qd}{2213}\,,
  \numberthis
\\
  \Sigma^K_{qd1} & =
   5.3\cdt{6}\bwc[(1)]{qd}{1212} +9.0\cdt{3}\bwc[(1)]{qd}{1312}
  -2.0\cdt{3} e^{i22^\circ}\bwc[(1)]{qd}{2321}
  +2.6\cdt{2} \bwc[(1)]{qd}{1213}
  \nline
  +8.1\cdt{1} e^{i24^\circ}\bwc[(1)]{qd}{2212}
  -3.4\cdt{1} e^{i26^\circ}\bwc[(1)]{qd}{1112}
  -1.7\cdt{1} e^{i23^\circ}\bwc[(1)]{qd}{3312}
  \nline
  -2.8 e^{i22^\circ}\bwc[(1)]{qd}{1232}
  -2.5 e^{i66^\circ}\bwc[(1)]{qd}{1211}
  +2.5e^{i66^\circ}\bwc[(1)]{qd}{1222}
  +7.2\cdt{-1} e^{i24^\circ}\bwc[(1)]{qd}{2312}
  \nline
   +4.4\cdt{-1} \bwc[(1)]{qd}{1313}
  -1.1\cdt{-1} e^{i45^\circ}\bwc[(1)]{qd}{1321}
  -9.6\cdt{-2} e^{i23^\circ}\bwc[(1)]{qd}{2331}
  \nline
  +1.0\cdt{-2} e^{i24^\circ}\bwc[(1)]{qd}{1322}
  -1.0\cdt{-2} e^{i24^\circ}\bwc[(1)]{qd}{1311}\,,
  \numberthis
\end{align*}
and
\begin{align*}
  \Sigma^{B_s}_{qd8} & =
   2.7\cdt{3}\bwc[(8)]{qd}{2323} +3.8\bwc[(8)]{qd}{3323}
  -2.1\bwc[(8)]{qd}{2223} +4.7\cdt{-1}e^{i22^\circ}\bwc[(8)]{qd}{1232}
  \nline
  -1.3\cdt{-1} \bwc[(8)]{qd}{2322} +1.3\cdt{-1}\bwc[(8)]{qd}{2333}
  +2.7\cdt{-2} \bwc[(8)]{qd}{2332}
  \nline
  -2.6\cdt{-2} e^{-i23^\circ}\bwc[(8)]{qd}{1323}
  -5.8\cdt{-3} e^{i21^\circ}\bwc[(8)]{qd}{1332}\,,
  \numberthis
\\
  \Sigma^{B_d}_{qd8} & =
   6.6\cdt{4} \bwc[(8)]{qd}{1313} -5.2\cdt{1}\bwc[(8)]{qd}{1213}
  -2.1\cdt{1} e^{i22^\circ}\bwc[(8)]{qd}{3313}
  \nline
  +1.1\cdt{1} e^{i22^\circ}\bwc[(8)]{qd}{1113} -3.2\bwc[(8)]{qd}{1312}
  +8.9\cdt{-1} e^{i23^\circ}\bwc[(8)]{qd}{2313}
  \nline
  -1.4\cdt{-1} e^{i21^\circ}\bwc[(8)]{qd}{2331}
  +7.5\cdt{-2} e^{i23^\circ}\bwc[(8)]{qd}{1323}
  +3.6\cdt{-2} e^{i21^\circ}\bwc[(8)]{qd}{1311}
  \nline
  -3.5\cdt{-2} e^{i21^\circ}\bwc[(8)]{qd}{1333}
  +3.1\cdt{-2} e^{i44^\circ}\bwc[(8)]{qd}{1331}
  +1.8\cdt{-2} e^{i22^\circ}\bwc[(8)]{qd}{2213}\,,
  \numberthis
\\
  \Sigma^K_{qd8} & =
   7.5\cdt{6} \bwc[(8)]{qd}{1212} +1.3\cdt{4}\bwc[(8)]{qd}{1312}
  -2.8\cdt{3} e^{i22^\circ}\bwc[(8)]{qd}{2321}
  +3.7\cdt{2} \bwc[(8)]{qd}{1213}
  \nline
  +1.1\cdt{2} e^{i24^\circ}\bwc[(8)]{qd}{2212}
  -4.7\cdt{1} e^{i27^\circ}\bwc[(8)]{qd}{1112}
  -2.3\cdt{1}  e^{i23^\circ}\bwc[(8)]{qd}{3312}
  \nline
  -4.0 e^{i22^\circ}\bwc[(8)]{qd}{1232}
  -3.4 e^{-i79^\circ}\bwc[(8)]{qd}{1222}
  +3.4e^{-i79^\circ}\bwc[(8)]{qd}{1211}
  \nline
  +9.7\cdt{-1} e^{i24^\circ}\bwc[(8)]{qd}{2312}
  +6.3\cdt{-1} \bwc[(8)]{qd}{1313}
  -1.5\cdt{-1} e^{i45^\circ}\bwc[(8)]{qd}{1321}
  \nline
  -1.4\cdt{-1} e^{i23^\circ}\bwc[(8)]{qd}{2331}
  +1.5\cdt{-2} e^{i24^\circ}\bwc[(8)]{qd}{1322}
  -1.4\cdt{-2} e^{i24^\circ}\bwc[(8)]{qd}{1311}
  \nline
  -6.7\cdt{-3} e^{i23^\circ}\bwc[(8)]{qd}{1332}
  +6.3\cdt{-3} e^{i46^\circ}\bwc[(8)]{qd}{1221}
  +6.2\cdt{-3} e^{i23^\circ}\bwc[(8)]{qd}{2213}\,,
  \numberthis
\end{align*}

For {\bf the up-basis} we have
\begin{align*}
  \hat\Sigma^{B_s}_{qd1} & =
    2.3\cdt{3}\bwcup[(1)]{qd}{2323} +5.2\cdt{2}\bwcup[(1)]{qd}{1323}
   +9.3\cdt{1}\bwcup[(1)]{qd}{2223} -9.3\cdt{1}\bwcup[(1)]{qd}{3323}
  \nline
  +2.2\cdt{1}\bwcup[(1)]{qd}{1223} +8.0e^{-i73^\circ}\bwcup[(1)]{qd}{1232}
  -3.8\bwcup[(1)]{qd}{2332} +1.8e^{-i73^\circ}\bwcup[(1)]{qd}{1123}
  \nline
  -3.3\cdt{-1} e^{-i74^\circ}\bwcup[(1)]{qd}{1332}
  -1.1\cdt{-1}\bwcup[(1)]{qd}{2322} +1.1\cdt{-1}\bwcup[(1)]{qd}{2333}
  \nline
  -2.5\cdt{-2}\bwcup[(1)]{qd}{1322} +2.5\cdt{-2}\bwcup[(1)]{qd}{1333}
  -5.4\cdt{-3} e^{-i11^\circ}\bwcup[(1)]{qd}{2313} \,,
  \numberthis
\\
  \hat\Sigma^{B_d}_{qd1} & =
   5.5\cdt{4}\bwcup[(1)]{qd}{1313} -1.3\cdt{4}\bwcup[(1)]{qd}{2313}
  +2.3\cdt{3}\bwcup[(1)]{qd}{1213} -5.3\cdt{2}\bwcup[(1)]{qd}{2213}
  \nline
  +5.0\cdt{2} e^{i22^\circ}\bwcup[(1)]{qd}{3313}
  +2.0\cdt{2} e^{-i73^\circ}\bwcup[(1)]{qd}{1113}
  -4.5\cdt{1} e^{-i73^\circ}\bwcup[(1)]{qd}{1231}
  \nline
  +2.0\cdt{1}e^{i22^\circ}\bwcup[(1)]{qd}{2331}
  -2.7\bwcup[(1)]{qd}{1312} +1.8e^{-i51^\circ}\bwcup[(1)]{qd}{1331}
  +6.2\cdt{-1}\bwcup[(1)]{qd}{2312}
  \nline
  -1.1\cdt{-1}\bwcup[(1)]{qd}{1212}
  +6.3\cdt{-2} e^{i23^\circ}\bwcup[(1)]{qd}{1323}
  -3.0\cdt{-2} e^{i21^\circ}\bwcup[(1)]{qd}{1333}
  \nline
  +3.0\cdt{-2}e^{i21^\circ}\bwcup[(1)]{qd}{1311}
  +2.6\cdt{-2}\bwcup[(1)]{qd}{2212}
  -2.4\cdt{-2} e^{i21^\circ}\bwcup[(1)]{qd}{3312}
  \nline
  -1.4\cdt{-2} e^{i23^\circ}\bwcup[(1)]{qd}{2323}
  -9.6\cdt{-3} e^{-i74^\circ}\bwcup[(1)]{qd}{1112}
  +6.6\cdt{-3} e^{i22^\circ}\bwcup[(1)]{qd}{2333}
  \nline
  -6.6\cdt{-3}e^{i22^\circ}\bwcup[(1)]{qd}{2311} \,,
  \numberthis
\\
  \hat\Sigma^K_{qd1} & =
   5.0\cdt{6}\bwcup[(1)]{qd}{1212} +1.2\cdt{6}\bwcup[(1)]{qd}{1112}
  -1.2\cdt{6}\bwcup[(1)]{qd}{2212} -2.7\cdt{5}\bwcup[(1)]{qd}{1221}
  \nline
  -2.0\cdt{5}\bwcup[(1)]{qd}{1312} +4.7\cdt{4}\bwcup[(1)]{qd}{2312}
  +4.5\cdt{4}e^{i22^\circ}\bwcup[(1)]{qd}{2321}
  +1.0\cdt{4}e^{i22^\circ}\bwcup[(1)]{qd}{1321}
  \nline
  -1.8\cdt{3} e^{i23^\circ}\bwcup[(1)]{qd}{3312} +2.5\cdt{2}\bwcup[(1)]{qd}{1213}
  +5.6\cdt{1}\bwcup[(1)]{qd}{1113} -5.6\cdt{1}\bwcup[(1)]{qd}{2213}
  \nline
  -1.3\cdt{1}\bwcup[(1)]{qd}{1231} -1.0\cdt{1}\bwcup[(1)]{qd}{1313}
  -2.6 e^{i22^\circ}\bwcup[(1)]{qd}{1232}
  -2.5 e^{i57^\circ}\bwcup[(1)]{qd}{1211}
  \nline
  +2.5e^{i57^\circ}\bwcup[(1)]{qd}{1222} +2.3\bwcup[(1)]{qd}{2313}
  +2.2e^{i23^\circ}\bwcup[(1)]{qd}{2331} -1.5 e^{i20^\circ}\bwcup[(1)]{qd}{2222}
  \nline
  +1.5e^{i20^\circ}\bwcup[(1)]{qd}{2211} +1.1e^{i27^\circ}\bwcup[(1)]{qd}{1122}
  -1.1 e^{i27^\circ}\bwcup[(1)]{qd}{1111}
  -6.1\cdt{-1}e^{i22^\circ}\bwcup[(1)]{qd}{1123}
  \nline
  +6.0\cdt{-1}e^{i22^\circ}\bwcup[(1)]{qd}{2223}
  +5.0\cdt{-1}e^{i23^\circ}\bwcup[(1)]{qd}{1331}
  -2.3\cdt{-1}e^{i22^\circ}\bwcup[(1)]{qd}{1322}
  \nline
  +2.3\cdt{-1}e^{i22^\circ}\bwcup[(1)]{qd}{1311}
  +1.4\cdt{-1}e^{i22^\circ}\bwcup[(1)]{qd}{1223}
  +1.1\cdt{-1}e^{i23^\circ}\bwcup[(1)]{qd}{1332}
  \nline
  +1.0\cdt{-1} e^{i34^\circ}\bwcup[(1)]{qd}{2322}
  -1.0\cdt{-1} e^{i34^\circ}\bwcup[(1)]{qd}{2311}
  -9.0\cdt{-2} e^{i24^\circ}\bwcup[(1)]{qd}{3313}
  \nline
  -2.5\cdt{-2} e^{i23^\circ}\bwcup[(1)]{qd}{2332}
  -2.3\cdt{-2} e^{i44^\circ}\bwcup[(1)]{qd}{2323}
  -5.4\cdt{-3} e^{i44^\circ}\bwcup[(1)]{qd}{1323} \,,
  \numberthis
\end{align*}
and
\begin{align*}
  \hat\Sigma^{B_s}_{qd8} & =
   2.6\cdt{3}\bwcup[(8)]{qd}{2323} +6.0\cdt{2}\bwcup[(8)]{qd}{1323}
  +1.1\cdt{2}\bwcup[(8)]{qd}{2223} -1.1\cdt{2}\bwcup[(8)]{qd}{3323}
  \nline
  +2.5\cdt{1}\bwcup[(8)]{qd}{1223} +9.3e^{-i73^\circ}\bwcup[(8)]{qd}{1232}
  -4.4\bwcup[(8)]{qd}{2332} +2.1e^{-i73^\circ}\bwcup[(8)]{qd}{1123}
  \nline
  -3.8\cdt{-1} e^{-i74^\circ}\bwcup[(8)]{qd}{1332}
  -1.3\cdt{-1}\bwcup[(8)]{qd}{2322}
  +1.3\cdt{-1}\bwcup[(8)]{qd}{2333}
  -2.9\cdt{-2}\bwcup[(8)]{qd}{1322}
  \nline
  +2.9\cdt{-2}\bwcup[(8)]{qd}{1333}
  -7.4\cdt{-3} e^{-i10^\circ}\bwcup[(8)]{qd}{2313}
  -6.5\cdt{-3}\bwcup[(8)]{qd}{2331}
  \nline
  +6.0\cdt{-3}e^{-i3^\circ}\bwcup[(8)]{qd}{3313}
  -5.3\cdt{-3}\bwcup[(8)]{qd}{2222}
  +5.3\cdt{-3}\bwcup[(8)]{qd}{2233}
  \nline
  +5.3\cdt{-3}\bwcup[(8)]{qd}{3322}
  -5.2\cdt{-3}\bwcup[(8)]{qd}{3333}
  +5.0\cdt{-3}\bwcup[(8)]{qd}{2213} \,,
  \numberthis
\\
  \hat\Sigma^{B_d}_{qd8} & =
   6.4\cdt{4}\bwcup[(8)]{qd}{1313} -1.5\cdt{4}\bwcup[(8)]{qd}{2313}
  +2.6\cdt{3}\bwcup[(8)]{qd}{1213} -6.1\cdt{2}\bwcup[(8)]{qd}{2213}
  \nline
  +5.7\cdt{2} e^{i22^\circ}\bwcup[(8)]{qd}{3313}
  +2.3\cdt{2} e^{-i73^\circ}\bwcup[(8)]{qd}{1113}
  -5.2\cdt{1} e^{-i73^\circ}\bwcup[(8)]{qd}{1231}
  \nline
  +2.4\cdt{1}e^{i22^\circ}\bwcup[(8)]{qd}{2331}
  -3.1\bwcup[(8)]{qd}{1312} +2.0e^{-i51^\circ}\bwcup[(8)]{qd}{1331}
  +7.2\cdt{-1}\bwcup[(8)]{qd}{2312}
  \nline
  -1.3\cdt{-1}\bwcup[(8)]{qd}{1212}
  +7.3\cdt{-2}e^{i23^\circ}\bwcup[(8)]{qd}{1323}
  +3.6\cdt{-2}e^{i21^\circ}\bwcup[(8)]{qd}{1311}
  \nline
  -3.5\cdt{-2} e^{i21^\circ}\bwcup[(8)]{qd}{1333}
  +3.0\cdt{-2}\bwcup[(8)]{qd}{2212}
  -2.8\cdt{-2} e^{i21^\circ}\bwcup[(8)]{qd}{3312}
  \nline
  -1.6\cdt{-2} e^{i23^\circ}\bwcup[(8)]{qd}{2323}
  -1.1\cdt{-2} e^{-i74^\circ}\bwcup[(8)]{qd}{1112}
  -7.8\cdt{-3} e^{i22^\circ}\bwcup[(8)]{qd}{2311}
  \nline
  +7.4\cdt{-3}e^{i22^\circ}\bwcup[(8)]{qd}{2333}\,,
  \numberthis
\\
  \hat\Sigma^{K}_{qd8} & =
   7.2\cdt{6}\bwcup[(8)]{qd}{1212} +1.6\cdt{6}\bwcup[(8)]{qd}{1112}
  -1.6\cdt{6}\bwcup[(8)]{qd}{2212} -3.8\cdt{5}\bwcup[(8)]{qd}{1221}
  \nline
  -2.9\cdt{5}\bwcup[(8)]{qd}{1312} +6.7\cdt{4}\bwcup[(8)]{qd}{2312}
  +6.4\cdt{4}e^{i22^\circ}\bwcup[(8)]{qd}{2321}
  +1.5\cdt{4}e^{i22^\circ}\bwcup[(8)]{qd}{1321}
  \nline
  -2.6\cdt{3} e^{i23^\circ}\bwcup[(8)]{qd}{3312}
  +3.5\cdt{2}\bwcup[(8)]{qd}{1213}
  +8.1\cdt{1}\bwcup[(8)]{qd}{1113} -8.0\cdt{1}\bwcup[(8)]{qd}{2213}
  \nline
  -1.9\cdt{1}\bwcup[(8)]{qd}{1231} -1.4\cdt{1}\bwcup[(8)]{qd}{1313}
  -3.8e^{i22^\circ}\bwcup[(8)]{qd}{1232} +3.3\bwcup[(8)]{qd}{2313}
  \nline
  +3.1 e^{i23^\circ}\bwcup[(8)]{qd}{2331}
  -3.0 e^{i87^\circ}\bwcup[(8)]{qd}{1211}
  +3.0 e^{i87^\circ}\bwcup[(8)]{qd}{1222}
  -2.3 e^{i19^\circ}\bwcup[(8)]{qd}{2222}
  \nline
  +2.3e^{i19^\circ}\bwcup[(8)]{qd}{2211} +1.5e^{i29^\circ}\bwcup[(8)]{qd}{1122}
  -1.5 e^{i29^\circ}\bwcup[(8)]{qd}{1111}
  -8.7\cdt{-1}e^{i22^\circ}\bwcup[(8)]{qd}{1123}
  \nline
  +8.6\cdt{-1}e^{i22^\circ}\bwcup[(8)]{qd}{2223}
  +7.2\cdt{-1}e^{i23^\circ}\bwcup[(8)]{qd}{1331}
  -3.3\cdt{-1}e^{i22^\circ}\bwcup[(8)]{qd}{1322}
  \nline
  +3.3\cdt{-1}e^{i22^\circ}\bwcup[(8)]{qd}{1311}
  +2.0\cdt{-1}e^{i22^\circ}\bwcup[(8)]{qd}{1223}
  +1.5\cdt{-1}e^{i23^\circ}\bwcup[(8)]{qd}{1332}
  \nline
  +1.4\cdt{-1} e^{i34^\circ}\bwcup[(8)]{qd}{2322}
  -1.4\cdt{-1} e^{i34^\circ}\bwcup[(8)]{qd}{2311}
  -1.3\cdt{-1} e^{i24^\circ}\bwcup[(8)]{qd}{3313}
  \nline
  -3.5\cdt{-2} e^{i23^\circ}\bwcup[(8)]{qd}{2332}
  -3.4\cdt{-2} e^{i44^\circ}\bwcup[(8)]{qd}{2323}
  -7.7\cdt{-3} e^{i44^\circ}\bwcup[(8)]{qd}{1323} \,.
  \numberthis
\end{align*}

%
%--------+---------+---------+---------+---------+---------+---------+---------+
\subsubsection{\boldmath $\Wc[(1)]{qu}$ and $\Wc[(8)]{qu}$}

We next present the results for $\wc[(1)]{qu}{ijkl}$
and $\wc[(8)]{qu}{ijkl}$ in the down and up bases. Here $\DDbar$ mixing
dominates the scene, in particular for 1212, but for certain elements like
1313 and 2323 the $\BBbarD$ and $\BBbarS$ mixings are more important.
Interestingly, the entries 1233, 1333 and 2333, can be relevant for $K^0$,
$B_d$ and $B_s$, respectively. The index ``3'' indicates the top Yukawa
at work. Yet, the constraints are much weaker than in previous cases.

This pattern can be summarized by the following rather accurate formulae for
the sum in \refeq{eq:master-M12SMEFT} in {\bf the down-basis}
\begin{align}
  \Sigma^{B_s}_{qu1} & =
  2.6\cdt{-1} \bwc[(1)]{qu}{2333} -3.2\cdt{-2} \bwc[(1)]{qu}{2323}\,,
\\
  \Sigma^{B_d}_{qu1} & =
  -1.3e^{i22^\circ} \bwc[(1)]{qu}{1333} +1.7\cdt{-1} \bwc[(1)]{qu}{1323}\,,
\\
  \Sigma^{K}_{qu1} & =
   2.2\cdt{-1} e^{i23^\circ}\bwc[(1)]{qu}{1233}
  -2.8\cdt{-2} \bwc[(1)]{qu}{1223}
  -2.6\cdt{-2} e^{i22^\circ}\bwc[(1)]{qu}{1232} \,,
\end{align}
and
\begin{align}
  \Sigma^{B_s}_{qu8} & =
  1.1\cdt{-1} \bwc[(8)]{qu}{2333} -1.4\cdt{-2} \bwc[(8)]{qu}{2323} \,,
\\
  \Sigma^{B_d}_{qu8} & =
  -5.2\cdt{-1} e^{i22^\circ}\bwc[(8)]{qu}{1333}
  +7.5\cdt{-2} \bwc[(8)]{qu}{1323} \,,
\\
  \Sigma^{K}_{qu8} & =
   9.2\cdt{-2} e^{i23^\circ}\bwc[(8)]{qu}{1233}
  -1.2\cdt{-2} \bwc[(8)]{qu}{1223}
  -1.1\cdt{-2} e^{i22^\circ}\bwc[(8)]{qu}{1232} \,.
\end{align}
For {\bf the up-basis} we have
\begin{align*}
  \hat\Sigma^{B_s}_{qu1} & =
   2.6\cdt{-1}\bwcup[(1)]{qu}{2333} +5.9\cdt{-2}\bwcup[(1)]{qu}{1333}
  -3.1\cdt{-2}\bwcup[(1)]{qu}{2323}
  \nline
  +1.1\cdt{-2}\bwcup[(1)]{qu}{2233} -1.0\cdt{-2}\bwcup[(1)]{qu}{3333}
  -7.2\cdt{-3}\bwcup[(1)]{qu}{1323} \,,
  \numberthis
\\
  \hat\Sigma^{B_d}_{qu1} & =
  -1.3e^{i22^\circ}\bwcup[(1)]{qu}{1333}
  +3.0\cdt{-1}e^{i22^\circ}\bwcup[(1)]{qu}{2333}
  +1.7\cdt{-1}\bwcup[(1)]{qu}{1323}
  \nline
  -5.5\cdt{-2} e^{i23^\circ}\bwcup[(1)]{qu}{1233}
  -3.8\cdt{-2}\bwcup[(1)]{qu}{2323}
  +1.3\cdt{-2}e^{i22^\circ}\bwcup[(1)]{qu}{2233}
  \nline
  -1.1\cdt{-2} e^{i44^\circ}\bwcup[(1)]{qu}{3333}
  +7.1\cdt{-3}\bwcup[(1)]{qu}{1223} \,,
  \numberthis
\\
  \hat \Sigma^K_{qu1} & =
   2.0\cdt{-1} e^{i23^\circ}\bwcup[(1)]{qu}{1233}
  +4.8\cdt{-2} e^{i23^\circ}\bwcup[(1)]{qu}{1133}
  -4.8\cdt{-2} e^{i23^\circ}\bwcup[(1)]{qu}{2233}
  \nline
  -2.5\cdt{-2}\bwcup[(1)]{qu}{1223}
  -2.4\cdt{-2} e^{i23^\circ}\bwcup[(1)]{qu}{1232}
  -1.2\cdt{-2} e^{i11^\circ}\bwcup[(1)]{qu}{1123}
  \nline
  +1.2\cdt{-2} e^{i11^\circ}\bwcup[(1)]{qu}{2223}
  -7.7\cdt{-3} e^{i23^\circ}\bwcup[(1)]{qu}{1333} \,,
  \numberthis
\end{align*}
and
\begin{align*}
  \hat\Sigma^{B_s}_{qu8} & =
   1.0\cdt{-1}\bwcup[(8)]{qu}{2333} +2.4\cdt{-2}\bwcup[(8)]{qu}{1333}
  -1.4\cdt{-2}\bwcup[(8)]{qu}{2323}\,,
  \numberthis
\\
  \hat\Sigma^{B_d}_{qu8} & =
  -5.1\cdt{-1}e^{i22^\circ}\bwcup[(8)]{qu}{1333}
  +1.2\cdt{-1}e^{i22^\circ}\bwcup[(8)]{qu}{2333}
  +7.3\cdt{-2}\bwcup[(8)]{qu}{1323}
  \nline
  -2.3\cdt{-2} e^{i23^\circ}\bwcup[(8)]{qu}{1233}
  -1.7\cdt{-2}\bwcup[(8)]{qu}{2323}
  +5.3\cdt{-3}e^{i22^\circ}\bwcup[(8)]{qu}{2233}\,,
  \numberthis
\\
  \hat \Sigma^{K}_{qu8} & =
   8.2\cdt{-2} e^{i23^\circ}\bwcup[(8)]{qu}{1233}
  +2.0\cdt{-2} e^{i23^\circ}\bwcup[(8)]{qu}{1133}
  -2.0\cdt{-2} e^{i23^\circ}\bwcup[(8)]{qu}{2233}
  \nline
  -1.1\cdt{-2}\bwcup[(8)]{qu}{1223}
  -1.0\cdt{-2} e^{i23^\circ}\bwcup[(8)]{qu}{1232}
  -5.0\cdt{-3} e^{i11^\circ}\bwcup[(8)]{qu}{1123}
  \nline
  +5.0\cdt{-3}e^{i11^\circ}\bwcup[(8)]{qu}{2223}\,.
  \numberthis
\end{align*}

%
%--------+---------+---------+---------+---------+---------+---------+---------+
\subsubsection{\boldmath $\Wc{dd}$, $\Wc{uu}$, $\Wc[(1)]{ud}$ and $\Wc[(8)]{ud}$}

We collect the results for these four operators in one section together, because
the corresponding $P_a^{ij}(\muNP)$ are the same in the down and up bases.

In the case of  $\wc{dd}{ijkl}$ large coefficients are only found in 1212,
1313 and 2323 entries from $\KKbar$, $\BBbarD$ and $\BBbarS$ mixings,
respectively. The corresponding coefficients in $\DDbar$ mixing are very small.

On the other hand in the case of $\wc{uu}{ijkl}$ only constraints from
$\DDbar$ mixing are relevant. Indeed, as seen in \eqref{eq:Cuu} the
contributions to down-quark mixings are eliminated by our constraints.

In the case of  $\wc[(1)]{ud}{ijkl}$ and $\wc[(8)]{ud}{ijkl}$ the numerical
coefficients are very small, in particular for charm, implying that $\DF=2$
transitions do not play any role in constraining these Wilson coefficients.
Yet, the entries 3312, 3313 and 3323 for $\KKbar$, $\BBbarD$ and $\BBbarS$
mixings, respectively, could play some role in specific models.

This pattern can be summarized by the following rather accurate formulae for
the sum in \refeq{eq:master-M12SMEFT} for both down and up basis
\begin{align*}
  \Sigma^{B_s}_{dd} & =
  -4.1\cdt{2}\bwc[]{dd}{2323} -2.2\cdt{-2}\bwc[]{dd}{2333}
  +1.9\cdt{-2}\bwc[]{dd}{2223} \,,
  \numberthis
\\
  \Sigma^{B_d}_{dd} & =
  -9.6\cdt{3}\bwc[]{dd}{1313} +4.7\cdt{-1}\bwc[]{dd}{1213}
  +1.4\cdt{-2}e^{i22^\circ}\bwc[]{dd}{1333}
  \nline
  -1.1\cdt{-2} e^{i23^\circ}\bwc[]{dd}{1323}
  +1.1\cdt{-2} e^{i22^\circ}\bwc[]{dd}{1223} \,,
  \numberthis
\\
  \Sigma^{K}_{dd} & =
  -3.7\cdt{4}\bwc[]{dd}{1212} -1.8\bwc[]{dd}{1213}
  -8.2\cdt{-2} e^{i23^\circ}\bwc[]{dd}{1222}
  \nline
  -4.8\cdt{-2}e^{i22^\circ}\bwc[]{dd}{1332}
  +1.9\cdt{-2}e^{i22^\circ}\bwc[]{dd}{1232}
  +5.5\cdt{-3}e^{i22^\circ}\bwc[]{dd}{1233} \,,
   \numberthis
\end{align*}
\begin{align}
  \label{eq:Cuu}
  \Sigma^{B_s}_{uu} & = 0 \,, &
  \Sigma^{B_d}_{uu} & = 0 \,, &
  \Sigma^{K}_{uu}   & = 0 \,,
\end{align}
\begin{align*}
  \Sigma^{B_s}_{ud1} & =
  -1.0 \bwc[(1)]{ud}{3323} +1.1\cdt{-1} \bwc[(1)]{ud}{2323} \,,
  \numberthis
\\
  \Sigma^{B_d}_{ud1} & =
   5.6 e^{i22^\circ}\bwc[(1)]{ud}{3313} -6.2\cdt{-1} \bwc[(1)]{ud}{2313}
  +5.4\cdt{-3} \bwc[(1)]{ud}{1313} \,,
  \numberthis
\\
  \Sigma^{K}_{ud1} & =
  -2.0\cdt{1} e^{i23^\circ}\bwc[(1)]{ud}{3312} +2.1 \bwc[(1)]{ud}{2312}
  +2.0 e^{i22^\circ}\bwc[(1)]{ud}{2321}
  \nline
  -1.9\cdt{-1} \bwc[(1)]{ud}{2212} -1.8\cdt{-2} \bwc[(1)]{ud}{1312}
  -8.1\cdt{-3} e^{i23^\circ}\bwc[(1)]{ud}{1112} \,,
  \numberthis
\end{align*}
and
\begin{align*}
  \Sigma^{B_s}_{ud8} & =
  -1.5 \bwc[(8)]{ud}{3323} +1.6\cdt{-1} \bwc[(8)]{ud}{2323} \,,
  \numberthis
\\
  \Sigma^{B_d}_{ud8} & =
   7.9 e^{i22^\circ}\bwc[(8)]{ud}{3313} -8.7\cdt{-1} \bwc[(8)]{ud}{2313}
  +7.6\cdt{-3} \bwc[(8)]{ud}{1313}
  \nline
  +5.5\cdt{-3} e^{i22^\circ}\bwc[(8)]{ud}{1113}
  +5.4\cdt{-3} e^{i22^\circ}\bwc[(8)]{ud}{2213}\,,
  \numberthis
\\
  \Sigma^{K}_{ud8} & =
  -3.8\cdt{1} e^{i23^\circ}\bwc[(8)]{ud}{3312}
  +4.2 \bwc[(8)]{ud}{2312} +4.0e^{i22^\circ}\bwc[(8)]{ud}{2321}
  \nline
  -3.9\cdt{-1} \bwc[(8)]{ud}{2212}
  -3.6\cdt{-2} \bwc[(8)]{ud}{1312}
  -2.5\cdt{-2} e^{i22^\circ}\bwc[(8)]{ud}{1112} \,.
  \numberthis
\end{align*}

%
%--------+---------+---------+---------+---------+---------+---------+---------+
\subsubsection{\boldmath $\Wc[(1,3)]{\phi q}$, $\Wc[]{\phi d}$
  and $\Wc[]{\phi u}$}

Here of particular interest are the entries 12, 13, 23 in $\wc[]{\phi d}{}$,
which in certain scenarios like VLQ models imply rather strong constraints
\cite{Bobeth:2016llm, Bobeth:2017xry}.

Explicitly we have in {\bf the down basis} for $\Wc[(1,3)]{\phi q}$
\begin{align}
  \Sigma^{B_s}_{\phi q1} & = -2.4\cdt{-1} \bwc[(1)]{\phi q}{23} \,,
& %\\
  \Sigma^{B_d}_{\phi q1} & = 1.2 e^{i22^\circ}\bwc[(1)]{\phi q}{13} \,,
& %\\
  \Sigma^{K}_{\phi q1}   & = -2.1\cdt{-1} e^{i23^\circ}\bwc[(1)]{\phi q}{12} \,,
\end{align}
and
\begin{align}
  \Sigma^{B_s}_{\phi q3} & =
   5.4\cdt{-1} \bwc[(3)]{\phi q}{23} +6.3\cdt{-3} \bwc[(3)]{\phi q}{33}
  +5.1\cdt{-3} \bwc[(3)]{\phi q}{22}\,,
\\
  \Sigma^{B_d}_{\phi q3} & =
  -2.7 e^{i22^\circ}\bwc[(3)]{\phi q}{13}
  -2.6\cdt{-2} e^{i21^\circ}\bwc[(3)]{\phi q}{12}
  \notag\nline
  +7.0\cdt{-3} e^{i44^\circ}\bwc[(3)]{\phi q}{33}
  +5.7\cdt{-3} e^{i44^\circ}\bwc[(3)]{\phi q}{11} \,,
\\
  \Sigma^{K}_{\phi q3} & =
   5.4\cdt{-1} e^{i23^\circ}\bwc[(3)]{\phi q}{12}
  +5.3\cdt{-3} e^{i24^\circ}\bwc[(3)]{\phi q}{13}\,,
\end{align}
and $\Wc{\phi d}$
\begin{align}
  \Sigma^{B_s}_{\phi d} & = 1.2 \bwc[]{\phi d}{23}\,,
&%\\
  \Sigma^{B_d}_{\phi d} & = -6.6 e^{i22^\circ}\bwc[]{\phi d}{13}\,,
&%\\
  \Sigma^K_{\phi d}     & = 2.5\cdt{1} e^{i23^\circ}\bwc[]{\phi d}{12}\,.
\end{align}
In the case $\Wc{\phi u}$ the corresponding contributions of $\Sigma_{\phi u}$
are too small to allow for a $5\%$ effect in $\big[M_{12}^{ij}\big]_\text{BSM}$
and hence are eliminated by our conditions.

For {\bf the up basis} we have
\begin{align*}
  \hat\Sigma^{B_s}_{\phi q1} & =
  -2.3\cdt{-1}\bwcup[(1)]{\phi q}{23} -5.4\cdt{-2}\bwcup[(1)]{\phi q}{13}
  -1.0\cdt{-2}\bwcup[(1)]{\phi q}{22} +9.4\cdt{-3}\bwcup[(1)]{\phi q}{33}\,,
  \numberthis
\\
  \hat\Sigma^{B_d}_{\phi q1} & =
   1.2e^{i22^\circ}\bwcup[(1)]{\phi q}{13}
  -2.7\cdt{-1}e^{i22^\circ}\bwcup[(1)]{\phi q}{23}
  +5.2\cdt{-2}e^{i23^\circ}\bwcup[(1)]{\phi q}{12}
  \nline
  -1.2\cdt{-2}e^{i22^\circ}\bwcup[(1)]{\phi q}{22}
  +1.0\cdt{-2}e^{i44^\circ}\bwcup[(1)]{\phi q}{33}\,,
  \numberthis
\\
  \hat\Sigma^K_{\phi q1} & =
  -1.9\cdt{-1} e^{i23^\circ}\bwcup[(1)]{\phi q}{12}
  -4.6\cdt{-2} e^{i23^\circ}\bwcup[(1)]{\phi q}{11}
  +4.6\cdt{-2} e^{i23^\circ}\bwcup[(1)]{\phi q}{22}
  \nline
  +7.3\cdt{-3}e^{i23^\circ}\bwcup[(1)]{\phi q}{13} \,,
  \numberthis
\end{align*}
and
\begin{align*}
  \hat\Sigma^{B_s}_{\phi q3} & =
   5.2\cdt{-1}\bwcup[(3)]{\phi q}{23} +1.2\cdt{-1}\bwcup[(3)]{\phi q}{13}
  +2.7\cdt{-2}\bwcup[(3)]{\phi q}{22}
  \nline
  -1.6\cdt{-2}\bwcup[(3)]{\phi q}{33}
  +7.3\cdt{-3}e^{-i19^\circ}\bwcup[(3)]{\phi q}{12} \,,
  \numberthis
\\
  \hat\Sigma^{B_d}_{\phi q3} & =
  -2.6e^{i22^\circ}\bwcup[(3)]{\phi q}{13}
  +6.1\cdt{-1}e^{i22^\circ}\bwcup[(3)]{\phi q}{23}
  -1.4\cdt{-1} e^{i23^\circ}\bwcup[(3)]{\phi q}{12}
 \nline
  +3.2\cdt{-2} e^{i22^\circ}\bwcup[(3)]{\phi q}{22}
  -1.8\cdt{-2} e^{i44^\circ}\bwcup[(3)]{\phi q}{33}
  -1.2\cdt{-2} e^{-i51^\circ}\bwcup[(3)]{\phi q}{11} \,,
  \numberthis
\\
  \hat\Sigma^K_{\phi q3} & =
   4.9\cdt{-1} e^{i23^\circ}\bwcup[(3)]{\phi q}{12}
  +1.2\cdt{-1} e^{i23^\circ}\bwcup[(3)]{\phi q}{11}
  -1.2\cdt{-1} e^{i23^\circ}\bwcup[(3)]{\phi q}{22}
  \nline
  -1.6\cdt{-2} e^{i23^\circ}\bwcup[(3)]{\phi q}{13}
  +7.4\cdt{-3} e^{i34^\circ}\bwcup[(3)]{\phi q}{23} \,,
  \numberthis
\end{align*}
whereas results for $\Wcup{\phi d}$ and $\Wcup{\phi u}$ are the
same as in the down basis.

%
%--------+---------+---------+---------+---------+---------+---------+---------+
\subsubsection{\boldmath $\Wc[(1,8)]{quqd}$}

The contributions from these coefficients are very small. However, what is
interesting is the dominance of $\DF = 1$ transitions in all meson systems,
demonstrating top Yukawa at work.

Explicitly we have in {\bf the down basis}
\begin{align*}
  \Sigma^{B_s}_{quqd1} & =
   1.0\cdt{-1}\bwc[(1)]{quqd}{3332} -5.0\cdt{-2}\bwc[(1)]{quqd}{2323}
  -9.2\cdt{-3}\bwc[(1)]{quqd}{3232} \,,
  \numberthis
\\
  \Sigma^{B_d}_{quqd1} & =
  -5.4\cdt{-1} e^{i22^\circ}\bwc[(1)]{quqd}{3331}
  -6.0\cdt{-2} \bwc[(1)]{quqd}{1313}
  +5.2\cdt{-2} \bwc[(1)]{quqd}{3231} \,,
  \numberthis
\\
  \Sigma^{K}_{quqd1} & =
  -2.8 \bwc[(1)]{quqd}{3321} -2.8 \bwc[(1)]{quqd}{3312}
  -1.2e^{i22^\circ} \bwc[(1)]{quqd}{2321}
  \nline
  +2.7\cdt{-1}\bwc[(1)]{quqd}{1312}
  +1.1\cdt{-1}\bwc[(1)]{quqd}{2221}
  -4.3\cdt{-2}\bwc[(1)]{quqd}{2331}
  \nline
  -4.1\cdt{-2}\bwc[(1)]{quqd}{1332}
  -2.4\cdt{-2}\bwc[(1)]{quqd}{1212} \,,
  \numberthis
\end{align*}
and
\begin{align*}
  \Sigma^{B_s}_{quqd8} &
  =  1.8\cdt{-2}\bwc[(8)]{quqd}{3332} -8.8\cdt{-3}\bwc[(8)]{quqd}{2323} \,,
  \numberthis
\\
  \Sigma^{B_d}_{quqd8} & =
  -9.6\cdt{-2}e^{i22^\circ}\bwc[(8)]{quqd}{3331}
  -1.1\cdt{-2}\bwc[(8)]{quqd}{1313}
  +9.0\cdt{-3}\bwc[(8)]{quqd}{3231}\,,
  \numberthis
\\
  \Sigma^K_{quqd8} & =
  -1.9\cdt{-1}e^{i22^\circ}\bwc[(8)]{quqd}{2321}
  +4.2\cdt{-2}\bwc[(8)]{quqd}{1312}
  +1.8\cdt{-2}\bwc[(8)]{quqd}{2221}
  \nline
  -6.2\cdt{-3}\bwc[(8)]{quqd}{2331}
  -5.7\cdt{-3}\bwc[(8)]{quqd}{1332} \,.
  \numberthis
\end{align*}

In {\bf the up basis} we find
\begin{align*}
  \hat\Sigma^{B_s}_{quqd1} & =
   1.0\cdt{-1}\bwcup[(1)]{quqd}{3332} -4.7\cdt{-2}\bwcup[(1)]{quqd}{2323}
  -1.1\cdt{-2}\bwcup[(1)]{quqd}{2313}
  \nline
  -1.1\cdt{-2}\bwcup[(1)]{quqd}{1323}
  -9.2\cdt{-3}\bwcup[(1)]{quqd}{3232} \,,
  \numberthis
\\
  \hat\Sigma^{B_d}_{quqd1} & =
  -5.5\cdt{-1} e^{i20^\circ}\bwcup[(1)]{quqd}{3331}
  -5.7\cdt{-2}\bwcup[(1)]{quqd}{1313} +5.2\cdt{-2}\bwcup[(1)]{quqd}{3231}
  \nline
  -2.2\cdt{-2} e^{i23^\circ}\bwcup[(1)]{quqd}{2331}
  -1.7\cdt{-2} e^{i28^\circ}\bwcup[(1)]{quqd}{3321}
  +1.3\cdt{-2}\bwcup[(1)]{quqd}{2313}
  \nline
  +1.3\cdt{-2}\bwcup[(1)]{quqd}{1323} \,,
  \numberthis
\\
  \hat\Sigma^K_{quqd1} & =
   3.4\cdt{1}\bwcup[(1)]{quqd}{3321} +1.5\cdt{1}\bwcup[(1)]{quqd}{3311}
  +2.2\bwcup[(1)]{quqd}{3322}-1.7\bwcup[(1)]{quqd}{3312}
  \nline
  +1.2\bwcup[(1)]{quqd}{2331} -1.1e^{i22^\circ}\bwcup[(1)]{quqd}{2321}
  +2.7\cdt{-1}\bwcup[(1)]{quqd}{1331}
  \nline
  -2.6\cdt{-1}e^{i22^\circ}\bwcup[(1)]{quqd}{1321}
  -2.6\cdt{-1}e^{i22^\circ}\bwcup[(1)]{quqd}{2311}
  +2.5\cdt{-1}\bwcup[(1)]{quqd}{1312}
  \nline
  +1.0\cdt{-1}\bwcup[(1)]{quqd}{2221}
  -6.0\cdt{-2} e^{i20^\circ}\bwcup[(1)]{quqd}{1311}
  -5.7\cdt{-2}\bwcup[(1)]{quqd}{1322}
  \nline
  -5.5\cdt{-2}\bwcup[(1)]{quqd}{2312} +2.4\cdt{-2}\bwcup[(1)]{quqd}{2211}
  +2.4\cdt{-2}\bwcup[(1)]{quqd}{1221}
  \nline
  -2.2\cdt{-2}\bwcup[(1)]{quqd}{1212}
  +2.0\cdt{-2}\bwcup[(1)]{quqd}{2332}
  +2.0\cdt{-2}e^{i3^\circ}\bwcup[(1)]{quqd}{1332}
  \nline
  +1.4\cdt{-2}\bwcup[(1)]{quqd}{2322}
  -6.0\cdt{-3} e^{i21^\circ}\bwcup[(1)]{quqd}{3331}
  +5.6\cdt{-3}\bwcup[(1)]{quqd}{1211}
  \nline
  +5.1\cdt{-3}\bwcup[(1)]{quqd}{2212}
  +5.1\cdt{-3}\bwcup[(1)]{quqd}{1222} \,,
  \numberthis
\end{align*}
and
\begin{align*}
  \hat\Sigma^{B_s}_{quqd8} & =
  1.8\cdt{-2}\bwcup[(8)]{quqd}{3332} -8.4\cdt{-3}\bwcup[(8)]{quqd}{2323} \,,
  \numberthis
\\
  \hat\Sigma^{B_d}_{quqd8} & =
  -9.5\cdt{-2}e^{i22^\circ}\bwcup[(8)]{quqd}{3331}
  -1.0\cdt{-2}\bwcup[(8)]{quqd}{1313}
  +9.0\cdt{-3}\bwcup[(8)]{quqd}{3231} \,,
  \numberthis
\\
  \hat\Sigma^K_{quqd8} & =
   3.6\cdt{-1}\bwcup[(8)]{quqd}{2331}
  -1.8\cdt{-1}e^{i22^\circ}\bwcup[(8)]{quqd}{2321}
  -8.2\cdt{-2}\bwcup[(8)]{quqd}{1331}
  \nline
  -4.1\cdt{-2}e^{i22^\circ}\bwcup[(8)]{quqd}{2311}
  -4.1\cdt{-2}e^{i22^\circ}\bwcup[(8)]{quqd}{1321}
  +4.0\cdt{-2}\bwcup[(8)]{quqd}{1312}
  \nline
  +1.7\cdt{-2}\bwcup[(8)]{quqd}{2221}
  +1.3\cdt{-2}\bwcup[(8)]{quqd}{1332}
  -9.6\cdt{-3}e^{i22^\circ}\bwcup[(8)]{quqd}{1311}
  \nline
  -9.3\cdt{-3}\bwcup[(8)]{quqd}{2312}
  -9.2\cdt{-3}\bwcup[(8)]{quqd}{1322}
  +7.3\cdt{-3}e^{i23^\circ}\bwcup[(8)]{quqd}{3321}
  \nline
  +5.1\cdt{-3}\bwcup[(8)]{quqd}{2332} \,.
  \numberthis
\end{align*}

%
%--------+---------+---------+---------+---------+---------+---------+---------+
\subsubsection{\boldmath $\Wc[(1,3)]{lequ}$, $\Wc[(1,3)]{lq}$, $\Wc[]{ld}$
and $\Wc[]{qe}$}

The contributions of the semileptonic operators are even smaller than the
previous ones. Most of them give too small contributions and we show
here only those that meet our criteria.\footnote{For the
interested reader we refer to the tables in the ancillary files, which provide
all contributions.} The correlations between rare Kaon decays,
$\KKbar$ mixing and the $\DF=1$ process $\epe$ have been discussed in
the framework of leptoquark models in \cite{Bobeth:2017ecx}.
Explicitly under our exclusion principle stated above we find in the
down basis
\begin{align}
  \Sigma^{B_s}_{lq3} & = 0\,,
&
  \Sigma^{B_d}_{lq3} & =
    1.1\cdt{-2} e^{i22^\circ} \left(
    \bwc[(3)]{lq}{3313} +\bwc[(3)]{lq}{1113} +\bwc[(3)]{lq}{2213} \right)\,,
&
  \Sigma^K_{lq3} & = 0 \,,
\end{align}
and in the up-basis
\begin{align}
  \hat\Sigma^{B_s}_{lq3} &  =0 \,,
&
  \hat\Sigma^{B_d}_{lq3} &
  = 1.1\cdt{-2} e^{i22^\circ} \left(
    \bwcup[(3)]{lq}{1113} +\bwcup[(3)]{lq}{2213}
   +\bwcup[(3)]{lq}{3313} \right)\,,
&
  \hat\Sigma^K_{lq3} & =  0\,.
\end{align}
While the numerical coefficients in the sums are the same in
up and down bases, the WCs could be different in these two bases.

%
%--------+---------+---------+---------+---------+---------+---------+---------+
\subsubsection{\boldmath $\Wc[]{uW}$}

For the operator $\Op[]{uW}$ one finds:
\begin{align}
  \Sigma^{B_s}_{uW} &
  = -6.1\cdt{-2} \bwc[]{uW}{23} \,,
&
  \Sigma^{B_d}_{uW} &
  = 3.1\cdt{-1} e^{i22^\circ}\bwc[]{uW}{13}\,,
&
  \Sigma^K_{uW} &
  =  0 \,,
\end{align}
and the same for the up-basis.

%
%
%--------+---------+---------+---------+---------+---------+---------+---------+
\subsection[Probing large values of $\muNP$]
{\boldmath Probing large values of $\muNP$}
\label{sec:Lambda-bounds}

\begin{figure}[htb]
\centering
  \includegraphics[width=0.99\textwidth]{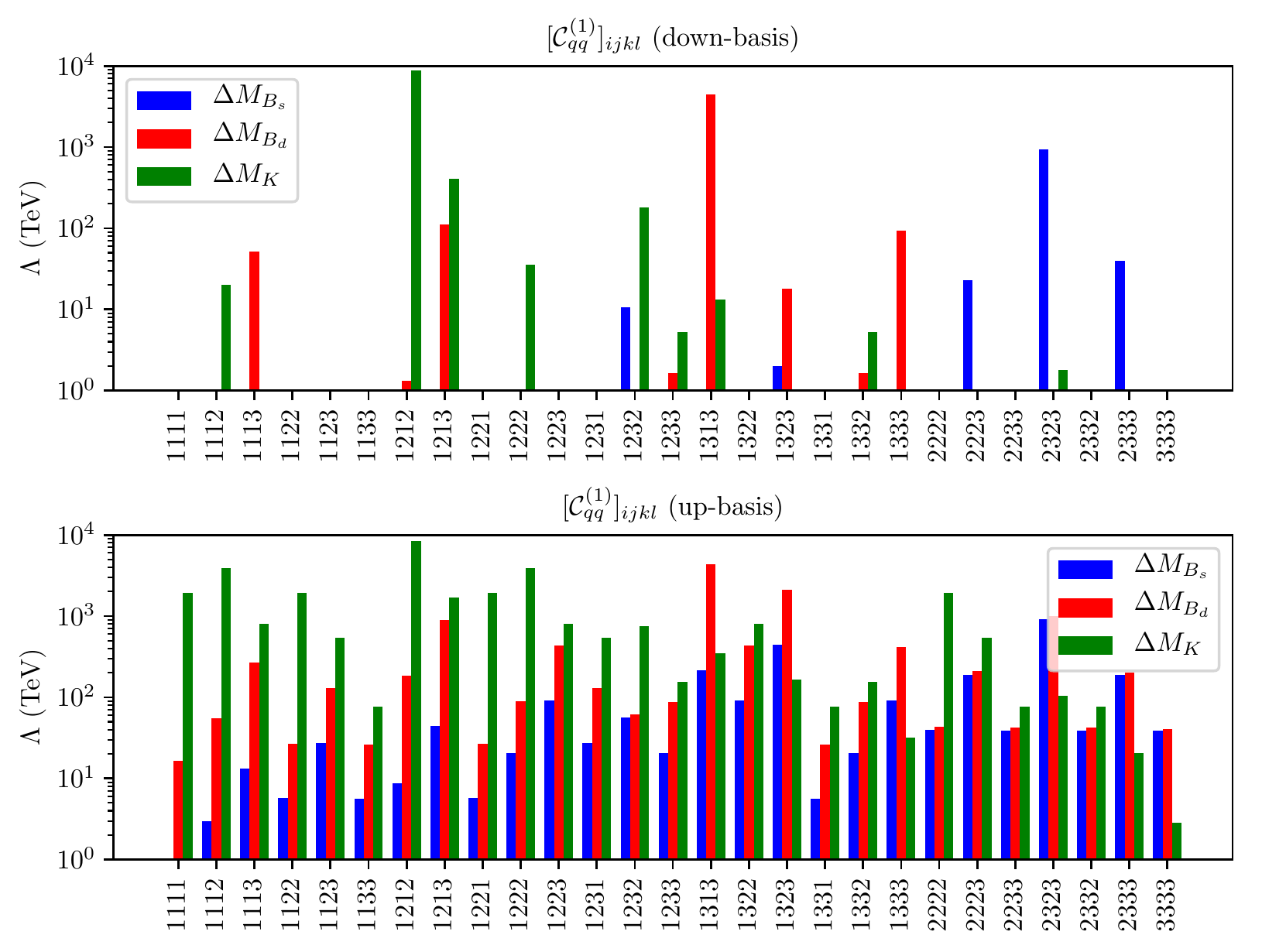}
\caption{\small
  The maximal NP scale $\muNP$ for $\bwc[(1)]{qq}{ijkl} = 10$ that
  corresponds to a $10\%$ effect in $2\big[M_{12}^{ij}\big]_\text{BSM}/
  (\Delta M_{ij})_\text{exp}$ for $B_s$ (blue), $B_d$ (red) and $K^0$
  (green), respectively.
}
\label{fig:lambda-qq1}
\end{figure}

In this section we present maximal values of the scale $\muNP$ at which
the NP contribution from a single Wilson coefficient could provide a shift
in the mass difference $\Delta M_{ij}$ of at least $10\%$ of
$(\Delta M_{ij})_\text{exp}$ when the dimensionless SMEFT coefficient becomes
$\bWc{a} = 10$. For lower values of $\bWc{a}$ these maximal values will be
smaller. On the other hand, keeping $\bWc{a} = 10$ and increasing $\muNP$
will imply NP effects below $10\%$ to $\Delta M_{ij}$. In view of the small
hadronic uncertainties in most of the $\DF=2$ matrix elements in
\reftab{tab:DF2-me-input}, NP effects below $10\%$ should be still of interest
when constraining BSM scenarios. The results are presented in
\reffigs{fig:lambda-qq1}{fig:lambda-lq3} for both down and up bases.
For an operator with a given flavour structure, strikingly different values
of $\muNP$ are found in the two bases. These results are self-explanatory.

\begin{figure}[htb]
\centering
  \includegraphics[width=0.99\textwidth]{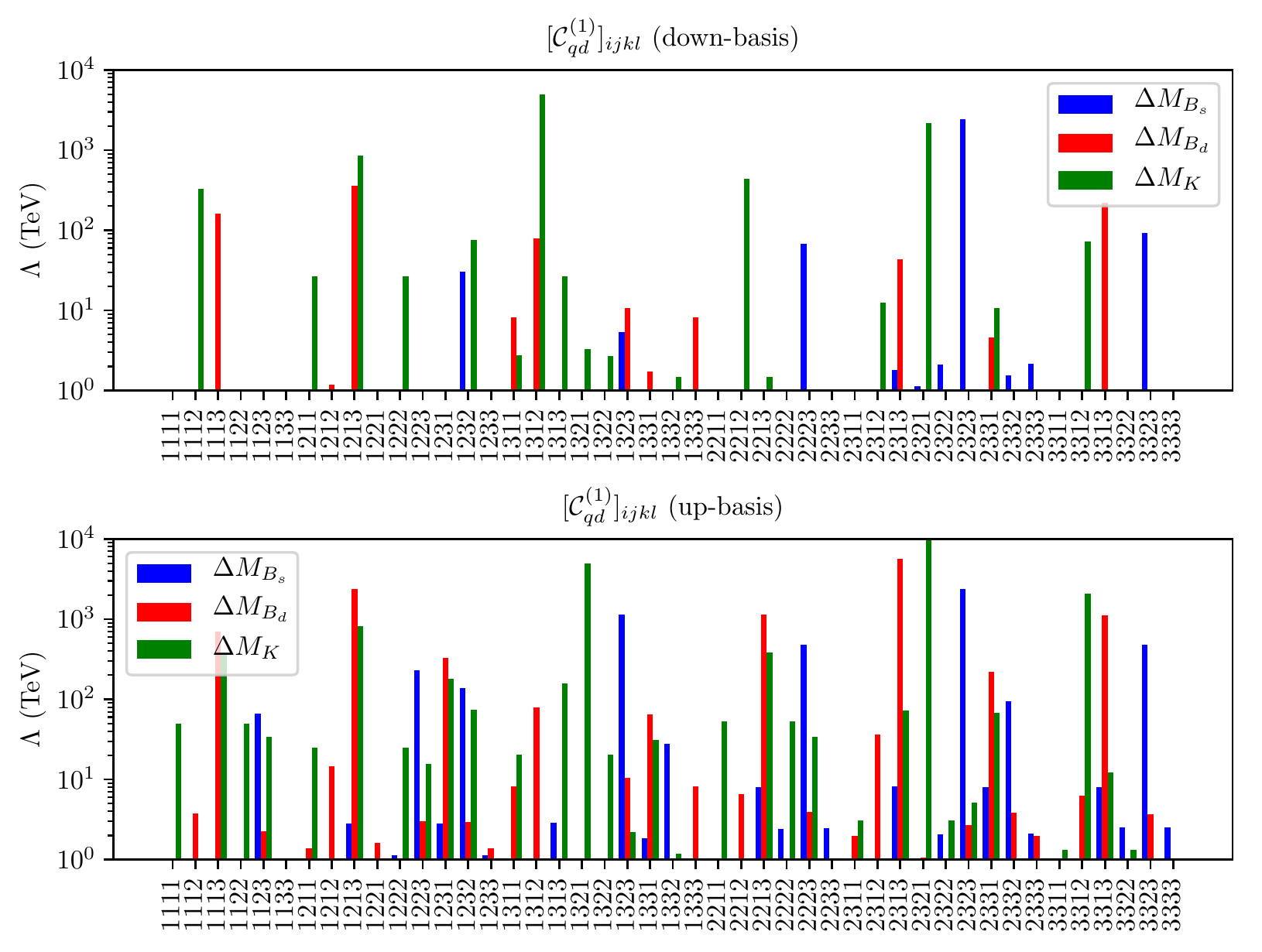}
\caption{\small
  The maximal NP scale $\muNP$ for $\bwc[(1)]{qd}{ijkl} = 10$ that
  corresponds to a $10\%$ effect in $2\big[M_{12}^{ij}\big]_\text{BSM}/
  (\Delta M_{ij})_\text{exp}$ for $B_s$ (blue), $B_d$ (red) and $K^0$
  (green), respectively.
}
\label{fig:lambda-qd1}
\end{figure}

\begin{figure}[htb]
\centering
  \includegraphics[width=0.99\textwidth]{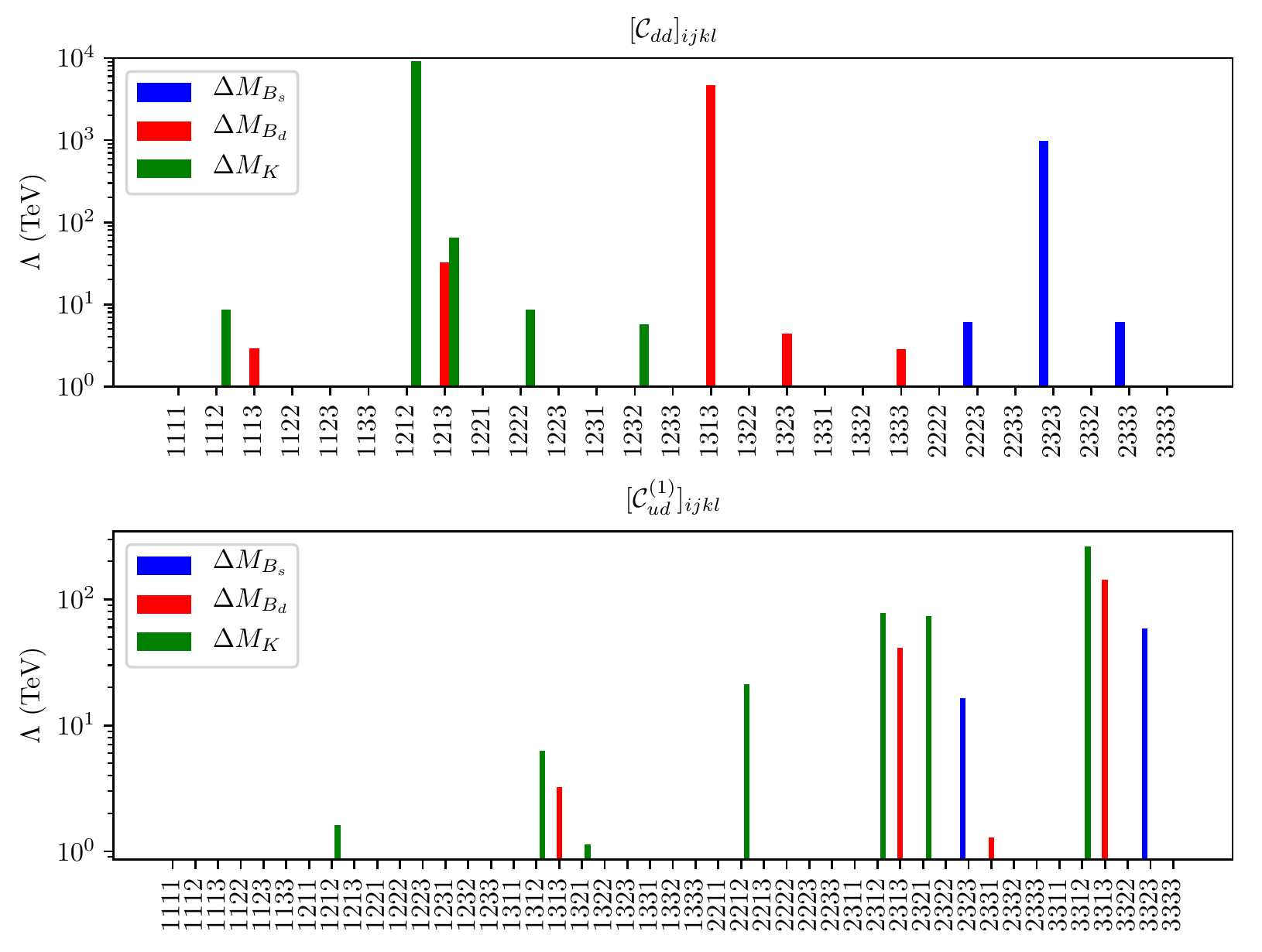}
\caption{\small
  The maximal NP scale $\muNP$ for $\bwc{dd}{ijkl} = 10$ (upper) and
  $\bwc[(1)]{ud}{ijkl} = 10$ (lower) that
  corresponds to a $10\%$ effect in $2\big[M_{12}^{ij}\big]_\text{BSM}/
  (\Delta M_{ij})_\text{exp}$ for $B_s$ (blue), $B_d$ (red) and $K^0$
  (green), respectively.
}
\label{fig:lambda-dd-ud1}
\end{figure}

\begin{figure}[htb]
\centering
  \includegraphics[width=0.99\textwidth]{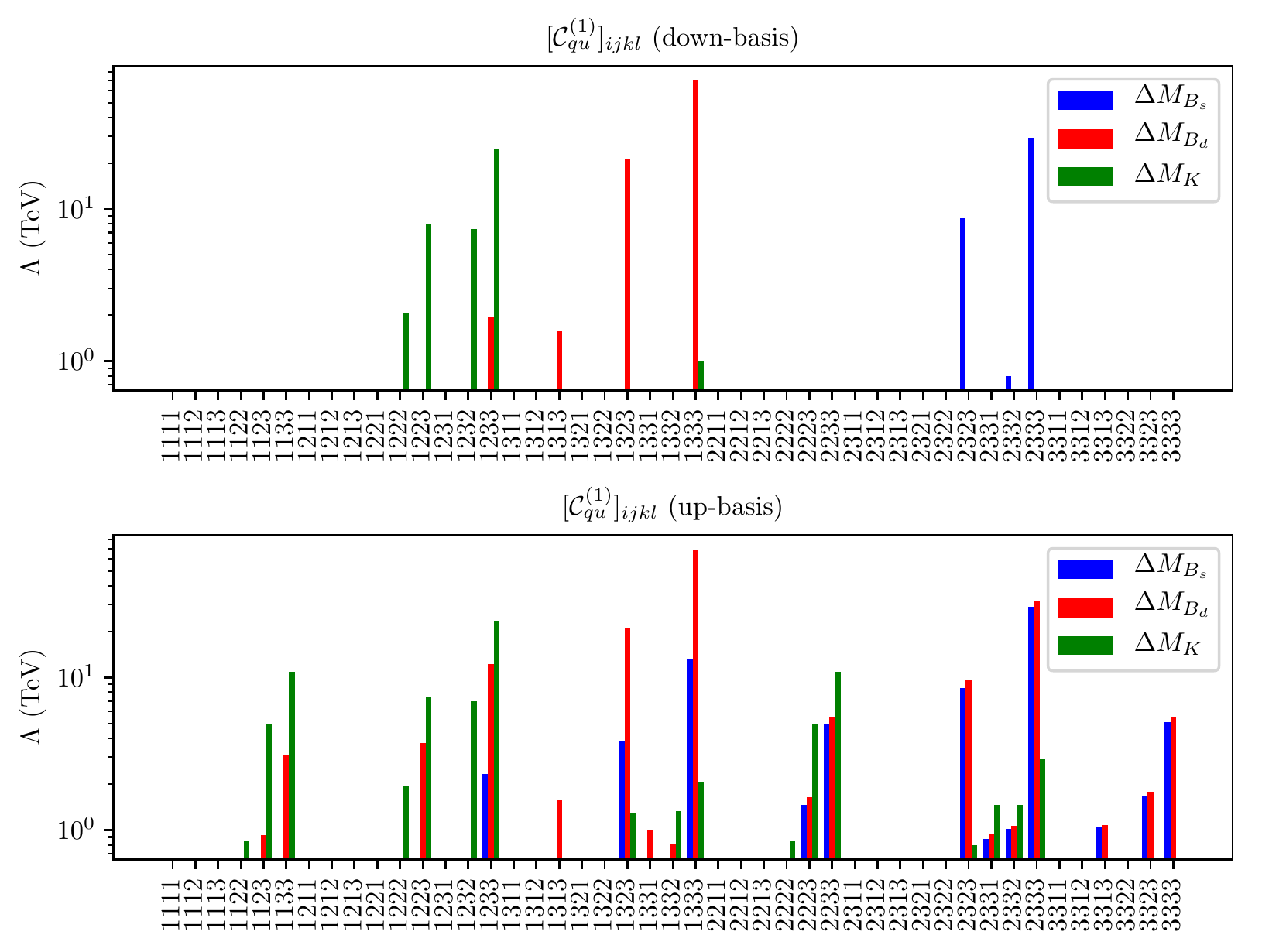}
\caption{\small
  The maximal NP scale $\muNP$ for $\bwc[(1)]{qu}{ijkl} = 10$ that
  corresponds to a $10\%$ effect in $2\big[M_{12}^{ij}\big]_\text{BSM}/
  (\Delta M_{ij})_\text{exp}$ for $B_s$ (blue), $B_d$ (red) and $K^0$
  (green), respectively.
}
\label{fig:lambda-qu1}
\end{figure}

\begin{figure}[htb]
\centering
  \includegraphics[width=0.47\textwidth]{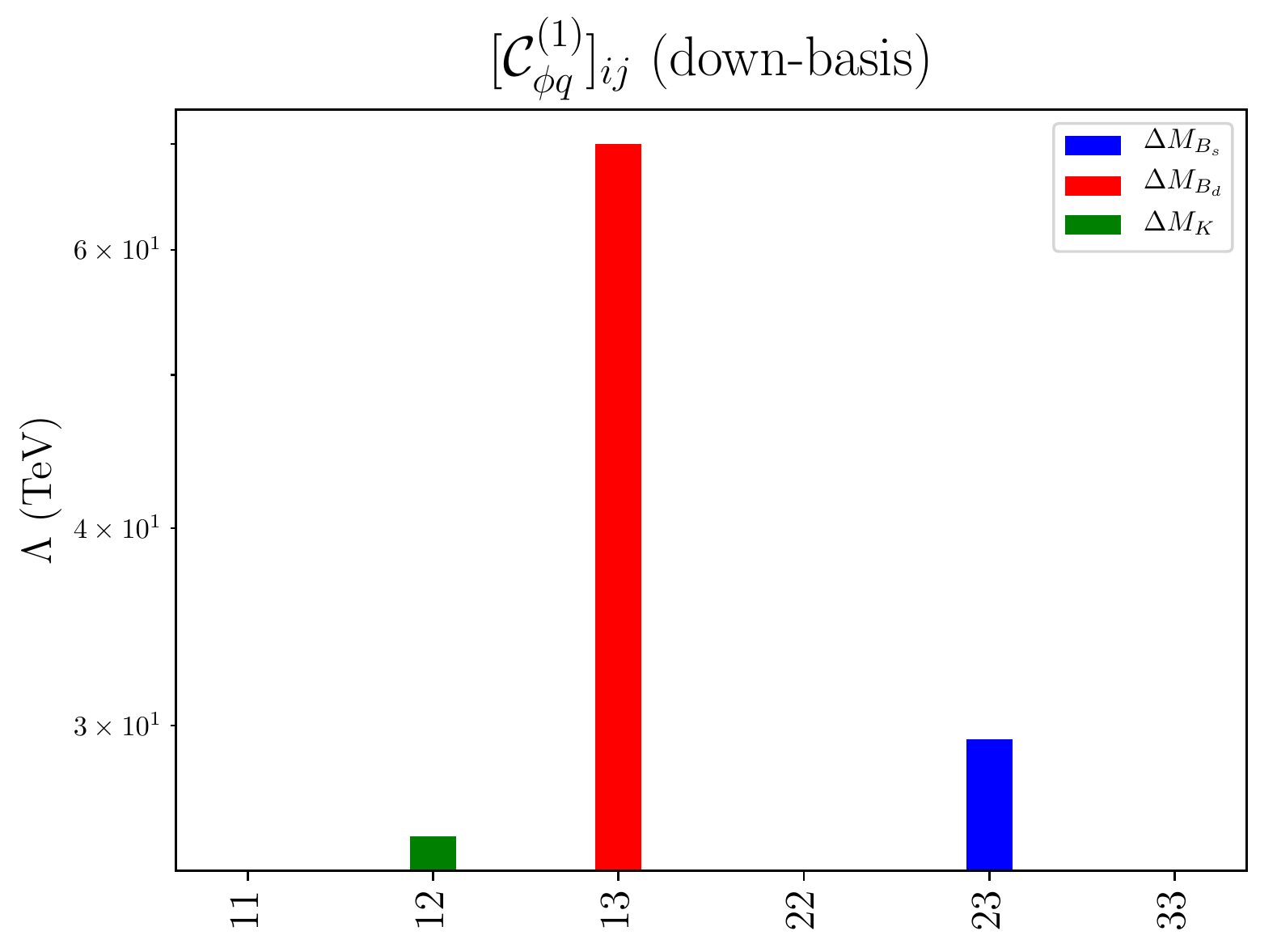}
  \hskip 0.01\textwidth
 \includegraphics[width=0.47\textwidth]{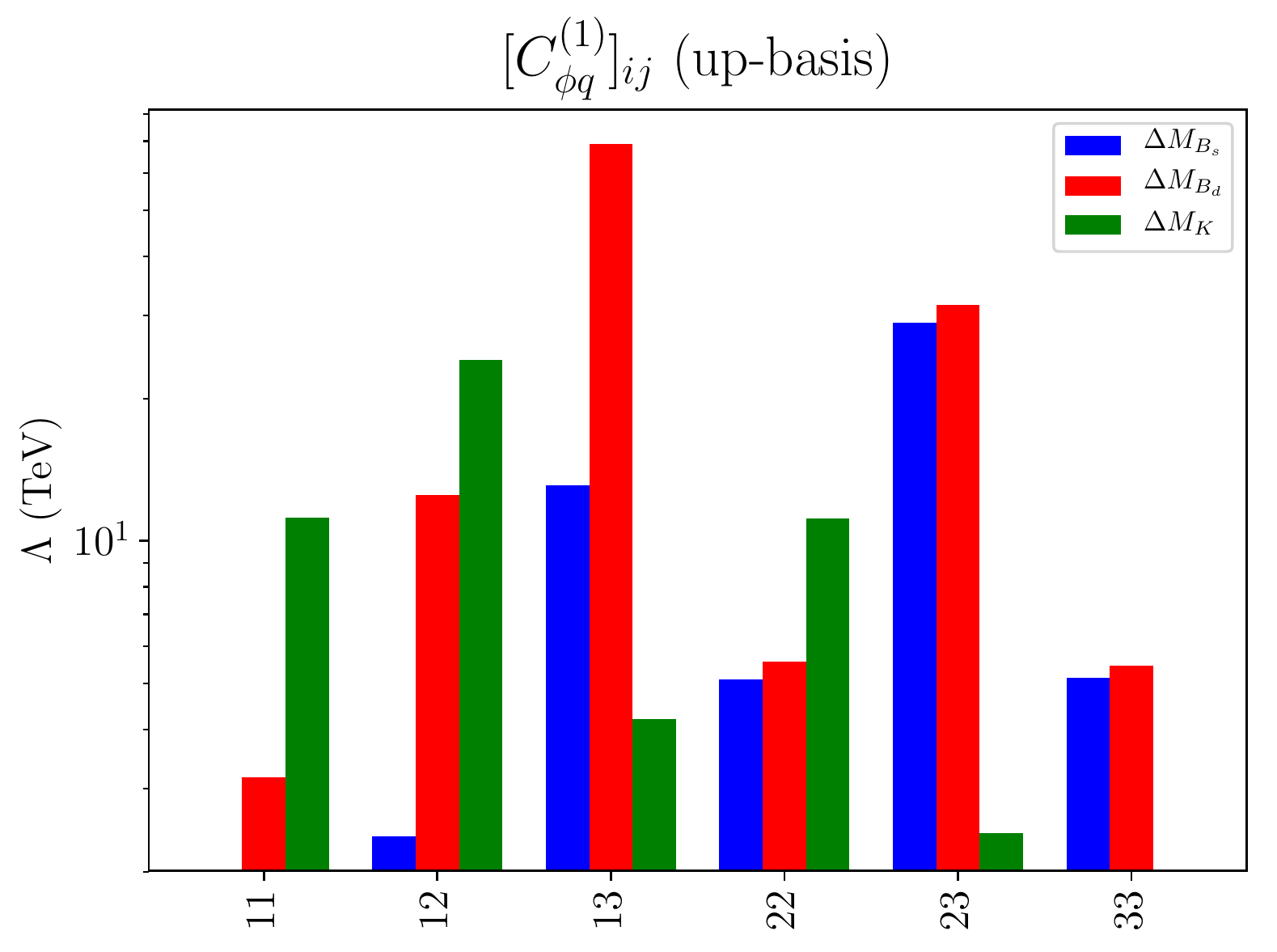}
\caption{\small
  The maximal NP scale $\muNP$ for $\bwc[(1)]{\phi q}{ij} = 10$ that
  corresponds to a $10\%$ effect in $2\big[M_{12}^{ij}\big]_\text{BSM}/
  (\Delta M_{ij})_\text{exp}$ for $B_s$ (blue), $B_d$ (red) and $K^0$
  (green), respectively.
}
\label{fig:lambda-phiq1}
\end{figure}

\begin{figure}[htb]
\centering
  \includegraphics[width=0.47\textwidth]{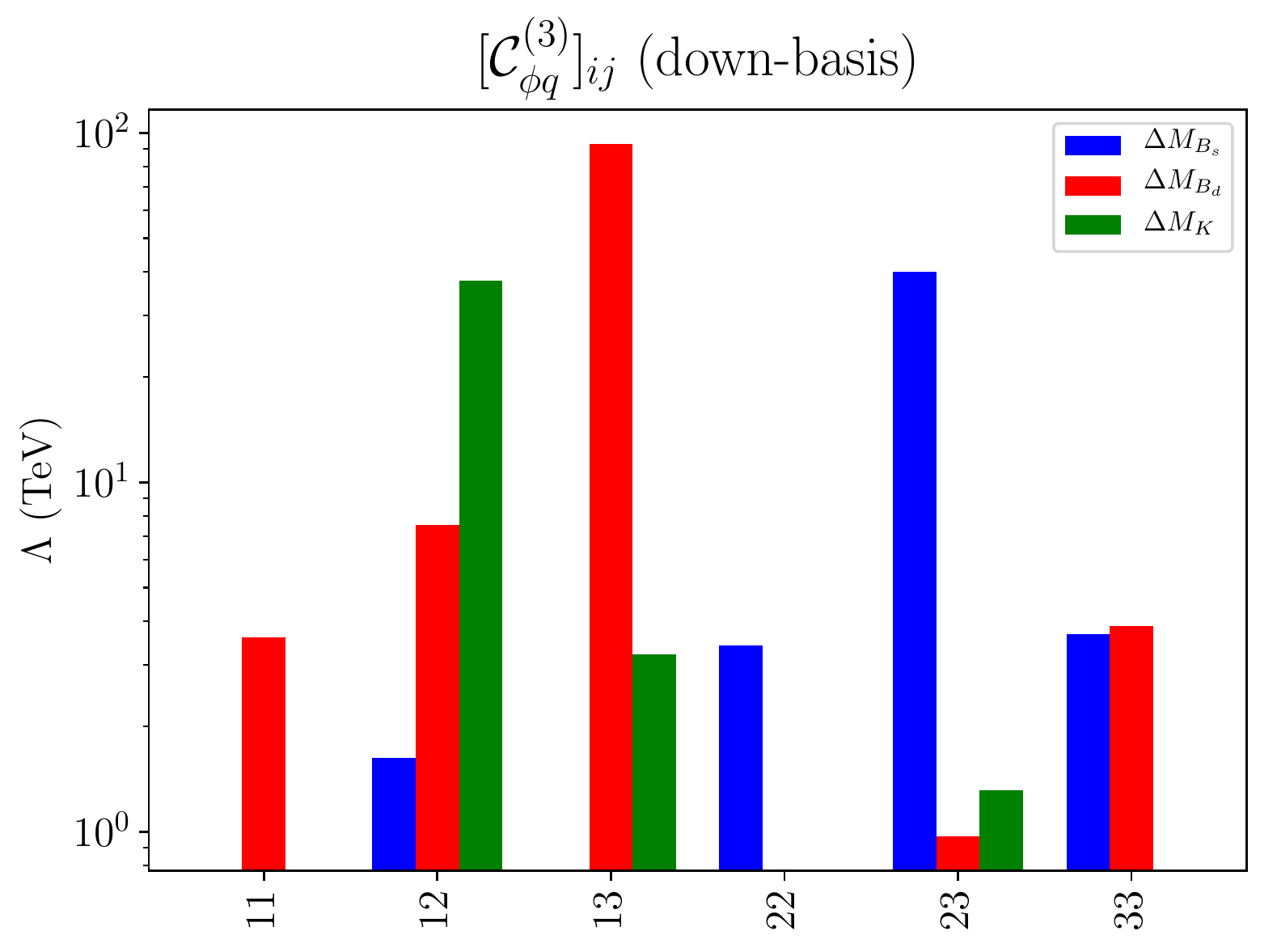}
  \hskip 0.01\textwidth
  \includegraphics[width=0.47\textwidth]{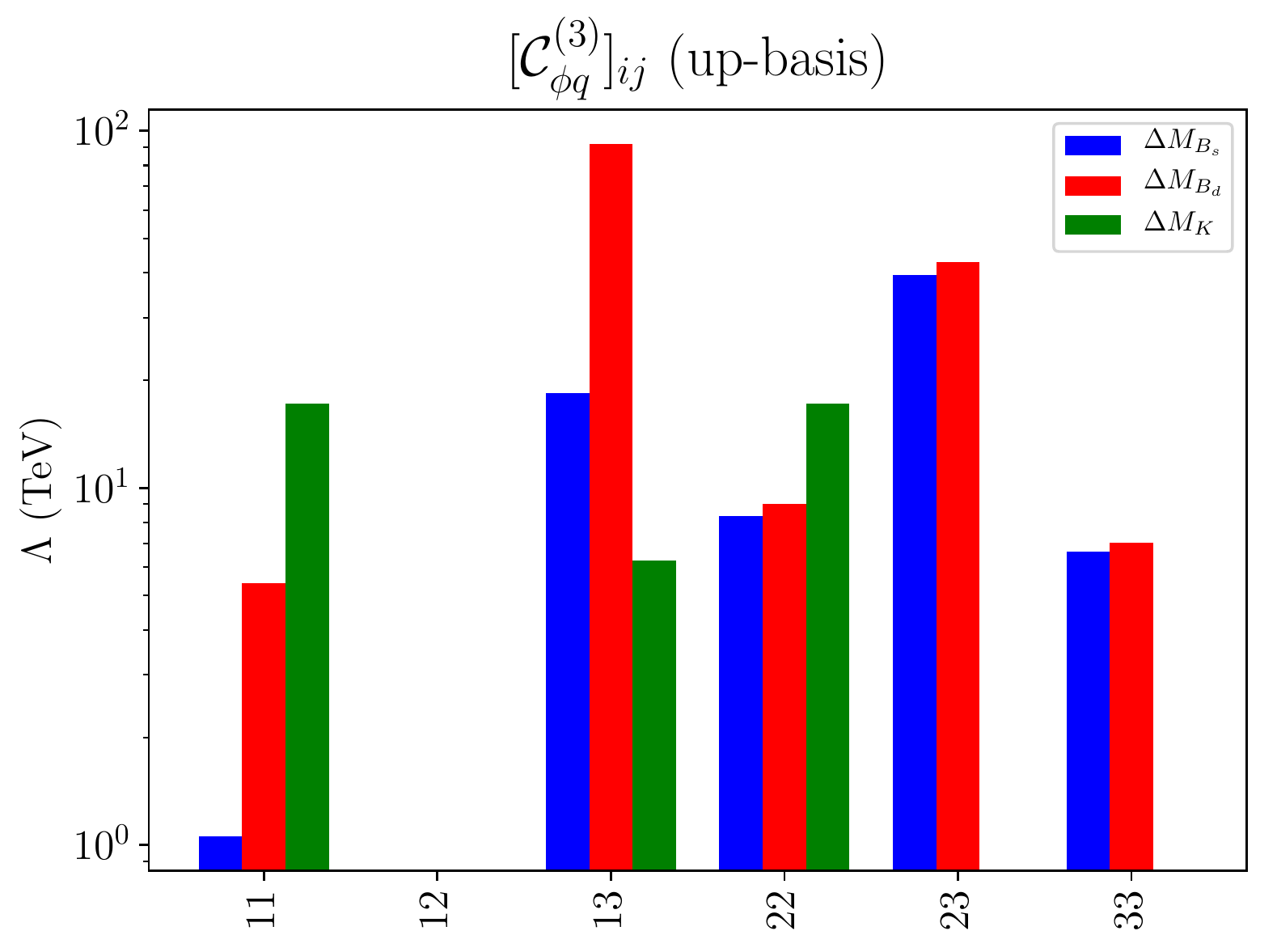}
\caption{\small
  The maximal NP scale $\muNP$ for $\bwc[(3)]{\phi q}{ij} = 10$ that
  corresponds to a $10\%$ effect in $2\big[M_{12}^{ij}\big]_\text{BSM}/
  (\Delta M_{ij})_\text{exp}$ for $B_s$ (blue), $B_d$ (red) and $K^0$
  (green), respectively.
}
\label{fig:lambda-phiq3}
\end{figure}

\begin{figure}[htb]
\centering
  \includegraphics[width=0.47\textwidth]{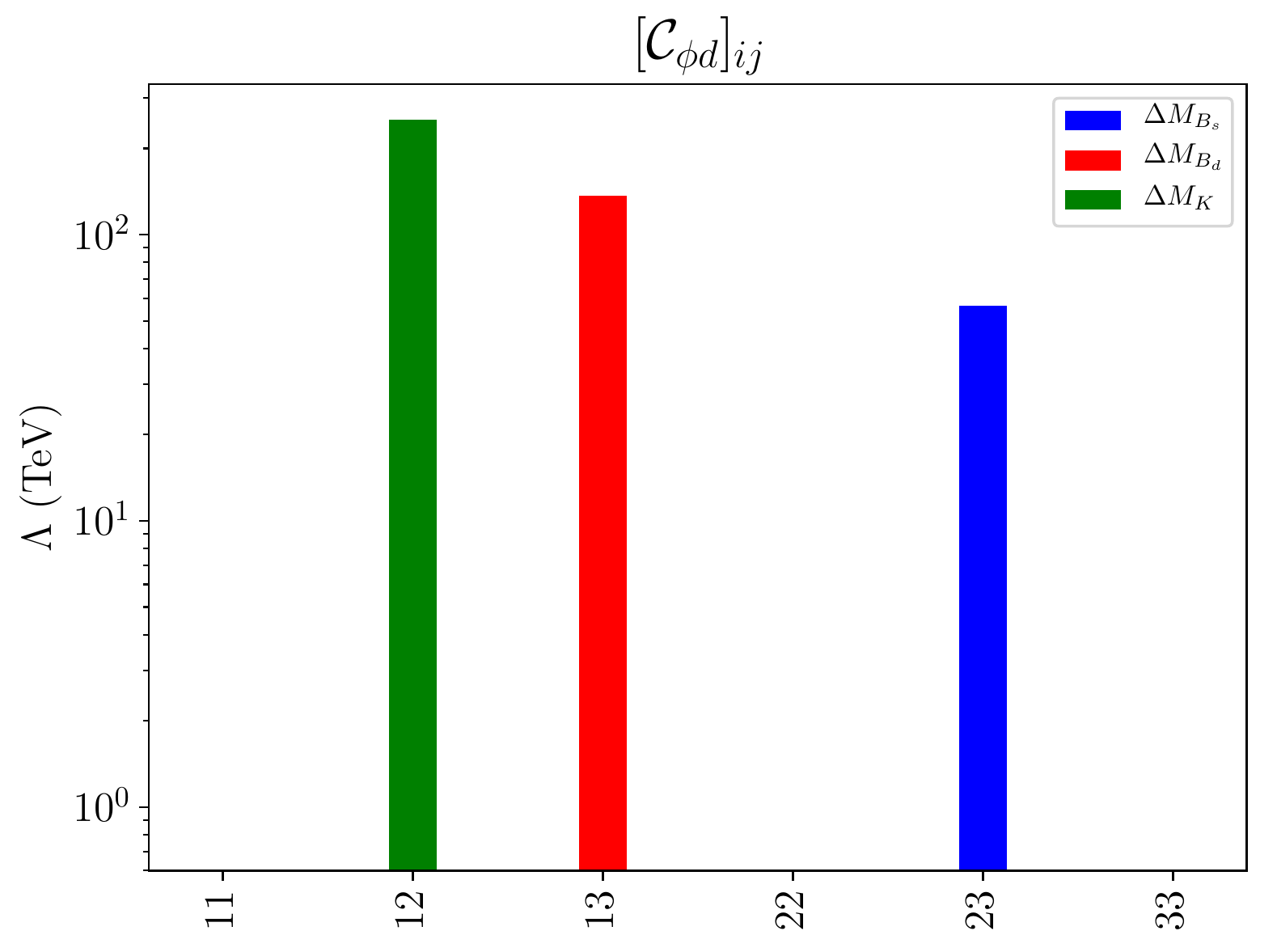}
\caption{\small
  The maximal NP scale $\muNP$ for $\bwc{\phi d}{ij} = 10$ that
  corresponds to a $10\%$ effect in $2\big[M_{12}^{ij}\big]_\text{BSM}/
  (\Delta M_{ij})_\text{exp}$ for $B_s$ (blue), $B_d$ (red) and $K^0$
  (green), respectively.
}
\label{fig:lambda-phid}
\end{figure}

\begin{figure}[htb]
\centering
  \includegraphics[width=0.99\textwidth]{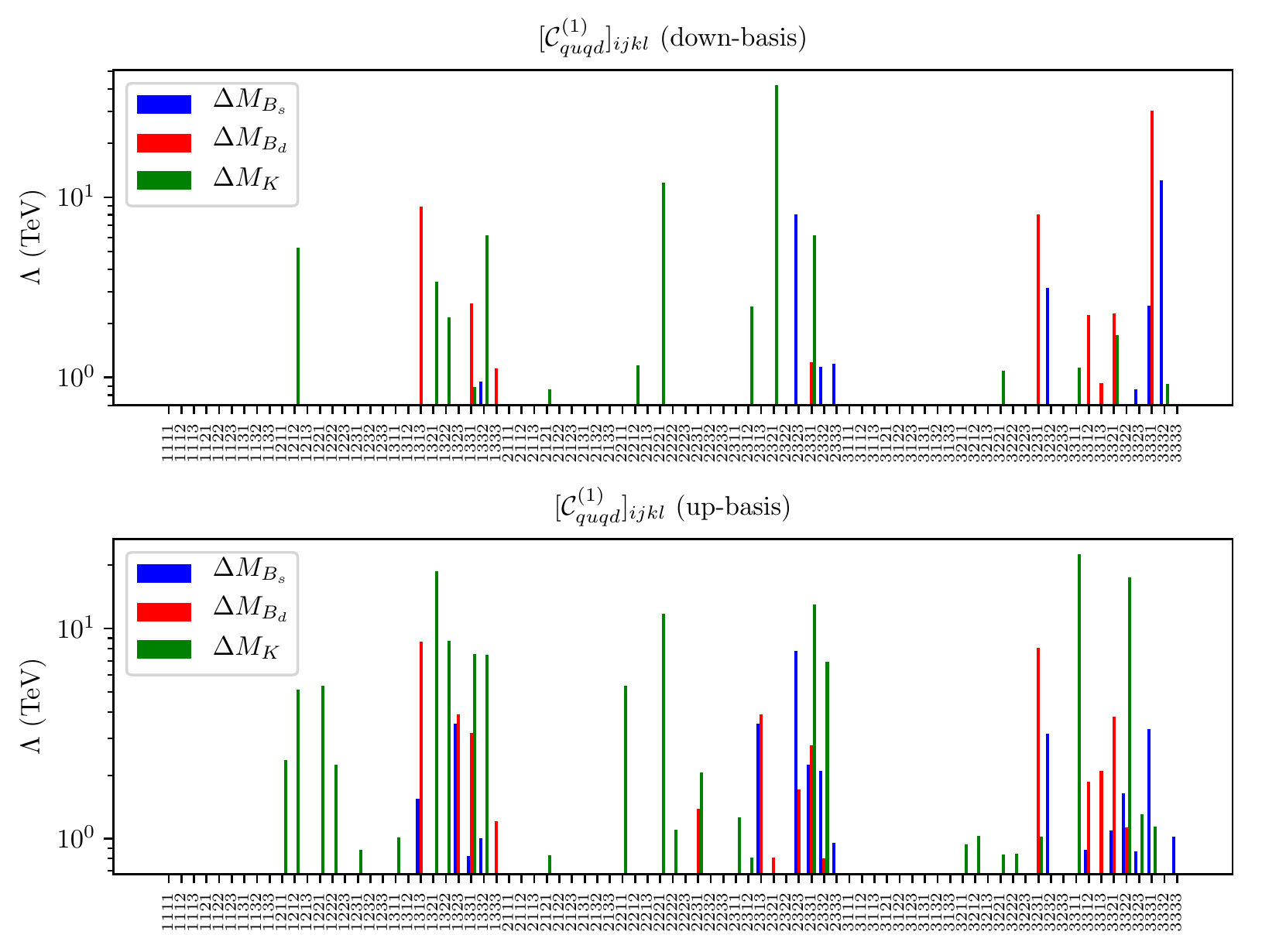}
\caption{\small
  The maximal NP scale $\muNP$ for $\bwc[(1)]{quqd}{ijkl} = 10$ that
  corresponds to a $10\%$ effect in $2\big[M_{12}^{ij}\big]_\text{BSM}/
  (\Delta M_{ij})_\text{exp}$ for $B_s$ (blue), $B_d$ (red) and $K^0$
  (green), respectively.
}
\label{fig:lambda-quqd1}
\end{figure}

\begin{figure}[htb]
\centering
  \includegraphics[width=0.99\textwidth]{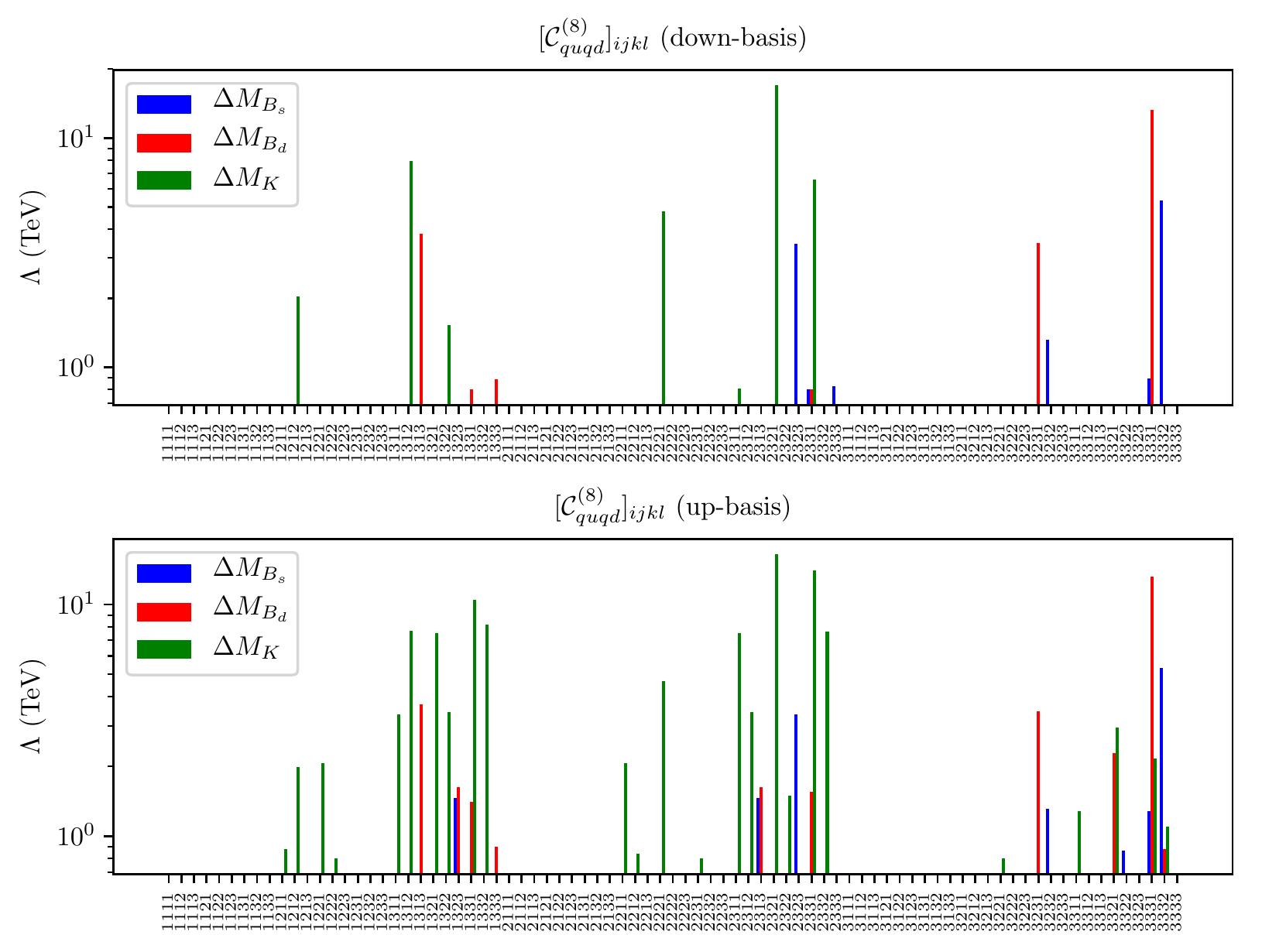}
\caption{\small
  The maximal NP scale $\muNP$ for $\bwc[(8)]{quqd}{ijkl} = 10$ that
  corresponds to a $10\%$ effect in $2\big[M_{12}^{ij}\big]_\text{BSM}/
  (\Delta M_{ij})_\text{exp}$ for $B_s$ (blue), $B_d$ (red) and $K^0$
  (green), respectively.
}
\label{fig:lambda-quqd8}
\end{figure}

\begin{figure}[htb]
\centering
  \includegraphics[width=0.99\textwidth]{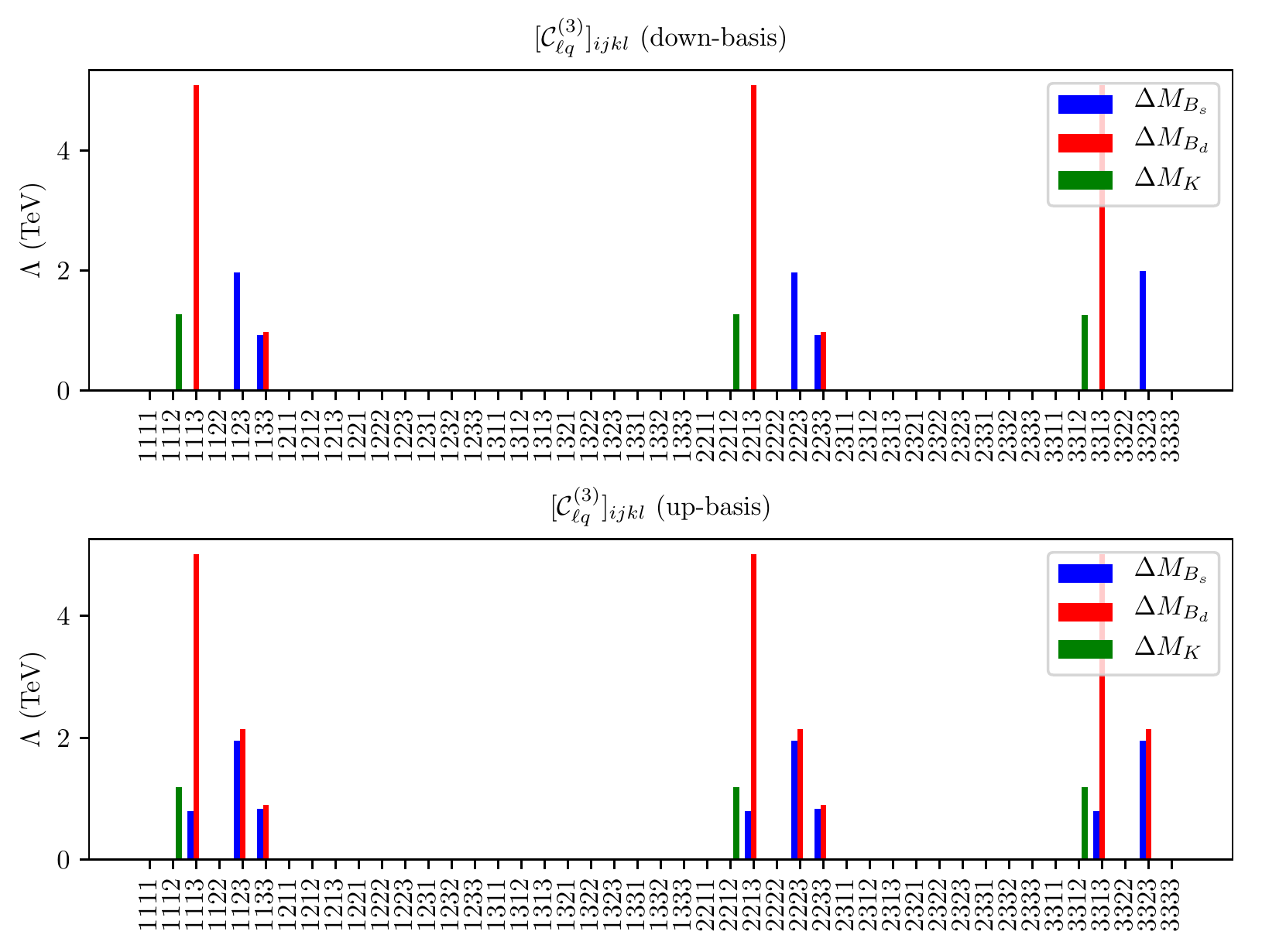}
\caption{\small
  The maximal NP scale $\muNP$ for $\bwc[(3)]{\ell q}{ijkl} = 10$ that
  corresponds to a $10\%$ effect in $2\big[M_{12}^{ij}\big]_\text{BSM}/
  (\Delta M_{ij})_\text{exp}$ for $B_s$ (blue), $B_d$ (red) and $K^0$
  (green), respectively.
}
\label{fig:lambda-lq3}
\end{figure}

%--------+---------+---------+---------+---------+---------+---------+---------+
%
%
%
%--------+---------+---------+---------+---------+---------+---------+---------+
\section{Simplified Models}
\label{sec:5}

In this section we discuss the tree-level models, which match onto one or several
operators relevant for the considered $\DF=2$ processes. Their complete matching
onto SMEFT can be found in \cite{deBlas:2017xtg}. \reftab{tab:scalarmodels} shows
all models with tree-level exchange of scalars that match onto relevant
four-quark operators. \reftab{tab:fermionmodels} lists the fermion and vector
models that match onto modified $Z$- and $W$ couplings or dipole operators at
tree-level. Finally, \reftab{tab:vectormodels} lists all tree-level mediated
vector models, which generate four-quark operators relevant for $\DF=2$
transitions. As examples we choose the models with vector bosons that are
singlets under $\text{SU(2)}_L \times \text{U(1)}_Y$ and with scalar bosons that
are doublets under $\text{SU(2)}_L$, i.e. for vector exchanges a colourless
heavy $Z'$ and a heavy coloured gluon $G'$ and similarly for scalar exchanges
a colourless $\varphi$ and a coloured $\Phi$.

\begin{table}[htb]
\centering
\begin{tabular}{clcccccccccc}
  \toprule
    Spin & Rep. &
    $\Op[(1)]{qq}$ &
    $\Op[(3)]{qq}$ &
    $\Op[(1)]{qd}$ &
    $\Op[(8)]{qd}$ &
    $\Op[(1)]{qu}$ &
    $\Op[(8)]{qu}$ &
    $\Op{dd}$ &
    $\Op{uu}$ &
    $\Op[(1)]{ud}$ &
    $\Op[(8)]{ud}$
    \\
    \midrule
    \multirow{9}{*}{$0$} & $\left(1,2\right)_{\frac 12}$ &&&$\times$&$\times$&$\times$&$\times$&&&  \\
    & $\left(3,1\right)_{-\frac 13}$ &$\times$&$\times$&&&&&&&$\times$& $\times$ \\
    & $\left(3,1\right)_{ \frac 23}$ &&&&&&&$\times$&&&  \\
    & $\left(3,1\right)_{-\frac 43}$ &&&&&&&&$\times$&&  \\
    & $\left(3,3\right)_{-\frac 13}$ &$\times$&$\times$&&&&&&&&  \\
    & $\left(6,1\right)_{ \frac 13}$ &$\times$&$\times$&&&&&&&&  \\
    & $\left(6,1\right)_{-\frac 23}$ &&&&&&&$\times$&&&  \\
    & $\left(6,1\right)_{ \frac 43}$ &&&&&&&&$\times$&&  \\
    & $\left(6,3\right)_{ \frac 13}$ &$\times$&$\times$&&&&&&&&  \\
    & $\left(8,2\right)_{ \frac 12}$ &&&$\times$&$\times$&$\times$&$\times$&&&&  \\
    \bottomrule
\end{tabular}
\caption{\small
  Four-quark $(\psi^4)$ operators generated from additional scalar fields.
}
\label{tab:scalarmodels}
\end{table}

\begin{table}[htb]
\centering
\begin{tabular}{clcccc}
  \toprule
    Spin & Rep. &
    $\Op[(1)]{\phi q}$ &
    $\Op[(3)]{\phi q}$ &
    $\Op{\phi d}$ &
    $\Op{uW}$
    \\
    \midrule
    \multirow{6}{*}{$\frac{1}{2}$} & $\left(3,1\right)_{\frac 23}$ &
    $\times$ & $\times$ \\
    & $\left(3,1\right)_{-\frac 13}$ &
    $\times$ & $\times$ \\
    & $\left(3,3\right)_{-\frac 13}$ &
    $\times$ & $\times$ \\
    & $\left(3,3\right)_{\frac 23}$ &
    $\times$ & $\times$ & & $\times$\\
    & $\left(3,2\right)_{\frac 16}$ &
    && $\times$ & $\times$ \\
    & $\left(3,2\right)_{-\frac 56}$ &
    && $\times$ & \\
    \midrule
    \multirow{2}{*}{$1$} & $\left(1,1\right)_{0}$ &$\times$&&$\times$  \\
    & $\left(1,3\right)_{0}$ &&$\times$& \\
    \bottomrule
\end{tabular}
\caption{\small
  $\psi^2\phi^2D$ and $\psi^2X\phi$ operators generated from additional
  fermion or vector fields.
}
\label{tab:fermionmodels}
\end{table}

\begin{table}[htb]
\centering
\begin{tabular}{clcccccccccc}
  \toprule
    Spin & Rep. &
    $\Op[(1)]{qq}$ &
    $\Op[(3)]{qq}$ &
    $\Op[(1)]{qd}$ &
    $\Op[(8)]{qd}$ &
    $\Op[(1)]{qu}$ &
    $\Op[(8)]{qu}$ &
    $\Op{dd}$ &
    $\Op{uu}$ &
    $\Op[(1)]{ud}$ &
    $\Op[(8)]{ud}$
    \\
    \midrule
    \multirow{9}{*}{$1$} & $\left(1,1\right)_{0}$ &$\times$&&$\times$&&$\times$&&$\times$&$\times$&$\times$&  \\
    & $\left(1,1\right)_{1}$ &&&&&&&&&$\times$&$\times$  \\
    & $\left(1,3\right)_{0}$ &&$\times$&&&&&&&&  \\
    & $\left(8,1\right)_{0}$ &$\times$&$\times$&&$\times$&&$\times$&$\times$&$\times$&& $\times$ \\
    & $\left(8,1\right)_{1}$ &&&&&&&&&$\times$&$\times$  \\
    & $\left(8,3\right)_{0}$ &$\times$&$\times$&&&&&&&&  \\
    & $\left(3,2\right)_{ \frac 16}$ &&&$\times$&$\times$&&&&&&  \\
    & $\left(3,2\right)_{-\frac 56}$ &&&&&$\times$&$\times$&&&&  \\
    & $\left(\bar 6,2\right)_{\frac 16}$ &&&$\times$&$\times$&&&&&&  \\
    & $\left(\bar 6,2\right)_{-\frac 56}$ &&&&&$\times$&$\times$&&&&  \\
    \bottomrule
\end{tabular}
\caption{\small
  Four-quark $(\psi^4)$ operators generated from additional vector fields.
}
\label{tab:vectormodels}
\end{table}

%
%
%
%--------+---------+---------+---------+---------+---------+---------+---------+
\subsection[$Z^\prime$ Model]
{\boldmath $Z^\prime$ Model}

The interaction Lagrangian of a $Z'=(1,1)_0$ field coupling to the quarks and
the SM Higgs doublet $\phi$ reads:
\begin{equation}\label{eq:Zplag}
  \mathcal{L}_{Z'} =
  - \Big[\, z_q^{ij} \, (\bar q^i \gamma^\mu q^j)
          + z_u^{ij} \, (\bar u^i \gamma^\mu u^j)
          + z_d^{ij} \, (\bar d^i \gamma^\mu d^j) \, \Big] \, Z'_{\mu}
    + z_\phi (\phi^\dag i \mathcal{D}^\mu \phi) Z'_{\mu}
    + \text{h.c.}\,.
\end{equation}
Matching this model onto the relevant SMEFT Wilson coefficients leads to the
following matching conditions at $\muNP$:
\begin{equation}
\begin{aligned}
  \wc[(1)]{qq}{ijkl}   & = -\frac{z_q^{ij} z_q^{kl}}{2 M_{Z'}^2} \,, &
  \wc[(1)]{qd}{ijkl}   & = -\frac{z_q^{ij} z_d^{kl}}{M_{Z'}^2}   \,, &
  \wc[(1)]{qu}{ijkl}   & = -\frac{z_q^{ij} z_u^{kl}}{M_{Z'}^2}   \,, \\
  \wc{dd}{ijkl}        & = -\frac{z_d^{ij} z_d^{kl}}{2 M_{Z'}^2} \,, &
  \wc{uu}{ijkl}        & = -\frac{z_u^{ij} z_u^{kl}}{2 M_{Z'}^2} \,, &
  \wc[(1)]{ud}{ijkl}   & = -\frac{z_u^{ij} z_d^{kl}}{M_{Z'}^2}   \,, \\
  \wc[(1)]{\phi q}{ij} & = -\frac{\re(z_\phi)z_q^{ij}}{M_{Z'}^2}\,, &
  \wc{\phi d}{ij}      & = -\frac{\re(z_\phi)z_d^{ij}}{M_{Z'}^2}\,.
\end{aligned}
\end{equation}
These equations imply the following tree-level relations between different
coefficients valid at the NP matching scale that are independent of $M_{Z'}$
and are universal in all meson systems considered ($a=u,d$)
\begin{align}
  \wc[(1)]{qa}{ijij}^2 & = 4 \wc[(1)]{qq}{ijij}\wc{aa}{ijij} \,, &
  \wc[(1)]{ud}{ijij}^2 & = 4 \wc{dd}{ijij}   \wc{uu}{ijij} \,.
\end{align}
Albeit they are generally modified through RG effects and matching
at one-loop level.

We note that the coefficients $\wc[(8)]{qa}{ijij}$
are absent in this list, which in the BMU basis  implies
$\WcL[ij]{\text{LR},2} = 0$, while $\WcL[ij]{\text{LR},1}$ are non-vanishing.
But including QCD corrections we find in the NDR scheme using the results
in \cite{Buras:2012fs} at $\muNP$
\begin{align}
  \wc[(1)]{qa}{ijij} &
  = - \frac{z_q^{ij}z_a^{ij}}{M_{Z'}^2}
    \left(1 - \frac{\alS}{4\pi} \left[
            2 \ln\frac{M_{Z'}^2}{\muNP^2} + \frac{1}{3} \right] \right) \,, &
  (a = u,d)
\\
  \wc[(8)]{qa}{ijij} &
  = \frac{\alS}{4\pi} \frac{z_q^{ij}z_a^{ij}}{M_{Z'}^2}
    \left[ 6 \ln\frac{M_{Z'}^2}{\muNP^2} + 1 \right] \,,
\end{align}
and this implies non-vanishing $\WcL[ij]{\text{LR},2}$.
The presence of the logarithms cancels the dependence on the
choice of the matching scale $\muNP$ present in the NLO RG evolution and the
constant terms remove the corresponding renormalization scheme dependence.

Using the master formulae of the previous section we find following master
formulae for the $Z^\prime$ model in the {\bf down-basis}
\begin{align*}
  \frac{M_{Z^\prime}^2\, \Sigma^{B_s}_{Z^\prime}}{(5\TeV)^2} & =
  -2.3\cdt{3}z_q^{23}z_d^{23} +2.1\cdt{2}z_d^{23}z_d^{23}
  +2.0\cdt{2}z_q^{23}z_q^{23} -3.3 z_q^{33}z_d^{23} +1.8z_q^{22}z_d^{23}
  \nlineS{-0.3cm}
  -1.2\re (z_\phi) z_d^{23} +1.0z_u^{33}z_d^{23}
  -4.0\cdt{-1}e^{i22^\circ}z_q^{12}z_d^{32} +2.7\cdt{-1}z_q^{23}z_q^{33}
  \nline
  -2.6\cdt{-1}z_q^{23}z_u^{33} +2.4\cdt{-1}\re (z_\phi) z_q^{23}
  -1.5\cdt{-1}z_q^{22}z_q^{23} +1.1\cdt{-1}z_q^{23}z_d^{22}
  \nline
  -1.1\cdt{-1}z_q^{23}z_d^{33} -1.1\cdt{-1}z_u^{23}z_d^{23}
  +3.4\cdt{-2}e^{i22^\circ}z_q^{12}z_q^{32} +3.2\cdt{-2}z_q^{23}z_u^{23}
  \nline
  -2.5\cdt{-2}z_q^{23}z_d^{32} +2.3\cdt{-2}e^{-i23^\circ}z_q^{13}z_d^{23}
  +1.1\cdt{-2}z_d^{23}z_d^{33} -9.3\cdt{-3}z_d^{22}z_d^{23}
  \nline
   +5.5\cdt{-3}e^{i21^\circ} z_q^{13}z_d^{32} \,,
  \numberthis
\\
  \frac{M_{Z^\prime}^2\, \Sigma^{B_d}_{Z^\prime}}{(5\TeV)^2} & =
  -5.7\cdt{4}z_q^{13}z_d^{13} +4.8\cdt{3}z_d^{13}z_d^{13}
  +4.5\cdt{3}z_q^{13}z_q^{13} +4.5\cdt{1}z_q^{12}z_d^{13}
  \nlineS{-0.3cm}
  +1.8\cdt{1}e^{i22^\circ}z_q^{33}z_d^{13} -9.8e^{i22^\circ}z_q^{11}z_d^{13}
  +6.6e^{i22^\circ}\re (z_\phi) z_d^{13} -5.6e^{i22^\circ}z_u^{33}z_d^{13}
  \nline
  -3.6z_q^{12}z_q^{13} +2.8z_q^{13}z_d^{12}
  -1.4e^{i22^\circ}z_q^{13}z_q^{33} +1.3e^{i22^\circ}z_q^{13}z_u^{33}
  -1.2e^{i22^\circ} \re (z_\phi) z_q^{13}
  \nline
  +7.8\cdt{-1}e^{i22^\circ}z_q^{11}z_q^{13}
  -7.6\cdt{-1} e^{i23^\circ} z_q^{23}z_d^{13}
  +6.2\cdt{-1}z_u^{23}z_d^{13} -2.4\cdt{-1}z_d^{12}z_d^{13}
  \nline -1.7\cdt{-1}z_q^{13}z_u^{23}
  +1.4\cdt{-1}e^{i21^\circ}z_q^{23}z_d^{31}
  -6.5\cdt{-2} e^{i23^\circ} z_q^{13}z_d^{23}
  +5.9\cdt{-2}e^{i23^\circ}z_q^{13}z_q^{23}
  \nline
  +3.1\cdt{-2}e^{i21^\circ}z_q^{13}z_d^{33}
  -3.1\cdt{-2} e^{i21^\circ}z_q^{13}z_d^{11}
  -3.0\cdt{-2} e^{i44^\circ}z_q^{13}z_d^{31}
  \nline
  -1.2\cdt{-2} e^{i21^\circ}z_q^{13}z_q^{32}
  -7.2\cdt{-3}e^{i22^\circ}z_q^{22}z_d^{13}
  -7.2\cdt{-3}e^{i22^\circ}z_d^{13}z_d^{33}
  \nline
  +5.5\cdt{-3}e^{i23^\circ}z_d^{13}z_d^{23}
  -5.4\cdt{-3} z_u^{13}z_d^{13}
  -5.3\cdt{-3}e^{i22^\circ}z_d^{12}z_d^{23} \,,
  \numberthis
\\
  \frac{M_{Z^\prime}^2\, \Sigma^{K}_{Z^\prime}}{(5\TeV)^2} & =
  -5.3\cdt{6}z_q^{12}z_d^{12} +1.8\cdt{4}z_d^{12}z_d^{12}
  +1.8\cdt{4}z_q^{12}z_q^{12} -9.0\cdt{3}z_q^{13}z_d^{12}
  \nlineS{-0.3cm}
  +2.0\cdt{3}e^{i22^\circ}z_q^{23}z_d^{21} -2.6\cdt{2}z_q^{12}z_d^{13}
  -8.1\cdt{1} e^{i24^\circ}z_q^{22}z_d^{12}
  +3.4\cdt{1}e^{i26^\circ}z_q^{11}z_d^{12}
  \nline
  +3.0\cdt{1}z_q^{12}z_q^{13}
  -2.5\cdt{1} e^{i23^\circ}\re (z_\phi) z_d^{12}
  +2.0\cdt{1}e^{i23^\circ}z_u^{33}z_d^{12}
  +1.7\cdt{1}e^{i23^\circ}z_q^{33}z_d^{12}
  \nline
  -6.6e^{i22^\circ}z_q^{12}z_q^{32} +2.8e^{i22^\circ}z_q^{12}z_d^{32}
  +2.5e^{i66^\circ}z_q^{12}z_d^{11} -2.5 e^{i66^\circ}z_q^{12}z_d^{22}
  -2.1z_u^{23}z_d^{12}
  \nline
  -2.0e^{i22^\circ}z_u^{23}z_d^{21} +9.0\cdt{-1}z_d^{12}z_d^{13}
  -7.2\cdt{-1} e^{i24^\circ}z_q^{23}z_d^{12} -4.4\cdt{-1}z_q^{13}z_d^{13}
  \nline
  +2.8\cdt{-1}e^{i23^\circ}z_q^{12}z_q^{22}
  -2.2\cdt{-1} e^{i23^\circ}z_q^{12}z_u^{33}
  +2.1\cdt{-1}e^{i23^\circ}\re (z_\phi) z_q^{12}
  \nline
  +1.9\cdt{-1}z_u^{22}z_d^{12}
  -1.3\cdt{-1} e^{i23^\circ}z_q^{11}z_q^{12}
  +1.1\cdt{-1}e^{i45^\circ}z_q^{13}z_d^{21}
  +9.6\cdt{-2}e^{i23^\circ}z_q^{23}z_d^{31}
  \nline
  -6.1\cdt{-2} e^{i23^\circ}z_q^{12}z_q^{33}
  +5.1\cdt{-2}z_q^{13}z_q^{13} +4.1\cdt{-2}e^{i23^\circ}z_d^{12}z_d^{22}
  +2.8\cdt{-2}z_q^{12}z_u^{23}
  \nline
  +2.6\cdt{-2}e^{i22^\circ}z_q^{12}z_u^{32}
  +2.4\cdt{-2}e^{i22^\circ}z_d^{13}z_d^{32}
  +1.8\cdt{-2}z_u^{13}z_d^{12}
  -1.3\cdt{-2} e^{i23^\circ}z_q^{13}z_q^{32}
  \nline
  -1.0\cdt{-2} e^{i24^\circ}z_q^{13}z_d^{22}
  +1.0\cdt{-2}e^{i24^\circ}z_q^{13}z_d^{11}
  -9.5\cdt{-3}e^{i22^\circ}z_d^{12}z_d^{32}
  \nline
  +8.1\cdt{-3}e^{i23^\circ}z_u^{11}z_d^{12} \,.
  \numberthis
\end{align*}
The corresponding expressions in {\bf the up-basis} are given in \refapp{NPMFup}.
Although the value of $M_{Z'}$ has been kept arbitrary, it should
be of the order of $\muNP$ to avoid the appearance of large logarithms
$\ln \muNP/M_{Z'}$. The same applies to the other simplified models.

%
%
%
%--------+---------+---------+---------+---------+---------+---------+---------+
\subsection[$G^\prime$ Model]
{\boldmath $G^\prime$ Model}

The interaction Lagrangian of a $G'=(8,1)_0$ field coupling to the quarks reads:
\begin{equation}
  \mathcal{L}_{G'}
  = - \Big[ \,
      g_q^{ij} \, (\bar q^i \gamma^\mu T^A q^j)
    + g_u^{ij} \, (\bar u^i \gamma^\mu T^A u^j)
    + g_d^{ij} \, (\bar d^i \gamma^\mu T^A d^j) \,\Big] \, G'^A_{\mu} \,.
\end{equation}
Matching this model onto the relevant SMEFT Wilson coefficients leads to the
following tree-level matching conditions at $\muNP$
\begin{align}
  \wc[(1)]{qq}{ijkl} &
  = \frac{g_q^{ij}g_q^{kl}}{12M_{G'}^2} -\frac{g_q^{il}g_q^{kj}}{8M_{G'}^2} \,,
&
  \wc[(3)]{qq}{ijkl} &
  = -\frac{g_q^{il} g_q^{kj}}{8M_{G'}^2} \,,
\notag \\
  \wc[(8)]{qd}{ijkl} &
  = -\frac{g_q^{ij}g_d^{kl}}{M_{G'}^2} \,,
&
  \wc[(8)]{qu}{ijkl} &
  = -\frac{g_q^{ij}g_u^{kl}}{M_{G'}^2}
\\
  \wc[]{dd}{ijkl} &
  = \frac{g_d^{ij} g_d^{kl}}{12 M_{G'}^2} - \frac{g_d^{il} g_d^{kj}}{4 M_{G'}^2} \,,
&
  \wc[]{uu}{ijkl} &
  = \frac{g_u^{ij} g_u^{kl}}{12 M_{G'}^2} - \frac{g_u^{il} g_u^{kj}}{4 M_{G'}^2} \,,
&
  \wc[(8)]{ud}{ijkl} &
  = -\frac{g_u^{ij}g_d^{kl}}{M_{G'}^2} \,.
\notag
\end{align}
These equations imply the following tree-level relations between different
coefficients that are independent of $M_{G'}$ and are universal in all meson
systems considered
\begin{equation}
\begin{aligned}
  \wc[(8)]{ud}{ijij}^2 & =  36 \wc[]{dd}{ijij}    \wc[]{uu}{ijij} \,, & \qquad\quad
  \wc[(8)]{qd}{ijij}^2 & = 144 \wc[(1)]{qq}{ijij} \wc[]{dd}{ijij} \,, \\
  \wc[(8)]{qu}{ijij}^2 & = 144 \wc[(1)]{qq}{ijij} \wc[]{uu}{ijij} \,, &
  \wc[(3)]{qq}{ijij}   & =   3 \wc[(1)]{qq}{ijij} \,.
\end{aligned}
\end{equation}

In this model we find in the {\bf down-basis}:
\begin{align*}
  \frac{M_{G'}^2 \Sigma^{B_s}_{G'}}{(5 \TeV)^2} & =
  -2.7\cdt{3}g_q^{23}g_d^{23} +6.9\cdt{1}g_d^{23}g_d^{23}
  +6.5\cdt{1}g_q^{23}g_q^{23} -3.8g_q^{33}g_d^{23}
  +2.1g_q^{22}g_d^{23}
  \nlineS{-0.3cm}
  +1.5g_u^{33}g_d^{23}
  -4.7\cdt{-1}e^{i22^\circ}g_q^{12}g_d^{32}
  -1.6\cdt{-1}g_u^{23}g_d^{23} +1.3\cdt{-1}g_q^{23}g_d^{22}
  \nline
  -1.3\cdt{-1}g_q^{23}g_d^{33} -1.1\cdt{-1}g_q^{23}g_u^{33}
  +8.7\cdt{-2}g_q^{23}g_q^{33} -5.2\cdt{-2}g_q^{23}g_q^{22}
  \nline
  -2.7\cdt{-2}g_q^{23}g_d^{32}
  +2.6\cdt{-2}e^{-i23^\circ}g_q^{13}g_d^{23}
  +1.4\cdt{-2}g_q^{23}g_u^{23}
  \nline
  +1.1\cdt{-2}e^{i22^\circ}g_q^{12}g_q^{32}
  +5.8\cdt{-3}e^{i21^\circ}g_q^{13}g_d^{32} \,,
 \numberthis
\\
  \frac{M_{G'}^2 \Sigma^{B_d}_{G'}}{(5 \TeV)^2} & =
  -6.6\cdt{4}g_q^{13}g_d^{13}
  +1.6\cdt{3}g_d^{13}g_d^{13}
  +1.5\cdt{3}g_q^{13}g_q^{13}
  +5.2\cdt{1}g_q^{12}g_d^{13}
  \nlineS{-0.3cm}
  +2.1\cdt{1}e^{i22^\circ}g_q^{33}g_d^{13}
  -1.1\cdt{1}e^{i22^\circ}g_q^{11}g_d^{13}
  -7.9e^{i22^\circ}g_u^{33}g_d^{13} +3.2g_q^{13}g_d^{12}
  \nline
  -1.2g_q^{13}g_q^{12}
  -8.9\cdt{-1}e^{i23^\circ}g_q^{23}g_d^{13}
  +8.7\cdt{-1}g_u^{23}g_d^{13}
  +5.2\cdt{-1}e^{i22^\circ}g_q^{13}g_u^{33}
  \nline
  -4.7\cdt{-1}e^{i21^\circ}g_q^{13}g_q^{33}
  +2.7\cdt{-1}e^{i22^\circ}g_q^{13}g_q^{11}
  +1.4\cdt{-1}e^{i21^\circ}g_q^{23}g_d^{31}
  \nline
  -7.8\cdt{-2}g_d^{13}g_d^{12}
  -7.5\cdt{-2}e^{i23^\circ}g_q^{13}g_d^{23}
  -7.5\cdt{-2}g_q^{13}g_u^{23}
  \nline
  -3.6\cdt{-2}e^{i21^\circ}g_q^{13}g_d^{11}
  +3.5\cdt{-2}e^{i21^\circ}g_q^{13}g_d^{33}
  -3.1\cdt{-2}e^{i44^\circ}g_q^{13}g_d^{31}
  \nline
  +1.9\cdt{-2}e^{i23^\circ}g_q^{13}g_q^{23}
  -1.8\cdt{-2}e^{i22^\circ}g_q^{22}g_d^{13}
  -7.6\cdt{-3}g_u^{13}g_d^{13}
  \nline
  -5.5\cdt{-3}e^{i22^\circ}g_u^{11}g_d^{13}
  -5.4\cdt{-3}e^{i22^\circ}g_u^{22}g_d^{13} \,,
  \numberthis
\\
  \frac{M_{G'}^2 \Sigma^{K}_{G'}}{(5 \TeV)^2} & =
  -7.5\cdt{6}g_q^{12}g_d^{12} -1.3\cdt{4}g_q^{13}g_d^{12}
  +6.1\cdt{3}g_d^{12}g_d^{12} +5.9\cdt{3}g_q^{12}g_q^{12}
  \nlineS{-0.3cm}
  +2.8\cdt{3}e^{i22^\circ}g_q^{23}g_d^{21} -3.7\cdt{2}g_q^{12}g_d^{13}
  -1.1\cdt{2}e^{i24^\circ}g_q^{22}g_d^{12} +4.7\cdt{1}e^{i27^\circ}g_q^{11}g_d^{12}
  \nline
  +3.8\cdt{1}e^{i23^\circ}g_u^{33}g_d^{12}
  +2.3\cdt{1}e^{i23^\circ}g_q^{33}g_d^{12}
  +1.0\cdt{1}g_q^{13}g_q^{12}
  -4.2g_u^{23}g_d^{12}
  \nline
  -4.0e^{i22^\circ}g_u^{23}g_d^{21}
  +4.0e^{i22^\circ}g_q^{12}g_d^{32}
  +3.4e^{-i79^\circ}g_q^{12}g_d^{22}
  -3.4e^{-i79^\circ}g_q^{12}g_d^{11}
  \nline
  -2.2e^{i22^\circ}g_q^{12}g_q^{32}
  -9.7\cdt{-1}e^{i24^\circ}g_q^{23}g_d^{12}
  -6.3\cdt{-1}g_q^{13}g_d^{13}
  \nline
  +3.9\cdt{-1}g_u^{22}g_d^{12}
  +3\cdt{-1}g_d^{12}g_d^{13}
  +1.5\cdt{-1}e^{i45^\circ}g_q^{13}g_d^{21}
  +1.4\cdt{-1}e^{i23^\circ}g_q^{23}g_d^{31}
  \nline
  +9.5\cdt{-2}e^{i23^\circ}g_q^{12}g_q^{22}
  -9.2\cdt{-2}e^{i23^\circ}g_q^{12}g_u^{33}
  -4.2\cdt{-2}e^{i23^\circ}g_q^{12}g_q^{11}
  \nline
  -3.9\cdt{-2}e^{i23^\circ}g_q^{13}g_q^{32}
  +3.6\cdt{-2}g_u^{13}g_d^{12}
  +2.5\cdt{-2}e^{i22^\circ}g_u^{11}g_d^{12}
  \nline
  +1.7\cdt{-2}g_q^{13}g_q^{13}
  -1.5\cdt{-2}e^{i24^\circ}g_q^{13}g_d^{22}
  +1.4\cdt{-2}e^{i24^\circ}g_q^{13}g_d^{11}
  \nline
  +1.4\cdt{-2}e^{i23^\circ}g_d^{12}g_d^{22}
  +1.2\cdt{-2}g_q^{12}g_u^{23}
  +1.2\cdt{-2}e^{i22^\circ}g_d^{12}g_d^{33}
  \nline
  +1.1\cdt{-2}e^{i22^\circ}g_q^{12}g_u^{32}
  +9.7\cdt{-3}e^{i23^\circ}g_q^{12}g_q^{33}
  +6.7\cdt{-3}e^{i23^\circ}g_q^{13}g_d^{32}
  \nline
  -6.3\cdt{-3}e^{i46^\circ}g_q^{12}g_d^{21}
  -6.2\cdt{-3}e^{i23^\circ}g_q^{22}g_d^{13}
-5.4\cdt{-3}e^{i22^\circ}g_d^{13}g_d^{32} \,.
  \numberthis
\end{align*}
The corresponding expressions in the up-basis are given in \refapp{NPMFup}.

%
%
%
%--------+---------+---------+---------+---------+---------+---------+---------+
\subsection{Colourless Scalar Model}

The interaction Lagrangian of a $\varphi=(1,2)_{1/2}$ scalar field coupling to
the quarks reads:
\begin{align}
  \label{eq:varphimatch}
  \mathcal L_\varphi &
  = - Y_d^{ij} (\bar q_i  d_j) \,\varphi
    - Y_u^{ij} (\bar q_i  u_j) \,\wT\varphi
    + \text{h.c.}
&
  \wT\varphi & \;\equiv i \sigma_2 \varphi^* \,.
\end{align}
Integrating out the heavy scalar leads to the following tree-level matching
conditions for the four-quark SMEFT operators \cite{deBlas:2017xtg} at $\muNP$
\begin{align}
  \label{matchscalar}
  \wc[(8)]{qu}{ijkl} &
  = 6 \wc[(1)]{qu}{ijkl}
  = - \, \frac{Y_{u}^{jk} Y_{u}^{il*}}{M_\varphi^{2}} \,,
&
  \wc[(8)]{qd}{ijkl} &
  = 6 \wc[(1)]{qd}{ijkl}
  = - \, \frac{Y_{d}^{li} Y_{d}^{kj*}}{M_\varphi^{2}} .
\end{align}

In the BMU basis these results imply $\WcL[ij]{\text{LR},1}=0$ and only
$\WcL[ij]{\text{LR},2}$ are non-vanishing. However, it is evident from the
charts in the previous section that QCD RG evolution will generate non-vanishing
$\WcL[ij]{\text{LR},1}$ at different scales. This is already seen when QCD
corrections to the matching in \eqref{matchscalar} are extracted from
\cite{Buras:2012fs}. In the NDR scheme we find
\begin{align}
 \wc[(1)]{qa}{ijij} &
 = - \frac{1}{6}\, \frac{Y_{a}^{ji} Y_{a}^{ij*}}{M_\varphi^{2}}
   \left(1 - \frac{5}{2}\frac{\alS}{4\pi} \right)\,,
&
 \wc[(8)]{qa}{ijij} &
 = - \, \frac{Y_{a}^{ji} Y_{a}^{ij*}}{M_\varphi^{2}}
   \left(1 - \frac{\alS}{4\pi} \right) \,,
\end{align}
so that the relation in \eqref{matchscalar} is violated and non-vanishing
$\WcL[ij]{\text{LR},1}$ are generated. These QCD corrections cancel the
renormalization scheme dependence present in two-loop anomalous dimensions
of the operators in question that enter the RG evolution at the NLO level.
One should note that no logarithms involving the NP scale are present in
these corrections. The reason for this is explained in \cite{Buras:2012fs}.

We find for the master formulae in the {\bf down-basis:}
\begin{align*}
  \frac{M_{\varphi}^2\, \Sigma^{B_s}_{\varphi}}{(5\TeV)^2} & =
  -3.1\cdt{3}Y_d^{32}Y_d^{23*} -4.4Y_d^{33}Y_d^{23*} +2.4Y_d^{32}Y_d^{22*}
  \nlineS{-0.3cm}
  -5.3\cdt{-1}e^{i22^\circ}Y_d^{21}Y_d^{32*}
  -1.5\cdt{-1}Y_u^{33}Y_u^{23*} +1.5\cdt{-1}Y_d^{22}Y_d^{23*}
  \nline
  -1.5\cdt{-1}Y_d^{32}Y_d^{33*} -3.1\cdt{-2}Y_d^{22}Y_d^{33*}
  +3.0\cdt{-2}e^{-i23^\circ}Y_d^{31}Y_d^{23*}
  \nline
  +1.9\cdt{-2}Y_u^{32}Y_u^{23*} +6.7\cdt{-3}e^{i21^\circ}Y_d^{21}Y_d^{33*} \,,
  \numberthis
\\
  \frac{M_{\varphi}^2\, \Sigma^{B_d}_{\varphi}}{(5\TeV)^2} & =
  -7.5\cdt{4}Y_d^{31}Y_d^{13*} +5.9\cdt{1}Y_d^{31}Y_d^{12*}
  +2.4\cdt{1}e^{i22^\circ}Y_d^{33}Y_d^{13*}
  \nlineS{-0.3cm}
  -1.3\cdt{1}e^{i22^\circ}Y_d^{31}Y_d^{11*}
  +3.7Y_d^{21}Y_d^{13*} -1.0 e^{i23^\circ}Y_d^{32}Y_d^{13*}
  \nline
  +7.4\cdt{-1}e^{i22^\circ}Y_u^{33}Y_u^{13*}
  +1.7\cdt{-1}e^{i21^\circ}Y_d^{12}Y_d^{33*}
  -1.0\cdt{-1}Y_u^{32}Y_u^{13*}
  \nline
  -8.5\cdt{-2} e^{i23^\circ}Y_d^{31}Y_d^{23*}
  -4.2\cdt{-2} e^{i21^\circ}Y_d^{11}Y_d^{13*}
  +4.1\cdt{-2}e^{i21^\circ}Y_d^{31}Y_d^{33*}
  \nline
  -3.6\cdt{-2} e^{i44^\circ}Y_d^{11}Y_d^{33*}
  -1.9\cdt{-2} e^{i22^\circ}Y_d^{32}Y_d^{12*} \,,
  \numberthis
\\
  \frac{M_{\varphi}^2\, \Sigma^{K}_{\varphi}}{(5\TeV)^2} & =
  -8.4\cdt{6}Y_d^{21}Y_d^{12*} -1.4\cdt{4}Y_d^{21}Y_d^{13*}
  +3.1\cdt{3}e^{i22^\circ}Y_d^{12}Y_d^{23*}
  \nlineS{-0.3cm}
  -4.1\cdt{2}Y_d^{31}Y_d^{12*}
  -1.3\cdt{2} e^{i24^\circ}Y_d^{22}Y_d^{12*}
  +5.2\cdt{1}e^{i27^\circ}Y_d^{21}Y_d^{11*}
  \nline
  +2.6\cdt{1}e^{i23^\circ}Y_d^{23}Y_d^{13*} +4.4e^{i22^\circ}Y_d^{21}Y_d^{32*}
  +3.7e^{-i82^\circ}Y_d^{21}Y_d^{22*}
  \nline
  -3.7e^{-i82^\circ}Y_d^{11}Y_d^{12*}
  -1.1 e^{i24^\circ}Y_d^{22}Y_d^{13*} -7.0\cdt{-1}Y_d^{31}Y_d^{13*}
  \nline
  +1.7\cdt{-1} e^{i45^\circ}Y_d^{11}Y_d^{23*}
  +1.5\cdt{-1} e^{i23^\circ}Y_d^{12}Y_d^{33*}
  -1.3\cdt{-1} e^{i23^\circ}Y_u^{23}Y_u^{13*}
  \nline
  +1.7\cdt{-2}Y_u^{22}Y_u^{13*}
  -1.6\cdt{-2} e^{i24^\circ}Y_d^{21}Y_d^{23*}
  +1.6\cdt{-2} e^{i24^\circ}Y_d^{11}Y_d^{13*}
  \nline
  +1.6\cdt{-2}e^{i22^\circ}Y_u^{23}Y_u^{12*}
  +7.5\cdt{-3}e^{i23^\circ}Y_d^{21}Y_d^{33*}
  -7.1\cdt{-3}e^{i46^\circ}Y_d^{11}Y_d^{22*}
  \nline
  -6.9\cdt{-3} e^{i23^\circ}Y_d^{32}Y_d^{12*} \,.
  \numberthis
\end{align*}
The corresponding expressions in the up-basis are given in \refapp{NPMFup}.

%
%
%
%--------+---------+---------+---------+---------+---------+---------+---------+
\subsection{Coloured Scalar Model}

The interaction Lagrangian of a $\Phi=(8,2)_{1/2}$ scalar field coupling to the
quarks reads:
\begin{align}
  \mathcal L_\Phi &
  = - X_d^{ij} (\bar q_i \,T^A d_j) \,\Phi^A
    - X_u^{ij} (\bar q_i \,T^A u_j) \,\wT \Phi^A
    + \text{h.c.} \,,
&
  \wT\Phi^A & \equiv i \sigma_2 (\Phi^A)^*  \,.
  \label{eq:LagPhi}
\end{align}
Integrating out the heavy scalar leads to the following tree-level matching
conditions for the four-quark SMEFT operators
\cite{deBlas:2017xtg} at $\muNP$
\begin{align}
  \frac{4}{3} \wc[(8)]{qu}{ijkl} &
  = - \wc[(1)]{qu}{ijkl}
  = \frac{2}{9} \frac{X_{u}^{jk} X_{u}^{il*}}{M_\Phi^{2}} \,,
&
  \frac{4}{3}  \wc[(8)]{qd}{ijkl} &
  = - \wc[(1)]{qd}{ijkl}
  = \frac{2}{9} \frac{X_{d}^{li} X_{d}^{kj*}}{M_\Phi^{2}} .
\end{align}
Note that this time the relation between the two coefficients differs from
\eqref{matchscalar} and  both $\WcL[ij]{\text{LR},1}$ and $\WcL[ij]{\text{LR},2}$
are non-vanishing already at tree-level.
For the master formula we find in {\bf the down-basis}:
\begin{align*}
  \frac{M_{\Phi}^2\, \Sigma^{B_s}_{\Phi}}{(5 \TeV)^2} & =
  -6.8\cdt{1}X_d^{32}X_d^{23*} -8.6\cdt{-2}X_d^{33}X_d^{23*}
  +5.3\cdt{-2}X_d^{32}X_d^{22*}
  \nlineS{-0.3cm}
  -4.1\cdt{-2}X_u^{33}X_u^{23*}
  -1.2\cdt{-2}e^{i22^\circ}X_d^{21}X_d^{32*} \,,
  \numberthis
\\
  \frac{M_{\Phi}^2\, \Sigma^{B_d}_{\Phi}}{(5 \TeV)^2} & =
  -1.7\cdt{3}X_d^{31}X_d^{13*} +1.3X_d^{31}X_d^{12*}
  +4.7\cdt{-1}e^{i22^\circ}X_d^{33}X_d^{13*}
  \nlineS{-0.3cm}
  -2.9\cdt{-1}e^{i22^\circ}X_d^{31}X_d^{11*}
  +2.0\cdt{-1}e^{i22^\circ}X_u^{33}X_u^{13*}
  +8.2\cdt{-2}X_d^{21}X_d^{13*}
  \nline
  -2.6\cdt{-2}X_u^{32}X_u^{13*}
  -2.0\cdt{-2} e^{i23^\circ}X_d^{32}X_d^{13*}
  +6.2\cdt{-3} e^{i21^\circ}X_d^{12}X_d^{33*} \,,
  \numberthis
\\
  \frac{M_{\Phi}^2\, \Sigma^{B_K}_{\Phi}}{(5 \TeV)^2} & =
  8.3\cdt{4}X_d^{21}X_d^{12*} +1.4\cdt{2}X_d^{21}X_d^{13*}
  -3.1\cdt{1}e^{i22^\circ}X_d^{12}X_d^{23*}
  \nlineS{-0.3cm}
  +4.1X_d^{31}X_d^{12*} +9.5\cdt{-1}e^{i33^\circ}X_d^{22}X_d^{12*}
  -3.4\cdt{-1}e^{-i6^\circ}X_d^{21}X_d^{22*}
  \nline
  +3.4\cdt{-1}e^{-i6^\circ}X_d^{11}X_d^{12*}
  -2.7\cdt{-1}e^{i61^\circ}X_d^{21}X_d^{11*}
  -4.4\cdt{-2}e^{i22^\circ}X_d^{21}X_d^{32*}
  \nline
  -4.0\cdt{-2} e^{i24^\circ}X_d^{23}X_d^{13*}
  -3.3\cdt{-2} e^{i23^\circ}X_u^{23}X_u^{13*}
  +6.9\cdt{-3}X_d^{31}X_d^{13*} \,.
  \numberthis
\end{align*}
The corresponding expressions in the up-basis are given in \refapp{NPMFup}.

%
%
%
%--------+---------+---------+---------+---------+---------+---------+---------+
\subsection{A Closer Look at NP Scenarios}

Let us next get a better insight into different NP scenarios by collecting in
\reftab{tab:model-couplings} the values of flavour violating couplings for
$M_{Z^\prime, G^\prime, \varphi, \Phi} = 5\TeV$ that give rise to $20\%$
NP corrections to $2 \big [M_{12}^{ij} \big]_\text{BSM}$ that is $0.2$ for the sum
entering \refeq{eq:master-M12SMEFT}. To this end we keep only the contribution
with the largest $P_a^{ij}(\muNP)$ in our results. These are the ones which come
directly from $\DF=2$ operators or from left-right operators, which for the
$Z^\prime$ and $G^\prime$ models involve the products $z_d^{ij} z_{q}^{ij}$
with $ij = 12,13,23$ for $K^0$, $B_d$ and $B_s$, respectively. In order
to get an idea of the size of the couplings we first take them to be real and
assume the relations
\begin{align}
  z_d^{ij} & = z_q^{ij} , &
  g_d^{ij} & = g_q^{ij} , &
  X_d^{ij} & = X_d^{ji} , &
  Y_d^{ij} & = Y_d^{ji} ,
\end{align}
that we will relax soon.

\begin{table}[htb]
\centering
\renewcommand{\arraystretch}{1.3}
\begin{tabular}{|c||ccc|cccccccccc|}
\hline
  Model & \multicolumn{3}{c|}{Couplings}
\\
\hline\hline
  $Z^\prime$ & $z_q^{12}$ & $z_q^{13}$ & $z_q^{23}$ \\
& $2.0\cdt{-4}$   &$1.9\cdt{-3}$& $9.3\cdt{-3}$     \\
\hline
  $G^\prime$ & $g_q^{12}$ & $g_q^{13}$ & $g_q^{23}$ \\
& $1.6\cdt{-4}$   &$1.7\cdt{-3}$& $8.6\cdt{-3}$     \\
\hline
  $\varphi$ &     $Y_{d}^{12}$ & $Y_{d}^{13}$ & $Y_{d}^{23}$ \\
& $1.5\cdt{-4}$   &$1.6\cdt{-3}$& $8.0\cdt{-3}$  \\
\hline
  $\Phi$ &  $X_{d}^{12}$ & $X_{d}^{13}$ & $X_{d}^{23}$ \\
& $1.6\cdt{-3}$   &$1.1\cdt{-2}$& $5.4\cdt{-2}$        \\
\hline
\end{tabular}
\renewcommand{\arraystretch}{1.0}
\caption{\small
  The values of flavour-violating couplings for $M_{Z^\prime,\, G^\prime,\,
  \varphi,\, \Phi} = 5\TeV$ that give rise to $20\%$ NP corrections to
  $2 \big [M_{12}^{ij} \big]_\text{BSM}$.
}
  \label{tab:model-couplings}
\end{table}

As the values of flavour-violating couplings turn out to be small the
question arises whether non-leading terms with smaller $P_a^{ij}(\muNP)$,
or equivalently smaller numerical coefficients multiplying the products of
couplings, could play a role, in particular those in which flavour-conserving
couplings are present. While a detailed analysis would require a simultaneous
study of $\DF=1$ transitions it is of interest to see whether other terms
generated by the RG evolution play eventually any role in the estimate of
NP contributions to $\DF=2$ observables. As we will see soon this indeed
can be the case provided flavour-conserving couplings are sufficiently large
but still in a perturbative regime. We will assume such couplings to be
at most $\sim 3$.

Evidently the outcome of such an analysis depends on the scenarios for
couplings considered and it is common in the case of $Z^\prime$ and
$G^\prime$ scenarios to investigate the following scenarios for couplings:
\begin{itemize}
\item
  {\bf Left-handed Scenario (LHS)} in which only coefficients involving
  the couplings $z_q^{ij}$ or $g_q^{ij}$ are kept non-zero.
\item
  {\bf Right-handed Scenario (RHS)} in which only coefficients involving
  the couplings $z_d^{ij}$ and $z_u^{ij}$ or $g_d^{ij}$ and $g_u^{ij}$
  are kept non-zero.
\item
  {\bf Left-Right-handed Scenario (LRS)} in which all coefficients
  are kept nonzero.
\end{itemize}

Inspecting the master formulae for the sums in  $Z^\prime$, $G^\prime$,
$\varphi$ and $\Phi$ scenarios listed above, we make the following observations:
\begin{itemize}
\item
  The pattern of various contributions depends on whether the model is
  formulated in the down basis or the up basis. While this could be at
  first sight surprizing it can be explained as the explicit breakdown
  of the $\text{U(3)}^5$ flavour symmetry in the NP scenarios
  considered. We will return to this point in \refsec{sec:basis-choice}.
\item
  The pattern also depends on the scenario of couplings as
  already mentioned above.
\item
  In the down basis the $\DF=2$ contributions seem to dominate by far
  in all the down-type meson systems $(K^0, B_d, B_s)$ in all four NP
  scenarios considered. This is in particular the case in LHS and RHS
  scenarios.
\item
  In the up basis the terms involving flavour-diagonal couplings, that
  are representing $\DF=1$ operators, can compete with direct $\DF=2$
  contributions. This is in particular the case for the LRS scenario with a
  hierarchy between left- and right-handed couplings, which, if
  necessary, can through cancellations between different important
  terms allow to suppress NP contributions to $\DF=2$ processes in the
  presence of significant NP contributions to $\DF=1$ transitions
  \cite{Buras:2014zga}.

  In particular the contributions $(23)(33)$ and $(23)(22)$ can
  compete with $(23)(23)$ in the $B_s$ system and $(12)(22)$ and
  $(12)(11)$ with $(12) (12)$ in the $K^0$ system, provided the
  flavour-conserving couplings are much larger than flavour-violating
  ones, but still being in the perturbative regime. Such competition
  seems to be less likely in the $B_d$ system.
\item
  We also find that the $\DF=1$ contributions are most relevant in the
  $G^\prime$ scenario followed by the $Z^\prime$ scenario. They can also
  be relevant in scalar scenarios but at most at the level of $30\%$ of
  the numerically leading contributions.
\item
  For the up basis we also find that contributions involving only
  flavour-violating couplings can compete with each other. In particular
  $(23)(13)$ can compete with $(23)(23)$ in the $B_s$ system, while
  $(13)(23)$ with $(13)(13)$ in the $B_d$ system. This also shows that
  the same  couplings enter $B_s$ and $B_d$ systems implying correlations
  between these two systems. Such correlations are missed if RG effects
  are not considered. Similar correlations are found between $K^0$ and
  $D^0$ systems.
\end{itemize}

These findings demonstrate that just keeping the direct $\DF=2$
contributions in the phenomenological analyses of $\DF=2$
observables can miss important dynamics of SMEFT.

%
%
%
%--------+---------+---------+---------+---------+---------+---------+---------+
\subsection{Comments on VLQ and LQ models}

We have seen that in the considered scenarios the role of
the operators
\begin{equation}
  \label{eq:ops-Hqd}
  \Op[(1)]{\phi q}, \qquad \Op[(3)]{\phi q}, \qquad \Op{\phi d} \,,
\end{equation}
was minor as they have been put under the shadow of the four-fermion
operators, in particular $\Op[(1)]{qd}$ and $\Op[(8)]{qd}$,
which are generated in the matching on SMEFT in these
scenarios already at tree-level.

The situation changes in VLQ models based on the SM group in which
operators in \eqref{eq:ops-Hqd} play the dominant role by generating the
operators  $\Op[(1)]{qd}$ and $\Op[(8)]{qd}$ through RG running dominated
by Yukawa couplings. In turn these operators give again important
contributions to $\DF=2$ transitions. The analysis of this scenario
with correlations between $\DF=2$ and $\DF=1$ transitions has been
already analyzed in \cite{Bobeth:2016llm} and we refer to this paper
for details. However, let us point out that in these VLQ
models the Wilson coefficients $\Wc[(1)]{qd}$ and $\Wc[(8)]{qd}$ receive
at the electroweak scale two contributions: firstly the one due to
tree-level VLQ exchange that generates $\Op{\phi d}$ at $\muNP$ and
enters via Yukawa mixing at $\muEW$ with a logarithmic enhancement.
And secondly due to a direct contribution from one-loop matching at
$\muNP$. Both contributions are loop-suppressed, but have a
different dependence on the VLQ Yukawa couplings $Y_\text{VLQ}$:
the former scales with $\propto (Y_\text{VLQ})^2 (V_\text{CKM})^2$,
whereas the latter scales with $\propto (Y_\text{VLQ})^4$. The
absolute numerical importance of both contributions depends strongly
on the allowed size of $Y_\text{VLQ}$ compared to $V_\text{CKM}$.
The physics behind these effects are FCNCs mediated by
the SM $Z$-boson that are generated through the mixing of VLQs and
the SM quarks in the process of electroweak symmetry breaking. A more
general discussion of this phenomenon in the context of the SMEFT can
be found in \cite{Bobeth:2017xry, Endo:2018gdn}.

Still different is the case of LQ models in which four-fermion operators
are generated at the electroweak scale through RG running from semileptonic
operators
\begin{equation}
  \Op[(1)]{lq} ,\quad \Op[(3)]{lq}{} ,\quad \Op[]{ld}{} ,\quad
  \Op[]{qe}{} ,\quad
  \Op[]{ledq}{} ,\quad \Op[(1)]{lequ}{} ,\quad \Op[(3)]{lequ}{} .
\end{equation}
Only electroweak interactions are involved here and the contributions
to $\DF=2$ processes in these models are small. Indeed we have seen that
the contributions of operators involving leptons in our master formulae
are strongly suppressed. We refer to \cite{Bobeth:2017ecx} for details.

%
%
%
%--------+---------+---------+---------+---------+---------+---------+---------+
\subsection{The issue of the basis choice}
\label{sec:basis-choice}

Having the set of linearly independent SMEFT operators, we had to
specify the weak-eigenstate basis in which we plan to perform
calculations including the RG evolution above the electroweak scale.
Performing the calculations in either the down basis or the up basis,
we found different results which could be surprizing because we are used to
basis-independent results within the SM. In order to understand better what
is going on let us repeat what is well know within the SM.

The gauge interactions in the SM are invariant under a [U(3)]$^5$ flavour symmetry
\begin{align}
 \label{equ:RotationSMquarks1}
  q_L & \rightarrow V_L^q\, q_L , &
  u_R & \rightarrow V_R^u\, u_R , &
  d_R & \rightarrow V_R^d\, d_R ,
\\
  \label{equ:RotationSMleptons1}
  \ell_L & \rightarrow V_L^e\, \ell_L , &
  e_R & \rightarrow V_R^e\, e_R ,
\end{align}
where $V_L^q, V_R^u, V_R^d, V_L^e$ and $V_R^e$ are unitary $3\times 3$
matrices. This is the consequence of the fact that there is the universality
of the gauge couplings for all fermion families of left- and right-handed fermions.
In the SM the Yukawa sector breaks this universality and
consequently [U(3)]$^5$ symmetry explicitly simply because the Yukawa
couplings to fermions are not subject to further symmetry constraints,
and in this way allows to account for the known mass spectrum of quarks and
leptons. The preferred basis for calculations is the mass-eigenstate basis in which
the Yukawa and consequently mass matrices are diagonalized as explicitly
given by
\begin{align}
  \label{equ:LYukRdiag1}
  (V_L^d)^\dagger \mathsf{Y}^D V_R^d & = \mathsf{\hat{Y}}^D , &
  (V_L^u)^\dagger \mathsf{Y}^U V_R^u & = \mathsf{\hat{Y}}^U , &
  (V_L^e)^\dagger \mathsf{Y}^E V_R^e & = \mathsf{\hat{Y}}^E ,
\end{align}
with $\mathsf{\hat{Y}}^{i}$ being diagonal. {Here $V_L^u$ and $V_L^d$
rotate the $\text{SU(2)}_L$ components of $q_L$ individually contrary
to $V_L^q$ in \eqref{equ:RotationSMquarks1}.

Now because of the universality of gauge couplings and the unitarity of
rotation matrices, FCNCs are absent and flavour changes appear
only in the charged currents
parametrized by CKM and PMNS matrices. It should be stressed that it is
irrelevant whether we rotate the down-quarks from flavour to mass eigenstates
and assume flavour and mass eigenstates in the up-quark system to be equal,
or vice versa. The interactions in the mass-eigenstate basis
remain unchanged. The same applies to the lepton sector.

Let us next assume that NP contributions, e.g. with non-universal but
generation-diagonal gauge couplings, break the [U(3)]$^5$
flavour symmetry explicitly. In order to see the consequences of this
breakdown let us consider a $Z^\prime$ model and choose the up basis,
i.e. $V^u_L = \mathbb{1}$ and $V^u_R = \mathbb{1}$.
This means that the Yukawa matrix or equivalently the mass matrix for
up-quarks is diagonal and the same applies to the
interactions of up-quarks with $Z^\prime$. There is no flavour violation
in the up-quark sector mediated by the $Z^\prime$ up to
contributions from matching and back-rotation in SMEFT. But with
$V^u_L = \mathbb{1}$ we have $V_L^d = V_\text{CKM}$. Therefore,
performing the usual rotations in the down sector from flavour- to
mass-eigenstate basis we find FCNC transitions in the down-quark sector
with
\begin{align}
  \label{Bijd}
  \Delta^{ij}_L(Z^{\prime}) &
  = g_{Z^\prime} \big[V_\text{CKM}^\dagger\, \hat Z^{d}_L\, V_\text{CKM} \big]_{ij}, &
  \Delta^{ij}_R(Z^{\prime}) &
  = g_{Z^\prime} \big[(V_R^d)^\dagger\, \hat Z^{d}_R\, V_R^d \big]_{ij} \,,
\end{align}
with $(i,j=d,s,b)$ and $\hat Z^{d}_{L,R}$ being diagonal matrices collecting
$\text{U(1)}^\prime$ charges of left- and right-handed down-quarks.

However, $V^u_L = \mathbb{1}$ and $V^u_R = \mathbb{1}$ is an assumption
which specifies our model. It assumes that in the basis in which Yukawa
matrices for up-quarks are diagonal also the interactions of the $Z^\prime$
with the up-quarks are flavour diagonal. In other words $\hat Y_u$ and
$Z^\prime$ interactions for the up-quarks are aligned with each other.
But we could as well choose $V^d_L = \mathbb{1}$ and $V^d_R = \mathbb{1}$
which would result in FCNCs mediated by the $Z^\prime$ in the up-quark sector
and no FCNCs in the down-quark sector again up to contributions from
matching and back-rotation.

These simple examples show that in the absence of a [U(3)]$^5$
flavour symmetry in the gauge sector
we have more freedom and the physics depends on how the Yukawa matrices
and matrices describing interactions are oriented in flavour space.
This also explains why the bounds on various coefficients found
in \cite{Silvestrini:2018dos} for the down-basis and up-basis differ
from each other and also implied different SMEFT master formulae in
our paper.

These findings underline the importance of the construction of UV completions
in which also a flavour theory is specified so that the orientation between
Yukawa matrices and the matrices describing the interactions are known.
Interesting model constructions in this direction can be found in
\cite{Bordone:2017bld, Gherardi:2019zil}.

%--------+---------+---------+---------+---------+---------+---------+---------+
%
%
%
%--------+---------+---------+---------+---------+---------+---------+---------+
\section{Summary and Outlook}\label{sec:6}

In the present paper we have worked out the model-independent anatomy
of the $\DF=2$ transitions $\KKbar$, $\BBbar$ and $\DDbar$ in the context of
SMEFT and LEFT. On the technical side the two most important novel results are two master formulae for the new physics contribution of
the mixing amplitude $\big[M_{12}^{ij}\big]_\text{BSM}$ with {$ij=ds,db,sb,cu$}.

The first Eq.~\eqref{eq:master-M12BSM} is given in terms of the Wilson coefficients (WCs) of the LEFT  operators evaluated at the electroweak scale $\muEW$. For each meson system there are eight WCs and
corresponding coefficients $P_a^{ij}(\muEW)$ that collect all the information
below the scales $\muEW$. This means the existing results for hadronic matrix
elements from LQCD combined with the presently known QCD renormalization group
evolution at the NLO level up to $\muEW$. The numerical values of
$P_a^{ij}(\muEW)$ in different operator bases are collected in
\reftab{tab:LEFT-Pa}. Calculating the WCs of the eight operators in question
at $\muEW$ in any BSM scenario this formula gives directly
$\big[M_{12}^{ij}\big]_\text{BSM}$ and consequently allows to calculate all
observables related to neutral-meson mixing. It is a modern
version of the formula presented already in \cite{Buras:2001ra}.
The advantage of this formula with respect to the second master formula
for SMEFT is the paucity of the terms entering it
and further it is not subject to constraints from $\text{SU(2)}_L
\times \text{U(1)}_Y$ gauge invariance present in the SMEFT master formula.
But the drawback is that it requires from the practitioners the calculation
of the RG evolution in the context of the SMEFT from the NP scale $\muNP$
down to the electroweak scale $\muEW$. Because of subtle and often important
RG effects related to the top-Yukawa coupling taking place on the route from
$\muNP$ down to $\muEW$, it is not evident from the structure of a given
extension of the SM that it is in agreement with the data at low energy
or not.

The second master formula, given in \eqref{eq:master-M12SMEFT}, although
containing many more terms than the LEFT one, is more powerful because it
allows right away to see whether a given BSM scenario has a chance to be
consistent with the data or not. It gives $\big[M_{12}^{ij}\big]_\text{BSM}$
directly in terms of the WCs of the SMEFT operators evaluated at the BSM
scale $\muNP$. The coefficients $P_a^{ij}(\muNP)$ entering this formula
generalize the information below the scale $\muEW$ present already in
$P_a^{ij}(\muEW)$ to include all RG effects between $\muEW$ and $\muNP$.
Therefore performing the matching of a given NP model to the SMEFT, using
in particular results of \cite{deBlas:2017xtg} at the scale $\muNP$, and
using the master formulae in \refsec{sec:SMEFT-ATLAS} allows to connect
NP at a very high scale with the observables measured at low energy scales.
While the numerical coefficients in these formulae are given for
the example of $\muNP = 5\TeV$, the dominant change in the
results for observables comes from the quadratic dependence on the masses
of gauge bosons and scalars and can be calculated right away. The dependence
of $P_a^{ij}(\muNP)$ on $\muNP$ is logarithmic and much weaker. One can
verify this by inspecting the corresponding master formulae for $100~\TeV$
in \refapp{app:100ATLAS}.

We stress that although the solution to the relevant RG equations
are collected in the first leading logarithmic approximation in
\refsec{sec:RGRunning}, the numerical values of $P_a^{ij}(\muNP)$ that
enter the master formulae are obtained by summing leading logarithms
to all orders in the coupling numerically and including the one-loop
matching from SMEFT onto LEFT \cite{Dekens:2019ept} at $\muEW$,
which is collected in \refapp{app:one-loop-SMEFT}.

Presenting our master formulae in the down and the up basis, we have
reemphasized their differences and the need for UV completions incorporating
a flavour theory in order to be able to understand the full dynamics of
the SMEFT. Whereas in the down basis the $\DF=2$ contributions
have the numerically largest coefficients compared to the $\DF=1$ ones
that enter mainly via Yukawa mixing, in the up basis this hierarchy
is absent. This implies very strong correlations between $\DF=2$ and
$\DF=1$ processes in the up basis, hence affecting many
phenomenological analysis of collider processes, in particular top-quark phenomenology.

We have also illustrated this technology by applying the SMEFT formula to
a number of simplified models  containing colourless heavy gauge bosons
($Z^\prime$) and scalars and models with coloured heavy gauge bosons
($G_a^\prime$) and scalars. Also the cases of vector-like quarks and
leptoquarks have been briefly discussed.

Our analysis demonstrates that RG effects of the running from $\muNP$ down
to $\muEW$, in particular those related to QCD interactions and top-Yukawa
couplings constitute an essential ingredient of any analysis of BSM scenarios.
However, it should be kept in mind that in some NP scenarios the role of the
model dependent one-loop matching at the NP scale $\muNP$ could also be
important if it generates the Wilson coefficients $\Wc[(1)]{qd}$ and
$\Wc[(8)]{qd}$ which are multiplied by very large $P_a^{ij}(\muNP)$ coefficients.

The corresponding analysis of $\DF=1$ transitions is expected to be much more
involved and subject to much larger hadronic uncertainties, but already our
$\DF=2$ ATLAS casts some doubts on the validity of many analyses present in
the literature that consider only one or two operators at the time and restrict
the analyses to a single meson system.

%
%
%--------+---------+---------+---------+---------+---------+---------+---------+
\section*{Acknowledgments}

J. A. acknowledges financial support from the Swiss National Science Foundation
(Project No. P400P2\_183838). A.J.B acknowledges financial support from the
Excellence Cluster ORIGINS, funded by the Deutsche Forschungsgemeinschaft
(DFG, German Research Foundation) under Germany's Excellence
Strategy – EXC-2094 – 390783311.
J.K. acknowledges financial support from NSERC of Canada.
The work of CB is supported by DFG under grant BO-4535/1-1.

%--------+---------+---------+---------+---------+---------+---------+---------+
%
%
%
%--------+---------+---------+---------+---------+---------+---------+---------+
\appendix

%--------+---------+---------+---------+---------+---------+---------+---------+
%
%
%
%--------+---------+---------+---------+---------+---------+---------+---------+

\section{\boldmath SMEFT ATLAS at $\muNP = 100\TeV$}
\label{app:100ATLAS}

Here we report the master formulae for the contributions of different operators,
this time generated at $\muNP = 100\TeV$. The dominant effect relative to the
$\muNP = 5\TeV$ case is suppression of these contributions by a factor of 400
originating in $1/\muNP^2$. RG effects, although visible, amount to shifts of
at most $50\%$ and this only for left-right operators.

%
%--------+---------+---------+---------+---------+---------+---------+---------+
\subsection[$\mathcal{C}^{(1)}_{qq}$ and $\mathcal{C}^{(3)}_{qq}$]
{\boldmath $\Wc[(1)]{qq}$ and $\Wc[(3)]{qq}$}

In the down-basis
\begin{align*}
  \Sigma^{B_s}_{qq1} & = -8.9\cdt{-1}\bwc[(1)]{qq}{2323} \,,
  \numberthis
\\
  \Sigma^{B_d}_{qq1} & =
  -2.1\cdt{1}\bwc[(1)]{qq}{1313} +2.6\cdt{-2}\bwc[(1)]{qq}{1213}
  +1.0\cdt{-2}e^{i22^\circ}\bwc[(1)]{qq}{1333}
  \nline
  -5.7\cdt{-3}e^{i22^\circ}\bwc[(1)]{qq}{1113} \,,
  \numberthis
\\
  \Sigma^K_{qq1} & =
  -8.1\cdt{1}\bwc[(1)]{qq}{1212} -2.2\cdt{-1}\bwc[(1)]{qq}{1213}
  +4.7\cdt{-2}e^{i22^\circ}\bwc[(1)]{qq}{1232} \,,
  \numberthis
\end{align*}
and up-basis
\begin{align*}
  \hat\Sigma^{B_s}_{qq1} & =
  -8.4\cdt{-1}\bwcup[(1)]{qq}{2323} -1.9\cdt{-1}\bwcup[(1)]{qq}{1323}
  -4.4\cdt{-2}\bwcup[(1)]{qq}{1313} -3.4\cdt{-2}\bwcup[(1)]{qq}{2223}
  \nline
  +3.4\cdt{-2}\bwcup[(1)]{qq}{2333} -7.9\cdt{-3}\bwcup[(1)]{qq}{1223}
  -7.9\cdt{-3}\bwcup[(1)]{qq}{1322} +7.8\cdt{-3}\bwcup[(1)]{qq}{1333} \,,
  \numberthis
\\
  \hat\Sigma^{B_d}_{qq1} & =
  -2.0\cdt{1}\bwcup[(1)]{qq}{1313} +4.5\bwcup[(1)]{qq}{1323}
  -1.0\bwcup[(1)]{qq}{2323} -8.0\cdt{-1}\bwcup[(1)]{qq}{1213}
  \nline
  +1.8\cdt{-1}\bwcup[(1)]{qq}{1223} +1.8\cdt{-1}\bwcup[(1)]{qq}{1322}
  -1.7\cdt{-1}e^{i22^\circ}\bwcup[(1)]{qq}{1333}
  \nline
  -6.8\cdt{-2} e^{-i73^\circ}\bwcup[(1)]{qq}{1113}
  -4.2\cdt{-2}\bwcup[(1)]{qq}{2223}
  +3.9\cdt{-2}e^{i22^\circ}\bwcup[(1)]{qq}{2333}
  \nline
  -3.3\cdt{-2}\bwcup[(1)]{qq}{1212}
  +1.6\cdt{-2}e^{-i73^\circ}\bwcup[(1)]{qq}{1231}
  +1.6\cdt{-2}e^{-i73^\circ}\bwcup[(1)]{qq}{1123}
  \nline
  +7.5\cdt{-3}\bwcup[(1)]{qq}{1222}
  -7.0\cdt{-3} e^{i23^\circ}\bwcup[(1)]{qq}{1233}
  -6.9\cdt{-3} e^{i23^\circ}\bwcup[(1)]{qq}{1332} \,,
  \numberthis
\\
  \hat\Sigma^K_{qq1} & =
  -7.3\cdt{1}\bwcup[(1)]{qq}{1212} -1.6\cdt{1}\bwcup[(1)]{qq}{1112}
  +1.6\cdt{1}\bwcup[(1)]{qq}{1222} -3.9\bwcup[(1)]{qq}{1111}
  \nline
  +3.9\bwcup[(1)]{qq}{1122} +3.9\bwcup[(1)]{qq}{1221}
  -3.9\bwcup[(1)]{qq}{2222} +2.9\bwcup[(1)]{qq}{1213}
  +6.4\cdt{-1}\bwcup[(1)]{qq}{1113}
  \nline
  -6.4\cdt{-1}\bwcup[(1)]{qq}{1223} -6.4\cdt{-1}\bwcup[(1)]{qq}{1322}
  -6.0\cdt{-1} e^{i23^\circ}\bwcup[(1)]{qq}{1232}
  \nline
  -3.0\cdt{-1} e^{i11^\circ}\bwcup[(1)]{qq}{1123}
  -3.0\cdt{-1} e^{i11^\circ}\bwcup[(1)]{qq}{1231}
  +3.0\cdt{-1}e^{i11^\circ}\bwcup[(1)]{qq}{2223}
  \nline
  -1.2\cdt{-1}\bwcup[(1)]{qq}{1313} +2.6\cdt{-2}\bwcup[(1)]{qq}{1323}
  +2.4\cdt{-2}e^{i23^\circ}\bwcup[(1)]{qq}{1233}
  \nline
  +2.4\cdt{-2}e^{i23^\circ}\bwcup[(1)]{qq}{1332}
  -1.1\cdt{-2}e^{i22^\circ}\bwcup[(1)]{qq}{2323}
  +5.9\cdt{-3}e^{i23^\circ}\bwcup[(1)]{qq}{1133}
  \nline
  -5.9\cdt{-3}e^{i23^\circ}\bwcup[(1)]{qq}{2233}
  +5.8\cdt{-3}e^{i23^\circ}\bwcup[(1)]{qq}{1331}
  -5.8\cdt{-3}e^{i23^\circ}\bwcup[(1)]{qq}{2332} \,,
  \numberthis
\end{align*}
and to a very good approximation the same expressions for $\Wc[(3)]{qq}$
and $\Wcup[(3)]{qq}$.

%
%--------+---------+---------+---------+---------+---------+---------+---------+
\subsection[$\mathcal{C}^{(1)}_{qd}$ and $\mathcal{C}^{(8)}_{qd}$]
{\boldmath $\Wc[(1)]{qd}$ and $\Wc[(8)]{qd}$}

In the down basis
\begin{align*}
  \Sigma^{B_s}_{qd1} & =
   7.3\bwc[(1)]{qd}{2323} +1.7\cdt{-2}\bwc[(1)]{qd}{3323}
  -9.3\cdt{-3}\bwc[(1)]{qd}{2223} \,,
  \numberthis
\\
  \Sigma^{B_d}_{qd1} & =
   1.8\cdt{2}\bwc[(1)]{qd}{1313} -2.3\cdt{-1}\bwc[(1)]{qd}{1213}
  -9.0\cdt{-2}e^{i22^\circ}\bwc[(1)]{qd}{3313}
  \nline
  +5.0\cdt{-2}e^{i22^\circ}\bwc[(1)]{qd}{1113}
  -1.4\cdt{-2}\bwc[(1)]{qd}{1312} \,,
  \numberthis
\\
  \Sigma^K_{qd1} & =
   1.7\cdt{4}\bwc[(1)]{qd}{1212} +4.6\cdt{1}\bwc[(1)]{qd}{1312}
  -1.0\cdt{1}e^{i22^\circ}\bwc[(1)]{qd}{2321} +1.4\bwc[(1)]{qd}{1213}
  \nline
  +4.4\cdt{-1}e^{i23^\circ}\bwc[(1)]{qd}{2212}
  -2.0\cdt{-1}e^{i23^\circ}\bwc[(1)]{qd}{1112}
  -7.8\cdt{-2}e^{i23^\circ}\bwc[(1)]{qd}{3312}
  \nline
  +3.2\cdt{-2}e^{i23^\circ}\bwc[(1)]{qd}{1222}
  -3.2\cdt{-2}e^{i23^\circ}\bwc[(1)]{qd}{1211}
  -1.5\cdt{-2}e^{i22^\circ}\bwc[(1)]{qd}{1232} \,,
  \numberthis
\end{align*}
and
\begin{align*}
  \Sigma^{B_s}_{qd8} & =
   8.7\bwc[(8)]{qd}{2323} +2.1\cdt{-2}\bwc[(8)]{qd}{3323}
  -1.1\cdt{-2}\bwc[(8)]{qd}{2223} \,,
  \numberthis
\\
  \Sigma^{B_d}_{qd8} & =
   2.1\cdt{2}\bwc[(8)]{qd}{1313} -2.7\cdt{-1}\bwc[(8)]{qd}{1213}
  -1.1\cdt{-1}e^{i22^\circ}\bwc[(8)]{qd}{3313}
  \nline
  +6.0\cdt{-2}e^{i22^\circ}\bwc[(8)]{qd}{1113}
  -1.7\cdt{-2}\bwc[(8)]{qd}{1312} \,,
  \numberthis
\\
  \Sigma^K_{qd8} & =
   2.4\cdt{4}\bwc[(8)]{qd}{1212} +6.5\cdt{1}\bwc[(8)]{qd}{1312}
  -1.4\cdt{1}e^{i22^\circ}\bwc[(8)]{qd}{2321} +1.9\bwc[(8)]{qd}{1213}
  \nline
  +6.2\cdt{-1}e^{i23^\circ}\bwc[(8)]{qd}{2212}
  -2.8\cdt{-1}e^{i23^\circ}\bwc[(8)]{qd}{1112}
  -1.0\cdt{-1}e^{i23^\circ}\bwc[(8)]{qd}{3312}
  \nline
  -4.4\cdt{-2}e^{i23^\circ}\bwc[(8)]{qd}{1211}
  +4.4\cdt{-2}e^{i23^\circ}\bwc[(8)]{qd}{1222}
  -2.0\cdt{-2}e^{i22^\circ}\bwc[(8)]{qd}{1232}
  \nline
  +5.1\cdt{-3}\bwc[(8)]{qd}{1313} \,.
  \numberthis
\end{align*}

For the up basis we have
\begin{align*}
  \hat\Sigma^{B_s}_{qd1} & =
   7.1\bwcup[(1)]{qd}{2323} +1.6\bwcup[(1)]{qd}{1323}
  +2.9\cdt{-1}\bwcup[(1)]{qd}{2223} -2.8\cdt{-1}\bwcup[(1)]{qd}{3323}
  +6.7\cdt{-2}\bwcup[(1)]{qd}{1223}
  \nline
  +2.5\cdt{-2}e^{-i73^\circ}\bwcup[(1)]{qd}{1232}
  -1.2\cdt{-2}\bwcup[(1)]{qd}{2332}
  +5.7\cdt{-3}e^{-i73^\circ}\bwcup[(1)]{qd}{1123} \,,
  \numberthis
\\
  \hat\Sigma^{B_d}_{qd1} & =
   1.7\cdt{2}\bwcup[(1)]{qd}{1313} -4.0\cdt{1}\bwcup[(1)]{qd}{2313}
  +7.1\bwcup[(1)]{qd}{1213} -1.6\bwcup[(1)]{qd}{2213}
  +1.5e^{i22^\circ}\bwcup[(1)]{qd}{3313}
  \nline
  +6.1\cdt{-1}e^{-i73^\circ}\bwcup[(1)]{qd}{1113}
  -1.4\cdt{-1}e^{-i73^\circ}\bwcup[(1)]{qd}{1231}
  +6.2\cdt{-2}e^{i22^\circ}\bwcup[(1)]{qd}{2331}
  \nline
  -1.4\cdt{-2}\bwcup[(1)]{qd}{1312}
  +5.3\cdt{-3}e^{-i51^\circ}\bwcup[(1)]{qd}{1331} \,,
  \numberthis
\\
  \hat\Sigma^K_{qd1} & =
   1.7\cdt{4}\bwcup[(1)]{qd}{1212} +3.8\cdt{3}\bwcup[(1)]{qd}{1112}
  -3.8\cdt{3}\bwcup[(1)]{qd}{2212} -8.8\cdt{2}\bwcup[(1)]{qd}{1221}
  \nline
  -6.6\cdt{2}\bwcup[(1)]{qd}{1312} +1.5\cdt{2}\bwcup[(1)]{qd}{2312}
  +1.4\cdt{2}e^{i22^\circ}\bwcup[(1)]{qd}{2321}
  +3.3\cdt{1}e^{i22^\circ}\bwcup[(1)]{qd}{1321}
  \nline
  -5.8 e^{i23^\circ}\bwcup[(1)]{qd}{3312} +1.3\bwcup[(1)]{qd}{1213}
  +3.0\cdt{-1}\bwcup[(1)]{qd}{1113} -3.0\cdt{-1}\bwcup[(1)]{qd}{2213}
  \nline
  -6.8\cdt{-2}\bwcup[(1)]{qd}{1231} -5.1\cdt{-2}\bwcup[(1)]{qd}{1313}
  +2.8\cdt{-2}e^{i23^\circ}\bwcup[(1)]{qd}{1222}
  \nline
  -2.8\cdt{-2}e^{i23^\circ}\bwcup[(1)]{qd}{1211}
  -1.4\cdt{-2}e^{i22^\circ}\bwcup[(1)]{qd}{1232}
  +1.2\cdt{-2}\bwcup[(1)]{qd}{2313}
  \nline
  +1.1\cdt{-2}e^{i23^\circ}\bwcup[(1)]{qd}{2331}
  +6.9\cdt{-3}e^{i23^\circ}\bwcup[(1)]{qd}{1122}
  -6.9\cdt{-3}e^{i23^\circ}\bwcup[(1)]{qd}{1111}
  \nline
  -6.9\cdt{-3}e^{i23^\circ}\bwcup[(1)]{qd}{2222}
  +6.9\cdt{-3}e^{i23^\circ}\bwcup[(1)]{qd}{2211} \,,
  \numberthis
\end{align*}
and
\begin{align*}
  \hat\Sigma^{B_s}_{qd8} & =
   8.5\bwcup[(8)]{qd}{2323} +2.0\bwcup[(8)]{qd}{1323}
  +3.5\cdt{-1}\bwcup[(8)]{qd}{2223} -3.4\cdt{-1}\bwcup[(8)]{qd}{3323}
  +8.0\cdt{-2}\bwcup[(8)]{qd}{1223}
  \nline
  +3.0\cdt{-2}e^{-i73^\circ}\bwcup[(8)]{qd}{1232}
  -1.4\cdt{-2}\bwcup[(8)]{qd}{2332}
  +6.8\cdt{-3}e^{-i73^\circ}\bwcup[(8)]{qd}{1123} \,,
  \numberthis
\\
  \hat\Sigma^{B_d}_{qd8} & =
   2.1\cdt{2}\bwcup[(8)]{qd}{1313} -4.8\cdt{1}\bwcup[(8)]{qd}{2313}
  +8.5\bwcup[(8)]{qd}{1213} -2.0\bwcup[(8)]{qd}{2213}
  +1.8e^{i22^\circ}\bwcup[(8)]{qd}{3313}
  \nline
  +7.3\cdt{-1}e^{-i73^\circ}\bwcup[(8)]{qd}{1113}
  -1.7\cdt{-1}e^{-i73^\circ}\bwcup[(8)]{qd}{1231}
  +7.4\cdt{-2}e^{i22^\circ}\bwcup[(8)]{qd}{2331}
  \nline
  -1.6\cdt{-2}\bwcup[(8)]{qd}{1312}
  +6.3\cdt{-3}e^{-i51^\circ}\bwcup[(8)]{qd}{1331} \,,
  \numberthis
\\
  \hat\Sigma^K_{qd8} & =
   2.3\cdt{4}\bwcup[(8)]{qd}{1212} +5.3\cdt{3}\bwcup[(8)]{qd}{1112}
  -5.3\cdt{3}\bwcup[(8)]{qd}{2212} -1.2\cdt{3}\bwcup[(8)]{qd}{1221}
  \nline
  -9.2\cdt{2}\bwcup[(8)]{qd}{1312} +2.1\cdt{2}\bwcup[(8)]{qd}{2312}
  +2.0\cdt{2}e^{i22^\circ}\bwcup[(8)]{qd}{2321}
  +4.7\cdt{1}e^{i22^\circ}\bwcup[(8)]{qd}{1321}
  \nline
  -8.1 e^{i23^\circ}\bwcup[(8)]{qd}{3312} +1.8\bwcup[(8)]{qd}{1213}
  +4.2\cdt{-1}\bwcup[(8)]{qd}{1113} -4.2\cdt{-1}\bwcup[(8)]{qd}{2213}
  \nline
  -9.6\cdt{-2}\bwcup[(8)]{qd}{1231} -7.2\cdt{-2}\bwcup[(8)]{qd}{1313}
  -4.0\cdt{-2} e^{i23^\circ}\bwcup[(8)]{qd}{1211}
  \nline
  +4.0\cdt{-2}e^{i23^\circ}\bwcup[(8)]{qd}{1222}
  -1.9\cdt{-2}e^{i22^\circ}\bwcup[(8)]{qd}{1232}
  +1.7\cdt{-2}\bwcup[(8)]{qd}{2313}
  \nline
  +1.6\cdt{-2}e^{i23^\circ}\bwcup[(8)]{qd}{2331}
  -9.7\cdt{-3}e^{i23^\circ}\bwcup[(8)]{qd}{1111}
  +9.7\cdt{-3}e^{i23^\circ}\bwcup[(8)]{qd}{1122}
  \nline
  +9.7\cdt{-3}e^{i23^\circ}\bwcup[(8)]{qd}{2211}
  -9.7\cdt{-3}e^{i23^\circ}\bwcup[(8)]{qd}{2222} \,.
  \numberthis
\end{align*}

%
%--------+---------+---------+---------+---------+---------+---------+---------+
\subsection[$\mathcal{C}^{(1)}_{qu}$ and $\mathcal{C}^{(8)}_{qu}$]
{\boldmath $\Wc[(1)]{qu}$ and $\Wc[(8)]{qu}$}

The only non-vanishing expression following our criteria is
in the down-basis
\begin{equation}
  \Sigma^{B_d}_{qu1} = -5.4\cdt{-3}e^{i22^\circ}\bwc[(1)]{qu}{1333} \,,
\end{equation}
and in the up basis
\begin{equation}
  \hat \Sigma^{B_d}_{qu1} = -5.2\cdt{-3}e^{i22^\circ}\bwcup[(1)]{qu}{1333} \,.
\end{equation}

%
%--------+---------+---------+---------+---------+---------+---------+---------+
\subsection[$\Wc{dd}$, $\Wc{uu}$, $\mathcal{C}^{(1)}_{ud}$ and $\mathcal{C}^{(8)}_{ud}$]
{\boldmath $\Wc{dd}$, $\Wc{uu}$, $\Wc[(1)]{ud}$ and $\Wc[(8)]{ud}$}

In both bases
\begin{align}
  \Sigma^{B_s}_{dd} & = -9.7\cdt{-1}\bwc{dd}{2323} \,, \\
  \Sigma^{B_d}_{dd} & = -2.2\cdt{1}\bwc{dd}{1313}  \,, \\
  \Sigma^K_{dd}     & = -8.6\cdt{1}\bwc{dd}{1212} -6.7\cdt{-3}\bwc{dd}{1213}\,,
\end{align}
\begin{align}
  \Sigma^{B_s}_{uu} & = 0 \,, &
  \Sigma^{B_d}_{uu} & = 0 \,, &
  \Sigma^{K}_{uu}   & = 0 \,,
\end{align}
\begin{align}
  \Sigma^{B_s}_{ud1} & =  0\,,
\\
  \Sigma^{B_d}_{ud1} & = 2.3\cdt{-2}e^{i22^\circ}\bwc[(1)]{ud}{3313} \,,
\\
  \Sigma^K_{ud1}     & =
   -7.8\cdt{-2}e^{i23^\circ}\bwc[(1)]{ud}{3312}
   +7.9\cdt{-3}\bwc[(1)]{ud}{2312} +7.6\cdt{-3}e^{i22^\circ}\bwc[(1)]{ud}{2321}\,,
\end{align}
\begin{align}
  \Sigma^{B_s}_{ud8} & = -7.4\cdt{-3}\bwc[(8)]{ud}{3323} \,,
\\
  \Sigma^{B_d}_{ud8} & = 4.0\cdt{-2}e^{i22^\circ}\bwc[(8)]{ud}{3313} \,,
\\
  \Sigma^{K}_{ud8}   & =
   -1.9\cdt{-1}e^{i23^\circ}\bwc[(8)]{ud}{3312}
   +1.9\cdt{-2}\bwc[(8)]{ud}{2312} +1.8\cdt{-2}e^{i22^\circ}\bwc[(8)]{ud}{2321}\,.
\end{align}

%
%--------+---------+---------+---------+---------+---------+---------+---------+
\subsection[$\mathcal{C}^{(1,3)}_{\phi q}$, $\Wc{\phi d}$ and $\Wc{\phi u}$]
{\boldmath $\Wc[(1,3)]{\phi q}$, $\Wc{\phi d}$ and $\Wc{\phi u}$}

The only non-vanishing expressions for these WCs in the down and up basis are
\begin{equation}
  \Sigma^{B_d}_{\phi q3} = -8.5\cdt{-3}e^{i22^\circ}\bwc[(3)]{\phi q}{13}\,,
\end{equation}
and
\begin{align}
  \Sigma^{B_s}_{\phi d} & =  5.7\cdt{-3}\bwc[]{\phi d}{23} \,, \\
  \Sigma^{B_d}_{\phi d} & = -3.1\cdt{-2}e^{i22^\circ}\bwc{\phi d}{13} \,, \\
  \Sigma^K_{\phi d}     & =  1.2\cdt{-1}e^{i23^\circ}\bwc{\phi d}{12} \,.
\end{align}

%
%--------+---------+---------+---------+---------+---------+---------+---------+
\subsection[$\mathcal{C}^{(1,8)}_{quqd}$]
{\boldmath $\Wc[(1,8)]{quqd}$}

For these operators the only non-zero contributions are in the down basis
\begin{equation}
  \Sigma^K_{quqd1} = -6.2\cdt{-3}e^{i22^\circ}\bwc[(1)]{quqd}{2321} \,,
\end{equation}
and in the up basis
\begin{equation}
  \hat\Sigma^{K}_{quqd1} = -5.9\cdt{-3}e^{i22^\circ}\bwcup[(1)]{quqd}{2321} \,.
\end{equation}

%
%--------+---------+---------+---------+---------+---------+---------+---------+
\subsection[$\mathcal{C}^{(1,3)}_{lequ}$, $\mathcal{C}^{(1,3)}_{lq}$, $\Wc{ld}$, $\Wc{qe}$ and $\Wc{uW}$]
{\boldmath $\Wc[(1,3)]{lequ}$, $\Wc[(1,3)]{lq}$, $\Wc{ld}$, $\Wc{qe}$ and $\Wc{uW}$}

The contributions of all the semileptonic Wilson coefficients as well as
$\Wc{uW}$ are negligible.

%--------+---------+---------+---------+---------+---------+---------+---------+
%
%
%
%--------+---------+---------+---------+---------+---------+---------+---------+
\section{NP Master Formulae in the up-basis}\label{NPMFup}
In this appendix we present the NP master formulae for the four scenarios
under consideration at 5 TeV in the up-basis.

%
%
%
%--------+---------+---------+---------+---------+---------+---------+---------+
\subsection{$Z'$}

\begin{align*}
  \frac{M_{Z^\prime}^2\, \hat\Sigma^{B_s}_{Z^\prime}}{(5\TeV)^2} & =
  -2.3\cdt{3}\hat{z}_q^{23}\hat{z}_d^{23} -5.2\cdt{2}\hat{z}_q^{13}\hat{z}_d^{23}
  +2.1\cdt{2}\hat{z}_d^{23}\hat{z}_d^{23} +1.9\cdt{2}\hat{z}_q^{23}\hat{z}_q^{23}
  \nlineS{-0.3cm}
  -9.3\cdt{1}\hat{z}_q^{22}\hat{z}_d^{23}
  +9.3\cdt{1}\hat{z}_q^{33}\hat{z}_d^{23}
  +4.3\cdt{1}\hat{z}_q^{13}\hat{z}_q^{23}
  -2.2\cdt{1}\hat{z}_q^{12}\hat{z}_d^{23}
  \nline
  +9.8\hat{z}_q^{13}\hat{z}_q^{13}
  -8.0 e^{-i73^\circ}\hat{z}_q^{12}\hat{z}_d^{32}
  +7.7\hat{z}_q^{22}\hat{z}_q^{23}
  -7.6\hat{z}_q^{23}\hat{z}_q^{33}
  +3.8\hat{z}_q^{23}\hat{z}_d^{32}
  \nline
  -1.8 e^{-i73^\circ}\hat{z}_q^{11}\hat{z}_d^{23}
  +1.8\hat{z}_q^{12}\hat{z}_q^{23}
  +1.8\hat{z}_q^{13}\hat{z}_q^{22}
  -1.8\hat{z}_q^{13}\hat{z}_q^{33}
  \nline
  -1.2\re (\hat{z}_\phi) \hat{z}_d^{23}
  +1.0\hat{z}_u^{33}\hat{z}_d^{23}
  +6.6\cdt{-1}e^{-i73^\circ}\hat{z}_q^{12}\hat{z}_q^{32}
  +4.1\cdt{-1}\hat{z}_q^{12}\hat{z}_q^{13}
  \nline
  +3.3\cdt{-1}e^{-i74^\circ}\hat{z}_q^{13}\hat{z}_d^{32}
  +3.2\cdt{-1}\hat{z}_q^{22}\hat{z}_q^{22}
  -3.2\cdt{-1}\hat{z}_q^{22}\hat{z}_q^{33}
  -3.1\cdt{-1}\hat{z}_q^{23}\hat{z}_q^{32}
  \nline
  +3.1\cdt{-1}\hat{z}_q^{33}\hat{z}_q^{33}
  -2.6\cdt{-1}\hat{z}_q^{23}\hat{z}_u^{33}
  +2.3\cdt{-1}\re (\hat{z}_\phi) \hat{z}_q^{23}
  \nline
  +1.5\cdt{-1}e^{-i73^\circ}\hat{z}_q^{12}\hat{z}_q^{31}
  +1.5\cdt{-1}e^{-i73^\circ}\hat{z}_q^{11}\hat{z}_q^{23}
  -1.1\cdt{-1}\hat{z}_u^{23}\hat{z}_d^{23}
  \nline
  +1.1\cdt{-1}\hat{z}_q^{23}\hat{z}_d^{22}
  -1.1\cdt{-1}\hat{z}_q^{23}\hat{z}_d^{33}
  +8.5\cdt{-2}e^{-i18^\circ}\hat{z}_q^{12}\hat{z}_q^{22}
  \nline
  -8.5\cdt{-2}e^{-i19^\circ}\hat{z}_q^{12}\hat{z}_q^{33}
  -8.4\cdt{-2}e^{-i19^\circ}\hat{z}_q^{13}\hat{z}_q^{32}
  -5.9\cdt{-2}\hat{z}_q^{13}\hat{z}_u^{33}
  \nline
  +5.4\cdt{-2}\re (\hat{z}_\phi) \hat{z}_q^{13}
  +3.5\cdt{-2}e^{-i73^\circ}\hat{z}_q^{11}\hat{z}_q^{13}
  +3.1\cdt{-2}\hat{z}_q^{23}\hat{z}_u^{23}
  \nline
  +2.5\cdt{-2}\hat{z}_q^{13}\hat{z}_d^{22}
  -2.5\cdt{-2}\hat{z}_q^{13}\hat{z}_d^{33}
  +1.5\cdt{-2}e^{-i5^\circ}\hat{z}_q^{12}\hat{z}_q^{12}
  \nline
  +1.1\cdt{-2}\hat{z}_d^{23}\hat{z}_d^{33}
  -1.1\cdt{-2}\hat{z}_q^{22}\hat{z}_u^{33}
  +1.0\cdt{-2}\re (\hat{z}_\phi) \hat{z}_q^{22}
  \nline
  +1.0\cdt{-2}\hat{z}_q^{33}\hat{z}_u^{33}
  -9.4\cdt{-3}\re (\hat{z}_\phi) \hat{z}_q^{33}
  -9.3\cdt{-3}\hat{z}_d^{22}\hat{z}_d^{23}
  \nline
  +7.2\cdt{-3}\hat{z}_q^{13}\hat{z}_u^{23}
  +6.3\cdt{-3}e^{-i73^\circ}\hat{z}_q^{12}\hat{z}_q^{21}
  +6.3\cdt{-3}e^{-i73^\circ}\hat{z}_q^{11}\hat{z}_q^{22}
  \nline
  -6.2\cdt{-3} e^{-i74^\circ}\hat{z}_q^{11}\hat{z}_q^{33}
  -6.2\cdt{-3} e^{-i74^\circ}\hat{z}_q^{13}\hat{z}_q^{31}
  +5.4\cdt{-3}e^{-i11^\circ}\hat{z}_q^{23}\hat{z}_d^{13} \,,
  \numberthis
\\
  \frac{M_{Z^\prime}^2\, \hat\Sigma^{B_d}_{Z^\prime}}{(5\TeV)^2} & =
  -5.5\cdt{4}\hat{z}_q^{13}\hat{z}_d^{13} +1.3\cdt{4}\hat{z}_q^{23}\hat{z}_d^{13}
  +4.8\cdt{3}\hat{z}_d^{13}\hat{z}_d^{13} +4.3\cdt{3}\hat{z}_q^{13}\hat{z}_q^{13}
  \nlineS{-0.3cm}
  -2.3\cdt{3}\hat{z}_q^{12}\hat{z}_d^{13} -9.9\cdt{2}\hat{z}_q^{13}\hat{z}_q^{23}
  +5.3\cdt{2}\hat{z}_q^{22}\hat{z}_d^{13}
  -5.0\cdt{2}e^{i22^\circ}\hat{z}_q^{33}\hat{z}_d^{13}
  \nline
  +2.3\cdt{2}\hat{z}_q^{23}\hat{z}_q^{23}
  -2.0\cdt{2} e^{-i73^\circ}\hat{z}_q^{11}\hat{z}_d^{13}
  +1.8\cdt{2}\hat{z}_q^{12}\hat{z}_q^{13}
  +4.5\cdt{1}e^{-i73^\circ}\hat{z}_q^{12}\hat{z}_d^{31}
  \nline
  -4.1\cdt{1}\hat{z}_q^{12}\hat{z}_q^{23} -4.1\cdt{1}\hat{z}_q^{13}\hat{z}_q^{22}
  +3.9\cdt{1}e^{i22^\circ}\hat{z}_q^{13}\hat{z}_q^{33}
  -2.0\cdt{1}e^{i22^\circ}\hat{z}_q^{23}\hat{z}_d^{31}
  \nline
  +1.5\cdt{1}e^{-i73^\circ}\hat{z}_q^{11}\hat{z}_q^{13}
  +9.4\hat{z}_q^{22}\hat{z}_q^{23}
  -8.9e^{i22^\circ}\hat{z}_q^{23}\hat{z}_q^{33}
  +7.4\hat{z}_q^{12}\hat{z}_q^{12}
  \nline
  +6.6e^{i22^\circ}\re (\hat{z}_\phi) \hat{z}_d^{13}
  -5.6 e^{i22^\circ}\hat{z}_u^{33}\hat{z}_d^{13}
  -3.5 e^{-i73^\circ}\hat{z}_q^{12}\hat{z}_q^{31}
  -3.5 e^{-i73^\circ}\hat{z}_q^{11}\hat{z}_q^{23}
  \nline
  +2.7\hat{z}_q^{13}\hat{z}_d^{12}
  -1.8 e^{-i51^\circ}\hat{z}_q^{13}\hat{z}_d^{31}
  -1.7\hat{z}_q^{12}\hat{z}_q^{22}
  +1.6e^{i23^\circ}\hat{z}_q^{12}\hat{z}_q^{33}
  +1.6e^{i23^\circ}\hat{z}_q^{13}\hat{z}_q^{32}
  \nline
  +1.3e^{i22^\circ}\hat{z}_q^{13}\hat{z}_u^{33}
  -1.2e^{i22^\circ}\re (\hat{z}_\phi) \hat{z}_q^{13}
  +8.1\cdt{-1}e^{-i68^\circ}\hat{z}_q^{12}\hat{z}_q^{32}
  \nline
  +6.3\cdt{-1}e^{-i72^\circ}\hat{z}_q^{11}\hat{z}_q^{12}
  +6.2\cdt{-1}\hat{z}_u^{23}\hat{z}_d^{13}
  -6.2\cdt{-1}\hat{z}_q^{23}\hat{z}_d^{12}
  \nline
  +3.9\cdt{-1}\hat{z}_q^{22}\hat{z}_q^{22}
  -3.7\cdt{-1}e^{i22^\circ}\hat{z}_q^{22}\hat{z}_q^{33}
  -3.7\cdt{-1}e^{i22^\circ}\hat{z}_q^{23}\hat{z}_q^{32}
  \nline
  +3.5\cdt{-1}e^{i44^\circ}\hat{z}_q^{33}\hat{z}_q^{33}
  -3.0\cdt{-1}e^{i22^\circ}\hat{z}_q^{23}\hat{z}_u^{33}
  +2.7\cdt{-1}e^{i22^\circ}\re (\hat{z}_\phi) \hat{z}_q^{23}
  \nline
  -2.4\cdt{-1}\hat{z}_d^{12}\hat{z}_d^{13}
  -1.7\cdt{-1}\hat{z}_q^{13}\hat{z}_u^{23}
  -1.5\cdt{-1} e^{-i73^\circ}\hat{z}_q^{12}\hat{z}_q^{21}
  \nline
  -1.5\cdt{-1} e^{-i73^\circ}\hat{z}_q^{11}\hat{z}_q^{22}
  +1.4\cdt{-1}e^{-i51^\circ}\hat{z}_q^{11}\hat{z}_q^{33}
  +1.4\cdt{-1}e^{-i51^\circ}\hat{z}_q^{13}\hat{z}_q^{31}
  \nline
  +1.1\cdt{-1}\hat{z}_q^{12}\hat{z}_d^{12}
  -6.3\cdt{-2} e^{i23^\circ}\hat{z}_q^{13}\hat{z}_d^{23}
  +5.5\cdt{-2}e^{i23^\circ}\hat{z}_q^{12}\hat{z}_u^{33}
  \nline
  -5.4\cdt{-2} e^{i34^\circ}\hat{z}_q^{11}\hat{z}_q^{11}
  -5.2\cdt{-2} e^{i23^\circ}\re (\hat{z}_\phi) \hat{z}_q^{12}
  +3.8\cdt{-2}\hat{z}_q^{23}\hat{z}_u^{23}
  \nline
  +3.0\cdt{-2}e^{i21^\circ}\hat{z}_q^{13}\hat{z}_d^{33}
  -3.0\cdt{-2} e^{i21^\circ}\hat{z}_q^{13}\hat{z}_d^{11}
  -2.6\cdt{-2}\hat{z}_q^{22}\hat{z}_d^{12}
  \nline
  +2.4\cdt{-2}e^{i21^\circ}\hat{z}_q^{33}\hat{z}_d^{12}
  +1.4\cdt{-2}e^{i23^\circ}\hat{z}_q^{23}\hat{z}_d^{23}
  -1.3\cdt{-2}e^{i22^\circ}\hat{z}_q^{22}\hat{z}_u^{33}
  \nline
  +1.2\cdt{-2}e^{i22^\circ}\re (\hat{z}_\phi) \hat{z}_q^{22}
  +1.1\cdt{-2}e^{i44^\circ}\hat{z}_q^{33}\hat{z}_u^{33}
  -1.0\cdt{-2} e^{i44^\circ}\re (\hat{z}_\phi) \hat{z}_q^{33}
  \nline
  +9.6\cdt{-3}e^{-i74^\circ}\hat{z}_q^{11}\hat{z}_d^{12}
  -7.2\cdt{-3}e^{i22^\circ}\hat{z}_d^{13}\hat{z}_d^{33}
  -7.1\cdt{-3}\hat{z}_q^{12}\hat{z}_u^{23}
  \nline
  -6.6\cdt{-3}e^{i22^\circ}\hat{z}_q^{23}\hat{z}_d^{33}
  +6.6\cdt{-3}e^{i22^\circ}\hat{z}_q^{23}\hat{z}_d^{11}
  +5.5\cdt{-3}e^{i23^\circ}\hat{z}_d^{13}\hat{z}_d^{23}
  \nline
  -5.4\cdt{-3}\hat{z}_u^{13}\hat{z}_d^{13}
  -5.3\cdt{-3}e^{i22^\circ}\hat{z}_d^{12}\hat{z}_d^{23} \,,
  \numberthis
\\
  \frac{M_{Z^\prime}^2\, \hat\Sigma^{K}_{Z^\prime}}{(5\TeV)^2} & =
  -5.0\cdt{6}\hat{z}_q^{12}\hat{z}_d^{12} -1.2\cdt{6}\hat{z}_q^{11}\hat{z}_d^{12}
  +1.2\cdt{6}\hat{z}_q^{22}\hat{z}_d^{12} +2.7\cdt{5}\hat{z}_q^{12}\hat{z}_d^{21}
  \nlineS{-0.3cm}
  +2.0\cdt{5}\hat{z}_q^{13}\hat{z}_d^{12} -4.7\cdt{4}\hat{z}_q^{23}\hat{z}_d^{12}
  -4.5\cdt{4}e^{i22^\circ}\hat{z}_q^{23}\hat{z}_d^{21}
  +1.8\cdt{4}\hat{z}_d^{12}\hat{z}_d^{12}
  \nline
  +1.6\cdt{4}\hat{z}_q^{12}\hat{z}_q^{12}
  -1.0\cdt{4}e^{i22^\circ}\hat{z}_q^{13}\hat{z}_d^{21}
  +3.5\cdt{3}\hat{z}_q^{11}\hat{z}_q^{12} -3.5\cdt{3}\hat{z}_q^{12}\hat{z}_q^{22}
  \nline
  +1.8\cdt{3}e^{i23^\circ}\hat{z}_q^{33}\hat{z}_d^{12}
  +8.5\cdt{2}\hat{z}_q^{11}\hat{z}_q^{11}
  -8.5\cdt{2}\hat{z}_q^{12}\hat{z}_q^{21} -8.5\cdt{2}\hat{z}_q^{11}\hat{z}_q^{22}
  \nline
  +8.5\cdt{2}\hat{z}_q^{22}\hat{z}_q^{22} -6.5\cdt{2}\hat{z}_q^{12}\hat{z}_q^{13}
  -2.5\cdt{2}\hat{z}_q^{12}\hat{z}_d^{13} -1.4\cdt{2}\hat{z}_q^{11}\hat{z}_q^{13}
  \nline
  +1.4\cdt{2}\hat{z}_q^{13}\hat{z}_q^{22} +1.4\cdt{2}\hat{z}_q^{12}\hat{z}_q^{23}
  +1.4\cdt{2}e^{i23^\circ}\hat{z}_q^{12}\hat{z}_q^{32}
  +6.6\cdt{1}e^{i11^\circ}\hat{z}_q^{11}\hat{z}_q^{23}
  \nline
  +6.6\cdt{1}e^{i11^\circ}\hat{z}_q^{12}\hat{z}_q^{31}
  -6.6\cdt{1} e^{i11^\circ}\hat{z}_q^{22}\hat{z}_q^{23}
  -5.6\cdt{1}\hat{z}_q^{11}\hat{z}_d^{13} +5.6\cdt{1}\hat{z}_q^{22}\hat{z}_d^{13}
  \nline
  +2.7\cdt{1}\hat{z}_q^{13}\hat{z}_q^{13}
  -2.5\cdt{1} e^{i23^\circ}\re (\hat{z}_\phi) \hat{z}_d^{12}
  +2.0\cdt{1}e^{i23^\circ}\hat{z}_u^{33}\hat{z}_d^{12}
  +1.3\cdt{1}\hat{z}_q^{12}\hat{z}_d^{31}
  \nline
  +1.0\cdt{1}\hat{z}_q^{13}\hat{z}_d^{13} -5.9\hat{z}_q^{13}\hat{z}_q^{23}
  -5.6 e^{i23^\circ}\hat{z}_q^{12}\hat{z}_q^{33}
  -5.5 e^{i23^\circ}\hat{z}_q^{13}\hat{z}_q^{32}
  +2.6e^{i22^\circ}\hat{z}_q^{12}\hat{z}_d^{32}
  \nline
  +2.5e^{i22^\circ}\hat{z}_q^{23}\hat{z}_q^{23}
  +2.5e^{i57^\circ}\hat{z}_q^{12}\hat{z}_d^{11}
  -2.5 e^{i57^\circ}\hat{z}_q^{12}\hat{z}_d^{22}
  -2.3\hat{z}_q^{23}\hat{z}_d^{13}
  \nline
  -2.2 e^{i23^\circ}\hat{z}_q^{23}\hat{z}_d^{31}
  -2.1\hat{z}_u^{23}\hat{z}_d^{12}
  -2.0e^{i22^\circ}\hat{z}_u^{23}\hat{z}_d^{21}
  +1.5e^{i20^\circ}\hat{z}_q^{22}\hat{z}_d^{22}
  \nline
  -1.5 e^{i20^\circ}\hat{z}_q^{22}\hat{z}_d^{11}
  -1.4 e^{i23^\circ}\hat{z}_q^{11}\hat{z}_q^{33}
  +1.3e^{i23^\circ}\hat{z}_q^{22}\hat{z}_q^{33}
  -1.3 e^{i23^\circ}\hat{z}_q^{13}\hat{z}_q^{31}
  \nline
  +1.3e^{i23^\circ}\hat{z}_q^{23}\hat{z}_q^{32}
  -1.1 e^{i27^\circ}\hat{z}_q^{11}\hat{z}_d^{22}
  +1.1e^{i27^\circ}\hat{z}_q^{11}\hat{z}_d^{11}
  +9.0\cdt{-1}\hat{z}_d^{12}\hat{z}_d^{13}
  \nline
  +6.1\cdt{-1}e^{i22^\circ}\hat{z}_q^{11}\hat{z}_d^{23}
  -6.0\cdt{-1} e^{i22^\circ}\hat{z}_q^{22}\hat{z}_d^{23}
  -5.0\cdt{-1} e^{i23^\circ}\hat{z}_q^{13}\hat{z}_d^{31}
  \nline
  +2.3\cdt{-1}e^{i23^\circ}\hat{z}_q^{13}\hat{z}_q^{33}
  +2.3\cdt{-1}e^{i22^\circ}\hat{z}_q^{13}\hat{z}_d^{22}
  -2.3\cdt{-1} e^{i22^\circ}\hat{z}_q^{13}\hat{z}_d^{11}
  \nline
  -2.0\cdt{-1} e^{i23^\circ}\hat{z}_q^{12}\hat{z}_u^{33}
  +1.9\cdt{-1}\hat{z}_u^{22}\hat{z}_d^{12}
  +1.9\cdt{-1}e^{i23^\circ} \re(\hat{z}_\phi) \hat{z}_q^{12}
  \nline
  -1.4\cdt{-1}e^{i22^\circ}\hat{z}_q^{12}\hat{z}_d^{23}
  -1.1\cdt{-1} e^{i23^\circ}\hat{z}_q^{13}\hat{z}_d^{32}
  -1.1\cdt{-1} e^{i34^\circ}\hat{z}_q^{23}\hat{z}_q^{33}
  \nline
  -1.0\cdt{-1} e^{i34^\circ}\hat{z}_q^{23}\hat{z}_d^{22}
  +1.0\cdt{-1}e^{i34^\circ}\hat{z}_q^{23}\hat{z}_d^{11}
  +9.0\cdt{-2}e^{i24^\circ}\hat{z}_q^{33}\hat{z}_d^{13}
  \nline
  -4.8\cdt{-2} e^{i23^\circ}\hat{z}_q^{11}\hat{z}_u^{33}
  +4.8\cdt{-2}e^{i23^\circ}\hat{z}_q^{22}\hat{z}_u^{33}
  +4.6\cdt{-2}e^{i23^\circ}\re (\hat{z}_\phi) \hat{z}_q^{11}
  \nline
  -4.6\cdt{-2} e^{i23^\circ}\re (\hat{z}_\phi) \hat{z}_q^{22}
  +4.1\cdt{-2}e^{i23^\circ}\hat{z}_d^{12}\hat{z}_d^{22}
  +2.5\cdt{-2}\hat{z}_q^{12}\hat{z}_u^{23}
  \nline
  +2.5\cdt{-2}e^{i23^\circ}\hat{z}_q^{23}\hat{z}_d^{32}
  +2.4\cdt{-2}e^{i22^\circ}\hat{z}_d^{13}\hat{z}_d^{32}
  +2.4\cdt{-2}e^{i23^\circ}\hat{z}_q^{12}\hat{z}_u^{32}
  \nline
  +2.3\cdt{-2}e^{i44^\circ}\hat{z}_q^{23}\hat{z}_d^{23}
  +1.8\cdt{-2}\hat{z}_u^{13}\hat{z}_d^{12}
  +1.2\cdt{-2}e^{i11^\circ}\hat{z}_q^{11}\hat{z}_u^{23}
  \nline
  -1.2\cdt{-2} e^{i11^\circ}\hat{z}_q^{22}\hat{z}_u^{23}
  -9.5\cdt{-3}e^{i22^\circ}\hat{z}_d^{12}\hat{z}_d^{32}
  +8.1\cdt{-3}e^{i23^\circ}\hat{z}_u^{11}\hat{z}_d^{12}
  \nline
  +7.7\cdt{-3}e^{i23^\circ}\hat{z}_q^{13}\hat{z}_u^{33}
  -7.3\cdt{-3} e^{i23^\circ}\re (\hat{z}_\phi) \hat{z}_q^{13}
  +5.4\cdt{-3}e^{i44^\circ}\hat{z}_q^{13}\hat{z}_d^{23} \,.
  \numberthis
\end{align*}

%
%
%
%--------+---------+---------+---------+---------+---------+---------+---------+
\subsection{$G'$}

\begin{align*}
  \frac{M_{G'}^2\, \hat\Sigma^{B_s}_{G'}}{(5 \TeV)^2} & =
  -2.6\cdt{3}\hat{g}_q^{23}\hat{g}_d^{23}
  -6.0\cdt{2}\hat{g}_q^{13}\hat{g}_d^{23}
  -1.1\cdt{2}\hat{g}_q^{22}\hat{g}_d^{23}
  +1.1\cdt{2}\hat{g}_q^{33}\hat{g}_d^{23}
  \nline
  +6.9\cdt{1}\hat{g}_d^{23}\hat{g}_d^{23}
  +6.2\cdt{1}\hat{g}_q^{23}\hat{g}_q^{23}
  -2.5\cdt{1}\hat{g}_q^{12}\hat{g}_d^{23}
  +1.4\cdt{1}\hat{g}_q^{13}\hat{g}_q^{23}
  \nline
  -9.3e^{-i73^\circ}\hat{g}_q^{12}\hat{g}_d^{32}
  +4.4\hat{g}_q^{23}\hat{g}_d^{32}
  +3.3\hat{g}_q^{13}\hat{g}_q^{13}
  -2.5\hat{g}_q^{23}\hat{g}_q^{33}
  +2.5\hat{g}_q^{23}\hat{g}_q^{22}
  \nline
  -2.1e^{-i73^\circ}\hat{g}_q^{11}\hat{g}_d^{23}
  +1.5\hat{g}_u^{33}\hat{g}_d^{23}
  +5.8\cdt{-1}\hat{g}_q^{12}\hat{g}_q^{23}
  +5.8\cdt{-1}\hat{g}_q^{13}\hat{g}_q^{22}
  \nline
  -5.9\cdt{-1}\hat{g}_q^{13}\hat{g}_q^{33}
  +3.8\cdt{-1}e^{-i74^\circ}\hat{g}_q^{13}\hat{g}_d^{32}
  +2.2\cdt{-1}e^{-i73^\circ}\hat{g}_q^{12}\hat{g}_q^{32}
  \nline
  -1.6\cdt{-1}\hat{g}_u^{23}\hat{g}_d^{23}
  +1.4\cdt{-1}\hat{g}_q^{12}\hat{g}_q^{13}
  +1.3\cdt{-1}\hat{g}_q^{23}\hat{g}_d^{22}
  -1.3\cdt{-1}\hat{g}_q^{23}\hat{g}_d^{33}
  \nline
  +1.1\cdt{-1}\hat{g}_q^{22}\hat{g}_q^{22}
  -1.1\cdt{-1}\hat{g}_q^{22}\hat{g}_q^{33}
  -1.1\cdt{-1}\hat{g}_q^{23}\hat{g}_q^{32}
  +1.0\cdt{-1}\hat{g}_q^{33}\hat{g}_q^{33}
  \nline
  -1.0\cdt{-1}\hat{g}_q^{23}\hat{g}_u^{33}
  +7.5\cdt{-2}e^{-i73^\circ}\hat{g}_q^{11}\hat{g}_q^{32}
  +7.5\cdt{-2}e^{-i73^\circ}\hat{g}_q^{13}\hat{g}_q^{21}
  +2.9\cdt{-2}\hat{g}_q^{13}\hat{g}_d^{22}
  \nline
  -2.9\cdt{-2}\hat{g}_q^{13}\hat{g}_d^{33}
  -2.8\cdt{-2}e^{-i19^\circ}\hat{g}_q^{12}\hat{g}_q^{33}
  -2.8\cdt{-2}e^{-i19^\circ}\hat{g}_q^{13}\hat{g}_q^{32}
  \nline
  +2.8\cdt{-2}e^{-i18^\circ}\hat{g}_q^{12}\hat{g}_q^{22}
  -2.5\cdt{-2}e^{-i73^\circ}\hat{g}_q^{12}\hat{g}_q^{31}
  -2.5\cdt{-2}e^{-i73^\circ}\hat{g}_q^{11}\hat{g}_q^{23}
  \nline
  -2.4\cdt{-2}\hat{g}_q^{13}\hat{g}_u^{33}
  +1.4\cdt{-2}\hat{g}_q^{23}\hat{g}_u^{23}
  +1.2\cdt{-2}e^{-i73^\circ}\hat{g}_q^{11}\hat{g}_d^{13}
  +7.4\cdt{-3}e^{-i10^\circ}\hat{g}_q^{23}\hat{g}_d^{13}
  \nline
  +6.5\cdt{-3}\hat{g}_q^{23}\hat{g}_d^{31}
  -6.0\cdt{-3}e^{-i3^\circ}\hat{g}_q^{33}\hat{g}_d^{13}
  +5.3\cdt{-3}\hat{g}_q^{22}\hat{g}_d^{22}
  -5.3\cdt{-3}\hat{g}_q^{22}\hat{g}_d^{33}
  \nline
  -\!5.3\cdt{-3}\hat{g}_q^{33}\hat{g}_d^{22}
  +\!5.2\cdt{-3}\hat{g}_q^{33}\hat{g}_d^{33}
  \!-\!5.0\cdt{-3}\hat{g}_q^{22}\hat{g}_d^{13}
  \!+\!5.0\cdt{-3}e^{-i5^\circ}\hat{g}_q^{12}\hat{g}_q^{12} \,,
  \numberthis
\\
  \frac{M_{G'}^2\, \hat\Sigma^{B_d}_{G'}}{(5 \TeV)^2} & =
  -6.4\cdt{4}\hat{g}_q^{13}\hat{g}_d^{13} +1.5\cdt{4}\hat{g}_q^{23}\hat{g}_d^{13}
  -2.6\cdt{3}\hat{g}_q^{12}\hat{g}_d^{13} +1.6\cdt{3}\hat{g}_d^{13}\hat{g}_d^{13}
  \nlineS{-0.3cm}
  +1.4\cdt{3}\hat{g}_q^{13}\hat{g}_q^{13} +6.1\cdt{2}\hat{g}_q^{22}\hat{g}_d^{13}
  -5.7\cdt{2}e^{i22^\circ}\hat{g}_q^{33}\hat{g}_d^{13}
  -3.3\cdt{2}\hat{g}_q^{13}\hat{g}_q^{23}
  \nline
  -2.3\cdt{2}e^{-i73^\circ}\hat{g}_q^{11}\hat{g}_d^{13}
  +7.6\cdt{1}\hat{g}_q^{23}\hat{g}_q^{23}
  +6.0\cdt{1}\hat{g}_q^{12}\hat{g}_q^{13}
  +5.2\cdt{1}e^{-i73^\circ}\hat{g}_q^{12}\hat{g}_d^{31}
  \nline
  -2.4\cdt{1}e^{i22^\circ}\hat{g}_q^{23}\hat{g}_d^{31}
  -1.4\cdt{1}\hat{g}_q^{12}\hat{g}_q^{23}
  -1.4\cdt{1}\hat{g}_q^{13}\hat{g}_q^{22}
  +1.3\cdt{1}e^{i22^\circ}\hat{g}_q^{13}\hat{g}_q^{33}
  \nline
  -7.9e^{i22^\circ}\hat{g}_u^{33}\hat{g}_d^{13}
  +5.2e^{-i73^\circ}\hat{g}_q^{13}\hat{g}_q^{11}
  +3.2\hat{g}_q^{22}\hat{g}_q^{23}
  +3.1\hat{g}_q^{13}\hat{g}_d^{12}
  -3.0e^{i22^\circ}\hat{g}_q^{23}\hat{g}_q^{33}
  \nline
  +2.5\hat{g}_q^{12}\hat{g}_q^{12}
  -2.0e^{-i51^\circ}\hat{g}_q^{13}\hat{g}_d^{31}
  -1.8e^{-i73^\circ}\hat{g}_q^{11}\hat{g}_q^{32}
  -1.8e^{-i73^\circ}\hat{g}_q^{13}\hat{g}_q^{21}
  \nline
  +8.7\cdt{-1}\hat{g}_u^{23}\hat{g}_d^{13}
  -7.2\cdt{-1}\hat{g}_q^{23}\hat{g}_d^{12}
  +5.8\cdt{-1}e^{-i73^\circ}\hat{g}_q^{12}\hat{g}_q^{31}
  +5.8\cdt{-1}e^{-i73^\circ}\hat{g}_q^{11}\hat{g}_q^{23}
  \nline
  -5.6\cdt{-1}\hat{g}_q^{12}\hat{g}_q^{22}
  +5.3\cdt{-1}e^{i23^\circ}\hat{g}_q^{13}\hat{g}_q^{32}
  +5.3\cdt{-1}e^{i23^\circ}\hat{g}_q^{12}\hat{g}_q^{33}
  \nline
  +5.1\cdt{-1}e^{i22^\circ}\hat{g}_q^{13}\hat{g}_u^{33}
  +2.7\cdt{-1}e^{-i68^\circ}\hat{g}_q^{12}\hat{g}_q^{32}
  +2.2\cdt{-1}e^{-i72^\circ}\hat{g}_q^{12}\hat{g}_q^{11}
  \nline
  +1.3\cdt{-1}\hat{g}_q^{22}\hat{g}_q^{22}
  +1.3\cdt{-1}\hat{g}_q^{12}\hat{g}_d^{12}
  -1.2\cdt{-1}e^{i22^\circ}\hat{g}_q^{23}\hat{g}_q^{32}
  \nline
  -1.2\cdt{-1}e^{i22^\circ}\hat{g}_q^{22}\hat{g}_q^{33}
  -1.2\cdt{-1}e^{i22^\circ}\hat{g}_q^{23}\hat{g}_u^{33}
  +1.2 \cdt{-1}e^{i44^\circ}\hat{g}_q^{33}\hat{g}_q^{33}
  \nline
  -7.8\cdt{-2}\hat{g}_d^{12}\hat{g}_d^{13}
  -7.3\cdt{-2}\hat{g}_q^{13}\hat{g}_u^{23}
  -7.3\cdt{-2}e^{i23^\circ}\hat{g}_d^{23}\hat{g}_q^{13}
  -4.8\cdt{-2}e^{-i73^\circ}\hat{g}_q^{11}\hat{g}_q^{22}
  \nline
  -4.8\cdt{-2}e^{-i73^\circ}\hat{g}_q^{12}\hat{g}_q^{21}
  +4.7\cdt{-2}e^{-i51^\circ}\hat{g}_q^{13}\hat{g}_q^{31}
  +4.4\cdt{-2}e^{-i51^\circ}\hat{g}_q^{11}\hat{g}_q^{33}
  \nline
  -3.6\cdt{-2}e^{i21^\circ}\hat{g}_d^{11}\hat{g}_q^{13}
  +3.5\cdt{-2}e^{i21^\circ}\hat{g}_d^{33}\hat{g}_q^{13}
  -3.0\cdt{-2}\hat{g}_d^{12}\hat{g}_q^{22}
  \nline
  +2.8\cdt{-2}e^{i21^\circ}\hat{g}_d^{12}\hat{g}_q^{33}
  +2.3\cdt{-2}e^{i23^\circ}\hat{g}_q^{12}\hat{g}_u^{33}
  -1.8\cdt{-2}e^{i34^\circ}\hat{g}_q^{11}\hat{g}_q^{11}
  \nline
  +1.7\cdt{-2}\hat{g}_q^{23}\hat{g}_u^{23}
  +1.6\cdt{-2}e^{i23^\circ}\hat{g}_d^{23}\hat{g}_q^{23}
  +1.1\cdt{-2}e^{-i74^\circ}\hat{g}_d^{12}\hat{g}_q^{11}
  \nline
  +7.8\cdt{-3}e^{i22^\circ}\hat{g}_q^{23}\hat{g}_d^{11}
  -7.6\cdt{-3}\hat{g}_u^{13}\hat{g}_d^{13}
  -7.4\cdt{-3}e^{i22^\circ}\hat{g}_q^{23}\hat{g}_d^{33}
  \nline
  -5.5\cdt{-3}e^{i22^\circ}\hat{g}_u^{11}\hat{g}_d^{13}
  -5.4\cdt{-3}e^{i22^\circ}\hat{g}_u^{22}\hat{g}_d^{13}
  -5.3\cdt{-3}e^{i22^\circ}\hat{g}_q^{22}\hat{g}_u^{33} \,,
  \numberthis
\\
  \frac{M_{G'}^2\, \hat\Sigma^{K}_{G'}}{(5 \TeV)^2} & =
  -7.2\cdt{6}\hat{g}_q^{12}\hat{g}_d^{12} -1.6\cdt{6}\hat{g}_q^{11}\hat{g}_d^{12}
  +1.6\cdt{6}\hat{g}_q^{22}\hat{g}_d^{12} +3.8\cdt{5}\hat{g}_q^{12}\hat{g}_d^{21}
  \nlineS{-0.3cm}
  +2.9\cdt{5}\hat{g}_q^{13}\hat{g}_d^{12} -6.7\cdt{4}\hat{g}_q^{23}\hat{g}_d^{12}
  -6.4\cdt{4}e^{i22^\circ}\hat{g}_q^{23}\hat{g}_d^{21}
  -1.5\cdt{4}e^{i22^\circ}\hat{g}_q^{13}\hat{g}_d^{21}
  \nline
  +6.1\cdt{3}\hat{g}_d^{12}\hat{g}_d^{12} +5.3\cdt{3}\hat{g}_q^{12}\hat{g}_q^{12}
  +2.6\cdt{3}e^{i23^\circ}\hat{g}_q^{33}\hat{g}_d^{12}
  -1.2\cdt{3}\hat{g}_q^{12}\hat{g}_q^{22}
  \nline
  +1.2\cdt{3}\hat{g}_q^{11}\hat{g}_q^{12}
  -3.5\cdt{2}\hat{g}_d^{13}\hat{g}_q^{12}
  -2.8\cdt{2}\hat{g}_q^{11}\hat{g}_q^{22}
  -2.8\cdt{2}\hat{g}_q^{12}\hat{g}_q^{21}
  \nline
  +2.8\cdt{2}\hat{g}_q^{22}\hat{g}_q^{22}
  +2.8\cdt{2}\hat{g}_q^{11}\hat{g}_q^{11}
  -2.2\cdt{2}\hat{g}_q^{12}\hat{g}_q^{13}
  -8.1\cdt{1}\hat{g}_d^{13}\hat{g}_q^{11}
  \nline
  +8.0\cdt{1}\hat{g}_d^{13}\hat{g}_q^{22}
  +4.8\cdt{1}\hat{g}_q^{12}\hat{g}_q^{23}
  +4.8\cdt{1}\hat{g}_q^{13}\hat{g}_q^{22}
  -4.8\cdt{1}\hat{g}_q^{11}\hat{g}_q^{13}
  \nline
  +4.5\cdt{1}e^{i23^\circ}\hat{g}_q^{12}\hat{g}_q^{32}
  +3.8\cdt{1}e^{i23^\circ}\hat{g}_d^{12}\hat{g}_u^{33}
  +3.3\cdt{1}e^{i11^\circ}\hat{g}_q^{11}\hat{g}_q^{32}
  \nline
  +3.3\cdt{1}e^{i11^\circ}\hat{g}_q^{13}\hat{g}_q^{21}
  -2.2\cdt{1}e^{i11^\circ}\hat{g}_q^{22}\hat{g}_q^{23}
  +1.9\cdt{1}\hat{g}_d^{31}\hat{g}_q^{12}
  +1.4\cdt{1}\hat{g}_d^{13}\hat{g}_q^{13}
  \nline
  -1.1\cdt{1}e^{i11^\circ}\hat{g}_q^{12}\hat{g}_q^{31}
  -1.1\cdt{1}e^{i11^\circ}\hat{g}_q^{11}\hat{g}_q^{23}
  +8.8\hat{g}_q^{13}\hat{g}_q^{13}
  -4.2\hat{g}_d^{12}\hat{g}_u^{23}
  \nline
  -4.2e^{i22^\circ}\hat{g}_d^{21}\hat{g}_u^{23}
  +3.8e^{i22^\circ}\hat{g}_d^{32}\hat{g}_q^{12}
  -3.3\hat{g}_d^{13}\hat{g}_q^{23}
  -3.1e^{i23^\circ}\hat{g}_d^{31}\hat{g}_q^{23}
  -3.2e^{i87^\circ}\hat{g}_d^{22}\hat{g}_q^{12}
  \nline
  +3.2e^{i87^\circ}\hat{g}_d^{11}\hat{g}_q^{12}
  +2.3e^{i19^\circ}\hat{g}_d^{22}\hat{g}_q^{22}
  -2.3e^{i19^\circ}\hat{g}_d^{11}\hat{g}_q^{22}
  -2.0\hat{g}_q^{13}\hat{g}_q^{23}
  -1.8e^{i23^\circ}\hat{g}_q^{12}\hat{g}_q^{33}
  \nline
  -1.8e^{i23^\circ}\hat{g}_q^{13}\hat{g}_q^{32}
  -1.5e^{i29^\circ}\hat{g}_d^{22}\hat{g}_q^{11}
  +1.5e^{i29^\circ}\hat{g}_d^{11}\hat{g}_q^{11}
  +8.7\cdt{-1}e^{i22^\circ}\hat{g}_d^{23}\hat{g}_q^{11}
  \nline
  -8.6\cdt{-1}e^{i22^\circ}\hat{g}_d^{23}\hat{g}_q^{22}
  +8.3\cdt{-1}e^{i22^\circ}\hat{g}_q^{23}\hat{g}_q^{23}
  -7.2\cdt{-1}e^{i23^\circ}\hat{g}_d^{31}\hat{g}_q^{13}
  \nline
  +4.5\cdt{-1}e^{i23^\circ}\hat{g}_q^{22}\hat{g}_q^{33}
  -4.5\cdt{-1}e^{i23^\circ}\hat{g}_q^{11}\hat{g}_q^{33}
  +4.5\cdt{-1}e^{i23^\circ}\hat{g}_q^{23}\hat{g}_q^{32}
  \nline
  -4.5\cdt{-1}e^{i23^\circ}\hat{g}_q^{13}\hat{g}_q^{31}
  +3.9\cdt{-1}\hat{g}_d^{12}\hat{g}_u^{22}
  +3.3\cdt{-1}e^{i22^\circ}\hat{g}_d^{22}\hat{g}_q^{13}
  \nline
  -3.3\cdt{-1}e^{i22^\circ}\hat{g}_d^{11}\hat{g}_q^{13}
  +3.0\cdt{-1}\hat{g}_d^{12}\hat{g}_d^{13}
  -2.0\cdt{-1}e^{i22^\circ}\hat{g}_d^{23}\hat{g}_q^{12}
  \nline
  -1.5\cdt{-1}e^{i23^\circ}\hat{g}_d^{32}\hat{g}_q^{13}
  -1.4\cdt{-1}e^{i34^\circ}\hat{g}_d^{22}\hat{g}_q^{23}
  +1.4\cdt{-1}e^{i34^\circ}\hat{g}_d^{11}\hat{g}_q^{23}
  \nline
  +1.3\cdt{-1}e^{i24^\circ}\hat{g}_d^{13}\hat{g}_q^{33}
  -8.2\cdt{-2}e^{i23^\circ}\hat{g}_q^{12}\hat{g}_u^{33}
  +7.7\cdt{-2}e^{i23^\circ}\hat{g}_q^{13}\hat{g}_q^{33}
  \nline
  +3.6\cdt{-2}\hat{g}_d^{12}\hat{g}_u^{13}
  -3.5\cdt{-2}e^{i34^\circ}\hat{g}_q^{23}\hat{g}_q^{33}
  +3.5\cdt{-2}e^{i23^\circ}\hat{g}_d^{32}\hat{g}_q^{23}
  \nline
  +3.4\cdt{-2}e^{i44^\circ}\hat{g}_d^{23}\hat{g}_q^{23}
  +2.5\cdt{-2}e^{i22^\circ}\hat{g}_d^{12}\hat{g}_u^{11}
  +2.0\cdt{-2}e^{i23^\circ}\hat{g}_q^{22}\hat{g}_u^{33}
  \nline
  -2.0\cdt{-2}e^{i23^\circ}\hat{g}_q^{11}\hat{g}_u^{33}
  +1.4\cdt{-2}e^{i23^\circ}\hat{g}_d^{12}\hat{g}_d^{22}
  +1.2\cdt{-2}e^{i22^\circ}\hat{g}_d^{12}\hat{g}_d^{33}
  \nline
  +1.1\cdt{-2}\hat{g}_q^{12}\hat{g}_u^{23}
  +1.0\cdt{-2}e^{i23^\circ}\hat{g}_q^{12}\hat{g}_u^{32}
  +7.7\cdt{-3}e^{i44^\circ}\hat{g}_d^{23}\hat{g}_q^{13}
  \nline
  -5.4\cdt{-3}e^{i22^\circ}\hat{g}_d^{13}\hat{g}_d^{32}
  -5.0\cdt{-3}e^{i11^\circ}\hat{g}_q^{22}\hat{g}_u^{23}
  +5.0\cdt{-3}e^{i11^\circ}\hat{g}_q^{11}\hat{g}_u^{23} \,.
  \numberthis
\end{align*}

%
%
%
%--------+---------+---------+---------+---------+---------+---------+---------+
\subsection{$\varphi$}

\begin{align*}
  \frac{M_{\varphi}^2\, \hat\Sigma^{B_s}_{\varphi}}{(5\TeV)^2} & =
  -3.0\cdt{3}\hat{Y}_d^{32}\hat{Y}_d^{23*}
  -6.9\cdt{2}\hat{Y}_d^{31}\hat{Y}_d^{23*}
  -1.2\cdt{2}\hat{Y}_d^{32}\hat{Y}_d^{22*}
  +1.2\cdt{2}\hat{Y}_d^{33}\hat{Y}_d^{23*}
  \nlineS{-0.3cm}
  -2.9\cdt{1}\hat{Y}_d^{31}\hat{Y}_d^{22*}
  -1.1\cdt{1} e^{-i73^\circ}\hat{Y}_d^{21}\hat{Y}_d^{32*}
  +5.0\hat{Y}_d^{22}\hat{Y}_d^{33*}
  -2.4 e^{-i73^\circ}\hat{Y}_d^{31}\hat{Y}_d^{21*}
  \nline
  +4.3\cdt{-1}e^{-i74^\circ}\hat{Y}_d^{21}\hat{Y}_d^{33*}
  -1.5\cdt{-1}\hat{Y}_u^{33}\hat{Y}_u^{23*}
  +1.5\cdt{-1}\hat{Y}_d^{22}\hat{Y}_d^{23*}
  \nline
  -1.4\cdt{-1}\hat{Y}_d^{32}\hat{Y}_d^{33*}
  +3.4\cdt{-2}\hat{Y}_d^{21}\hat{Y}_d^{23*}
  -3.4\cdt{-2}\hat{Y}_d^{31}\hat{Y}_d^{33*}
  \nline
  -3.4\cdt{-2}\hat{Y}_u^{33}\hat{Y}_u^{13*}
  +1.9\cdt{-2}\hat{Y}_u^{32}\hat{Y}_u^{23*}
  +8.3\cdt{-3}e^{-i10^\circ}\hat{Y}_d^{32}\hat{Y}_d^{13*}
  \nline
  +7.2\cdt{-3}\hat{Y}_d^{12}\hat{Y}_d^{33*}
  -6.7\cdt{-3}e^{-i3^\circ}\hat{Y}_d^{33}\hat{Y}_d^{13*}
  -6.4\cdt{-3}\hat{Y}_u^{23}\hat{Y}_u^{23*}
  \nline
  +6.1\cdt{-3}\hat{Y}_d^{22}\hat{Y}_d^{22*}
  -6.1\cdt{-3}\hat{Y}_d^{32}\hat{Y}_d^{32*}
  -6.0\cdt{-3}\hat{Y}_d^{23}\hat{Y}_d^{23*}
  \nline
  +6.0\cdt{-3}\hat{Y}_d^{33}\hat{Y}_d^{33*}
  +5.7\cdt{-3}\hat{Y}_u^{33}\hat{Y}_u^{33*}
  -5.6\cdt{-3}\hat{Y}_d^{32}\hat{Y}_d^{12*} \,,
  \numberthis
\\
  \frac{M_{\varphi}^2\, \hat\Sigma^{B_d}_{\varphi}}{(5\TeV)^2} & =
  -7.3\cdt{4}\hat{Y}_d^{31}\hat{Y}_d^{13*} +1.7\cdt{4}\hat{Y}_d^{32}\hat{Y}_d^{13*}
  -3.0\cdt{3}\hat{Y}_d^{31}\hat{Y}_d^{12*}
  +7.0\cdt{2}\hat{Y}_d^{32}\hat{Y}_d^{12*}
  \nlineS{-0.3cm}
  -6.6\cdt{2} e^{i22^\circ}\hat{Y}_d^{33}\hat{Y}_d^{13*}
  -2.6\cdt{2} e^{-i73^\circ}\hat{Y}_d^{31}\hat{Y}_d^{11*}
  +6.0\cdt{1}e^{-i73^\circ}\hat{Y}_d^{11}\hat{Y}_d^{32*}
  \nline
  -2.7\cdt{1}e^{i22^\circ}\hat{Y}_d^{12}\hat{Y}_d^{33*}
  +3.6\hat{Y}_d^{21}\hat{Y}_d^{13*}
  -2.3 e^{-i51^\circ}\hat{Y}_d^{11}\hat{Y}_d^{33*}
  -8.2\cdt{-1}\hat{Y}_d^{22}\hat{Y}_d^{13*}
  \nline
  +7.2\cdt{-1}e^{i22^\circ}\hat{Y}_u^{33}\hat{Y}_u^{13*}
  -1.7\cdt{-1}e^{i22^\circ}\hat{Y}_u^{33}\hat{Y}_u^{23*}
  +1.5\cdt{-1}\hat{Y}_d^{21}\hat{Y}_d^{12*}
  \nline
  -1.0\cdt{-1}\hat{Y}_u^{32}\hat{Y}_u^{13*}
  -8.3\cdt{-2} e^{i23^\circ}\hat{Y}_d^{31}\hat{Y}_d^{23*}
  -4.0\cdt{-2} e^{i21^\circ}\hat{Y}_d^{11}\hat{Y}_d^{13*}
  \nline
  +4.0\cdt{-2}e^{i21^\circ}\hat{Y}_d^{31}\hat{Y}_d^{33*}
  -3.4\cdt{-2}\hat{Y}_d^{22}\hat{Y}_d^{12*}
  +3.2\cdt{-2}e^{i23^\circ}\hat{Y}_u^{23}\hat{Y}_u^{13*}
  \nline
  +3.2\cdt{-2}e^{i21^\circ}\hat{Y}_d^{23}\hat{Y}_d^{13*}
  +2.3\cdt{-2}\hat{Y}_u^{32}\hat{Y}_u^{23*}
  +1.9\cdt{-2}e^{i23^\circ}\hat{Y}_d^{32}\hat{Y}_d^{23*}
  \nline
  +1.3\cdt{-2}e^{-i74^\circ}\hat{Y}_d^{21}\hat{Y}_d^{11*}
  +9.0\cdt{-3}e^{i22^\circ}\hat{Y}_d^{12}\hat{Y}_d^{13*}
  -8.5\cdt{-3}e^{i22^\circ}\hat{Y}_d^{32}\hat{Y}_d^{33*}
  \nline
  -7.4\cdt{-3}e^{i22^\circ}\hat{Y}_u^{23}\hat{Y}_u^{23*}
  +6.3\cdt{-3}e^{i44^\circ}\hat{Y}_u^{33}\hat{Y}_u^{33*} \,,
  \numberthis
\\
  \frac{M_{\varphi}^2\, \hat\Sigma^{K}_{\varphi}}{(5\TeV)^2} & =
  -8.0\cdt{6}\hat{Y}_d^{21}\hat{Y}_d^{12*} -1.8\cdt{6}\hat{Y}_d^{21}\hat{Y}_d^{11*}
  +1.8\cdt{6}\hat{Y}_d^{22}\hat{Y}_d^{12*}
  +4.2\cdt{5}\hat{Y}_d^{11}\hat{Y}_d^{22*}
  \nlineS{-0.3cm}
  +3.3\cdt{5}\hat{Y}_d^{21}\hat{Y}_d^{13*}
  -7.5\cdt{4}\hat{Y}_d^{22}\hat{Y}_d^{13*}
  -7.1\cdt{4}e^{i22^\circ}\hat{Y}_d^{12}\hat{Y}_d^{23*}
  \nline
  -1.6\cdt{4}e^{i22^\circ}\hat{Y}_d^{11}\hat{Y}_d^{23*}
  +2.9\cdt{3}e^{i23^\circ}\hat{Y}_d^{23}\hat{Y}_d^{13*}
  -3.9\cdt{2}\hat{Y}_d^{31}\hat{Y}_d^{12*}
  \nline
  -9.0\cdt{1}\hat{Y}_d^{31}\hat{Y}_d^{11*} +9.0\cdt{1}\hat{Y}_d^{32}\hat{Y}_d^{12*}
  +2.1\cdt{1}\hat{Y}_d^{11}\hat{Y}_d^{32*} +1.6\cdt{1}\hat{Y}_d^{31}\hat{Y}_d^{13*}
  \nline
  +4.2e^{i22^\circ}\hat{Y}_d^{21}\hat{Y}_d^{32*}
  -3.7\hat{Y}_d^{32}\hat{Y}_d^{13*} -3.5 e^{i23^\circ}\hat{Y}_d^{12}\hat{Y}_d^{33*}
  +3.3e^{i83^\circ}\hat{Y}_d^{11}\hat{Y}_d^{12*}
  \nline
  -3.3e^{i83^\circ}\hat{Y}_d^{21}\hat{Y}_d^{22*}
  +2.5e^{i19^\circ}\hat{Y}_d^{22}\hat{Y}_d^{22*}
  -2.5 e^{i19^\circ}\hat{Y}_d^{12}\hat{Y}_d^{12*}
  -1.7 e^{i29^\circ}\hat{Y}_d^{21}\hat{Y}_d^{21*}
  \nline
  +1.7e^{i29^\circ}\hat{Y}_d^{11}\hat{Y}_d^{11*}
  +9.7\cdt{-1}e^{i22^\circ}\hat{Y}_d^{31}\hat{Y}_d^{21*}
  -9.6\cdt{-1}e^{i22^\circ}\hat{Y}_d^{32}\hat{Y}_d^{22*}
  \nline
  -8.0\cdt{-1}e^{i23^\circ}\hat{Y}_d^{11}\hat{Y}_d^{33*}
  +3.6\cdt{-1}e^{i22^\circ}\hat{Y}_d^{21}\hat{Y}_d^{23*}
  -3.6\cdt{-1}e^{i22^\circ}\hat{Y}_d^{11}\hat{Y}_d^{13*}
  \nline
  -2.2\cdt{-1}e^{i22^\circ}\hat{Y}_d^{31}\hat{Y}_d^{22*}
  -1.7\cdt{-1}e^{i23^\circ}\hat{Y}_d^{21}\hat{Y}_d^{33*}
  -1.6\cdt{-1}ee^{i34^\circ}\hat{Y}_d^{22}\hat{Y}_d^{23*}
  \nline
  +1.6\cdt{-1}e^{i34^\circ}\hat{Y}_d^{12}\hat{Y}_d^{13*}
  +1.4\cdt{-1}e^{i24^\circ}\hat{Y}_d^{33}\hat{Y}_d^{13*}
  -1.2\cdt{-1}e^{i23^\circ}\hat{Y}_u^{23}\hat{Y}_u^{13*}
  \nline
  +3.9\cdt{-2}e^{i23^\circ}\hat{Y}_d^{22}\hat{Y}_d^{33*}
  +3.7\cdt{-2}e^{i44^\circ}\hat{Y}_d^{32}\hat{Y}_d^{23*}
  -2.8\cdt{-2} e^{i23^\circ}\hat{Y}_u^{13}\hat{Y}_u^{13*}
  \nline
  +2.8\cdt{-2}e^{i23^\circ}\hat{Y}_u^{23}\hat{Y}_u^{23*}
  +1.5\cdt{-2}\hat{Y}_u^{22}\hat{Y}_u^{13*}
  +1.4\cdt{-2}e^{i23^\circ}\hat{Y}_u^{23}\hat{Y}_u^{12*}
  \nline
  +8.6\cdt{-3}e^{i44^\circ}\hat{Y}_d^{31}\hat{Y}_d^{23*}
  +7.0\cdt{-3}e^{i11^\circ}\hat{Y}_u^{12}\hat{Y}_u^{13*}
  -7.0\cdt{-3} e^{i11^\circ}\hat{Y}_u^{22}\hat{Y}_u^{23*} \,.
  \numberthis
\end{align*}

%
%
%
%--------+---------+---------+---------+---------+---------+---------+---------+
\subsection{$\Phi^A$}

\begin{align*}
  \frac{M_{\Phi}^2\, \hat\Sigma^{B_s}_{\Phi}}{(5 \TeV)^2} & =
  -6.6\cdt{1}\hat{X}_d^{32}\hat{X}_d^{23*}
  -1.5\cdt{1}\hat{X}_d^{31}\hat{X}_d^{23*}
  -2.7\hat{X}_d^{32}\hat{X}_d^{22*} +2.7\hat{X}_d^{33}\hat{X}_d^{23*}
  \nlineS{-0.3cm}
  -6.3\cdt{-1}\hat{X}_d^{31}\hat{X}_d^{22*}
  -2.3\cdt{-1} e^{-i73^\circ}\hat{X}_d^{21}\hat{X}_d^{32*}
  +1.1\cdt{-1}\hat{X}_d^{22}\hat{X}_d^{33*}
  \nline
  -5.4\cdt{-2} e^{-i72^\circ}\hat{X}_d^{31}\hat{X}_d^{21*}
  -4.0\cdt{-2}\hat{X}_u^{33}\hat{X}_u^{23*}
  +9.5\cdt{-3}e^{-i74^\circ}\hat{X}_d^{21}\hat{X}_d^{33*}
  \nline
  -9.1\cdt{-3}\hat{X}_u^{33}\hat{X}_u^{13*} \,,
  \numberthis
\\
  \frac{M_{\Phi}^2\, \hat\Sigma^{B_d}_{\Phi}}{(5 \TeV)^2} & =
  -1.6\cdt{3}\hat{X}_d^{31}\hat{X}_d^{13*} +3.8\cdt{2}\hat{X}_d^{32}\hat{X}_d^{13*}
  -6.8\cdt{1}\hat{X}_d^{31}\hat{X}_d^{12*}
  +1.6\cdt{1}\hat{X}_d^{32}\hat{X}_d^{12*}
  \nlineS{-0.3cm}
  -1.5\cdt{1}e^{i22^\circ}\hat{X}_d^{33}\hat{X}_d^{13*}
  -5.8e^{-i73^\circ}\hat{X}_d^{31}\hat{X}_d^{11*}
  +1.3e^{-i73^\circ}\hat{X}_d^{11}\hat{X}_d^{32*}
  \nline
  -6.0\cdt{-1}e^{i22^\circ}\hat{X}_d^{12}\hat{X}_d^{33*}
  +2.0\cdt{-1}e^{i22^\circ}\hat{X}_u^{33}\hat{X}_u^{13*}
  +8.0\cdt{-2}\hat{X}_d^{21}\hat{X}_d^{13*}
  \nline
  -5.2\cdt{-2} e^{-i51^\circ}\hat{X}_d^{11}\hat{X}_d^{33*}
  -4.7\cdt{-2} e^{i22^\circ}\hat{X}_u^{33}\hat{X}_u^{23*}
  -2.5\cdt{-2}\hat{X}_u^{32}\hat{X}_u^{13*}
  \nline
  -1.8\cdt{-2}\hat{X}_d^{22}\hat{X}_d^{13*}
  +8.4\cdt{-3}e^{i23^\circ}\hat{X}_u^{23}\hat{X}_u^{13*}
  +5.8\cdt{-3}\hat{X}_u^{32}\hat{X}_u^{23*} \,,
  \numberthis
\\
  \frac{M_{\Phi}^2\, \hat\Sigma^{K}_{\Phi}}{(5 \TeV)^2} & =
   7.9\cdt{4}\hat{X}_d^{21}\hat{X}_d^{12*} +1.8\cdt{4}\hat{X}_d^{21}\hat{X}_d^{11*}
  -1.8\cdt{4}\hat{X}_d^{22}\hat{X}_d^{12*}
  \nlineS{-0.3cm}
  -4.2\cdt{3}\hat{X}_d^{11}\hat{X}_d^{22*} -3.2\cdt{3}\hat{X}_d^{21}\hat{X}_d^{13*}
  +7.4\cdt{2}\hat{X}_d^{22}\hat{X}_d^{13*}
  \nline
  +7.1\cdt{2}e^{i22^\circ}\hat{X}_d^{12}\hat{X}_d^{23*}
  +1.6\cdt{2}e^{i22^\circ}\hat{X}_d^{11}\hat{X}_d^{23*}
  -2.9\cdt{1} e^{i23^\circ}\hat{X}_d^{23}\hat{X}_d^{13*}
  \nline
  +3.9\hat{X}_d^{31}\hat{X}_d^{12*} +8.9\cdt{-1}\hat{X}_d^{31}\hat{X}_d^{11*}
  -8.9\cdt{-1}\hat{X}_d^{32}\hat{X}_d^{12*}
  \nline
  -2.7\cdt{-1}e^{-i7^\circ}\hat{X}_d^{21}\hat{X}_d^{22*}
  +2.7\cdt{-1}e^{-i7^\circ}\hat{X}_d^{11}\hat{X}_d^{12*}
  -2.1\cdt{-1}\hat{X}_d^{11}\hat{X}_d^{32*}
  \nline
  -1.6\cdt{-1}\hat{X}_d^{31}\hat{X}_d^{13*}
  -4.2\cdt{-2} e^{i22^\circ}\hat{X}_d^{21}\hat{X}_d^{32*}
  -4.0\cdt{-2} e^{i11^\circ}\hat{X}_d^{22}\hat{X}_d^{22*}
  \nline
  +4.0\cdt{-2}e^{i11^\circ}\hat{X}_d^{12}\hat{X}_d^{12*}
  +3.6\cdt{-2}\hat{X}_d^{32}\hat{X}_d^{13*}
  +3.5\cdt{-2}e^{i23^\circ}\hat{X}_d^{12}\hat{X}_d^{33*}
  \nline
  -3.0\cdt{-2}e^{i23^\circ}\hat{X}_u^{23}\hat{X}_u^{13*}
  -9.6\cdt{-3}e^{i22^\circ}\hat{X}_d^{31}\hat{X}_d^{21*}
  +9.6\cdt{-3}e^{i22^\circ}\hat{X}_d^{32}\hat{X}_d^{22*}
  \nline
  -8.1\cdt{-3}e^{-i80^\circ}\hat{X}_d^{21}\hat{X}_d^{21*}
  +8.1\cdt{-3}e^{-i80^\circ}\hat{X}_d^{11}\hat{X}_d^{11*}
  +8.0\cdt{-3}e^{i23^\circ}\hat{X}_d^{11}\hat{X}_d^{33*}
  \nline
  -7.3\cdt{-3}e^{i23^\circ}\hat{X}_u^{13}\hat{X}_u^{13*}
  +7.3\cdt{-3}e^{i23^\circ}\hat{X}_u^{23}\hat{X}_u^{23*} \,.
  \numberthis
\end{align*}

%--------+---------+---------+---------+---------+---------+---------+---------+
%
%
%
%--------+---------+---------+---------+---------+---------+---------+---------+
\section{LEFT bases changes}
\label{Relations}

The transformation of the Wilson coefficients between the various
LEFT bases read:\\[0.2cm]
SUSY $\to$ BMU:
\begin{align}
  \WcL{\text{VLL}}  & = \WcL{1} \,, &
  \WcL{\text{LR},1} & = - \frac{1}{2} \WcL{5} \,,
&
  \WcL{\text{LR},2} & = \WcL{4} \,,
  \label{eq:susy-bmu}
\intertext{JMS $\to$ BMU: }
  \WcL{\text{VLL}}  & = \WcL[V,LL]{dd} , &
  \WcL{\text{LR},1} & = \WcL[V1,LR]{dd} - \frac{1}{6} \WcL[V8,LR]{dd} ,
&
  \WcL{\text{LR},2} & = -\WcL[V8,LR]{dd} ,
  \label{eq:jms-bmu}
\intertext{BMU $\to$ JMS:}
  \WcL[V,LL]{dd}  & = \WcL{\text{VLL}} , &
  \WcL[V1,LR]{dd} & = \WcL{\text{LR},1} - \frac{1}{6} \WcL{\text{LR},2} ,
&
  \WcL[V8,LR]{dd} & = -\WcL{\text{LR},2} ,
  \label{eq:bmu-jms}
\intertext{SUSY $\to$ JMS:}
  \WcL[V,LL]{dd}  & = \WcL{1} \,, &
  \WcL[V1,LR]{dd} & = \frac{1}{6} \WcL{4} + \frac{1}{2} \WcL{5} \,,
&
  \WcL[V8,LR]{dd} & = \WcL{4} \,,
  \label{eq:susy-jms}
\intertext{JMS $\to$ SUSY:}
  \WcL{1} & = \WcL[V,LL]{dd} , &
  \WcL{4} & = \WcL[V1,LR]{dd} ,
&
  \WcL{5} & = 2 \WcL[V1,LR]{dd} - \frac{1}{3} \WcL[V8,LR]{dd} ,
  \label{eq:jms-susy}
\end{align}
where we have omitted flavour indices $\wcL{a}{ijij}$ on all Wilson coefficients.

Here we report the basis transformation from the JMS basis to the SUSY
operator basis introduced in \cite{Jenkins:2017jig}. The basis change
can be written in a compact form as:
\begin{equation}
  \label{eq:SusyinJMS-general}
\begin{aligned}
  \OpL[ij]{1} &
  = \opL[V,LL]{aa}{ijij} \,,
&
  \OpL[ij]{5} &
  = -\frac{1}{2} \opL[V1,LR]{aa}{ijij} \,,
\\[1mm]
  \OpL[ij]{2} &
  = \opL[S1,RR\, \dag]{aa}{jiji} \,,
&
  \OpL[ij]{1'} &
  = \opL[V,RR]{aa}{ijij} \,,
\\[1mm]
  \OpL[ij]{3} & = \frac{1}{3} \opL[S1,RR\, \dag]{aa}{jiji}
           + 2 \, \opL[S8,RR\, \dag]{aa}{jiji} \,, \qquad
&
  \OpL[ij]{2'} &
  = \opL[S1,RR]{aa}{ijij} \,,
\\[1mm]
  \OpL[ij]{4} & =-\frac{1}{6} \opL[V1,LR]{aa}{ijij}
           - \opL[V8,LR]{aa}{ijij} \,,
&
  \OpL[ij]{3'} & = \frac{1}{3} \opL[S1,RR]{aa}{ijij}
           + 2\, \opL[S8,RR]{aa}{ijij}   \,,
\end{aligned}
\end{equation}
where we have for the down-($aa=dd$) and up-sector ($aa=uu$), respectively.

%
%
%
%--------+---------+---------+---------+---------+---------+---------+---------+
\section{SMEFT one-loop matching}
\label{app:one-loop-SMEFT}

We report the matching of the SMEFT Wilson coefficient onto LEFT (JMS basis)
derived in Ref.~\cite{Dekens:2019ept}. The SMEFT Wilson coefficients are given in the
down-basis as opposed to \cite{Dekens:2019ept}. Furthermore, the matching is
performed in a redundant basis containing operators which are related to each
other. For our purpose we adopt the non-redundant basis defined in
\cite{Aebischer:2017ugx} and find for $\KKbar$ and $\BBbar$ down-type
meson mixing in {\bf the down-basis}
\begin{align}
  \wcL[V,LL]{dd}{ijij} &
  = -\frac{\alpha}{\pi s_W^2} \left(\wc[(1)]{qq}{ijij} + \wc[(3)]{qq}{ijij}
    \right) I_3(m_W, m_Z, \muEW)  \notag
\\ &
  - \frac{2\alpha \lambda_t^{im} \lambda_t^{nj}}{\pi s_W^2}
  \left( \wc[(1)]{qq}{ijmn} + \wc[(1)]{qq}{mnij}
       - \wc[(3)]{qq}{ijmn} - \wc[(3)]{qq}{mnij} \right. \notag
\\ & \qquad\qquad\;\;
  + \left. 2\wc[(3)]{qq}{inmj} + 2\wc[(3)]{qq}{mjin} \right) J(x_t) \notag
\\ &
  - \frac{\alpha }{\pi s_W^2} \bigg[ \lambda_t^{im}
    \left( \wc[(1)]{qq}{mjij} + \wc[(1)]{qq}{ijmj}
         + \wc[(3)]{qq}{mjij} + \wc[(3)]{qq}{ijmj} \right) \notag
\\ & \qquad\quad
  + \lambda_t^{mj} \left( \wc[(1)]{qq}{imij} + \wc[(1)]{qq}{ijim}
                        + \wc[(3)]{qq}{imij} + \wc[(3)]{qq}{ijim} \right) \bigg]
   K(x_t, \muEW) \notag
\\ &
  + \frac{\alpha \lambda_t^{ij}}{\pi s_W^2} \wc[(1)]{\phi q}{ij} I_1(x_t, \muEW)
  - \frac{\alpha \lambda_t^{ij}}{\pi s_W^2} \wc[(3)]{\phi q}{ij} I_2(x_t, \muEW) \notag
\\ &
  + \frac{\alpha \lambda_t^{ij}}{4\pi s_W^2} \left(
    \lambda_t^{im} \wc[(3)]{\phi q}{mj} + \lambda_t^{mj} \wc[(3)]{\phi q}{im} \right) S_0(x_t) \notag
\\ &
  - \frac{\alpha \lambda_t^{ij}}{\pi s_W^2} \left(
     \wc[(1)]{qu}{ij33} + \frac{N_c-1}{2N_c} \wc[(8)]{qu}{ij33} \right) I_1(x_t, \muEW) \notag
\\ &
  + \frac{\alpha \lambda_t^{ij}}{\pi s_W^2} \left(
       \wc[]{uW}{i3}V_{3j} + \wc[]{uW}{j3}^* V_{3i}^* \right) I_4(x_t, \muEW) \,,
  \label{eq:1loop-match1}
\\[2mm]
  \wcL[V1,LR]{dd}{ijij} &
  = -\frac{\alpha}{\pi s_W^2} \wc[(1)]{qd}{ijij} I_5(m_W, m_Z, \muEW)
  + \frac{\alpha \lambda_t^{ij}}{\pi s_W^2} \wc[]{\phi d}{ij} I_1(x_t, \muEW)
  \notag
\\ &
  - \frac{\alpha \lambda_t^{ij}}{\pi s_W^2} \wc[(1)]{ud}{33ij} I_1(x_t,\muEW)
  - \frac{\alpha}{\pi s_W^2} \left(
      \lambda_t^{im} \wc[(1)]{qd}{mjij}
  +   \lambda_t^{mj} \wc[(1)]{qd}{imij} \right) K(x_t, \muEW) \notag
\\ &
  - \frac{2\alpha \lambda_t^{ij}\lambda_t^{mn}}{\pi s_W^2} \wc[(1)]{qd}{nmij} J(x_t) \,,
\label{eq:1loop-match2}
\\[2mm]
\label{eq:1loop-match3}
  \wcL[V8,LR]{dd}{ijij} &
  = -\frac{\alpha}{\pi s_W^2} \wc[(8)]{qd}{ijij} I_5(m_W, m_Z, \muEW)
  - \frac{2\alpha \lambda_t^{ij}\lambda_t^{mn}}{\pi s_W^2}  \wc[(8)]{qd}{nmij} J(x_t)
\\ &
  - \frac{\alpha \lambda_t^{ij}}{\pi s_W^2} \wc[(8)]{ud}{33ij} I_1(x_t, \muEW)
  - \frac{\alpha}{\pi s_W^2} \left(
      \lambda_t^{im} \wc[(8)]{qd}{mjij}
    + \lambda_t^{mj} \wc[(8)]{qd}{imij} \right) K(x_t, \muEW) \,, \notag
\\[2mm]
\label{eq:1loop-match4}
  \wcL[V,RR]{dd}{ijij} &
  = -\frac{\alpha}{\pi s_W^2} \wc[]{dd}{ijij} I_6(m_W, m_Z, \muEW) \,,
\end{align}
where we use $x_t = m_t^2/m_W^2$ and the loop-functions:
\begin{align}
  K(x, \muEW) &
  = -\frac{3 x (1 + x)}{64 (x-1)} + \frac{x (4 - 2 x + x^2) }{32 (x-1)^2}\ln x
  - \frac{x}{16} \ln \frac{\muEW}{m_W} \,,
\\
  I_1(x,\muEW) &
  = -\frac{x(x-7)}{32(x-1)} - \frac{x(x^2-2x+4)}{16(x-1)^2} \ln x
  + \frac{x}{8} \ln \frac{\muEW}{m_W} \,,
\\
  I_2(x,\muEW) &
  = \frac{7x^2-25x}{32(x-1)} - \frac{x(x^2-14x+4)}{16(x-1)^2} \ln x
  + \frac{x}{8} \ln \frac{\muEW}{m_W} \,,
\\
  I_4(x) &
  = \frac{3 x^{3/2}(x+1)}{8\sqrt{2} (x-1)^2}
  - \frac{3 \sqrt{x} x^2}{4\sqrt{2} (x-1)^3} \ln x \,,
\\
  J(x) & = \frac{x}{16} \,,
\\
  S_0(x) &
  = \frac{x(x^2-11x+4)}{4(x-1)^2} + \frac{3x^3}{2(x-1)^3} \ln x\,,
\\
  I_3(m_W, m_Z, \muEW) &
  = -\frac{11(2m_W^2+m_Z^2)^2}{144m_W^2 m_Z^2}
  - \frac{(2m_W^2+m_Z^2)^2}{12 m_W^2 m_Z^2} \ln\frac{\muEW}{m_Z} \,,
\\
  I_5(m_W, m_Z, \muEW) &
  = \frac{-2m_W^4+m_W^2m_Z^2+m_Z^4}{72m_W^2 m_Z^2}
  + \frac{-m_Z^4-m_W^2 m_Z^2+2m_W^4}{6m_W^2 m_Z^2} \ln\frac{\muEW}{m_Z} \,,
\\
  I_6(m_W, m_Z, \muEW) &
  = -\frac{11 (m_W^2 - m_Z^2)^2 }{36 m_W^2 m_Z^2}
  - \frac{(m_W^2 - m_Z^2)^2 }{3 m_W^2 m_Z^2}\ln \frac{\muEW}{m_Z} \,.
\end{align}
The remaining Wilson coefficients $\WcL[S1,RR]{dd}$ and $\WcL[S8,RR]{dd}$ do
not get a matching contribution neither at tree-level nor at one-loop.

For the up-sector one finds at one-loop for $\DDbar$
up-type meson mixing in {\bf the up-basis}:
\begin{align}
  \wcL[V,LL]{uu}{ijij} &
  = -\frac{\alpha}{\pi s_W^2} \left(\wcup[(1)]{qq}{ijij} + \wcup[(3)]{qq}{ijij} \right)
    I_7(m_W, m_Z, \muEW)  \,,
\\
  \wcL[V1,LR]{uu}{ijij} &
  = -\frac{\alpha}{\pi s_W^2} \wcup[(1)]{qu}{ijij}\, I_8(m_W, m_Z, \muEW)\,,
\\
  \wcL[V8,LR]{uu}{ijij} &
  = -\frac{\alpha}{\pi s_W^2} \wcup[(8)]{qu}{ijij}\, I_8(m_W, m_Z, \muEW)\,,
\\
  \wcL[V,RR]{uu}{ijij} &
  = -\frac{4\alpha}{\pi s_W^2} \wcup{uu}{ijij}\, I_6(m_W, m_Z, \muEW)\,,
\end{align}
with the loop functions
\begin{align}
  I_7(m_W, m_Z, \muEW) &
  = -\frac{11 (m_Z^2-4 m_W^2)^2}{144 m_W^2 m_Z^2}
    -\frac{(m_Z^2-4 m_W^2)^2}{12 m_W^2 m_Z^2} \ln \frac{\muEW}{m_Z} \,,
\\
  I_8(m_W, m_Z, \muEW) &
  = \frac{(4 m_W^4 - 5 m_W^2 m_Z^2 + m_Z^4)}{36 m_W^2 m_Z^2}
    \left(-1 + 12 \ln \frac{\muEW}{m_Z} \right) \,.
\end{align}

%--------+---------+---------+---------+---------+---------+---------+---------+
%
%
%
%--------+---------+---------+---------+---------+---------+---------+---------+
\section{The Issue of Evanescent Operators}
\label{EVO}

The one-loop matching conditions discussed in our paper constitute a part of an
NLO calculation. It is well known that in the process of NLO calculations in the
NDR-\MSbar{} scheme, where ultraviolet divergences are regulated dimensionally,
the so-called evanescent operators that vanish in $D = 4$ dimensions have to be
considered \cite{Buras:1989xd, Buras:2000if}. They arise in particular when
complicated Dirac structures are projected onto the chosen basis of physical
operators and can also arise when different operator bases are related by
performing usual Fierz transformations that generally are not valid in
$D \neq 4$ dimensions.

The treatment of these operators in the process of one-loop matching must be
consistent with the one used in the calculation of two-loop anomalous dimensions.
In performing the NLO QCD evolutions we have used the $P_a^{ij}(\muEW)$
from \cite{Buras:2001ra}, see \eqref{eq:P_a-BJU}, which are based
on the two-loop anomalous dimensions
of operators calculated in \cite{Buras:2000if}. Therefore it is mandatory for
us to treat evanescent operators appearing in the one-loop matching in the same
manner as done in \cite{Buras:2000if}. Now, the latter paper used the treatment
of evanescent operators as proposed in the context of the formulation of the
NDR-\MSbar{} scheme introduced in \cite{Buras:1989xd}. The virtue of this
treatment is that the evanescent operators defined in this scheme influence
only two-loop anomalous dimensions. Indeed
 \begin{itemize}
\item By definition they do not contribute to the matching and to the finite
  corrections to the matrix elements of renormalized physical
  operators. They are simply
  subtracted away in the process of renormalization. This issue is summarized
  in Section 6.9.4 of \cite{Buras:1998raa}, where further references can be
  found. Important papers in this context are also \cite{Dugan:1990df,
  Herrlich:1994kh}.
\item Similarly using the projections of products of gamma matrices consistent
  with \cite{Buras:1989xd} also the usual Fierz transformations are not affected
  by the so-called Fierz-vanishing evanescent operators as long
  as the infrared divergences are not regulated dimensionally
  \cite{Buras:2000if}. Moreover, this issue is absent in the operators
  considered by us.
\end{itemize}

Even if these issues have been discussed in the literature it is instructive
to demonstrate in our case that indeed the usual Fierz identities can be used
to relate different bases by simply calculating the matrix elements of the
involved operators and checking that the $D = 4$ relations between different
bases do  not receive any $\ord{\alS}$ corrections.

In the case of the relation between the operators $\OpL{\text{LR},1}$ and
$\OpL{5}$ such a test  can be performed by calculating the matrix elements
of these operators inserting them in the current-current topologies. Master
formulae for these one-loop matrix elements can be found in (6.64)--(6.69) of
\cite{Buras:1998raa}. There, these formulae have been used for the calculation
of one-loop anomalous dimensions. Strictly speaking, when finite parts of
one-loop diagrams are considered these formulae should include a universal
factor $(1 + 2\epsilon)$ that results from the calculation of the integrals.
We will keep this factor but it is irrelevant for our test.

According to the procedure outlined above, in doing the reduction of products
of $\gamma_\mu$ matrices one can omit the usual evanescent operators that are
relevant for two-loop anomalous dimensions but are chosen not to contribute
to one-loop matchings. They are simply subtracted in the process of
renormalization \cite{Buras:1989xd}. Consequently the reductions to be used
are as follows. In the case of the $\OpL{\text{LR},1}$ operator we have
\begin{align}
  \gamma_\mu\gamma_\rho\gamma_\tau P_L \gamma^\rho\gamma^\mu\otimes\gamma^\tau P_R &
  = 4(1-2\epsilon) \gamma_\tau P_L\otimes\gamma^\tau P_R \,,
\\
  \gamma_\tau P_L\gamma_\rho\gamma_\mu\otimes\gamma^\tau P_R\gamma^\rho\gamma^\mu &
  = 4(1 \;+ \,\epsilon\,) \gamma_\tau P_L\otimes\gamma^\tau P_R \,,
\\
  \gamma_\tau P_L\gamma_\rho\gamma_\mu\otimes\gamma^\mu\gamma^\rho \gamma^\tau P_R &
  = 16(1-\epsilon) \gamma_\tau P_L\otimes\gamma^\tau P_R\,.
\end{align}
In the case $\OpL{5}$ we have
\begin{align}
  \gamma_\mu\gamma_\rho P_L \gamma^\rho\gamma^\mu\otimes P_R &
  = 16(1-\epsilon)  P_L \otimes P_R,
\\
  P_L\gamma_\rho\gamma_\mu\otimes P_R\gamma^\rho\gamma^\mu &
  = 4(1 \;+ \,\epsilon\,) P_L \otimes P_R,
\\
  P_L\gamma_\rho\gamma_\mu\otimes\gamma^\mu\gamma^\rho P_R &
  = 4(1-2\epsilon)  P_L \otimes P_R\,.
\end{align}

We find then the following $\ord{\alS}$ contributions to the matrix
elements in question:
\begin{align}
  \langle \OpL{\text{LR},1} \rangle &
  = \frac{1}{\epsilon}\frac{\alpha_s}{16\pi}(1+2\epsilon) \left[
      \frac{(20-44\epsilon)}{3} \langle \OpL{\text{LR},1} \rangle_0
    + (12-20\epsilon)\langle \OpLt{\text{LR},1}\rangle_0 \right] ,
\\
  \langle \OpL{5} \rangle &
  = \frac{1}{\epsilon}\frac{\alpha_s}{16\pi}(1+2\epsilon) \left[
      \frac{(20-44\epsilon)}{3} \langle \OpL{5} \rangle_0
    + (12-20\epsilon)\langle \OpL{4} \rangle_0 \right] ,
\end{align}
where
\begin{equation}
  \OpLt[ij]{\text{LR},1} =
  (\bar d_i^\alpha \gamma_\mu P_L d_j^\beta)
  (\bar d_i^\beta \gamma^\mu P_R d_j^\alpha),
\end{equation}
and $''0''$ indicates that these are tree-level matrix elements.
We can use for the latter $D=4$ Fierz relations
\begin{align}
  \label{FR}
   \OpL{\text{LR},1} & = -2 \OpL{5} \,, &
  \OpLt{\text{LR},1} & = -2 \OpL{4} \,.
\end{align}
Using these results we indeed find that the $D=4$ Fierz relation
between $\OpL{\text{LR},1}$ and $\OpL{5}$
\begin{equation}
  \OpL{\text{LR},1} = - 2 \OpL{5} + \ord{\alS^2}
\end{equation}
is satisfied.

For the transformation from the BMU to the JMS basis we still need one-loop
matrix elements of  $\OpLt{\text{LR},1}$ and $\OpL{\text{LR},2}$.
The reason is that
\begin{equation}
 \opL[V8,LR]{dd}{ijij}
  = [\bar{d}_i \gamma_\mu P_L T^A d_j][\bar{d}_i \gamma^\mu P_R T^A d_j]
  = \frac{1}{2} \left(\OpLt[ij]{\text{LR},1} -\frac{1}{3} \OpL[ij]{\text{LR},1} \right)
\end{equation}
and $\OpLt{\text{LR},1}$ do not belong to the BMU basis. We find
\begin{align}
  \langle \OpLt{\text{LR},1} \rangle &
  = \frac{1}{\epsilon}\frac{\alpha_s}{16\pi}(1+2\epsilon) \left[
      \frac{4}{3} (32-29\epsilon) \langle \OpLt{\text{LR},1} \rangle_0
    - 12\epsilon \langle \OpL{\text{LR},1} \rangle_0 \right] ,
\\
  \langle \OpL{\text{LR},2} \rangle &
  = \frac{1}{\epsilon}\frac{\alpha_s}{16\pi}(1+2\epsilon) \left[
      \frac{4}{3}(32-29\epsilon)\langle \OpL{\text{LR},2} \rangle^0
    - 12\epsilon \langle \OpL{5} \rangle_0 \right] .
\end{align}
Using these results and the Fierz identities for tree matrix elements in
\eqref{FR} we indeed find
\begin{equation}
  \OpLt{\text{LR},1} = -2 \, \OpL{\text{LR},2} + \ord{\alS^2} ,
\end{equation}
which implies the relation between the JMS and the BMU basis given in the text.

Our results for the matrix elements of $\OpL{\text{LR},1}$ and
$\OpL{\text{LR},2} = \OpL{4}$ confirm the ones obtained in \cite{Buras:2012fs}
but to test the Fierz relations we had to calculate the ones of $\OpL{5}$
and $\OpLt{\text{LR},1}$ as well.

%--------+---------+---------+---------+---------+---------+---------+---------+
%
%
%
%--------+---------+---------+---------+---------+---------+---------+---------+
\addcontentsline{toc}{section}{References}

\small

\bibliographystyle{JHEP}
\bibliography{Bookallrefs}

\end{document}